\documentclass[a4paper]{article}
\usepackage{lineno,hyperref}
\usepackage{graphicx} 
\usepackage[toc,page]{appendix} 
\usepackage{pifont}
\usepackage{array}
\usepackage[usenames,dvipsnames]{color}
\usepackage{epstopdf}
\usepackage{epsfig}
\usepackage{authblk}

\def\softd{{\leavevmode\setbox1=\hbox{d}%
          \hbox to 1.05\wd1{d\kern-0.4ex{\char039}\hss}}}

\modulolinenumbers[10] 

\newcommand{\fr}{\displaystyle \frac} 
\newcommand{\del}{\partial} 
\newcommand{\be}{\begin{equation}} 
\newcommand{\ee}{\end{equation}} 
\newcommand{\bea}{\begin{eqnarray}} 
\newcommand{\eea}{\end{eqnarray}} 
\newcommand{\ba}{\begin{array}} 
\newcommand{\ea}{\end{array}}  

\newcommand{\gre}{\color{Green}}
\newcommand{\rr}{\color{red}}
\newcommand{\br}{\color{Sepia}}
\newcommand{\cmark}{\ding{52}}
\newcommand{\xmark}{\ding{54}}

\providecommand{\keywords}[1]
{
  \small	
  \textbf{\textit{Keywords---}} #1
}

\date{} 
     
\begin{document} 

\title{A Kinetic Flux Difference Splitting Method for Compressible Flows} 

\author{Shrinath.K.S\thanks{Research Scholar, Department of Aerospace Engineering, Indian Institute of Science, Bangalore, India; HAL, Bangalore, India. email:{\tt \href{shrinath.k.s@gmail.com}{shrinath.k.s@gmail.com} }} \   N.H.Maruthi\thanks{SankhyaSutra Labs, Manyata Embassy Business Park, Bangalore, India; email:{\tt \href{maruthinh@gmail.com}{maruthinh@gmail.com}}} \  Raghurama Rao.S.V\thanks{ Corresponding Author, Department of Aerospace Engineering, Indian Institute of Science, Bangalore, India. email:{\tt \href{ raghu@ero.iisc.ernet.in}{raghu@ero.iisc.ernet.in}}} \ and Veeredhi Vasudeva Rao\thanks{ Department of Mechanical and Industrial Engineering, University of South Africa (UNISA), Johannesburg, South Africa. email:{\tt \href{vasudvr@unisa.ac.za}{vasudvr@unisa.ac.za}}}

}

\maketitle

%
%
%

\begin{abstract} 
A  low diffusive  flux difference splitting based kinetic scheme is developed based on a discrete velocity Boltzmann equation, with a novel three velocity model. While two discrete velocities are used for upwinding, the third discrete velocity is utilized to introduce appropriate additional numerical diffusion only in the expansion regions, identified using relative entropy  (Kullback-Liebler divergence) at the cell-interface, along with the estimation of physical entropy. This strategy provides an  interesting alternative to entropy fix, which is typically needed for low diffusive schemes.  Grid-aligned steady discontinuities are captured exactly by fixing the primary numerical diffusion such that flux equivalence leads to zero numerical diffusion across discontinuities. Results for bench-mark test problems are presented for inviscid and viscous compressible flows.    
\end{abstract} 
 

\section{\bf Introduction}  

Kinetic or Boltzmann schemes have been interesting alternatives to the traditional Riemann solvers for the numerical simulation of fluid flows and this topic has been the focus of research for the past several decades.  The basic framework for this class of numerical methods is the connection between the Boltzmann equation of the kinetic theory of gases and the macroscopic equations of gas dynamics, which can be obtained through the well-known {\em moment method strategy}.  The advantage of this strategy is the linearity of the convection terms in the Boltzmann equation, though the collision term is nonlinear.  The linearity of the convection terms makes the application of upwinding easier.  Thus, this strategy avoids the complications of Riemann solvers, field-by-field decompositions and the strong dependence on eigen-structure. In this paper, a new upwind Boltzmann scheme named as {\em Kinetic Flux Difference Splitting} (KFDS) method is presented, utilizing the framework of a {\em Discrete Velocity Boltzmann Equation} (DVBE).  The discrete velocity Boltzmann equation is first introduced based on the classical Boltzmann equation with a BGK model for the collision term, by replacing the Maxwellian distribution function by a set of Dirac delta functions.  This formulation leads to simpler expressions for the equilibrium distribution and the moment relations.  Applying upwinding to the discrete velocity Boltzmann equation leads to a macroscopic flux difference splitting method with simple expressions for the split fluxes.  The discrete velocities are chosen based on physical and numerical considerations and one of the resulting schemes leads to exact capturing of steady grid-aligned discontinuities, a feature shared by a few of the well-known macroscopic schemes.  Many of the existing kinetic schemes, like the flux vector splitting methods, suffer from high numerical diffusion and hence are not accurate enough compared to the Riemann solvers. The scheme presented in this paper overcomes this drawback,  captures steady grid-aligned discontinuities exactly and is less diffusive otherwise. An interesting alternative to entropy fix, which is typically needed for less diffusive schemes, is introduced in this paper based on switching over to a three velocity model instead of a two velocity model in one dimension.  The multi-dimensional extension is based on the standard finite volume method.  The component of the discrete distribution corresponding to the additional discrete velocity is chosen based on the variation of relative entropy (also known as Kullback-Liebler divergence, directed divergence or $D^{2}$-distance) and entropy.  
 
The first Boltzmann scheme was presented by Chu \cite{Chu} in which a finite difference method is applied to the Boltzmann equation with the BGK model.  Sanders \& Prendergast \cite{Sanders_Prendergast} presented the {\em Beam scheme} in which the Maxwellian is replaced by a set of weighted Dirac delta functions (called beams) and the propagation of beams across cells is accounted for, in an algorithm which ensures conservation.  van Albada {\em et al.} \cite{vanAlbada} demonstrated the upwinding property of Beam scheme.  Pullin \cite{Pullin} introduced an {\em Equilibrium Flux Method} (EFM) in which the fluxes across cells are calculated based on half-Maxwellians, which is equivalent to an upwind scheme.  Reitz \cite{Reitz}  proposed a Boltzmann scheme in which operator-splitting is used together with an appropriate velocity discretization for solving the Boltzmann equation.  Harten, Lax and van Leer \cite{HLL} gave a general description of kinetic schemes and suggested the equilibrium  distribution function as a weighted sum of Dirac delta functions, connecting them to the eigen-structure of macroscopic Euler equations, thus generalizing the approach of Sanders and Prendergast.  Deshpande \cite{Deshpande_KNM} presented a second order accurate kinetic numerical method based on utilizing the Chapman-Enskog distribution to provide an anti-diffusive contribution to an otherwise inviscid scheme.  Deshpande \cite{Deshpande_KFVS} and Mandal \& Deshpande \cite{Mandal_Deshpande} introduced the Kinetic Flux Vector Splitting (KFVS) method which introduces upwinding directly to the convection terms of the Boltzmann equation, together with operator splitting and instantaneous relaxation to equilibrium, and this leads to a macroscopic upwind scheme with identical split flux expressions as in EFM, though the approach is different.  Boltzmann schemes in which the Maxwellian is replaced by a compactly supported hat functions were introduced by Kaniel \cite{Kaniel} and Perthame \cite{Perthame}.  Prendergast and Kun Xu \cite{Prendergast_Kun_Xu}  developed a Boltzmann scheme without operator splitting, by using the solution of the BGK equation with appropriate approximations and a pressure sensor to detect the shocks.  Raghurama Rao \& Deshpande \cite{SVRRao_SMD,SVRRao_PVU} introduced a peculiar velocity based upwind method which leads to a convection-pressure splitting at the macroscopic level, with simpler split flux expressions avoiding error functions and exponentials.  It is worth noting that the macroscopic Flux Vector Splitting (FVS) method of Steger \& Warming \cite{Steger_Warming} can be recovered from the Beam scheme for a particular value of $\gamma$ and similarly van Leer's FVS method \cite{vanLeer} can be recovered from Perthame's kinetic scheme.    While there are several other developments based on the above strategies, a slightly different approach is introduced with the framework of {\em discrete velocity based kinetic schemes}, by Natalini \cite{Natalini} and later by Aregba-Driollet \& Natalini \cite{Driollet_Natalini}.  They introduce efficient kinetic schemes based discrete velocity Boltzmann equation and also establish the connection between the {\em relaxation systems} of Jin \& Xin \cite{Jin_Xin} and {\em Discrete Velocity Boltzmann systems}. While there is an extensive volume of research based on discrete velocity based kinetic schemes, some of the schemes which have the flavour of the traditional continuous molecular velocity based kinetic schemes are due to Bouchut {\em et al.} \cite{Bouchut_Guarguaglini_Natalini}, Raghurama Rao \& Balakrishna \cite{SVRRao_Balakrishna}, Raghurama Rao \& Subba Rao \cite{SVRRao_SubbaRao}, Bajpayi and Raghurama Rao \cite{Mayank_SVRRao}, Arun {\em et al.} \cite{Arun_SVRRao,Arun_SVRRao_KRS,Arun_Maria},  Raghavendra \cite{Venkat_Thesis}, Raghavendra and Raghurama Rao \cite{Venkat_SVRRao},  Abgrall \& Torlo \cite{Abgrall_Torlo}. An interesting connection of this discrete velocity Boltzmann framework to Lattice Boltzmann Method is introduced by Raghurama Rao {\em et al.} \cite{SVRRao_Rohan_Kotnala} and Deshmukh \cite{Rohan_Thesis} by developing a Lattice Boltzmann Relaxation Scheme for compressible flows. \\

 The discrete velocity based kinetic schemes are simpler than the the continuous molecular velocity based kinetic schemes, as the integrals are replaced by summations and the Maxwellians are simple algebraic functions of conserved variables and fluxes.  While some of the discrete velocity kinetic schemes are quite efficient, none of the above mentioned kinetic or discrete kinetic schemes can capture steady grid-aligned discontinuities exactly, a feat achieved by some of the macroscopic schemes.  In this paper, an exact discontinuity capturing kinetic scheme is presented based on discrete velocity formulation, by choosing the discrete velocity magnitudes in such a way that steady grid-aligned discontinuities are captured without any numerical diffusion. While the design of this kinetic scheme requires just two discrete velocities, the three-velocity model is utilized in a novel way  such that the third component is adjusted to avoid any entropy condition violation, which is typical of low diffusive algorithms, by utilizing the  relative entropy. The relative entropy is also known as Kullback-Liebler divergence, the directed divergence or $D^{2}$-distance (see Kullback \cite{Kullback}).   It represents the information theoretic distance between two bivariate normal distributions and its moment is an efficient indicator to distinguish among different nonlinear waves. It is also closely connected to the concept of Mahalanobis distance, introduced in 1936 \cite{Mahalanobis}. Raghavendra {\em et al.} \cite{Venkat_Varma_Biju_SMD} utilized the $D^{2}$-Distance successfully as a tool for mesh adaptation.  In the next section, the basics of moment method strategy are introduced and in the following section the discrete velocity Boltzmann equation and its moments to recover the macroscopic equations are introduced.

\section{Moment Method Strategy for Kinetic Schemes} 

The kinetic or Boltzmann schemes are based on the fact that Euler equations of gas dynamics can be obtained as moments of the classical Boltzmann equation.  Thus, the Euler equations can be written in the form 

\be \label{EE_as_moments}
\left\langle \Psi \left( \underbrace{ \fr{\del f}{\del t} + \vec{v} \cdot \fr{\del f}{\del \vec{x}} = J(f,f)}_{\rm Boltzmann \ equation} \right) \right\rangle 
\ee 
where the moments are defined by 
\be 
\left \langle \Psi f \right \rangle = \int_{0}^{\infty} \int_{-\infty}^{\infty} \int_{-\infty}^{\infty} \int_{-\infty}^{\infty} \Psi f \ dv_{1} dv_{2} dv_{3} \ dI 
\ee 
Here, $f$ is the molecular velocity distribution function, $v$ is the molecular velocity, $I$ is the internal energy variable corresponding to non-translational degrees of freedom and $J(f,f)$ is the collision term. $\Psi$ is the moment function vector, representing the mass, momenta and energy of the molecules (which are conserved during collisions). 
\bea 
\Psi = \left[ \ba{c} 1 \\ v_{1} \\ v_{2} \\ v_{3} \\ I + \fr{1}{2} \left( v_{1}^{2} + v_{2}^{2} + v_{3}^{2} \right) \ea \right]
\eea
The Euler equations (\ref{EE_as_moments}), after completing the moments, are given in their macroscopic form by 
\be 
\fr{\del U}{\del t} + \sum_{i=1}^{3} \fr{\del G_{i}}{\del x_{i}} = 0  
\ee 
where $U$ is the conserved variable vector and $G_{i}$ are the inviscid flux vectors with standard definitions representing conservation of mass, momenta and energy.  
\bea 
U = \left[ \ba{c} \rho \\ \rho u_{1} \\ \rho u_{2} \\ \rho u_{3} \\ \rho E \ea \right]; \ 
G_{1} = \left[\ba{c} \rho u_{1} \\ p + \rho u_{1}^{2} \\ \rho u_{1} u_{2} \\ \rho u_{1} u_{3} \\ p u_{1} + \rho u_{1} E \ea \right]; \ G_{2} = \left[\ba{c} \rho u_{2} \\ \rho u_{2} u_{1} \\ p + \rho u_{2}^{2} \\ \rho u_{2} u_{3} \\ p u_{2} + \rho u_{2} E \ea \right]; \ G_{3} = \left[\ba{c} \rho u_{3} \\ \rho u_{3} u_{1} \\ \rho u_{3} u_{2} \\ p + \rho u_{3}^{2} \\ p u_{3} + \rho u_{3} E \ea \right]   
\eea 
Here, the total energy is the sum of internal and kinetic energies. 
\be 
E = \fr{p}{\rho \left(\gamma - 1\right)} + \fr{1}{2} \left(u_{1}^{2} + u_{2}^{2} + u_{3}^{2} \right) 
\ee 
$\gamma$ is the ratio of specific heats. A simpler model for the collision term ($J(f,f)$) in the Boltzmann equation is given by the popular  B-G-K model \cite{Bhatnagar_Gross_Krook}, which reduces the otherwise integro-differential equation to a PDE.  
\be 
J(f,f) = - \fr{1}{\epsilon} \left[f - f^{eq} \right] 
\ee 
Here, $\epsilon$ is the relaxation time and $f^{eq}$ is the distribution representing local thermodynamic equilibrium, given by a Maxwellian as 
\be 
f^{eq} = \fr{\rho}{I_{0}} \left(\fr{\beta}{\pi}\right)^{\frac{3}{2}} e^{- \beta \left(v_{1} - u_{1} \right)^{2} 
- \beta \left(v_{2} - u_{2} \right)^{2} 
- \beta \left(v_{3} - u_{3} \right)^{2}} e^{-\frac{I}{I_{0}}} 
\ee 
where 
\be 
\beta = \fr{1}{2RT} \ \textrm{and} \ I_{0} = \fr{\left(2+D\right) - \gamma D}{2\left(\gamma - 1\right)} RT 
\ee  
with $D$ being the degrees of freedom. 

\subsection{Strategy of Kinetic Schemes} 

\subsubsection{Kinetic Schemes for Euler Equations} 
Utilizing the B-G-K model and operator splitting, the solution of the Boltzmann equation can be split into two steps as 
\be   
\textrm{\bf Convection Step:} \ \fr{\del f}{\del t} + \fr{\del \vec{h}}{\del \vec{x}} = 0 
\ee  
\be   
\textrm{\bf Collision Step:} \ \fr{d f}{d t} = - \fr{1}{\epsilon} \left[f - f^{eq} \right]  
\ee 
where $ \vec{h} =  \vec{v}  f$ is the flux, with the Boltzmann equation being written in conservation form.   
Choosing an {\em instantaneous relaxation to equilibrium} ($\epsilon=0$), the collision step becomes a simple relaxation step as $f = f^{eq}$.  Thus, the Euler equations can be written in an intriguing form as 

\be 
\left \langle \Psi \left( \underbrace{\fr{\del f}{\del t} +  \fr{\del \vec{h}}{\del \vec{x}}=0}_{\rm convection}, \ \underbrace{f = f^{eq}}_{\rm relaxation} \right) \right \rangle
\ee 
The essential advantage of using this representation for  macroscopic Euler equations is the linearity of the convection terms, which makes introducing upwinding easier in a kinetic or Boltzmann scheme, automatically leading to an upwind scheme for Euler equations.  Note that the nonlinearity is present only in the last step of taking moments.  This strategy avoids dependence on the eigen-structure of the nonlinear Euler equations and hence is a good alternative to the Riemann solvers. 
\subsubsection{Kinetic Schemes for Navier-Stokes Equations} 
In the above formulation, $f^{eq}$ represents local thermodyanmic equilibrium and hence is insufficient to obtain Navier-Stokes equations.  For this purpose, Chapman-Enskog distribution, $f^{CE}$, (derived using Chapman-Enskog expansion) can be used instead of $f^{eq}$.  
Then, we can write the Navier-Stokes equations in the moment form as 

\be 
\left \langle \Psi \left( \fr{\del f}{\del t} +  \fr{\del \vec{h}}{\del \vec{x}}=0, \ f = f^{CE} \right) \right \rangle
\ee 
where the instantaneous relaxation in the collision step is to Chapman-Enskog distribution (see Junk and  Raghurama Rao  \cite{Junk_SVRRao}).  The Chapman-Enskog distribution, for the general case of poly-atomic gases, is derived by Deshpande \cite{Deshpande_KNM} for 1-D case and by Mano Kumar {\em et al.} \cite{Manoj_SVRR_SMD} for multi-dimensions (see also Chou \& Baganoff \cite{Chou_Baganoff} and  Deshpande {\em et al.} \cite{Ramanan_SVRR_SMD}).  The numerical schemes in this work begin with the above distributions for introducing the discrete velocity versions.   
\subsubsection{Discrete Velocity Boltzmann Schemes}  
Here, we utilize the above formulations and introduce new versions of the equilibrium and Chapman-Enskog distribution functions based on discrete velocities. These discrete distributions are then utilized to introduce novel kinetic schemes for solving Euler and Navier-Stokes equations.  Compared to traditional kinetic schemes, the expressions here are much simpler, as integrations are replaced by summations and Maxwellians are replaced by simpler algebraic expressions.  Unlike in the Lattice Boltzmann Method, further Chapman-Enskog expansion (to fit the coefficients of viscosity and thermal conductivity) is not needed here as this formulation is based instantaneous relaxation to Maxwellian or Chapman-Enskog distributions in the collision step.  One disadvantage with the B-G-K model is the limitation of unit Prandtl number, which is overcome here by employing a special strategy which introduces a correction, in a later section.   It is worth noting here that, unlike in the Lattice Boltzmann Method, the numerical schemes developed in this work begin with the conservation form of the Boltzmann equation.  Further, the framework of finite volume method and the flux difference splitting form of numerical diffusion ensure the preservation of the conservation form.  These frameworks ensure conservation even if we make the discrete velocities functions of conserved variables. Such a modification only makes the numerical diffusion part in the flux difference splitting framework nonlinear but yet  preserves conservation.  Further, the essential numerical diffusion is fixed based on {\em flux equivalence across discontinuities}, which amounts to enforcing Rankine-Hugoniot (R-H) conditions, and R-H conditions represent the quintessential conservation of fluxes across discontinuities.  This is similar to the modification of diffusion terms in expressions for the cell-interface fluxes in the macroscopic flux difference splitting methods, without the loss of conservation. In the next section, the new discrete velocity Boltzmann equation is introduced.

\section{\bf Equilibrium distribution function and its representations}  
The discrete velocity model is first introduced in 1-D in this section.   Consider 1-D Euler equations given by  
\be \label{1D_EE} \ \ \ \ \ \ 
\fr{\del U}{\del t} + \fr{\del G(U)}{\del x} = 0 
\ee 
where $U$ is the conserved variable vector and $G(U)$ its nonlinear flux vector, given by  
\bea \label{1D_EE_Vectors}
U = \left[ \ba{c} \rho \\ \rho u \\ \rho E \ea \right] \ \textrm{and} \ 
G(U) = \left[ \ba{c} \rho u \\ p + \rho u^{2} \\ p u + \rho u E \ea \right] 
\eea
Here, $\rho$ is the density, $u$ is the fluid velocity, $p$ is the pressure and $E$ is the total (internal + kinetic) energy, given by $E = \fr{p}{\rho(\gamma -1)} +
\fr{1}{2} u^{2}$.   
The above equations can be obtained as moments of the 1-D   Boltzmann equation, with the BGK model, given by 

\be \label{1D_BE}
\fr{\del f}{\del t} + \fr{\del h}{\del x} = - \fr{1}{\epsilon} \left[ f - f^{eq} \right] 
\ee  
where $h = vf$. 
 The 1-D Maxwellian is given by 
\be 
\label{1D_Maxwellian}
f^{eq} = \rho \fr{\sqrt{\beta}}{\sqrt{\pi}} 
e^{- \beta \left( v - u \right)^{2} } e^{-\frac{I}{I_{0}}} \ \textrm{with} \ 
\beta = \fr{1}{2RT} \ \textrm{and} \ I_{0} = \fr{\left(3-\gamma\right)}{2\left(\gamma -1\right)}RT  
\ee    
where $R$ is the gas constant and $T$ is the temperature, with the ideal gas equation of state being given by $p = \rho R T$.  The internal energy variable corresponding to non-translational degrees of freedom, $I$, takes the role of providing the right value of $\gamma$  for poly-atomic case.  The moments to obtain the macroscopic variables are defined by 
\be \label{1D_moments} 
U_{i} = \int_{0}^{\infty} \int_{-\infty}^{\infty} {\psi}_{i} f \ dvdI \ \textrm{and} \ 
G_{i}(U) = \int_{0}^{\infty}\int_{-\infty}^{\infty} {\psi}_{i} v f \ dvdI, \ i=1,2,3  
\ee  
with 
\be 
\psi = \left[ \ba{c} 1 \\ v \\ I + \fr{1}{2} v^{2} \ea \right] 
\ee 
Introducing a truncated  distribution as 
\be 
\tilde{f} = \int_{0}^{\infty} f \ dI 
\ee 
we can redefine the moment relations as  
\be \label{1D_moments_modified_1} 
U_{i} = \int_{-\infty}^{\infty}  {\psi}_{i} \tilde{f} \ dv \  \textrm{and} \ 
G(U) = \int_{-\infty}^{\infty} {\psi}_{i} v \tilde{f} \ dv  
\ee 
\subsection{\bf Two Velocity Model}
In order to obtain a discrete velocity model, let us start with the moment relations for the equilibrium distribution function given by  
\be \label{equil_Euler}
U_{i} = \int_{- \infty}^{\infty} {\psi}_{i} \tilde{f}^{eq} dv \ \textrm{and} \ 
G(U)_{i} = \int_{- \infty}^{\infty}  {\psi}_{i} v \tilde{f}^{eq} dv 
\ee
Introducing a set of Dirac delta functions comprising of two discrete velocities $\lambda^{+}$ and $\lambda^{-}$ (for replacing the molecular velocity $v$) and also two corresponding components of $\tilde{f}^{eq}$ as  $f^{eq}_{+}$ and $f^{eq}_{-}$ (which absorb the effect of $\psi$), we write 
\be
 {\psi}_{i} \tilde{f}^{eq} = { \left\{ f^{eq}_{+} \delta (v - \lambda^{+}) +   f^{eq}_{-} \delta (v - \lambda^{-}) \right\} }_{i} 
\ee  
where $\delta\left(v-\lambda^{\pm}\right)$ is the Dirac delta function.  
Thus, the conserved variable vector becomes 
\be  
U_{i} ={ \left[ \int_{-\infty}^{\infty} f^{eq}_{+} \delta (v - \lambda^{+})  dv 
+ \int_{-\infty}^{\infty} f^{eq}_{-} \delta (v - \lambda^{-})  dv \right] }_{i} 
\ee 
Let us further assume, for simplicity, that the discrete velocities,  $\lambda^{+}$ and $\lambda^{-}$ for each $i$  are given by 
\be \label{simple_2DVs_1D}
\lambda^{+}_{i} = \lambda_{i} \ \textrm{ and } \  \lambda^{-}_{i} = - \lambda_{i} 
\ee  
Thus, we have three unknowns, namely, $f^{eq}_{+}$,  $f^{eq}_{-}$ and $\lambda$ to be fixed for the equilibrium distribution, for each $i$.  
The moment relations in (\ref{equil_Euler}) are utilized to fix $f^{eq}_{\pm}$ here.  The fixing of $\lambda$ is explained in later sections,  to develop a low diffusion scheme.  \\
\\    
$$ \ba{rcl}   
U_{i} & = & \displaystyle \int_{-\infty}^{\infty} {\psi}_{i} \tilde{f}^{eq} dv \\ 
 & = & \displaystyle \int_{-\infty}^{\infty} \left\{ f^{eq}_{+} \delta (v - \lambda^{+}) + f^{eq}_{-} \delta (v - \lambda^{-}) \right\}_{i} dv \\ 
 & = & \left\{ f^{eq}_{+} + f^{eq}_{-} \right\}_{i} 
\ea $$  
Thus 
\be  
\left\{f^{eq}_{+} + f^{eq}_{-}\right\}_{i} = U_{i} 
\ee  
$$ \ba{rcl}  
G(U)_{i} & = & \displaystyle \int_{-\infty}^{\infty} v {\psi}_{i} \tilde{f}^{eq} dv \\ [3mm]
 & = & \displaystyle \int_{-\infty}^{\infty} v 
\left\{ f^{eq}_{+} \delta (v - \lambda^{+}) + f^{eq}_{-} \delta (v - \lambda^{-}) \right\}_{i} dv \\ [2mm]
 & = & \left\{ f^{eq}_{+} \displaystyle \int_{-\infty}^{\infty} v \delta (v - \lambda^{+}) dv +  f^{eq}_{-} \displaystyle \int_{-\infty}^{\infty} v \delta (v - \lambda^{-}) dv  \right\}_{i}  \\
 & = & \left\{  f^{eq}_{+} \lambda^{+} +  f^{eq}_{-} \lambda^{-}  \right\}_{i}
\ea $$  
Thus  
\be 
 \left\{ f^{eq}_{+} \lambda^{+} +   f^{eq}_{-} \lambda^{-} \right\}_{i}  = G(U)_{i}
\ee   
Solving the above two equations and simplifying  of discrete velocities for each $i$ using (\ref{simple_2DVs_1D}), we get
\be \label{2_Vel_equilibria}
{f^{eq}_{+}}_{i} = \fr{1}{2} U_{i} + \fr{1}{2 {\lambda}_{i}} G(U)_{i} \ \textrm{and} \ {f^{eq}_{-}}_{i} = \fr{1}{2} U_{i}- \fr{1}{2 {\lambda}_{i}} G(U)_{i}
\ee  
Therefore, the {\em Discrete Velocity Boltzmann Equation} (DVBE) based on the two discrete velocity model can be written as    
\be \label{DVBE_1D_2_Vel} 
\left\{\fr{\del \bf f}{\del t} + \bf  \fr{\del \bf h}{\del x} = - \fr{1}{\epsilon} \left[ \bf f - \bf f^{eq} \right] \right\}_{i} \ i=1,2,3
\ee 
where $\bf {h}_{i} = {\Lambda}_{i}{f}_{i} $  
and
\be \label{1D_2Vel_Feq} 
\bf f_{i} = \left[ \ba{c} f_{+} \\ f_{-} \ea \right]_{i}, \ 
\Lambda_{i} = \left[ \ba{cc} \lambda^{+} & 0 \\ 0 & \lambda^{-} \ea \right]_{i} \ \textrm{and} \ 
\bf {f^{eq}}_{i} = \left[ \ba{c} f^{eq}_{+} \\ f^{eq}_{-} \ea \right]_{i}
= \left[ \ba{c} \fr{1}{2} U + \fr{1}{2 \lambda} G(U) \\[4mm]  
\fr{1}{2} U - \fr{1}{2 \lambda} G(U) \ea \right]_{i}  
\ee
The equation (\ref{DVBE_1D_2_Vel}) can be written in compact form as

\be \label{DVBE_1D_2_Vel_Comp} 
\fr{\del \bf F}{\del t} + \bf  \fr{\del \bf H}{\del x} = - \fr{1}{\epsilon} \left[ \bf F - \bf F^{eq} \right]
\ee
where $ \bf{H} = \bf{\tilde \Lambda F} $ is the flux and  
\be
\bf F = \left[ \ba{c}\bf f_{1} \\ \bf f_{2} \\ \bf f_{3} \ea \right] =  \left[ \ba{c}{f_{+}}_{1} \\  {f_{-}}_{1} \\{f_{+}}_{2} \\ {f_{-}}_{2} \\ {f_{+}}_{3} \\ {f_{-}}_{3} \ea \right], \  \ \textrm{and} \ 
\bf {F^{eq}} = \left[ \ba{c} \bf{f^{eq}}_{1} \\ \bf{f^{eq}}_{2} \\\bf{f^{eq}}_{3} \ea \right] = \left[ \ba{c} {f_{+}^{eq}}_{1}\\ {f_{-}^{eq}}_{1} \\ {f_{+}^{eq}}_{2} \\ {f_{-}^{eq}}_{2} \\ {f_{+}^{eq}}_{3} \\ {f_{-}^{eq}}_{3} \ea \right] 
\ee
$ \bf {\tilde \Lambda}$ is given by
\be
\bf {\tilde \Lambda} = \left[ \ba{ccc} \bf \Lambda_{1} & \  \bf 0  & \ \bf 0 \\ \bf  0  & \ \bf \Lambda_{2} &  \  \bf 0 \\ \bf 0 & \  \bf 0 & \ \bf  \Lambda_{3}  \ea \right]  = \left[ \ba{cccccc}  \lambda^{+}_{1} & \ 0  & \ 0  & \ 0  & \ 0  & \ 0 \\  0 & \lambda^{-}_{1} & \ 0  & \ 0  & \ 0  & \ 0 \\  0  & \ 0  & \ \lambda^{+}_{2} &  \ 0 &  \ 0 &  \ 0 \\ 0 & 0  & \ 0  & \ \lambda^{-}_{2} &  \ 0 &  \ 0  \\ 0 & \ 0 & 0 & \ 0 & \  \lambda^{+}_{3} & \ 0 \\ 0 & 0 & \ 0 & 0 & \ 0 & \  \lambda^{+}_{3}  \ea \right]
\ee
where  $\bf f_{1,2,3}$ , $\bf \Lambda_{1,2,3}$ and $ \bf{f^{eq}}_{1,2,3}$ are obtained from (\ref{1D_2Vel_Feq}).
It is worth noting that moments of the above discrete velocity Boltzmann equation, for the 2-velocity model, yield the relaxation system of Jin and Xin \cite{Jin_Xin}, as noted by Aregba-Driollet \& Natalini \cite{Driollet_Natalini}.  To recover the macroscopic system of equations, we  take moments, which in this case will be premultiplying the system with $\bf P$, which is defined as 
\be \label{2_Vel_P} 
\bf{P} =  \left[ \ba{cccccc} {1} & {1} & \ 0 & \ 0 & \ 0 & \ 0 \\  0 & \ 0 & {1} & {1} & \ 0 & \ 0 \\   0 & \ 0 & \ 0 & \ 0 & {1} & {1} \ea  \right]
\ee  
\subsection{\bf Three Velocity Model}
The basic motivation for this part is to keep the additional discrete velocity and its corresponding discrete distribution as free parameters, to be fixed for avoiding entropy condition violation in a novel way.  Let us replace the equilibrium distribution by another combination of Dirac delta functions as 
\be
{\psi}_{i} \tilde{f}^{eq} =\left\{ f^{eq}_{+} \delta (v - \lambda^{+}) +  f^{eq}_{o} \delta (v - \lambda_{o})  + f^{eq}_{-} \delta (v - \lambda^{-}) \right\}_{i} 
\ee 
Let us further assume, for simplicity, that the discrete velocities, $\lambda^{+}$ and $\lambda^{-}$ for each $i$ are given by 
\be \label{New_3DV_1D} 
\lambda^{+}_{i} = \lambda_{i} \ \textrm{and} \  \lambda^{-}_{i} = - \lambda_{i}  
\ee 
Let us assume that $\lambda^{0}$ and $f^{eq}_{0}$ for each $i$ are known (which will be fixed later).  Then, using the two moment relations in (\ref{1D_moments_modified_1})  we obtain the following.  
$$ \ba{rcl}  
U_{i} & = & \displaystyle \int_{-\infty}^{\infty} {\psi}_{i} \tilde{f}^{eq} dv \\ 
 & = & \displaystyle \int_{-\infty}^{\infty} \left\{ f^{eq}_{+} \delta (v - \lambda^{+}) + f^{eq}_{o}\delta (v - \lambda_{o})  + f^{eq}_{-} \delta (v - \lambda^{-}) \right\}_{i} dv \\ 
 & = &\left\{  f^{eq}_{+}  + f^{eq}_{o}  + f^{eq}_{-} \right\}_{i}  
\ea $$  
or 
\be 
\left\{f^{eq}_{+} + f^{eq}_{-}\right\}_{i} = U_{i} - { f^{eq}_{o}}_{i} 
\ee 
$$ \ba{rcl} 
G(U)_{i} & = & \displaystyle \int_{-\infty}^{\infty} v {\psi}_{i} \tilde{f}^{eq} dv \\ [3mm]
 & = & \displaystyle \int_{-\infty}^{\infty} v 
\left\{ f^{eq}_{+} \delta (v - \lambda^{+}) + f^{eq}_{o} \delta (v -\lambda_{o}) + f^{eq}_{-} \delta (v - \lambda^{-}) \right\}_{i} dv \\ [2mm]
 & = &\left\{  f^{eq}_{+} \displaystyle \int_{-\infty}^{\infty} \phi(v) \delta (v - \lambda^{+}) dv + f^{eq}_{o} \displaystyle \int_{-\infty}^{\infty} \phi(v) \delta (v - \lambda_{o}) dv + 
        f^{eq}_{-} \displaystyle \int_{-\infty}^{\infty} \phi(v) \delta (v - \lambda^{-}) dv\right\}_{i}, \ \   (\phi(v) = v) \\ 
 & = &\left\{ f^{eq}_{+} \lambda^{+} +  f^{eq}_{o} \lambda_{o}  + f^{eq}_{-} \lambda^{-}\right\}_{i}
\ea $$ 
or 
\be 
\left\{f^{eq}_{+} \lambda^{+}  + f^{eq}_{-} \lambda^{-}\right\}_{i}  = G(U)_{i} - {\left\{ f^{eq}_{o}\lambda_{o}\right\}}_{i}
\ee 
Solving the above two equations  and simplifying the discrete velocities using  (\ref{New_3DV_1D}), we get
\be \label{3_Vel_equilibria} 
{f^{eq}_{+}}_{i} = \fr{1}{2} U_{i} + \fr{1}{2 {\lambda}_{i}} G(U)_{i} - \left\{\fr{\lambda + \lambda_{o}}{2 \lambda} f^{eq}_{o}\right\}_{i} \ \textrm{and} \ {f^{eq}_{-}}_{i} = \fr{1}{2} U_{i} - \fr{1}{2 {\lambda}_{i}} G(U)_{i} -   \left\{\fr{\lambda - \lambda_{o}}{2 \lambda} f^{eq}_{o} \right\}_{i}
\ee 
This is similar to the equilibria derived using the two velocity model (\ref{2_Vel_equilibria})  with the respective additional  $-\fr{\lambda + \lambda_{o}}{2 \lambda} f^{eq}_{o}$ and $-\fr{\lambda - \lambda_{o}}{2 \lambda} f^{eq}_{o}$ terms.  These additional terms will later be utilized in the numerical scheme to avoid possible entropy violations.  

The {\em Discrete Velocity Boltzmann Equation} (DVBE) for three velocity model can now be written as    
\be \label{DVBE_1D_3_Vel} 
\left\{\fr{\del \bf f}{\del t} + \fr{\del \bf h}{\del x} = - \fr{1}{\epsilon} \left[ \bf f - \bf f^{eq} \right] \right\}_{i} \ i=1,2,3
\ee
where  ${\bf h_{i}=\Lambda_{i} f_{i}}$, with the above Boltzmann equation being in conservation form.    Here 
\be \label{3V_DVBEq} 
\bf f_{i} = \left[ \ba{c} f_{+} \\ f_{o} \\ f_{-} \ea \right]_{i}, \ 
{\Lambda}_{i} = \left[ \ba{ccc} \lambda^{+} & 0 & 0 \\ 0 &  \lambda_{o} & 0 \\  0 & 0 &   \lambda^{-} \ea \right]_{i} \ \textrm{and} \ 
\bf{ f^{eq}}_{i} = \left[ \ba{c} f^{eq}_{+} \\  f^{eq}_{o} \\ f^{eq}_{-} \ea \right]_{i} 
= \left[ \ba{c} \fr{1}{2} U + \fr{1}{2 \lambda} G(U)  - \fr{\lambda + \lambda_{o}}{2 \lambda} f^{eq}_{o}\\
f^{eq}_{o} \\
\fr{1}{2} U - \fr{1}{2 \lambda} G(U)  - \fr{\lambda - \lambda_{o}}{2 \lambda} f^{eq}_{o} \ea \right]_{i}  
\ee
$f^{eq}_{0}$ and $\lambda^{0}$ for each $i$ will be chosen later in such a way that entropy condition violation is avoided.  
The equation (\ref{DVBE_1D_3_Vel}) can be written in compact form as

\be \label{DVBE_1D_3_Vel_Comp} 
\fr{\del \bf F}{\del t} +  \fr{\del \bf H}{\del x} = - \fr{1}{\epsilon} \left[ \bf F - \bf F^{eq} \right]
\ee
where 
\be \\[1 mm]
\bf F = \left[ \ba{c}\bf f_{1} \\[1 mm] \bf f_{2} \\[1 mm] \bf f_{3} \ea \right] =  \left[ \ba{c}{f_{+}}_{1} \\[1 mm]  {f_{o}}_{1} \\[1 mm]  {f_{-}}_{1} \\[1 mm] {f_{+}}_{2}\\[1 mm] {f_{o}}_{2} \\[1 mm]  {f_{-}}_{2} \\[1 mm] {f_{+}}_{3}  \\[1 mm] {f_{o}}_{3} \\[1 mm] {f_{-}}_{3} \ea \right] \ \textrm{and} \  
\bf {F^{eq}} = \left[ \ba{c} \bf{f^{eq}}_{1} \\[1 mm] \bf{f^{eq}}_{2} \\[1 mm] \bf{f^{eq}}_{3} \ea \right] = \left[ \ba{c} {f_{+}^{eq}}_{1} \\[1 mm] {f_{o}^{eq}}_{1} \\[1 mm] {f_{-}^{eq}}_{1} \\[1 mm] {f_{+}^{eq}}_{2}\\[1 mm] {f_{o}^{eq}}_{2} \\[1 mm] {f_{-}^{eq}}_{2} \\[1 mm] {f_{+}^{eq}}_{3} \\[1 mm] {f_{o}^{eq}}_{3} \\[1 mm] {f_{-}^{eq}}_{3} \ea \right] 
\ee
$\bf {\tilde \Lambda}$ is given by
\be
\bf {\tilde \Lambda} = \left[ \ba{ccc} \bf \Lambda_{1} & \  \bf 0  & \ \bf 0 \\ \bf  0  & \ \bf \Lambda_{2} &  \  \bf 0 \\ \bf 0 & \  \bf 0 & \ \bf  \Lambda_{3}  \ea \right]  = \left[ \ba{ccccccccc}  \lambda^{+}_{1} & \ 0  & \ 0  & \ 0  & \ 0  & \ 0 & \ 0  & \ 0  & \ 0 \\  0 & \lambda^{o}_{1} & \ 0  & \ 0  & \ 0  & \ 0 & \ 0  & \ 0  & \ 0 \\  0  & \ 0  & \ \lambda^{-}_{1} &  \ 0 &  \ 0 &  \ 0 & \ 0  & \ 0  & \ 0 \\  0 & \ 0 & \ 0 & \lambda^{+}_{2} & \ 0 & \ 0 & \ 0  & \ 0  & \ 0 \\  0 & \ 0  & \ 0  & \ 0 & \lambda^{o}_{2} & \ 0  & \ 0  & \ 0  & \ 0 \\  0  & \ 0 &  \ 0 &  \ 0 &  \ 0  & \ \lambda^{-}_{2} &  \ 0 &  \ 0 &  \ 0 \\ 0 & \ 0 & \ 0  & \ 0 & \ 0  & \ 0  & \ \lambda^{+}_{3} &  \ 0 &  \ 0  \\  0 & \ 0 & \ 0 & \ 0 & \ 0 & \ 0 & \ 0 & \  \lambda^{o}_{3} & \ 0  \\ 0 & \ 0 & \ 0 & \ 0 & \ 0 & \ 0 & \ 0 & \ 0 &  \ \lambda^{-}_{3} \ea \right] 
\ee
where $\bf f_{1,2,3}$ , $\bf \Lambda_{1,2,3}$ and $ \bf{f^{eq}}_{1,2,3}$ are obtained from (\ref{3V_DVBEq}).
To recover the macroscopic system of equations, we  take moments, which in this case will be by  premultiplying the system with $\bf P$, which is defined as 
\be \label{3_Vel_P} 
\bf{P} =  \left[ \ba{ccccccccc} {1} & {1} & {1} & \ 0 & \ 0 & \ 0 & \ 0  & \ 0 & \ 0 \\  0 & \ 0 & \ 0 & {1} & {1} & {1} & \ 0 & \ 0 & \ 0 \\   0 & \ 0 & \ 0 & \ 0 & \ 0 & \ 0 & {1} & {1} & {1} \ea  \right]
\ee 
Let us now develop a kinetic scheme for the discrete velocity models thus developed using a flux difference splitting approach.  

\section{Kinetic Flux Difference Splitting (KFDS) Scheme}
\subsection{KFDS method with two velocity model} \label{2KFDS_Section}

Let us write  (\ref{DVBE_1D_2_Vel})  in a finite volume framework 
\be \label{K_update_FVM} 
\left \{ {\bf f}_{j}^{n+1} = {\bf f}_{j}^{n} - \fr{\Delta t}{\Delta x} \left[ {\bf h}_{j+\frac{1}{2}}^{n} - {\bf h}_{j-\frac{1}{2}}^{n}  \right]  \right \}_{i}
\ee
where $ {\bf h}_{i} = {\bf \Lambda}_{i} {\bf f}_{i}$.  
To introduce upwinding in two velocity model based DVBE, let us split the discrete velocity matrix ${\bf \Lambda}_{i}$ in to two parts, separating the positive and negative velocities.  
\bea \label{DV_splitting_1D_2VM} 
{\bf \Lambda}_{i} = \left[ \ba{cc} \lambda^{+} & 0 \\ 0 & \lambda^{-} \ea \right]_{i}
= \left[ \ba{cc} \lambda^{+} & 0 \\ 0 & 0 \ea \right] + \left[ \ba{cc} 0 & 0 \\ 0 & \lambda^{-} \ea \right]_{i}
= {\bf \Lambda}^{+}_{i} + {\bf \Lambda}^{-}_{i}
\eea
It is possible to write
\be \label{mod_lambda} 
 {|\bf \Lambda|}_{i}={\bf  \Lambda}^{+}_{i} - {\bf \Lambda}^{-}_{i}
\ee
Thus the upwind fluxes for each $i$, applied on a three point stencil, can be written as 
\bea 
\left \{ {\bf h}_{j+\frac{1}{2}}^{n}= [{\bf \Lambda}^{+} {\bf f}_{eq}]_{j} 
+  [ {\bf \Lambda}^{-} {\bf f}_{eq}]_{j+1}\right \}_{i}\\
\left \{ {\bf h}_{j-\frac{1}{2}}^{n}= [ {\bf \Lambda}^{+} {\bf f}_{eq}]_{j-1} 
+  [ {\bf \Lambda}^{-} {\bf f}_{eq}]_{j} \right \}_{i}
\eea
Using (\ref {DV_splitting_1D_2VM}) and (\ref{mod_lambda}) we can write 
 
\bea 
\left \{ {\bf h}_{j+\frac{1}{2}}^{n}= \underbrace{\fr{1}{2}[ {\bf h}_{j+1}^{n}+ {\bf h}_{j}^{n}]}_{\rm average \ flux} 
- \underbrace{\fr{1}{2}|\Lambda|[{\bf f}^{eq}_{j+1}-[{\bf f}^{eq}_{j}]}_{\rm diffusive \ flux}  \right \}_{i}\\ 
\left \{ {\bf h}_{j-\frac{1}{2}}^{n}= \fr{1}{2}[ {\bf h}_{j}^{n}+ {\bf h}_{j-1}^{n}] 
- \fr{1}{2}|\Lambda|[{\bf f}^{eq}_{j}-[{\bf f}^{eq}_{j-1}]  \right \}_{i}
\eea 
Clearly, $\Lambda$ represents the coefficients of numerical diffusion.  Choosing the values of $\lambda_{i}$ for numerical considerations is an efficient strategy to control the numerical diffusion without losing conservation, as the above expressions for cell-interface fluxes in the flux difference splitting form enforce conservation in the basic finite volume framework.  
The above expressions can further be rewritten in flux difference splitting form as 
\bea \label{2vel_KineticUpdate} 
\left \{ {\bf h}_{j+\frac{1}{2}}^{n}= \fr{1}{2}[ {\bf h}_{j+1}^{n}+ {\bf h}_{j}^{n}] 
- \fr{1}{2}[\Delta {\bf h}^{+}_{j+\frac{1}{2}}-\Delta {\bf h}^{-}_{j+\frac{1}{2}}]  \right \}_{i}\\ 
\left \{ {\bf h}_{j-\frac{1}{2}}^{n}= \fr{1}{2}[ {\bf h}_{j}^{n}+ {\bf h}_{j-1}^{n}] 
- \fr{1}{2}[\Delta {\bf h}^{+}_{j-\frac{1}{2}}-\Delta {\bf h}^{-}_{j-\frac{1}{2}}]  \right \}_{i}
\eea
where $ \Delta {\bf h}_{j\pm\frac{1}{2}}^{\pm} = [{\bf \Lambda}^{\pm}\Delta {\bf f}_{eq}]_{j\pm\frac{1}{2}}$.  In the next sub-section, the coefficient of numerical diffusion, which corresponds to $|\bf \Lambda|$, is chosen a function of both the left and right states.  Thus, the flux difference splitting is the appropriate choice rather than flux vector splitting.  

To recover the macroscopic update formula, let is take moments by multiplying with $P$ as given in (\ref{2_Vel_P}).  Therefore the macroscopic update formula for the Kinetic Flux Difference Splitting (KFDS) scheme thus developed using 2 velocity model based  DVBE can be written as 
\be 
\left \{ U_{j}^{n+1} = U_{j}^{n} - \fr{\Delta t}{\Delta x}\left[ G(U)_{j+\frac{1}{2}}^{n} - G(U)_{j-\frac{1}{2}}^{n}  \right]  \right \}_{i}
\ee
where the interface fluxes are given by
\bea 
\left \{ G(U)_{j+\frac{1}{2}}^{n}= \fr{1}{2}[  G(U)_{j+1}^{n}+ G(U)_{j}^{n}]- \fr{1}{2}[\Delta G(U)^{+,n}_{j+\frac{1}{2}}-\Delta G(U)^{-,n}_{j+\frac{1}{2}}]\right \}_{i}\\ 
 \left \{G(U)_{j-\frac{1}{2}}^{n}= \fr{1}{2}[  G(U)_{j}^{n}+ G(U)_{j-1}^{n}]- \fr{1}{2}[\Delta G(U)^{+,n}_{j-\frac{1}{2}}-\Delta G(U)^{-,n}_{j-\frac{1}{2}}] \right \}_{i}
\eea
\bea \label{2Vel_NKFDS} 
 \left \{\Delta G(U)_{j+\frac{1}{2}}^{\pm} = \fr{1}{2}[  G(U)_{j+1}- G(U)_{j}]\pm \fr{1}{2}|\lambda|[U_{j+1}-U_{j}] \right \}_{i}\\ 
 \left \{\Delta G(U)_{j-\frac{1}{2}}^{\pm} = \fr{1}{2}[  G(U)_{j}- G(U)_{j-1}]\pm \fr{1}{2}|\lambda|[U_{j}-U_{j-1}] \right \}_{i}
\eea  
Clearly, $\lambda$ controls the  numerical diffusion.  
\subsection{Fixing $\lambda$} 
For developing an accurate shock capturing scheme, the value of $\lambda$ is fixed such that numerical diffusion vanishes for a steady discontinuity, as suggested in \cite{HLL}.  As a consequence, though there is a jump in the conserved variables  across a steady discontinuity, the fluxes on the left and right side of the discontinuity are equal. This principle is referred to as {\em flux equivalence across a steady discontinuity} here.   Equating the cell-interface flux separately to the left flux ($G(U)_{j+\frac{1}{2}} = G(U)_{j}$) and the right flux ($G(U)_{j+\frac{1}{2}} = G(U)_{j+1}$) and generalizing leads to 
\be 
\Delta G(U) = |\lambda| \Delta U 
\ee
which is nothing but the Rankine-Hugnoiot (R-H) jump conditions.  Thus, we can choose 
\be 
|\lambda| = \left|\fr{\Delta G(U)}{\Delta U}\right|
\ee 
However, as $U$ and $G(U)$ are vectors, choosing a scalar numerical diffusion is not easy.  We choose a diagonal matrix for representing $\lambda$ (as in \cite{Jaisankar}) and this leads to  
\be 
|\lambda|_{i} = \left|\fr{\Delta G(U)}{\Delta U}\right|_{i}, \ i=1,2,3  
\ee   
Choosing this value for $\lambda_{i}$, which represent the coefficients of numerical diffusion, will enforce R-H conditions directly in the discretization process.  As R-H conditions represent the quintessential conservation of fluxes across discontinuities, conservation is enforced, as is also ensured by beginning with the conservative form of equations and the utilization of the finite volume framework. As a consequence of the above choice, this 2-velocity model based KFDS scheme can capture steady discontinuities exactly, without numerical diffusion.  However, as this variant is low in numerical diffusion, entropy condition violation is likely to occur.  A 3-velocity model based KFDS is introduced in the next section in which the additional velocity and the corresponding component of its distribution function are chosen in a novel way to avoid any possible entropy condition violation.  

   It is worth noting here that though the numerical scheme intially begins with the discrete velocity form of the Boltzmann equation, as $\lambda$ turns out to be the coefficient of numerical diffusion, a suitable choice for the value of $\lambda$ (without losing conservation, in the finite volume and FDS framework, as noted before) changes the discrete nature of the velocities, thus leading to an essentially new framework.  Yet, the name DVBE is retained here, for the sake of convenience.  

\subsection{\bf KFDS method with three Velocity model} \label{3KFDS_Section}
To derive the finite volume scheme for the three velocity based DVBE, we follow the same steps as in the previous section and obtain the update formula at kinetic level to be same as that of two velocity DVBE scheme. The definitions of $\Lambda$ are as  given below.  
\be \label{3velocityLambda}  
{\bf \Lambda}_{i} = \left[ \ba{ccc} \lambda^{+} & \ 0 & \  0 \\[1 mm]   0  &   \ \lambda_{o} &  \  0 \\[1 mm]  0 &  \ 0  & \   \lambda^{-} \ea \right]_{i} = \left[ \ba{ccc} \lambda^{+} & \ 0  &  \  0 \\[1 mm]  0  &   \ \lambda_{o}^{+} & \  0 \\[1 mm]  0 &  \ 0   &  \   0 \ea \right]_{i}+ \left[ \ba{ccc} 0 & \ 0  & \  0 \\[1 mm]  0  &   \ \lambda_{o}^{-} & \  0 \\[1 mm]  0 &  \ 0   & \   \lambda^{-} \ea \right]_{i} = {\bf \Lambda}_{i}^{+} + {\bf \Lambda}_{i}^{-}
\ee
Again it is possible to define $|\bf \Lambda|$ in a similar way as  
\be \label{mod_lambda3} 
 |\bf \Lambda|_{i}= {\bf \Lambda}_{i}^{+} - {\bf \Lambda}_{i}^{-} 
\ee

Following the same finite volume procedure and applying it to (\ref{K_update_FVM}), together with  (\ref{3velocityLambda}), (\ref{mod_lambda3}) and taking moments using $P$ as given in (\ref{3_Vel_P}), we get  
\bea 
 \left \{ G(U)_{j+\frac{1}{2}}^{n}= \fr{1}{2}[  G(U)_{j+1}^{n}+ G(U)_{j}^{n}]- \fr{1}{2}[\Delta G(U)^{+,n}_{j+\frac{1}{2}}-\Delta G(U)^{-,n}_{j+\frac{1}{2}}] \right \}_{i}\\ 
 \left \{ G(U)_{j-\frac{1}{2}}^{n}= \fr{1}{2}[  G(U)_{j}^{n}+ G(U)_{j-1}^{n}]- \fr{1}{2}[\Delta G(U)^{+,n}_{j-\frac{1}{2}}-\Delta G(U)^{-,n}_{j-\frac{1}{2}}] \right \}_{i}
\eea
\bea \label{3Vel_NKFDS} 
 \left \{ \Delta G(U)_{j+\frac{1}{2}}^{\pm} = \fr{1}{2}[  G(U)_{j+1}- G(U)_{j}]]\pm \fr{1}{2}|\lambda|[U_{j+1}-U_{j} ] \mp \fr{1}{2}(\lambda -|\lambda_{o}|)[{f^{o}_{eq}}_{j+1}-{f^{o}_{eq}}_{j} ] \right \}_{i}\\ 
 \left \{ \Delta G(U)_{j-\frac{1}{2}}^{\pm} = \fr{1}{2}[  G(U)_{j}- G(U)_{j-1}]\pm \fr{1}{2}|\lambda|[U_{j}-U_{j-1}]\mp \fr{1}{2}(\lambda -|\lambda_{o}|)[{f^{o}_{eq}}_{j}-{f^{o}_{eq}}_{j-1} ] \right \}_{i}
\eea 
The split flux differences in the interface flux for the three velocity DVBE model based KFDS derived in (\ref{3Vel_NKFDS}) have taken an interesting form in comparison with those in the two velocity DVBE model based KFDS derived in (\ref{2Vel_NKFDS}). We can write
\be \label{mod_lambda2} 
 \left\{ G(U)_{j \pm \frac{1}{2}}^{3V-KFDS}\right\}_{i}= \left \{ G(U)_{j \pm \frac{1}{2}}^{2V-KFDS} \mp  \fr{1}{2}(\lambda -|\lambda_{o}|)(\Delta f^{o}_{eq})\right \}_{i}
\ee  
 wherein the interface flux for 3-Velocity Model is essentially that of 2-Velocity Model with additional terms, which are just functions of $\lambda_{0}$ and $f^{0}_{eq}$.   In the next subsection, $\lambda$, which represents the essential contribution to the coefficient of numerical diffusion of the scheme, is fixed in such a way that numerical diffusion vanishes for a steady discontinuity, leading to the enforcement of R-H conditions, as described in the 2-velocity model.  $\lambda_{0}$ and $f^{0}_{eq}$ are chosen in such a way that entropy condition violation is avoided and hence are activated only in the expansion regions.    This happens automatically in the formulation, based on the utilization of relative entropy or the $D^{2}$-distance.  Since $\lambda_{0}$ and $f^{0}_{eq}$ are activated only in the smooth regions, their contribution to the numerical diffusion in the scheme will be low.  Here, for simplicity, we reduce the unknowns ($\lambda_{0}$ and $f^{0}_{eq}$) from two to one, by expressing one in terms of the other. Thus, to determine $f^{o}_{eq}$ for each $i$, we equate the diffusion flux term of the equation (\ref{3Vel_NKFDS}) to zero. 
\bea 
 \fr{1}{2}|\lambda|\Delta U_{j+\frac{1}{2}} - \fr{1}{2}(\lambda -|\lambda_{o}|)  \Delta {f^{o}_{eq}}_{j+\frac{1}{2}}= 0\\ 
or \ \Delta {f^{o}_{eq}}_{j+\frac{1}{2}} = \fr{|\lambda|}{\lambda - |\lambda_{o}|}\Delta U_{j+\frac{1}{2}}\\ 
\textrm{Thus we infer,} \ {f^{o}_{eq}}_{i} =  \fr{|\lambda|_{i}}{\lambda_{i} - |\lambda_{o}|_{i}}U_{i}
\eea
For convenience, together with $f^{o}_{eq}$, a coefficient which enables a control on the additional diffusion flux is introduced.   Thus we fix $f^{o}_{eq}$ for each $i$ as 

\be
  {f^{o}_{eq}}_{i} =  \left( \fr{k|\lambda|}{\lambda - |\lambda_{o}|}U \right )_{i} =  \left ( \fr{\lambda_{A}}{\lambda - |\lambda_{o}|}U \right )_{i}
\ee 
where $\lambda_{A}$ is the wave speed for the additional diffusion term, which is required to avoid any possible violation of entropy condition.  The flux differences in (\ref{3Vel_NKFDS}) can be rewritten as

\bea \label{3Vel_NKFDS_updated} 
 \left \{ \Delta G(U)_{j+\frac{1}{2}}^{\pm} = \fr{1}{2}[  G(U)_{j+1}^{n}- G(U)_{j}^{n}]\pm \fr{1}{2}|\lambda|[U_{j+1}-U_{j} ] \mp \fr{1}{2}\lambda_{A} [U_{j+1}-U_{j} ] \right \}_{i}\\  \left \{ \Delta G(U)_{j-\frac{1}{2}}^{\pm} = \fr{1}{2}[  G(U)_{j}^{n}- G(U)_{j-1}^{n}]\pm \fr{1}{2}|\lambda|[U_{j}-U_{j-1}]\mp \fr{1}{2}\lambda_{A} [U_{j}-U_{j-1} ] \right \}_{i}
\eea 
It can be observed from the above expressions that two parameters, $\lambda$ and $\lambda_{A}$ are now required to be fixed, to control numerical diffusion.  Out of these two, $\lambda$ is assigned the role of controlling the primary numerical diffusion to capture steady discontinuities exactly, which is described in the next subsection.  $\lambda_{A}$ is assigned the role of introducing additional numerical diffusion to avoid any possible entropy condition violation, which is described in a later subsection. 
\subsection{Fixing the primary numerical diffusion through $\lambda$}
In the construction of the KFDS scheme, fixing of the coefficients $\lambda$ and $\lambda_{A}$ will be critical in determining the fundamental capabilities of the scheme. As mentioned earlier, the idea behind introducing $\lambda_{A}$ is to provide a means to introduce additional diffusion in specific regions of expansion in order to prevent the entropy violation which is typical of low diffusion schemes. The basic definition of $\lambda$ comes with the assumption that $\lambda_{A}$ is zero. $\lambda$ is fixed to obtain a steady shock with zero numerical diffusion.  For a steady shock to be captured exactly, when the left and right fluxes are equal the numerical diffusion must vanish, according to Harten, Lax and van Leer [\cite {HLL}].  This {\em flux equivalence across a steady shock} is the consequence of flux conservation.  Thus, in the expression for the cell-interface flux $G_{j+\frac{1}{2}}$ for each $i$, we can take 
$G_{j} = G_{j+\frac{1}{2}} = G_{j+1}$ to obtain an expression for $\lambda$. First we substitute $G_{j+\frac{1}{2}}=G_{j}$.  Then, we get  
\be 
|{\lambda}_{i}| \Delta U_{i} = \left(G_{j+1} - G_{j}\right)_{i} = {\Delta G}_{i}
\ee
Similarly, $G_{j+\frac{1}{2}} = G_{j+1}$ leads to 
\be 
|{\lambda}|_{i}\Delta U_{i} = - \Delta G_{i}
\ee
Generalizing from both the above expressions, we write   
\be
|\lambda|_{i} = \left|\frac{\Delta G_{i}}{\Delta U_{i}}\right| 
\ee 
Therefore, the values of $|{\lambda}|$ for the continuity, momentum and energy equations in a one dimensional flow will take the form as given below.  
\be
|{\lambda}|_{1} = \left |\fr{\Delta (\rho u)}{\Delta \rho}\right|; \ \ \ |{\lambda}|_{2} = \left |\fr{\Delta (p + \rho {u}^{2})}{\Delta \rho u} \right|; \ \ \ |{\lambda}|_{3} = \left |\fr{\Delta( p u + \rho u E)}{\Delta  \rho u E}\right| 
\ee
As the coefficient of numerical diffusion is a function of both the left and right states across a cell-interface, flux difference splitting approach chosen is appropriate. In order to avoid numerical overflow, ${|\lambda|}_{i}$ is restricted to the minimum eigenvalue when $\Delta {U}_{i} \le \epsilon$ where $\epsilon = {10}^{-10}$.  For $G_{j} = G_{j+1}$ or $\Delta G = 0$, the coefficient of diffusion $\lambda$ becomes zero, thus capturing the steady shocks and contact discontinuities exactly.    
\subsection{Need for additional diffusion}
 
The requirement of the additional diffusion arises in smooth regions of the flow. In particular, sonic points can occur in regions of expansions in a flow field and the low diffusive numerical schemes would require an entropy fix to prevent the occurrence of unphysical expansion shocks.  Such an entropy fix introduces additional non-zero numerical diffusion near the sonic points.  However, if this is done everywhere, the exact shock capturing ability will be lost. It is therefore desirable to retain the low diffusive nature of the basic scheme and add additional numerical diffusion only near the expansive sonic points.  The identification of smooth regions, in contrast with the regions of discontinuity, is done in a novel way utilizing the relative entropy or the $D^{2}$-distance.
\subsubsection{$D^{2}$-distance and entropy}

\begin{figure} 
\begin{center} 
\includegraphics[width=15.8cm,angle=0]{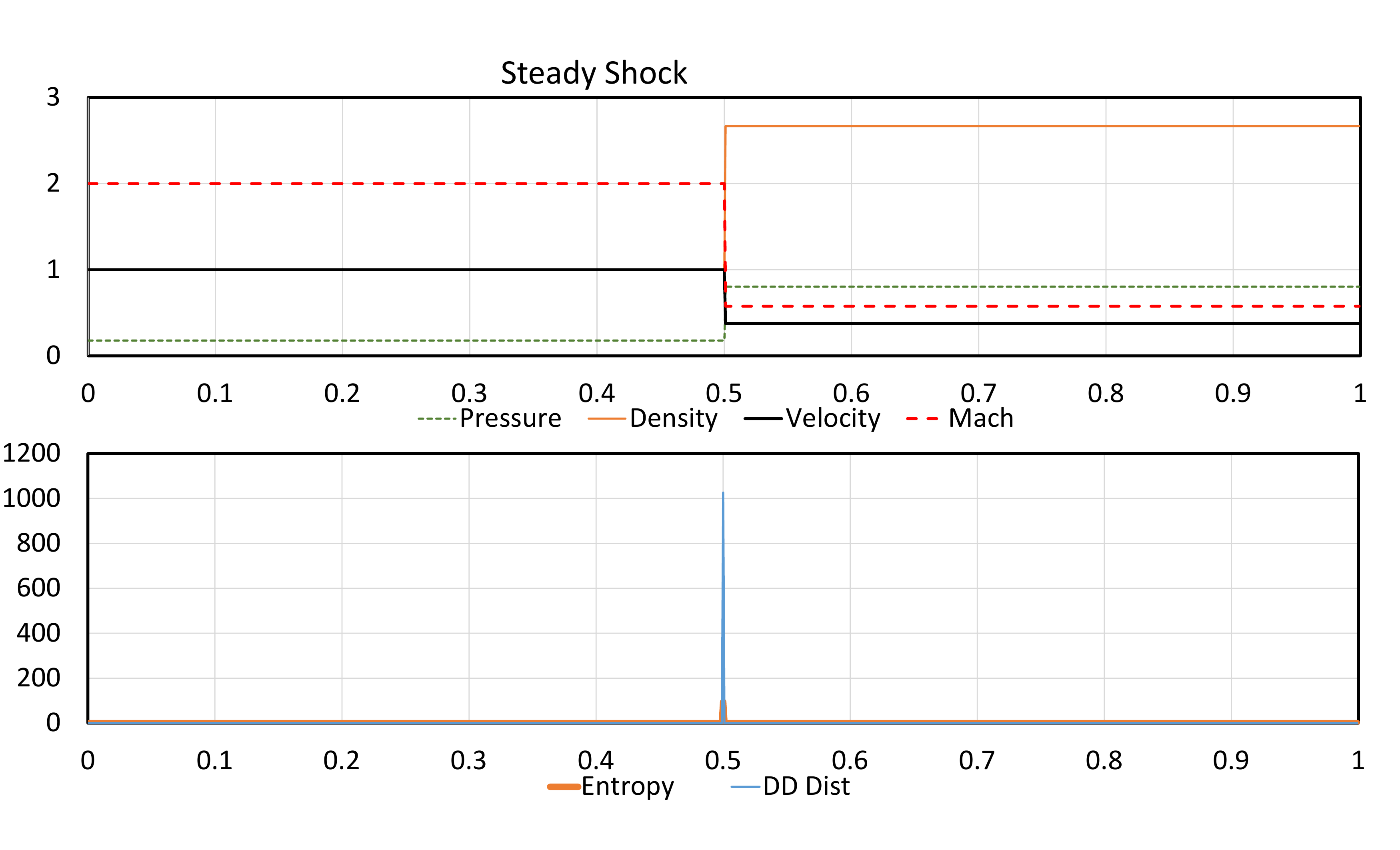}
\includegraphics[width=15.8cm,angle=0]{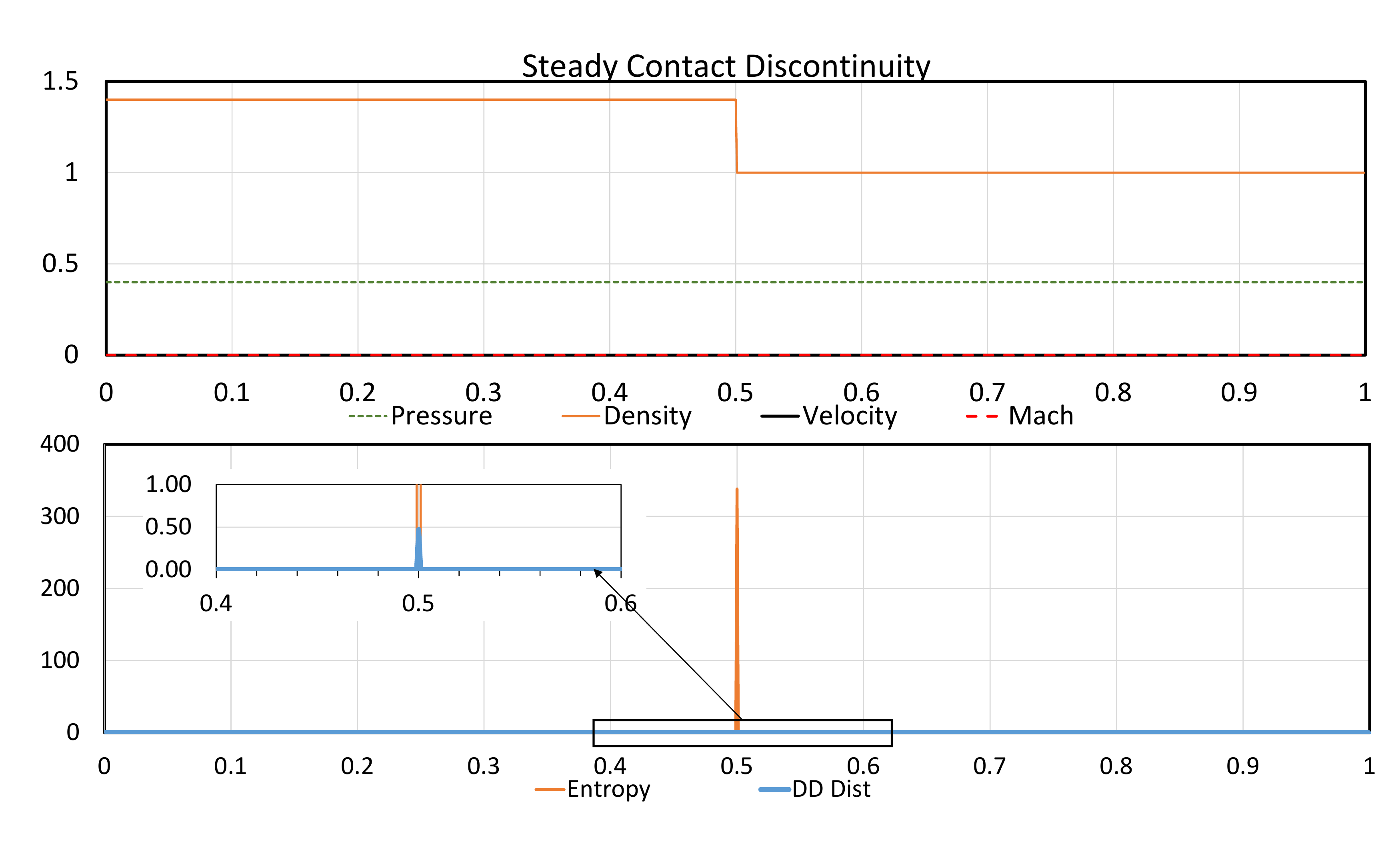}
\end{center} 
\caption{Entropy and $D^{2}$-distance (relative entropy) for steady shock and steady contact discontinuity} 
\label{MahalanobisStShkCt}  
\end{figure}

\begin{figure} 
\begin{center} 
\includegraphics[width=15.8cm,angle=0]{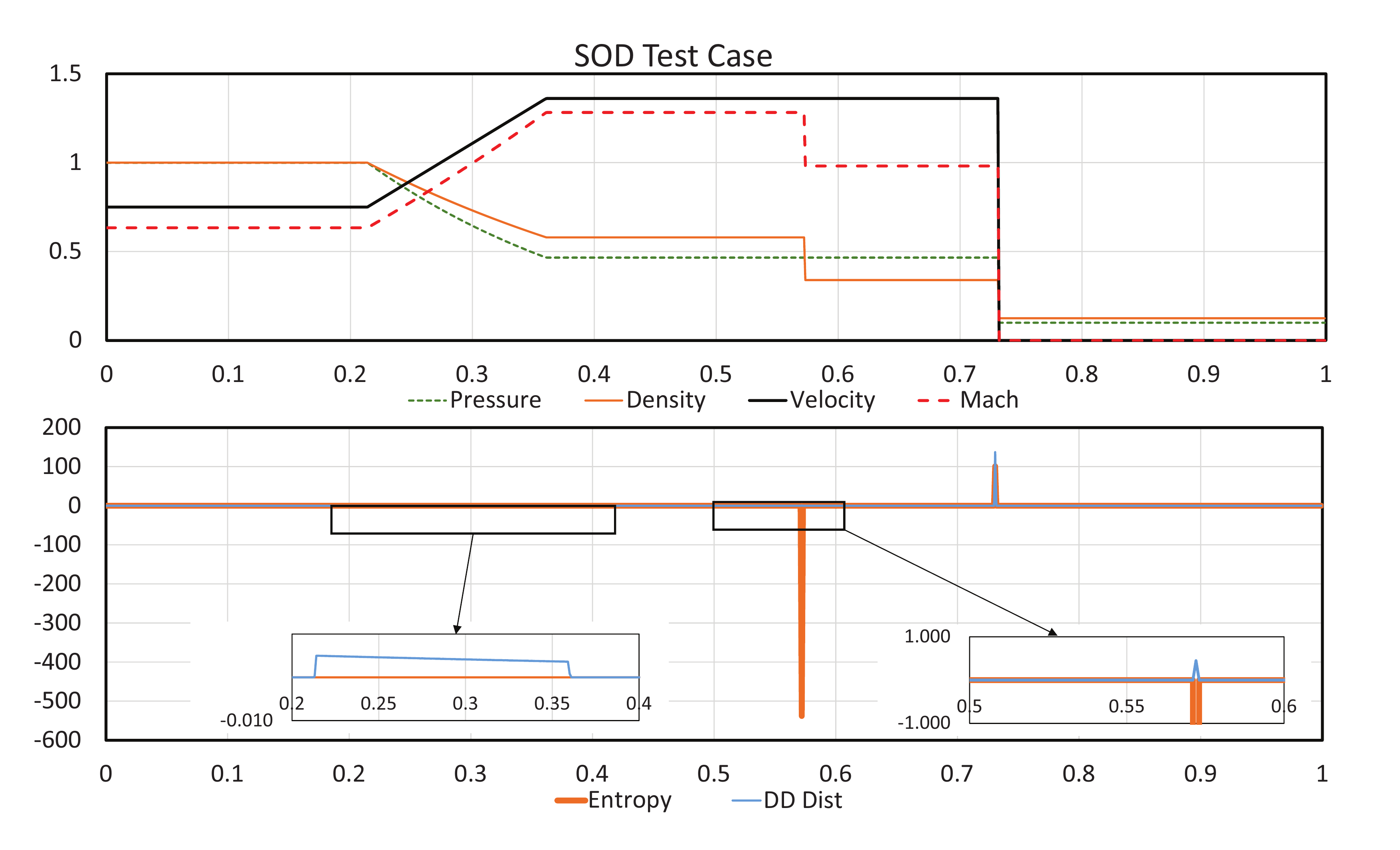}
\end{center} 
\caption{Entropy and $D^{2}$-distance (relative entropy) for SOD test case} 
\label{MahalanobisSOD}  
\end{figure}

Relative entropy, also known as Kullback-Liebler divergence, directed divergence or $D^{2}$-distance,  is a measure of variation between two distribution functions. This measure and its variants are popularly used in statistics to identify the distinguishing features between two sets of statistical data (discrete samples) or continuous functions.  It is worth noting that the form of $D^{2}$-distance resembles the Boltzmann H-function and thereby is a comprehensive representation of entropy at the kinetic level.   
The relative entropy to measure divergence between two distributions is given by (\ref{Mahal-Kinetic}).  
\be \label{Mahal-Kinetic}
D\left(f^{eq}_{k}, f^{eq}_{j}\right)=\int_{0}^{\infty} \int_{-\infty}^{\infty}\left(f^{eq}_{j}-f^{eq}_{k}\right) \ln \left(\frac{f^{eq}_{j}}{f^{eq}_{k}}\right) d v d I
\ee
Here, $f^{eq}_{k,j}$ represents the equilibrium distribution and  $k,j$ represent two locations, which represent in the present framework the left and right state of the interface.   Utilizing the definition of classical Maxwellians for the equilibrium distribution, the moment of the above $D^{2}$-distance yields \cite{Venkat_ME_Thesis}   
\be \label{Mahal-DD}
 D^{2}\left(f^{eq}_{k}, f^{eq}_{j}\right)= \left(\rho_{j}-\rho_{k}\right) \ln \left[\frac{\rho_{j}}{\rho_{k}}\left(\frac{T_{k}}{T_{j}}\right)^{\frac{5}{2}}\right] + \frac{\rho_{k}}{2 R T_{k}}\left[\frac{\rho_{j}}{\rho_{k}}+\frac{T_{k}}{T_{j}}\right]\left(u_{k}-u_{j}\right)^{2}+\frac{5}{2}\left[\frac{\rho_{k}\left(T_{k}-T_{j}\right)}{T_{j}}+\frac{\rho_{j}\left(T_{j}-T_{k}\right)}{T_{k}}\right]
\ee
The thermodynamic entropy used in this work is obtained from gas dynamic relations as  
\be
S_{j}=-R\left(\ln \rho_{j}+ \frac{\ln \beta_{j}}{\gamma-1} + {\rm constant} \right)
\ee  
To evaluate the capabilities of the  relative entropy as a sensor function, numerical tests were performed on the exact solutions of several one dimensional Euler test cases. The behaviour of the relative entropy and the corresponding  entropy for the some of most representative test cases are shown in Figures [\ref{MahalanobisStShkCt}] and [\ref{MahalanobisSOD}].  

It can be seen that the relative entropy gives a positive signal of distinct magnitude for shocks and contact discontinuities.  Even in case of expansion waves the function produces a positive signal whose magnitude is one order lower than that of strong shocks and contact discontinuities.  It is evident that this function (\ref{Mahal-DD}) senses gradients arising in pressure, temperature, density and velocity and can be effective in identifying flow gradients and discontinuities of many kinds.  While identification of shocks, contact discontinuities or expansions individually is challenging, the $D^{2}$-distance along with the entropy function seem to be a good indicator for this task. In the present work, the $D^{2}$- distance along with the entropy function is used to identify expansion regions and we introduce additional diffusion only at these regions to avoid unphysical expansion shocks.  The rendering of the discrete version of relative entropy to fit the discrete Boltzmann system presented here, with the study of the required mathematical properties, is beyond the scope of this paper and will be pursued elsewhere.    
\subsection{Fixing the additional numerical diffusion through $\lambda_{A}$} \label{Fixing_Lambda_A}  
For fixing the additional numerical diffusion, $\lambda_{A}$ 
 is defined as follows.    
\bea \label{Add_Diffusion}
\lambda_{A} = \left\{ \ba{l}  \lambda_{DD} \ \textrm{for } \  D^{2} > 0 \ \textrm{  } \  \& \ \textrm{  } \ \Delta S =0  \\ 
                                              0 \ \ \textrm{Otherwise }  \ea \right. 
\eea 
where $D^{2}$ refers to the relative entropy or $D^{2}$-distance and $\Delta S$ is the estimated difference in thermodynamic entropy across the cell interface. 
Two possibilities are presented here for fixing $\lambda_{DD}$.  

\subsubsection{Maximum eigenvalue based $\lambda_{DD}$}
A simple and yet robust strategy to capture expansion region without violating entropy conditions is to fix the coefficient of numerical diffusion as the maximum of the eigenvalues, as in Rusanov of LLF method.  It it is done only at expansions, as it would yet be possible retain the steady discontinuity capturing ability of the scheme.   Thus the first possible definition of $\lambda_{DD}$ will be 

\bea
\lambda_{DD} = \lambda_{max} - \lambda \\ {\rm where} \ 
\lambda_{\max }=\max \left\{\max (|u+a|,|u|,|u-a|)_{L}, \max (|u+a|,|u|,|u-a|)_{R}\right\}
\eea 
This makes the coefficient of the total numerical diffusion as $\lambda_{max}$ at the expansions.  

\pagebreak

\subsubsection{Fluid velocity based $\lambda_{DD}$}  
Another choice for $\lambda_{A}$ is to select a coefficient of numerical diffusion based on normal fluid velocity across the cell interface.  Raghavendra \cite{Venkat_Thesis}, Ramesh \cite{Ramesh_Thesis} and Ramesh {\em et al.} \cite{Ramesh_Venkat_SVRRao_Sekhar} have demonstrated that the numerical diffusion can be fixed based on Riemann invariants and have reported stable capturing of the expansion regions without the need for an entropy fix. This approach automatically leads to the normal fluid velocity as the coefficient of numerical diffusion.  We define
 
\bea
\lambda_{DD} = \lambda_{RI} - \lambda \\ 
{\rm where} \ \lambda_{RI }=\frac{|u_{L}| + |u_{R}|}{2}
\eea 
This makes the coefficient of total numerical diffusion in the scheme as $\lambda_{RI}$ in expansions.

\subsubsection{Activation strategy of additional numerical diffusion}
To evaluate the efficiency of (\ref{Add_Diffusion}), preliminary numerical tests were performed on shock tube problems. The capability of the basic scheme with no additional diffusion is presented in Table 1 for comparison. 

\begin{table}[htb] 
\centering
\begin{tabular}{|l|c|c|c|c|c|}
\hline \bf TEST CASE & \bf KFDS & \bf   KFDS  &\bf    KFDS & \bf KFDS   &\bf KFDS  \\
&  \bf No additional & + DD $\bf>0$ & + DD $\bf>0$ & + DD $\bf>0$&+ DD $\bf>0$ \\ 
&  \bf Diffusion & & $ +\bf \Delta S = 0$ &  +$\bf \Delta S \le {S}_{max}$ &  + $\bf \Delta S \le {S}_{mean}$\\ 
\hline & \bf \gre\cmark & \bf \gre\cmark &  \bf \gre\cmark &  \bf \gre\cmark &  \bf \gre\cmark \\
\bf Steady C.D. & Exact & Diffused & Exact & Exact & Exact \\
\hline & \bf \gre\cmark & \bf \gre\cmark &  \bf \gre\cmark &  \bf \gre\cmark &  \bf \gre\cmark \\
\bf Steady Shock & Exact & Diffused & Exact & Diffused & Exact \\
\hline &\bf \rr\xmark & \bf \gre\cmark &  \bf \rr\xmark &  \bf \gre\cmark &  \bf \gre\cmark \\
\bf SOD case & Expansion Shocks & smooth & Expansion Shocks & smooth & smooth \\
\hline &\bf \rr\xmark & \bf \gre\cmark &  \bf \gre\cmark &  \bf \gre\cmark &  \bf \gre\cmark \\
\bf Overheating & Blows up & smooth & smooth  & smooth & smooth \\
\hline &\bf \br\cmark & \bf \gre\cmark &  \bf \br\cmark &  \bf \gre\cmark &  \bf \gre\cmark \\
\bf Toro Case 3 & Not smooth & smooth & Not smooth & smooth & smooth \\
\hline &\bf \br\cmark & \bf \gre\cmark &  \bf \br\cmark &  \bf \gre\cmark &  \bf \gre\cmark \\
\bf Shock Collision & Not smooth & smooth & Not smooth & smooth & smooth \\
\hline &\bf \rr\xmark & \bf \gre\cmark &  \bf \rr\xmark &  \bf \gre\cmark &  \bf \gre\cmark \\
\bf Toro Case 5 & Blows up & smooth & Blows up & smooth & smooth \\
\hline
\end{tabular}
\caption{Numerical experiments to deduce activation strategy for additional numerical diffusion} 
\end{table}
Upon introducing additional diffusion in regions where directed divergence is nonzero, the scheme is stable but loses its ability to capture steady shocks and steady contact  discontinuities exactly.  When the condition in (\ref{Add_Diffusion}) is strictly enforced, the diffusion required for the solution to evolve in the regions of shocks and contact discontinuities is found to be not adequate.  As a result, expansion shocks appear in SOD test case and the solution blows up for Toro case 5.  It is observed that additional diffusion is required in regions of both shocks and expansions in the unsteady test cases for the solution to evolve. This means that the condition $\Delta$S=0 has to be revised to $\Delta S \le {\rm a \ parameter}$. The numerical experiments further confirmed that the parameter cannot be a constant and its magnitude scales up according to the maximum energy / entropy level of the system.  Therefore, the parameter has to be a function of the primitive variables defining the system. After careful numerical experiments, maximum entropy and mean entropy of the system were identified as the limits, upto which entropy change ($\Delta S$) is acceptable to permit additional diffusion.  While both the limits provided positive solutions, the use of mean entropy limit for regulating additional diffusion zone is capable of retaining the exact steady shock and steady contact discontinuity capturing ability of the numerical scheme. Therefore, the condition to introduce additional diffusion as given in (\ref{Add_Diffusion}) is revised to the following. 

\bea \label{Add_Diffusion_New}
\lambda_{A} = \left\{ \ba{l}  \lambda_{DD} \ \textrm{for } \  D^{2} > 0 \ \textrm{  } \  \& \ \textrm{  } \ \Delta S \le {S}_{mean}  \\ 
                                              0 \ \ \textrm{Otherwise }  \ea \right. 
\eea 
It is clear that the additional numerical diffusion to get all the typical and different test cases working well is quite sensitive to the involved parameters, a fact well-known to the macroscopic CFD algorithm developers.  The optimal numerical diffusion to avoid entropy violation in smooth regions still remains an active research topic.    


\subsection{Final expressions for KFDS scheme}
The final form of the numerical scheme is given as
\be 
\left\{U_{j}^{n+1} = U_{j}^{n} - \fr{\Delta t}{\Delta x}\left[ G(U)_{j+\frac{1}{2}}^{n} - G(U)_{j-\frac{1}{2}}^{n}  \right]  \right\}_{i}
\ee
where the interface fluxes are given by
\bea 
\left\{ G(U)_{j+\frac{1}{2}}^{n}= \fr{1}{2}[  G(U)_{j+1}^{n}+ G(U)_{j}^{n}]- \fr{1}{2}[\Delta G(U)^{+,n}_{j+\frac{1}{2}}-\Delta G(U)^{-,n}_{j+\frac{1}{2}}]\right\}_{i}\\ 
\left\{ G(U)_{j-\frac{1}{2}}^{n}= \fr{1}{2}[  G(U)_{j}^{n}+ G(U)_{j-1}^{n}]- \fr{1}{2}[\Delta G(U)^{+,n}_{j-\frac{1}{2}}-\Delta G(U)^{-,n}_{j-\frac{1}{2}}]\right\}_{i}
\eea
The split flux differences are given by
\bea \label{3Vel_NKFDS_updated_2} 
\left\{ \Delta G(U)_{j+\frac{1}{2}}^{\pm} = \fr{1}{2}[  G(U)_{j+1}^{n}- G(U)_{j}^{n}]]\pm \fr{1}{2}|\lambda|[U_{j+1}-U_{j} ] \mp \fr{1}{2}\lambda_{A} [U_{j+1}-U_{j} ]\right\}_{i}\\ 
\left\{ \Delta G(U)_{j-\frac{1}{2}}^{\pm} = \fr{1}{2}[  G(U)_{j}^{n}- G(U)_{j-1}^{n}]\pm \fr{1}{2}|\lambda|[U_{j}-U_{j-1}]\mp \fr{1}{2}\lambda_{A} [U_{j}-U_{j-1} ]\right\}_{i}
\eea
where  $|{\lambda}_{i}| = \left|\frac{\Delta G(U)_{i}}{\Delta U_{i}}\right|$ and $\lambda_{A}$ for each $i$ is defined by
\bea 
\lambda_{A} = \left\{ \ba{l}  \lambda_{DD} \ \textrm{for } \  D^{2} > 0 \ \textrm{  } \  \& \ \textrm{  } \ \Delta S \le {S}_{mean} \\ 
                                              0 \ \ \textrm{Otherwise }  \ea \right. 
\eea 
$\lambda_{A}$ is determined by the choice of the $\lambda_{DD}$ and correspondingly the scheme will be named as KFDS-A for the use of maximum eigenvalue based wave speed and KFDS-B for the use of normal fluid velocity based wave speed.  
 
\subsection{\bf Higher order accuracy} 
There are many ways to obtain higher order accuracy \cite{Leveque1,Laney,Torobook,AbgrallHb}.  One of the simplest of the ways is to reconstruct the primitive or the conserved variables by a linear combination of the solution variables from the  appropriate neighbouring cells. It is important to note that a mere linear combination can result in spurious oscillations and can lead to unphysical results. Thus such linear combinations are generally accompanied  with moderation of the slopes or fluxes using limiter functions.  In this paper, second order accuracy is obtained by means of a piece-wise linear reconstruction of the distribution functions from neighbouring cells, at a cell-interface.  
We take the same approach as in section \ref{3KFDS_Section} for discrete kinetic upwinding, with linear reconstructions of the distributions. Although this section is presented without suffix $i$ for simplicity, each of the expressions does have a valid suffix $i$, representing the respective conservation equation. 

\bea
{\bf h}_{j+\frac{1}{2}}^{n}= \left[ \left(\Lambda^{+} \mathbf{f}_{e q}\right)_{L}+\left(\Lambda^{-} \mathbf{f}_{e q}\right)_{R} \right]_{j+\frac{1}{2}}  \\ 
{\bf h}_{j-\frac{1}{2}}^{n} =  \left[ \left(\Lambda^{+} \mathbf{f}_{e q}\right)_{L}+\left(\Lambda^{-} \mathbf{f}_{e q}\right)_{R} \right]_{j-\frac{1}{2}}
\eea

In order to attain second order accuracy, we do a piece-wise linear reconstruction of the conserved variable (in this case, the distribution function $f$) to obtain the equivalent left and right state variables for the given cell-interface.

\be 
{\bf f}_{j}\left(x, t^{n}\right) =\left[{\bf f}_{j}^{n}+\left(\frac{\partial {\bf f}}{\partial x}\right)_{j}\left(x-x_{j}\right)\right]
\ee
The slope can be limited by a minmod limiter to obtain a non-oscillatory behaviour. 
\be
\left(\frac{\partial {\bf f}}{\partial x}\right)_{j}={\bf minmod}\left[\frac{{\bf f}_{j+1}-{\bf f}_{j}}{\Delta x }, \frac{{\bf f}_{j}-{\bf f}_{j-1}}{\Delta x }\right]
\ee  
Using such piece-wise linear reconstructions at appropriate locations, the left and right states  ${\bf f}_{L}$ and  ${\bf f}_{R}$ at the cell-interface $I=j+\frac{1}{2}$ turn out to be 
\bea
{\bf f}_{R,j+\frac{1}{2}} = {\bf f}_{j+1} - \fr{1}{2}{\bf minmod} \left[ ({\bf f}_{j+2}-{\bf f}_{j+1}),({\bf f}_{j+1}-{\bf f}_{j})\right]  \\\
{\bf f}_{L,j+\frac{1}{2}} = {\bf f}_{j}     + \fr{1}{2}{\bf minmod} \left[ ({\bf f}_{j+1}-{\bf f}_{j}),({\bf f}_{j}-{\bf f}_{j-1})\right] 
\eea

Therefore the interface flux in the finite volume update formula can be written  in flux difference split form as 
\bea \label{interfaceflux_3kfds}
{\bf h}_{j+\frac{1}{2}}^{n}= \fr{1}{2}\left[ {\bf h}_{R}^{n}+ {\bf h}_{L}^{n}\right]_{j+\frac{1}{2}} 
- \fr{1}{2}[\Delta {\bf h}^{+}_{j+\frac{1}{2}}-\Delta {\bf h}^{-}_{j+\frac{1}{2}}] \\
{\bf h}_{j-\frac{1}{2}}^{n}= \fr{1}{2}\left[ {\bf h}_{R}^{n}+ {\bf h}_{L}^{n}\right]_{j-\frac{1}{2}} 
- \fr{1}{2}[\Delta {\bf h}^{+}_{j-\frac{1}{2}}-\Delta {\bf h}^{-}_{j-\frac{1}{2}}]  
\eea
where 
\be \label{FluxDiffSplits}
\Delta {\bf h}_{j\pm\frac{1}{2}}^{\pm} = [\Lambda^{\pm}\left({\bf f}_{R}-{\bf f}_{L}\right)]_{j\pm\frac{1}{2}}
\ee
Let us evaluate the moments of the above equations to recover the corresponding macroscopic update formula. Equation (\ref{K_update_FVM}) gives
\be 
U_{j}^{n+1} = U_{j}^{n} - \fr{\Delta t}{\Delta x}\left[ G(U)_{j+\frac{1}{2}}^{n} - G(U)_{j-\frac{1}{2}}^{n}  \right]   
\ee 
where the interface fluxes are obtained by taking moments of (\ref{interfaceflux_3kfds}).  
\bea \label {Right_Interface_Flux}
\ba{ll}{\bf P}{\bf h}_{j+\frac{1}{2}} & =    G(U)_{j+\frac{1}{2}}\\[6mm]
& =   \fr{1}{2}\left[G(U)_{R}+ G(U)_{L}\right]_{j+\frac{1}{2}} - \fr{1}{2}[\Delta G(U)^{+}_{j+\frac{1}{2}}-\Delta G(U)^{-}_{j+\frac{1}{2}}]\\[6mm]
& =   \fr{1}{2}\left[G(U)_{j}+ G(U)_{j+1}\right] -\fr{1}{4}{\bf minmod}\left[ (G(U)_{j+2}-G(U)_{j+1}),(G(U)_{j+1}-G(U)_{j}) \right] \\[4mm]
&    +\fr{1}{4}{\bf minmod}\left[ (G(U)_{j+1}-G(U)_{j}),(G(U)_{j}-G(U)_{j-1})\right]- \fr{1}{2}[\Delta G(U)^{+}_{j+\frac{1}{2}}-\Delta G(U)^{-}_{j+\frac{1}{2}}]\\[6mm]
\ea
\eea
Similarly 
\bea \label {Left_Interface_Flux}
\ba{ll}{\bf P}{\bf h}_{j-\frac{1}{2}} & =    G(U)_{j-\frac{1}{2}}\\[6mm]
& =   \fr{1}{2}\left[G(U)_{R}+ G(U)_{L}\right]_{j-\frac{1}{2}} - \fr{1}{2}[\Delta G(U)^{+}_{j-\frac{1}{2}}-\Delta G(U)^{-}_{j-\frac{1}{2}}]\\[6mm]
& =   \fr{1}{2}\left[G(U)_{j-1}+ G(U)_{j}\right] -\fr{1}{4}{\bf minmod}\left[ (G(U)_{j+1}-G(U)_{j}),(G(U)_{j}-G(U)_{j-1}) \right] \\[4mm]
&    +\fr{1}{4}{\bf minmod}\left[ (G(U)_{j}-G(U)_{j-1}),(G(U)_{j-1}-G(U)_{j-2})\right]- \fr{1}{2}[\Delta G(U)^{+}_{j-\frac{1}{2}}-\Delta G(U)^{-}_{j-\frac{1}{2}}]\\[6mm]
\ea
\eea
Here, the split flux differences, namely $\Delta G(U)^{\pm}_{j\pm\frac{1}{2}}$, can be evaluated by taking moments of (\ref{FluxDiffSplits}).  
\bea
\ba{ll}{\bf P}\Delta {\bf h}_{j+\frac{1}{2}}^{+} &= \Delta {G(U)}^{+}_{j+\frac{1}{2}}\\[4mm]
&= {\bf P}[\Lambda^{+}\left({\bf f}_{R}-{\bf f}_{L}\right)]_{j+\frac{1}{2}}\\[4mm]
& = {\bf P}\Lambda^{+}\left[ {\bf f}_{j+1} -\fr{1}{2}{\bf minmod} \left[ ({\bf f}_{j+2}-{\bf f}_{j+1}),({\bf f}_{j+1}-{\bf f}_{j})\right] \right]\\[4mm]
& - {\bf P}\Lambda^{+}\left[ {\bf f}_{j} +\fr{1}{2}{\bf minmod} \left[ ({\bf f}_{j+1}-{\bf f}_{j}),({\bf f}_{j}-{\bf f}_{j-1})\right] \right]\\[4mm]
\ea
\eea
Upon expanding and rearranging which we get
$$
\ba{ll}
 \Delta {G(U)}^{+}_{j+\frac{1}{2}} &= \fr{1}{2}\left( {G(U)}_{j+1}-{G(U)}_{j}\right) + \fr{1}{2}|\lambda|\left( {U}_{j+1}-{U}_{j}\right)- \fr{(\lambda-|\lambda_{o}|)}{2}\left( { {\bf f}_{eq,}^{o}}_{j+1}-{ {\bf f}_{eq,}^{o}}_{j}\right)\\[4mm]
& - \fr{1}{2} {\bf minmod} \left[ \ba{c} \left\{ \fr{1}{2}\left( {G(U)}_{j+2}-{G(U)}_{j+1}\right) + \fr{1}{2}|\lambda|\left( {U}_{j+2}-{U}_{j+1}\right)
- \fr{(\lambda-|\lambda_{o}|)}{2}\left( { {\bf f}_{eq,}^{o}}_{j+2}-{ {\bf f}_{eq,}^{o}}_{j+1}\right) \right\}, \\[2mm]  
\left\{ \fr{1}{2}\left( {G(U)}_{j+1}-{G(U)}_{j}\right) + \fr{1}{2}|\lambda|\left( {U}_{j+1}-{U}_{j}\right)
- \fr{(\lambda-|\lambda_{o}|)}{2}\left( { {\bf f}_{eq,}^{o}}_{j+1}-{ {\bf f}_{eq,}^{o}}_{j}\right)  \right\} \ea \right]\\[10mm]
& - \fr{1}{2} {\bf minmod} \left[ \ba{c} \left\{ \fr{1}{2}\left( {G(U)}_{j+1}-{G(U)}_{j}\right) + \fr{1}{2}|\lambda|\left( {U}_{j+1}-{U}_{j}\right)
- \fr{(\lambda-|\lambda_{o}|)}{2}\left( { {\bf f}_{eq,}^{o}}_{j+1}-{ {\bf f}_{eq,}^{o}}_{j}\right) \right\}, \\[2mm]  
\left\{ \fr{1}{2}\left( {G(U)}_{j}-{G(U)}_{j-1}\right) + \fr{1}{2}|\lambda|\left( {U}_{j}-{U}_{j-1}\right)
- \fr{(\lambda-|\lambda_{o}|)}{2}\left( { {\bf f}_{eq,}^{o}}_{j}-{ {\bf f}_{eq,}^{o}}_{j-1}\right)  \right\} \ea \right]
\ea
$$
Similarly the $ \Delta {G(U)}^{-}_{j+\frac{1}{2}}$ term can be obtained as
$$
\ba{ll}
 \Delta {G(U)}^{-}_{j+\frac{1}{2}} &= \fr{1}{2}\left( {G(U)}_{j+1}-{G(U)}_{j}\right) - \fr{1}{2}|\lambda|\left( {U}_{j+1}-{U}_{j}\right)+ \fr{(\lambda-|\lambda_{o}|)}{2}\left( { {\bf f}_{eq,}^{o}}_{j+1}-{ {\bf f}_{eq,}^{o}}_{j}\right)\\[4mm]
& - \fr{1}{2} {\bf minmod} \left[ \ba{c} \left\{ \fr{1}{2}\left( {G(U)}_{j+2}-{G(U)}_{j+1}\right) - \fr{1}{2}|\lambda|\left( {U}_{j+2}-{U}_{j+1}\right)
+ \fr{(\lambda-|\lambda_{o}|)}{2}\left( { {\bf f}_{eq,}^{o}}_{j+2}-{ {\bf f}_{eq,}^{o}}_{j+1}\right) \right\}, \\[2mm]  
\left\{ \fr{1}{2}\left( {G(U)}_{j+1}-{G(U)}_{j}\right) - \fr{1}{2}|\lambda|\left( {U}_{j+1}-{U}_{j}\right)
+ \fr{(\lambda-|\lambda_{o}|)}{2}\left( { {\bf f}_{eq,}^{o}}_{j+1}-{ {\bf f}_{eq,}^{o}}_{j}\right)  \right\} \ea \right]\\[4mm]
& - \fr{1}{2} {\bf minmod} \left[ \ba{c} \left\{ \fr{1}{2}\left( {G(U)}_{j+1}-{G(U)}_{j}\right) - \fr{1}{2}|\lambda|\left( {U}_{j+1}-{U}_{j}\right)
+ \fr{(\lambda-|\lambda_{o}|)}{2}\left( { {\bf f}_{eq,}^{o}}_{j+1}-{ {\bf f}_{eq,}^{o}}_{j}\right) \right\}, \\[2mm]  
\left\{ \fr{1}{2}\left( {G(U)}_{j}-{G(U)}_{j-1}\right) - \fr{1}{2}|\lambda|\left( {U}_{j}-{U}_{j-1}\right)
+ \fr{(\lambda-|\lambda_{o}|)}{2}\left( { {\bf f}_{eq,}^{o}}_{j}-{ {\bf f}_{eq,}^{o}}_{j-1}\right)  \right\} \ea \right]
\ea
$$
In the same way, the flux difference split fluxes at $j-\fr{1}{2} $  interface can be obtained as
$$
\ba{ll}
 \Delta {G(U)}^{+}_{j-\frac{1}{2}} &= \fr{1}{2}\left( {G(U)}_{j}-{G(U)}_{j-1}\right) + \fr{1}{2}|\lambda|\left( {U}_{j}-{U}_{j-1}\right)- \fr{(\lambda-|\lambda_{o}|)}{2}\left( { {\bf f}_{eq,}^{o}}_{j}-{ {\bf f}_{eq,}^{o}}_{j-1}\right)\\[4mm]
& - \fr{1}{2} {\bf minmod} \left[ \ba{c} \left\{ \fr{1}{2}\left( {G(U)}_{j+1}-{G(U)}_{j}\right) + \fr{1}{2}|\lambda|\left( {U}_{j+1}-{U}_{j}\right)
- \fr{(\lambda-|\lambda_{o}|)}{2}\left( { {\bf f}_{eq,}^{o}}_{j+1}-{ {\bf f}_{eq,}^{o}}_{j}\right) \right\}, \\[2mm]  
\left\{ \fr{1}{2}\left( {G(U)}_{j}-{G(U)}_{j-1}\right) + \fr{1}{2}|\lambda|\left( {U}_{j}-{U}_{j-1}\right)
- \fr{(\lambda-|\lambda_{o}|)}{2}\left( { {\bf f}_{eq,}^{o}}_{j}-{ {\bf f}_{eq,}^{o}}_{j-1}\right)  \right\} \ea \right]\\[10mm]
& - \fr{1}{2} {\bf minmod} \left[ \ba{c} \left\{ \fr{1}{2}\left( {G(U)}_{j}-{G(U)}_{j-1}\right) + \fr{1}{2}|\lambda|\left( {U}_{j}-{U}_{j-1}\right)
- \fr{(\lambda-|\lambda_{o}|)}{2}\left( { {\bf f}_{eq,}^{o}}_{j}-{ {\bf f}_{eq,}^{o}}_{j-1}\right) \right\}, \\[2mm]  
\left\{ \fr{1}{2}\left( {G(U)}_{j-1}-{G(U)}_{j-2}\right) + \fr{1}{2}|\lambda|\left( {U}_{j-1}-{U}_{j-2}\right)
- \fr{(\lambda-|\lambda_{o}|)}{2}\left( { {\bf f}_{eq,}^{o}}_{j-1}-{ {\bf f}_{eq,}^{o}}_{j-2}\right)  \right\} \ea \right]
\ea
$$

$$
\ba{ll}
 \Delta {G(U)}^{-}_{j-\frac{1}{2}} &= \fr{1}{2}\left( {G(U)}_{j}-{G(U)}_{j-1}\right) - \fr{1}{2}|\lambda|\left( {U}_{j}-{U}_{j-1}\right)+ \fr{(\lambda-|\lambda_{o}|)}{2}\left( { {\bf f}_{eq,}^{o}}_{j}-{ {\bf f}_{eq,}^{o}}_{j-1}\right)\\[4mm]
& - \fr{1}{2} {\bf minmod} \left[ \ba{c} \left\{ \fr{1}{2}\left( {G(U)}_{j+1}-{G(U)}_{j}\right) - \fr{1}{2}|\lambda|\left( {U}_{j+1}-{U}_{j}\right)
+ \fr{(\lambda-|\lambda_{o}|)}{2}\left( { {\bf f}_{eq,}^{o}}_{j+1}-{ {\bf f}_{eq,}^{o}}_{j}\right) \right\}, \\[2mm]  
\left\{ \fr{1}{2}\left( {G(U)}_{j}-{G(U)}_{j-1}\right) - \fr{1}{2}|\lambda|\left( {U}_{j}-{U}_{j-1}\right)
+ \fr{(\lambda-|\lambda_{o}|)}{2}\left( { {\bf f}_{eq,}^{o}}_{j}-{ {\bf f}_{eq,}^{o}}_{j-1}\right)  \right\} \ea \right]\\[10mm]
& - \fr{1}{2} {\bf minmod} \left[ \ba{c} \left\{ \fr{1}{2}\left( {G(U)}_{j}-{G(U)}_{j-1}\right) - \fr{1}{2}|\lambda|\left( {U}_{j}-{U}_{j-1}\right)
+ \fr{(\lambda-|\lambda_{o}|)}{2}\left( { {\bf f}_{eq,}^{o}}_{j}-{ {\bf f}_{eq,}^{o}}_{j-1}\right) \right\}, \\[2mm]  
\left\{ \fr{1}{2}\left( {G(U)}_{j-1}-{G(U)}_{j-2}\right) - \fr{1}{2}|\lambda|\left( {U}_{j-1}-{U}_{j-2}\right)
+ \fr{(\lambda-|\lambda_{o}|)}{2}\left( { {\bf f}_{eq,}^{o}}_{j-1}-{ {\bf f}_{eq,}^{o}}_{j-2}\right)  \right\} \ea \right]
\ea
$$
The above set of expressions can be put in a compact form by absorbing the minmod terms in  (\ref{Right_Interface_Flux}) $ \& $ (\ref{Left_Interface_Flux}) into the respective flux difference split terms.  The resulting second order update formulation can be written as
\be
U_{j}^{n+1} = U_{j}^{n} - \fr{\Delta t}{\Delta x}\left[ G(U)_{j+\frac{1}{2}}^{n} - G(U)_{j-\frac{1}{2}}^{n}  \right]  
\ee
where in the interface flux are computed using
\bea
G(U)_{j+\frac{1}{2}} & =   \fr{1}{2}\left[G(U)_{j}+ G(U)_{j+1}\right] - \fr{1}{2}[{\bf \Delta G(U)}^{+}_{j+\frac{1}{2}}-{\bf \Delta G(U)}^{-}_{j+\frac{1}{2}}]\\[2mm]
G(U)_{j-\frac{1}{2}} & =   \fr{1}{2}\left[G(U)_{j-1}+ G(U)_{j}\right] - \fr{1}{2}[{\bf \Delta G(U)}^{+}_{j-\frac{1}{2}}-{\bf \Delta G(U)}^{-}_{j-\frac{1}{2}}]
\eea
Using the definition of $f^{o}_{eq}$, we can rewrite the split flux differences in the above equation gets redefined as 
 
\bea
\ba{ll}
{\bf \Delta {G(U)}}^{+}_{j+\frac{1}{2}} &= \fr{1}{2}\left( {G(U)}_{j+1}-{G(U)}_{j}\right) + \fr{1}{2}|\lambda|\left( {U}_{j+1}-{U}_{j}\right)- \fr{1}{2}\lambda_{A}\left( {U}_{j+1}-{U}_{j}\right)\\[4mm]
&+\fr{1}{2}{{\bf minmod}}\left[ (G(U)_{j+2}-G(U)_{j+1}),(G(U)_{j+1}-G(U)_{j}) \right] \\[4mm]
& - \fr{1}{2} {\bf minmod} \left[ \ba{c} \left\{ \fr{1}{2}\left( {G(U)}_{j+2}-{G(U)}_{j+1}\right) + \fr{1}{2}|\lambda|\left( {U}_{j+2}-{U}_{j+1}\right)
- \fr{1}{2}\lambda_{A}\left( {U}_{j+2}-{U}_{j+1}\right) \right\}, \\[2mm]  
\left\{ \fr{1}{2}\left( {G(U)}_{j+1}-{G(U)}_{j}\right) + \fr{1}{2}|\lambda|\left( {U}_{j+1}-{U}_{j}\right)
- \fr{1}{2}\lambda_{A}\left( {U}_{j+1}-{U}_{j}\right)  \right\} \ea \right]\\[10mm]
& - \fr{1}{2} {\bf minmod} \left[ \ba{c} \left\{ \fr{1}{2}\left( {G(U)}_{j+1}-{G(U)}_{j}\right) + \fr{1}{2}|\lambda|\left( {U}_{j+1}-{U}_{j}\right)
- \fr{1}{2}\lambda_{A}\left( {U}_{j+1}-{U}_{j}\right) \right\}, \\[2mm]  
\left\{ \fr{1}{2}\left( {G(U)}_{j}-{G(U)}_{j-1}\right) + \fr{1}{2}|\lambda|\left( {U}_{j}-{U}_{j-1}\right)
-\fr{1}{2}\lambda_{A}\left( {U}_{j}-{U}_{j-1}\right)  \right\} \ea \right]
\ea
\eea

\bea
\ba{ll}
 {\bf \Delta {G(U)}}^{-}_{j+\frac{1}{2}} &= \fr{1}{2}\left( {G(U)}_{j+1}-{G(U)}_{j}\right) - \fr{1}{2}|\lambda|\left( {U}_{j+1}-{U}_{j}\right)+ \fr{1}{2}\lambda_{A}\left( {U}_{j+1}-{U}_{j}\right)\\[4mm]
& +\fr{1}{2}{{\bf minmod}}\left[ (G(U)_{j+1}-G(U)_{j}),(G(U)_{j}-G(U)_{j-1})\right]\\[4mm]
& - \fr{1}{2} {\bf minmod} \left[ \ba{c} \left\{ \fr{1}{2}\left( {G(U)}_{j+2}-{G(U)}_{j+1}\right) - \fr{1}{2}|\lambda|\left( {U}_{j+2}-{U}_{j+1}\right)
+ \fr{1}{2}\lambda_{A}\left( {U}_{j+2}-{U}_{j+1}\right) \right\}, \\[2mm]  
\left\{ \fr{1}{2}\left( {G(U)}_{j+1}-{G(U)}_{j}\right) - \fr{1}{2}|\lambda|\left( {U}_{j+1}-{U}_{j}\right)
+ \fr{1}{2}\lambda_{A}\left( {U}_{j+1}-{U}_{j}\right)  \right\} \ea \right]\\[10mm]
& - \fr{1}{2} {\bf minmod} \left[ \ba{c} \left\{ \fr{1}{2}\left( {G(U)}_{j+1}-{G(U)}_{j}\right) - \fr{1}{2}|\lambda|\left( {U}_{j+1}-{U}_{j}\right)
+ \fr{1}{2}\lambda_{A}\left( {U}_{j+1}-{U}_{j}\right) \right\}, \\[2mm]  
\left\{ \fr{1}{2}\left( {G(U)}_{j}-{G(U)}_{j-1}\right) - \fr{1}{2}|\lambda|\left( {U}_{j}-{U}_{j-1}\right)
+\fr{1}{2}\lambda_{A}\left( {U}_{j}-{U}_{j-1}\right)  \right\} \ea \right]
\ea
\eea

\bea
\ba{ll}
 {\bf \Delta {G(U)}}^{+}_{j-\frac{1}{2}} &= \fr{1}{2}\left( {G(U)}_{j}-{G(U)}_{j-1}\right) + \fr{1}{2}|\lambda|\left( {U}_{j}-{U}_{j-1}\right)- \fr{1}{2}\lambda_{A}\left( {U}_{j}-{U}_{j-1}\right)\\[4mm]
&+ \fr{1}{2}{\bf minmod}\left[ (G(U)_{j+1}-G(U)_{j}),(G(U)_{j}-G(U)_{j-1}) \right] \\[4mm]
& - \fr{1}{2} {\bf minmod} \left[ \ba{c} \left\{ \fr{1}{2}\left( {G(U)}_{j+1}-{G(U)}_{j}\right) + \fr{1}{2}|\lambda|\left( {U}_{j+1}-{U}_{j}\right)
- \fr{1}{2}\lambda_{A}\left( {U}_{j+1}-{U}_{j}\right) \right\}, \\[2mm]  
\left\{ \fr{1}{2}\left( {G(U)}_{j}-{G(U)}_{j-1}\right) + \fr{1}{2}|\lambda|\left( {U}_{j}-{U}_{j-1}\right)
- \fr{1}{2}\lambda_{A}\left( {U}_{j}-{U}_{j-1}\right)  \right\} \ea \right]\\[10mm]
& - \fr{1}{2} {\bf minmod} \left[ \ba{c} \left\{ \fr{1}{2}\left( {G(U)}_{j}-{G(U)}_{j-1}\right) + \fr{1}{2}|\lambda|\left( {U}_{j}-{U}_{j-1}\right)
- \fr{1}{2}\lambda_{A}\left( {U}_{j}-{U}_{j-1}\right) \right\}, \\[2mm]  
\left\{ \fr{1}{2}\left( {G(U)}_{j-1}-{G(U)}_{j-2}\right) + \fr{1}{2}|\lambda|\left( {U}_{j-1}-{U}_{j-2}\right)
- \fr{1}{2}\lambda_{A}\left( {U}_{j-1}-{U}_{j-2}\right) \right\} \ea \right]
\ea
\eea

\bea
\ba{ll}
 {\bf \Delta {G(U)}}^{-}_{j-\frac{1}{2}} &= \fr{1}{2}\left( {G(U)}_{j}-{G(U)}_{j-1}\right) - \fr{1}{2}|\lambda|\left( {U}_{j}-{U}_{j-1}\right)+ \fr{1}{2 }\lambda_{A}\left( {U}_{j}-{U}_{j-1}\right)\\[4mm]
& +\fr{1}{2}{\bf minmod}\left[ (G(U)_{j}-G(U)_{j-1}),(G(U)_{j-1}-G(U)_{j-2})\right]\\[4mm]
& - \fr{1}{2} {\bf minmod} \left[ \ba{c} \left\{ \fr{1}{2}\left( {G(U)}_{j+1}-{G(U)}_{j}\right) - \fr{1}{2}|\lambda|\left( {U}_{j+1}-{U}_{j}\right)
+ \fr{1}{2}\lambda_{A}\left( {U}_{j+1}-{U}_{j}\right) \right\}, \\[2mm]  
\left\{ \fr{1}{2}\left( {G(U)}_{j}-{G(U)}_{j-1}\right) + \fr{1}{2}|\lambda|\left( {U}_{j}-{U}_{j-1}\right)
+ \fr{1}{2}\lambda_{A}\left( {U}_{j}-{U}_{j-1}\right)  \right\} \ea \right]\\[10mm]
& - \fr{1}{2} {\bf minmod} \left[ \ba{c} \left\{ \fr{1}{2}\left( {G(U)}_{j}-{G(U)}_{j-1}\right) - \fr{1}{2}|\lambda|\left( {U}_{j}-{U}_{j-1}\right)
+ \fr{1}{2}\lambda_{A}\left( {U}_{j}-{U}_{j-1}\right) \right\}, \\[2mm]  
\left\{ \fr{1}{2}\left( {G(U)}_{j-1}-{G(U)}_{j-2}\right) - \fr{1}{2}|\lambda|\left( {U}_{j-1}-{U}_{j-2}\right)
+ \fr{1}{2}\lambda_{A}\left( {U}_{j-1}-{U}_{j-2}\right) \right\} \ea \right]
\ea
\eea
 
\subsection{\bf KFDS scheme for viscous flows} 
Consider 1-D Navier-Stokes equations given by
\be \label{1D_NS} \ \ \ \ \ \ 
\fr{\del U}{\del t} + \fr{\del G(U)}{\del x} = \fr{\del G_{v}(U)}{\del x}  
\ee 
where $U$ is the conserved variable vector, $G(U)$ its nonlinear inviscid flux vector and $G_{v}(U)$ is the viscous flux vector, given by  
\bea \label{1D_NS_Vectors}
U = \left[ \ba{c} \rho \\ \rho u \\ \rho E \ea \right] , \ 
G(U) = \left[ \ba{c} \rho u \\ p + \rho u^{2} \\ p u + \rho u E \ea \right] \ \textrm{and} \
G_{v}(U) = \left[ \ba{c} 0 \\ \tau \\ u\tau - q \ea \right] 
\eea
Here, $\tau$ is the one dimensional component of the stress tensor and $q$ is the corresponding component of the heat flux vector.   $\mu$ is the coefficient of fluid viscosity and $k$ is the coefficient of  thermal conductivity. 
The kinetic theory framework for this system of  equations as moments, based on the Boltzmann equation with the BGK model,  can in a similar way as in (\ref{1D_BE}), be given by 
\be \label{1D_VBE}
\fr{\del f}{\del t} +  \fr{\del h}{\del x} = - \fr{1}{\epsilon} \left[ f - f_{CE} \right]  
\ee 
and the 1-D Navier-Stokes equations can be recovered by taking moments as 
\be 
\left\langle \Psi 
\left(
\fr{\del f}{\del t} +  \fr{\del h}{\del x} = 0, f = f_{CE} 
\right) \right\rangle
\ee
Here, the distribution function instantaneously relaxes to $f_{CE}$, the Chapman-Enskog distribution function, in the collision step.  The derivation of this equilibrium distribution and its moments to recover the conservation equations are well documented in  \cite{Deshpande_KNM,Manoj_SVRR_SMD,Chou_Baganoff}.  The Chapman-Enskog distribution function for 1D is given by 
\bea
\label{1D_Visc_Max}
f_{CE} = f^{eq}(1 + P_{CE})\\
\textrm{where} \ \ f^{eq} = \rho \fr{\sqrt{\beta}}{\sqrt{\pi}} e^{- \beta \left( v - u \right)^{2} } e^{-\frac{I}{I_{0}}} 
\eea
The perturbation term $P_{CE}$ is given by 
\be
P_{CE}=\fr{\tau_{CE}}{p} Z_{\tau}-\fr{q_{CE}}{p \sqrt{2RT}} Z_{q}
\ee
where 
\be
\tau_{CE}=(3-\gamma) \epsilon p \fr{\del u}{\del x} \ \ and \ \ q_{{CE}}=\fr{\gamma}{\gamma-1} \epsilon p \fr{\del}{\del x}\left(\fr{p}{\rho}\right)
\ee
The coefficients $Z_{\tau}$ and $Z_{q}$ are given by
\bea
Z_{\tau}(v, I)=\fr{3 \gamma-5}{2(3-\gamma)}+\fr{(v-u)^{2}}{2RT}-\fr{4(\gamma-1)^{2}}{(3-\gamma)^{2}} \fr{I}{2 R T} \\[2 mm]
Z_{q}(v, I)=\frac{(\gamma-1)}{\gamma RT}\left[\fr{(v-u)^{3}}{\sqrt{2RT}}-\frac{5}{2}(v-u)\sqrt{2RT}+\frac{4(\gamma-1)}{3-\gamma} \frac{I(v-u)}{\sqrt{2 R T}}\right]
\eea
Thus perturbation term is a function of $(v-u)$ and $I$ with the analogous fluid viscosity and thermal conductivity coefficients. 
The moments to obtain the macroscopic variables [\cite{Deshpande_KNM}] are defined by 
\be \label{1D_Visc_moments} 
U_{i} = \int_{0}^{\infty} \int_{-\infty}^{\infty} {\psi}_{i} f_{CE} \ dvdI \ \textrm{and} \ 
G_{T,i}(U) = G(U)_{i} - G_{v,i}(U) = \int_{0}^{\infty}\int_{-\infty}^{\infty} {\psi}_{i} v f_{CE} \ dvdI 
\ee  
with 
\be 
\psi = \left[ \ba{c} 1 \\ v \\ I + \fr{1}{2} v^{2} \ea \right] \ \textrm{and} \   i = 1, 2,3 \ \ ({\rm for\ 1-D \ case}) 
\ee 
Introducing a truncated  distribution as 
\be 
\tilde{f}_{CE} = \int_{0}^{\infty} f_{CE} \ dI 
\ee 
we can redefine the moment relations as 
\be \label{1D_moments_modified_2} 
U_{i} = \int_{-\infty}^{\infty}  {\psi}_{i} \tilde{f}_{CE} \ dv \  \textrm{and} \ 
G_{T,i}(U) = \int_{-\infty}^{\infty} {\psi}_{i} v \tilde{f}_{CE} \ dv   
\ee  
\subsubsection{Prandtl number fix} 
The KFDS schemes are based on BGK model for the collision terms. The limitation of this model is the Prandtl number being fixed as unity.  In order to overcome this limitation, a Prandtl number fix is employed to the heat flux term at the macroscopic level as suggested in [\cite{ZhangShu}]. The Prandtl number fix for the viscous flux term in 1D is given by
\be
{G^{'}}_{v}(U) =  G_{v}(U) +  \left[ \ba{c} 0 \\ 0 \\ (1-Pr)q \ea \right]
\ee
where Pr refers to the actual Prandtl number.  
\subsubsection{\bf Discrete velocity model for viscous flow}

 To arrive at a discrete velocity model for the above framework,  the continuous Chapman-Enskog distribution $ \tilde{f}_{CE}$ is approximated by a combination of Dirac delta functions as  
\be
{\psi}_{i} \tilde{f}_{CE} =\left\{ {f_{CE}}_{+} \delta (v - \lambda^{+}) +  {f_{CE}}_{o} \delta (v - \lambda_{o})  +{f_{CE}}_{-} \delta (v - \lambda^{-}) \right\}_{i} 
\ee
Let us further assume, for simplicity, that the discrete velocities, $\lambda^{+}$ and $\lambda^{-}$ for each $i$ are given by 
\be \label{New_DVs_1D}
\lambda^{+}_{i} = \lambda_{i} \ \textrm{and} \  \lambda^{-}_{i} = - \lambda_{i}  
\ee 
Let us assume that $\lambda^{o}$ and ${f_{CE}^{eq}}_{o}$ for each $i$ are known  (which will be fixed later, as done in the inviscid case).  Then, using the two moment relations in (\ref{1D_moments_modified_2})  we obtain the following.  
$$ \ba{rcl}  
U_{i} & = & \displaystyle \int_{-\infty}^{\infty} {\psi}_{i} \tilde{f}_{CE} dv \\ 
 & = & \displaystyle \int_{-\infty}^{\infty} \left\{ {f_{CE}}_{+} \delta (v - \lambda^{+}) + {f_{CE}}_{o}\delta (v - \lambda_{o})  +{f_{CE}}_{-} \delta (v - \lambda^{-}) \right\}_{i} dv \\ 
 & = &\left\{  {f_{CE}}_{+}  + {f_{CE}}_{o}   + {f_{CE}}_{-} \right\}_{i}  
\ea $$ 
or 
\be 
\left\{{f_{CE}}_{+} + {f_{CE}}_{-}\right\}_{i} = U_{i} -{ {f_{CE}}_{o}}_{i} 
\ee 
$$ \ba{rcl} 
G_{T,i}(U) & = & \displaystyle \int_{-\infty}^{\infty} v {\psi}_{i} \tilde{f}_{CE} dv \\ [3mm]
 & = & \displaystyle \int_{-\infty}^{\infty} v 
\left\{ {f_{CE}}_{+} \delta (v - \lambda^{+}) + {f_{CE}}_{o} \delta (v -\lambda_{o}) + {f_{CE}}_{-} \delta (v - \lambda^{-}) \right\}_{i} dv \\ [2mm]
 & = &\left\{ \ba{l} {f_{CE}}_{+} \displaystyle \int_{-\infty}^{\infty} \phi(v) \delta (v - \lambda^{+}) dv + {f_{CE}}_{o} \displaystyle \int_{-\infty}^{\infty} \phi(v) \delta (v - \lambda_{o}) dv \\ + 
        {f_{CE}}_{-} \displaystyle \int_{-\infty}^{\infty} \phi(v) \delta (v - \lambda^{-}) dv \ea \right\}_{i}, 
        \ (\phi(v) = v) \\ 
 & = &\left\{ {f_{CE}}_{+} \lambda^{+} +  {f_{CE}}_{o} \lambda_{o}  + {f_{CE}}_{-} \lambda^{-}\right\}_{i}
\ea $$ 
or 
\be 
\left\{{f_{CE}}_{+} \lambda^{+}  + {f_{CE}}_{-} \lambda^{-}\right\}_{i}  = G_{T,i}(U) - {\left\{ {f_{CE}}_{o}\lambda_{o}\right\}}_{i}
\ee 
Solving the above two equations  and simplifying, we get
\be \label{3_Vel_equilibria_Visc}
{{f_{CE}}_{+}}_{i} = \fr{1}{2} U_{i} + \fr{1}{2 {\lambda}_{i}} G_{T,i}(U) - \left\{\fr{\lambda + \lambda_{o}}{2 \lambda} {f_{CE}}_{o}\right\}_{i} \ \textrm{and} \ {{f_{CE}}_{-}}_{i} = \fr{1}{2} U_{i} - \fr{1}{2 {\lambda}_{i}} G_{T,i}(U) -   \left\{\fr{\lambda - \lambda_{o}}{2 \lambda}{f_{CE}}_{o} \right\}_{i}
\ee 
The {\em Discrete Velocity Boltzmann Equation} (DVBE) for three velocity model thus derived can be written as    
\be \label{DVBE_1D_3_Vel_Visc} 
\left\{\fr{\del \bf f}{\del t} +  \fr{\del \bf h}{\del x} = - \fr{1}{\epsilon} \left[ \bf f - \bf f_{CE} \right] \right\}_{i}  \ \ \textrm{with} \ \ i=1,2,3
\ee
where 
\bea \label{3V_DVBEq_Visc}
\bf {f_{CE}}_{i} = \left[ \ba{c} {f_{CE}}_{+} \\ {f_{CE}}_{o} \\ {f_{CE}}_{-} \ea \right]_{i}, \ 
{\Lambda}_{i} = \left[ \ba{cc}  \lambda^{+} \  \ 0 \  \  \  0 \\ \ 0  \ \ \lambda_{o} \ \ \  0 \\  \ \  0\  \  \ 0  \ \ \   \lambda^{-} \ea \right]_{i} \ \textrm{and} \\ [4 mm]
\bf{ f_{CE}}_{i} = \left[ \ba{c}  {f_{CE}}_{+} \\ [2 mm] {f_{CE}}_{o} \\[2 mm] {f_{CE}^{eq}}_{-} \ea \right]_{i} 
= \left[ \ba{c} \fr{1}{2} U + \fr{1}{2 \lambda} \left(G(U)-G_{v}(U)\right)  - \fr{\lambda + \lambda_{o}}{2 \lambda}  {f_{CE}}_{o}\\[2 mm]
 {f_{CE}}_{o}\\[2 mm]
\fr{1}{2} U - \fr{1}{2 \lambda} \left(G(U)-G_{v}(U)\right)  - \fr{\lambda - \lambda_{o}}{2 \lambda}  {f_{CE}}_{o} \ea \right]_{i} 
\eea
It is interesting to note that the discrete velocity model for the inviscid flow is a subset of the model for viscous flow, can be obtained in the limit of viscous fluxes being set to zero and therefore the model is consistent.   
 \subsubsection{\bf KFDS scheme for viscous flows}
We follow the same procedures as in section \ref{3KFDS_Section} to arrive at the finite volume formulation for viscous flows along with the Prandtl number fix. We get the update formula as
\be
\left(U_{j}^{n+1} = U_{j}^{n} - \fr{\Delta t}{\Delta x}\left[ G(U)_{j+\frac{1}{2}}^{n} - G(U)_{j-\frac{1}{2}}^{n}  \right] + \fr{\Delta t}{\Delta x}\left[{G}_{v}(U)_{j+\frac{1}{2}}^{n} - {G}_{v}(U)_{j-\frac{1}{2}}^{n}  \right] \right)_{i}  
\ee where $i = \ 1,2,3$ for 1D system of equations.  The interface fluxes are given by
\bea
\left( G(U)_{j+\frac{1}{2}} =\frac{1}{2}\left[G(U)_{j+1}+G(U)_{j}\right]-\frac{1}{2}\left[\Delta G(U)_{j+\frac{1}{2}}^{+}-\Delta G(U)_{j+\frac{1}{2}}^{-}\right] \right)_{i} \\
\left(G(U)_{j-\frac{1}{2}} =\frac{1}{2}\left[G(U)_{j}+G(U)_{j-1}\right]-\frac{1}{2}\left[\Delta G(U)_{j-\frac{1}{2}}^{+}-\Delta G(U)_{j-\frac{1}{2}}^{-}\right] \right)_{i}
\eea
\\ and
\bea
\left( \Delta G(U)_{j+\frac{1}{2}}^{\pm} =\frac{1}{2}\left[G(U)_{j+1}-G(U)_{j}\right] \pm \frac{1}{2}|\lambda|\left[U_{j+1}-U_{j}\right] \mp \frac{1}{2} \lambda_{A}\left[U_{j+1}-U_{j}\right] \right)_{i}  \\ [2 mm]
\left( \Delta G(U)_{j-\frac{1}{2}}^{\pm} =\frac{1}{2}\left[G(U)_{j}-G(U)_{j-1}\right] \pm \frac{1}{2}|\lambda|\left[U_{j}-U_{j-1}\right] \mp \frac{1}{2} \lambda_{A}\left[U_{j}-U_{j-1}\right] \right)_{i}
\eea
The viscous interface fluxes are computed as 
\be
\left( {G}_{v}(U)_{j+\frac{1}{2}}=\frac{1}{2}\left({G}_{v}(U)_{j}+{G}_{v}(U)_{j+1}\right) \right)_{i} \  \& \ \left( {G}_{v}(U)_{j-\frac{1}{2}}=\frac{1}{2}\left({G}_{v}(U)_{j-1}+{G}_{v}(U)_{j}\right) \right)_{i}
\ee  
The values of $\lambda$ and $\lambda_{A}$ for each $i$ in the above update formulation can be fixed using the methodologies described earlier for the inviscid case. 
 
\section{KFDS scheme in 2-D}
The 2-D KFDS scheme is derived from a 2-D version of discrete velocity Boltzmann equation, which is based on the isotropic relaxation system introduced by Raghurama Rao and utilized in \cite{Jayaraj,Arun_SVRRao,Arun_SVRRao_KRS,Arun_Maria}.  
Consider a two dimensional hyperbolic system of conservation laws as given by
\be
\frac{\partial U}{\partial t}+\frac{\partial G_{1}(U)}{\partial x}+\frac{\partial G_{2}(U)}{\partial y}=0
\ee
A five velocity DVBE equation for the above system can be derived as 

\be  
\frac{\partial \mathbf{f}}{\partial t}+ \frac{\partial \mathbf{{h}_{1}}}{\partial x}+ \frac{\partial \mathbf{\bf {h}_{2}}}{\partial y}= - \fr{1}{\epsilon}\left[ \mathbf{f} -\mathbf{f}^{eq}\right] 
\ee
where $\bf {h}_{1} = {\Lambda}_{1}f , \bf {h}_{2} = {\Lambda}_{2}f $ and the discrete velocity matrices are given by 
\be
{\Lambda}_{1}=\left[\begin{array}{ccccc}{-\lambda} & {0} & {0} & {0} & {0} \\ {0} & {\lambda} & {0} & {0} & {0} \\ {0} & {0}  & {\lambda}_{o} & {0} & {0} \\ {0} & {0} & {0} & {\lambda} & {0} \\ {0} & {0} & {0} & {0} & {-\lambda}\end{array}\right]   \Lambda_{2}=\left[\begin{array}{ccccc}{-\lambda} & {0} & {0} & {0} & {0} \\ {0} & {-\lambda} & {0} & {0} & {0} \\ {0} & {0}  & {\lambda}_{o} & {0} & {0} \\ {0} & {0} & {0} & {\lambda} & {0} \\ {0} & {0} & {0} & {0} & {\lambda}\end{array}\right]
\ee
and the equilibria are given by 
\be
{\bf f}^{eq}=\left[\begin{array}{l}{f^{eq}_{1}} \\ {f^{eq}_{2}} \\ {f^{eq}_{3}} \\ {f^{eq}_{4}} \\ {f^{eq}_{5}}\end{array}\right]=\left[\begin{array}{c}{\frac{1}{4} U-\frac{1}{4 \lambda}G_{1}-\frac{1}{4 \lambda} G_{2}} - \fr{\lambda - 2\lambda_{o}}{4\lambda}f_{o}^{eq} \\ {\frac{1}{4}U+\frac{1}{4 \lambda} G_{1}-\frac{1}{4 \lambda} G_{2}} -\fr{1}{4}f_{o}^{eq} \\ f_{o}^{eq} \\ {\frac{1}{4} U+\frac{1}{4 \lambda} G_{1}+\frac{1}{4 \lambda} G_{2}} - \fr{\lambda + 2\lambda_{o}}{4\lambda}f_{o}^{eq} \\ {\frac{1}{4} U-\frac{1}{4 \lambda} G_{1}+\frac{1}{4} \lambda} G_{2} -\fr{1}{4}f_{o}^{eq}\end{array}\right]
\ee
Let us denote $\mathbf{h}_{x}=\mathbf{\Lambda}_{1} \mathbf{f}, \mathbf{h}_{y}=\mathbf{\Lambda}_{2} \mathbf{f}$ as the fluxes along the $x-$ and $y-$ directions.   
For a quadrilateral finite volume, the cell-interface fluxes are constructed be normal to the interfaces and can be obtained as 
\bea
\mathbf{h}_{ L}=\left(\mathbf{h}_{x} \cos \theta+\mathbf{h}_{y} \sin \theta\right)_{L}\\
\mathbf{h}_{ R}=\left(\mathbf{h}_{x} \cos \theta+\mathbf{h}_{y} \sin \theta\right)_{R}
\eea
where the suffix L and R represent the left and the right states of the cell-interface.  
A finite volume update formula for the 2D Euler system, for a quadrilateral mesh, can be derived as 
\be 
\mathbf{f}_{j,k}^{n+1}=\mathbf{f}_{j,k}^{n}-\frac{\Delta t}{A_{j,k}} \sum_{I_{c}=1}^{4} \mathbf{h}^{n}_{n I_{c}} \Delta s_{I_{c}}
\ee
where  $ A_{j,k}$ is the area of the cell centered at $(j,k)$ and $\Delta s_{I_{c}}$ is the length of the cell interface $I_{c}.$  
Applying the {\em Kinetic Flux Difference Splitting} across a finite volume cell-interface, we get  
\be
\mathbf{h}_{n, I_{c}}=\frac{1}{2}\left[h_{ R}+h_{ L}\right]-\frac{1}{2}\left[\Delta h_{I,n}^{+}-\Delta h_{I,n}^{-}\right]
\ee 
where 
\be
\Delta h_{I,n}^{\pm} = \left(\Lambda_{1 } \cos \theta\right)^{\pm}\Delta \mathbf{f}_{I,n}^{e q}+\left(\Lambda_{2 } \sin \theta\right)^{\pm}\Delta \mathbf{f}_{In}^{e q}
\ee 
Upon taking moments of the above equations and extending the same techniques that we had applied in one dimensional system in determining the  diffusion coefficients and $f_{o}^{eq}$, we recover the macroscopic update formula as 
\be 
\mathbf{U}_{j,k}^{ n+1}=\mathbf{U}_{j,k}^{ n}-\frac{\Delta t}{A_{j,k}} \sum_{I_{c}=1}^{4} \mathbf{G}^{n}_{n, I_{c}} \Delta s_{I_{c}}
\ee
where 
\bea
\mathbf{G}_{n, I_{c}} = \frac{1}{2}\left[G(U)_{n, R}+G(U)_{n, L}\right]-\frac{1}{2}\left[\Delta G(U)_{I_{c}}^{+}-\Delta G(U)_{I_{c}}^{-}\right]\\
\Delta G(U)_{I_{c}}^{\pm} = \left.\frac{1}{2}\left[G(U)_{n, R}-G(U)_{n, L}\right]\right] \pm \frac{1}{2}|\lambda|\left[U_{n, R}-U_{n, L}\right] \mp \frac{1}{2} \lambda_{A}\left[U_{n R}-U_{n L}\right].
\eea

Here again, the primary diffusion coefficient $|\lambda|$ is fixed based on the principle of flux equivalence across a steady discontinuity.  It is interesting to note that unlike in the one dimensional cases, there is inherent multidimensional diffusion in the 2D test version of the numerical scheme owing to the standard finite volume formulation which lacks truly multidimensional modeling.  Therefore the strategy for adding additional definition does not require an entropy scale to allow for additional diffusion. However, the numerical errors in calculating the change in thermodynamic entropy still need to be considered. Therefore the strategy for fixing additional diffusion is defined as follows. 

\bea \label{Add_Diffusion_2D}
\lambda_{A} = \left\{ \ba{l}  \lambda_{DD} \ \textrm{for } \  D^{2} > 0 \ \textrm{  } \  \& \ \textrm{  } \ \Delta S \le \epsilon  \\ 
                                              0 \ \ \textrm{Otherwise }  \ea \right. 
\eea
where $\epsilon$ is a small number and is taken as $1.0 \times 10^{-10}$.    As in 1-D, two variations are introduced for evaluating $\lambda_{DD}$ in the 2-D case, for the finite volume method.      

\subsection{Maximum eigenvalue based $\lambda_{DD}$}
 The definition of $\lambda_{DD}$ in 2D at the cell-interface is  
 
\be
\lambda_{DD} = \lambda_{max} - \lambda 
\ee 
where 
\be  
\lambda_{\max }=\max \left\{\max (|{u}_{n}+a|,|{u}_{n}|,|{u}_{n}-a|)_{L}, \max (|{u}_{n}+a|,|{u}_{n}|,|{u}_{n}-a|)_{R}\right\} 
\ee 
with 
\be  
{\left( {u}_{ n}={u}_{1} \cos \theta+{u}_{2} \sin \theta\right)}_{L,R}
\ee 
This makes the coefficient of the total numerical diffusion as $\lambda_{max}$ at the expansions. 

\subsection{Fluid velocity based  $\lambda_{DD}$}
			
\bea
\lambda_{DD} = \lambda_{RI} - \lambda \\ 
{\rm where} \ \lambda_{RI }=\frac{|u_{n,L}| + |u_{n,R}|}{2}
\eea 


\section{Results and Discussion}
The Kinetic Flux Difference Splitting (KFDS) method is tested systematically on various 1-D and 2-D benchmark test cases to evaluate and establish its capabilities.  Presented in this section are the one dimensional test cases [\cite{Torobook}] which are meant to evaluate the robustness of the scheme, involving nonlinear waves and their interactions.  Following them, the numerical scheme is tested for typical benchmark cases in two dimensional Eulerian flow fields, which present various shock waves, contact discontinuities, expansion waves and their interactions.  Finally, the scheme is tested for viscous fluid flows involving boundary layers and their interactions with nonlinear waves. Before presenting the results, the experimental order of convergence (EOC) studies are presented in the following subsection, demonstrating the order of accuracy of KFDS scheme.  
	
\subsection{Experimental Order of Convergence (EOC)} 
Here, a typical test is presented for evaluating the {\em Experimental Order of Convergence} (EOC) for the KFDS scheme.  Consider a one dimensional computational domain of size [0,2]. The flow variables in the domain are initialized with a sinusoidal variation in the density while keeping the pressure and velocity constant. The exact solution for such initial conditions is given below. 
\bea
\rho(x, t)=1.0+0.2 \sin (\pi(x-u t))\\
u(x, t)=0.1, \ p(x, t)=0.5
\eea

The numerical simulations are carried out by changing the number of computational cells methodically as 10, 20, 40, $\cdots$, 160 cells. The solutions are computed with each grid size for the final time of $0.5 s$. The exact solution is used for both initializing the computational domain and for enforcing the periodic boundary condition. The L1 and L2 errors represented by $\left\|\mathcal{E}_{K}\right\|_{L^{1}} and \left\|\mathcal{E}_{K}\right\|_{L^{2}}$  are calculated at $t^{n}=0.5 s$ using equation (\ref{L1Form}) and (\ref{L2Form}) as given below.

\be
\label{L1Form}
\left\|\mathcal{E}_{K}\left(t^{n}\right)\right\|_{L^{1}}=\Delta x \sum_{j=1}^{K}\left|\rho_{j}^{numerical}-\rho_{j}^{analytical}\right| 
\ee
\be
\label{L2Form}
\left\|\mathcal{E}_{K}\left(t^{n}\right)\right\|_{L^{2}}=\sqrt{\Delta x \sum_{j=1}^{K}\left(\rho_{j}^{numerical}-\rho_{j}^{analytical}\right)^{2}} \\
\ee
where $K$ is the number of cells.  As we expect the error to behave as $\left\|\mathcal{E}\right\| = C \Delta x^{p} + \mathcal{O}\left(\Delta x^{p+1}\right)$ for $p^{th}$ order accuracy, we can write 
\be 
\fr{\rho_{\Delta x} - \rho_{exact}}{\rho_{\frac{\Delta x}{2}} - \rho_{exact}} 
= \fr{C \Delta x^{p} + \mathcal{O}\left(\Delta x^{p+1}\right)}{C \left(\frac{\Delta x}{2}\right)^{p} + \mathcal{O}\left(\frac{\Delta x}{2}\right)^{p+1}}  
= 2^{p} + \mathcal{O}\left(\Delta x\right)
\ee  
Thus, 
\be 
{\rm log}_{2} \left(  \fr{\rho_{\Delta x} - \rho_{exact}}{\rho_{\frac{\Delta x}{2}} - \rho_{exact}} \right) = p + \mathcal{O}\left( \Delta x \right)
\ee
As $K \propto \fr{1}{\Delta x} $, the {\em Experimental Order of Convergence} (EOC) is computed using (\ref{EOC}) as follows.  
\be
\mathrm{EOC}=\log _{2}\left(\frac{\left\|\mathcal{E}_{K / 2}\right\|}{\left\|\mathcal{E}_{K}\right\|}\right)
\label{EOC}
\ee

\begin{figure}[h!]  
\begin{center} 
\includegraphics[width=12.5cm]{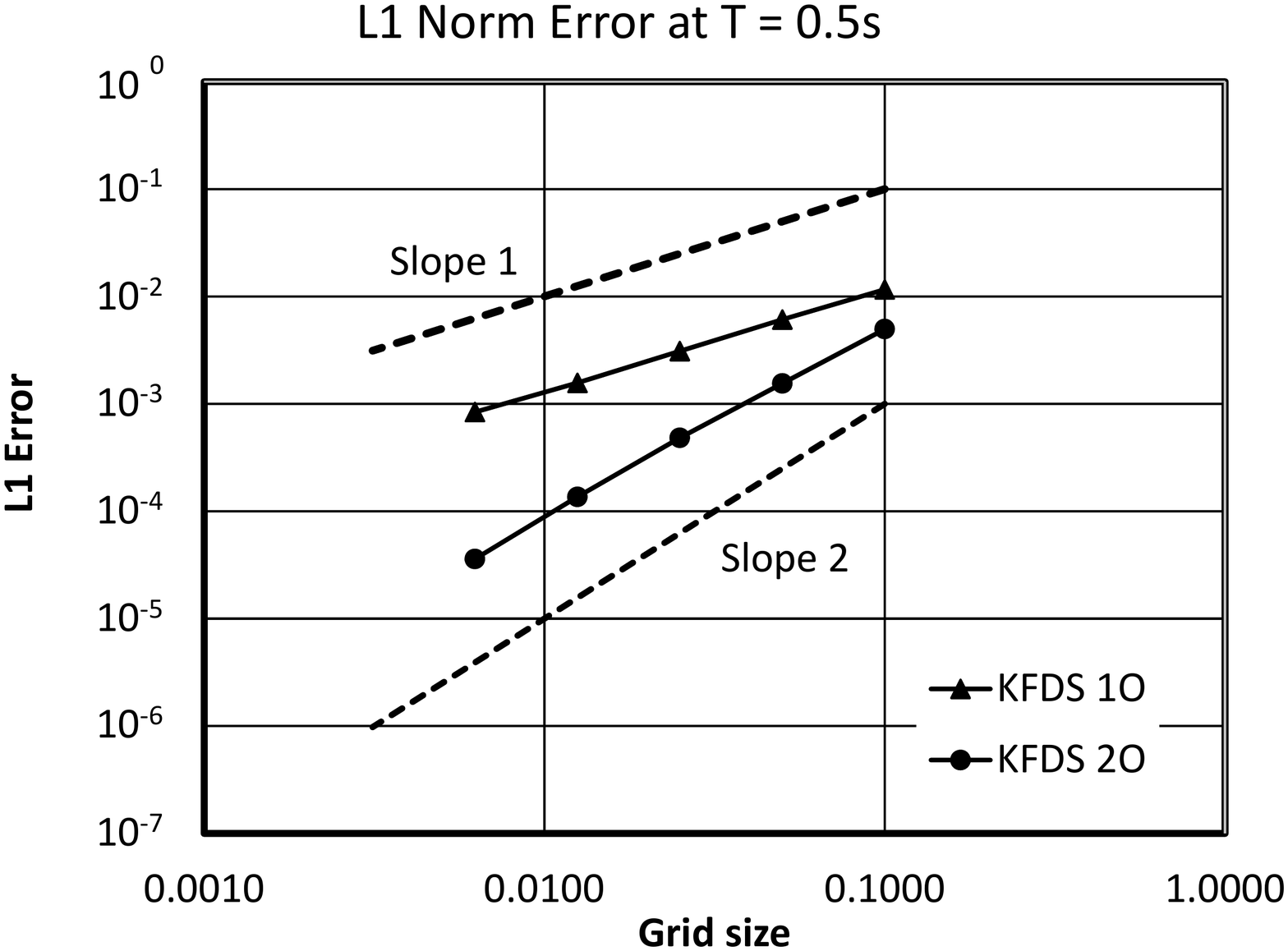}
\end{center} 
\caption{L1 norm errors for KFDS-1O and KFDS-2O schemes } 
\label{L1_Norm}  
\end{figure}

\begin{figure} 
\begin{center} 
\includegraphics[width=12.5cm]{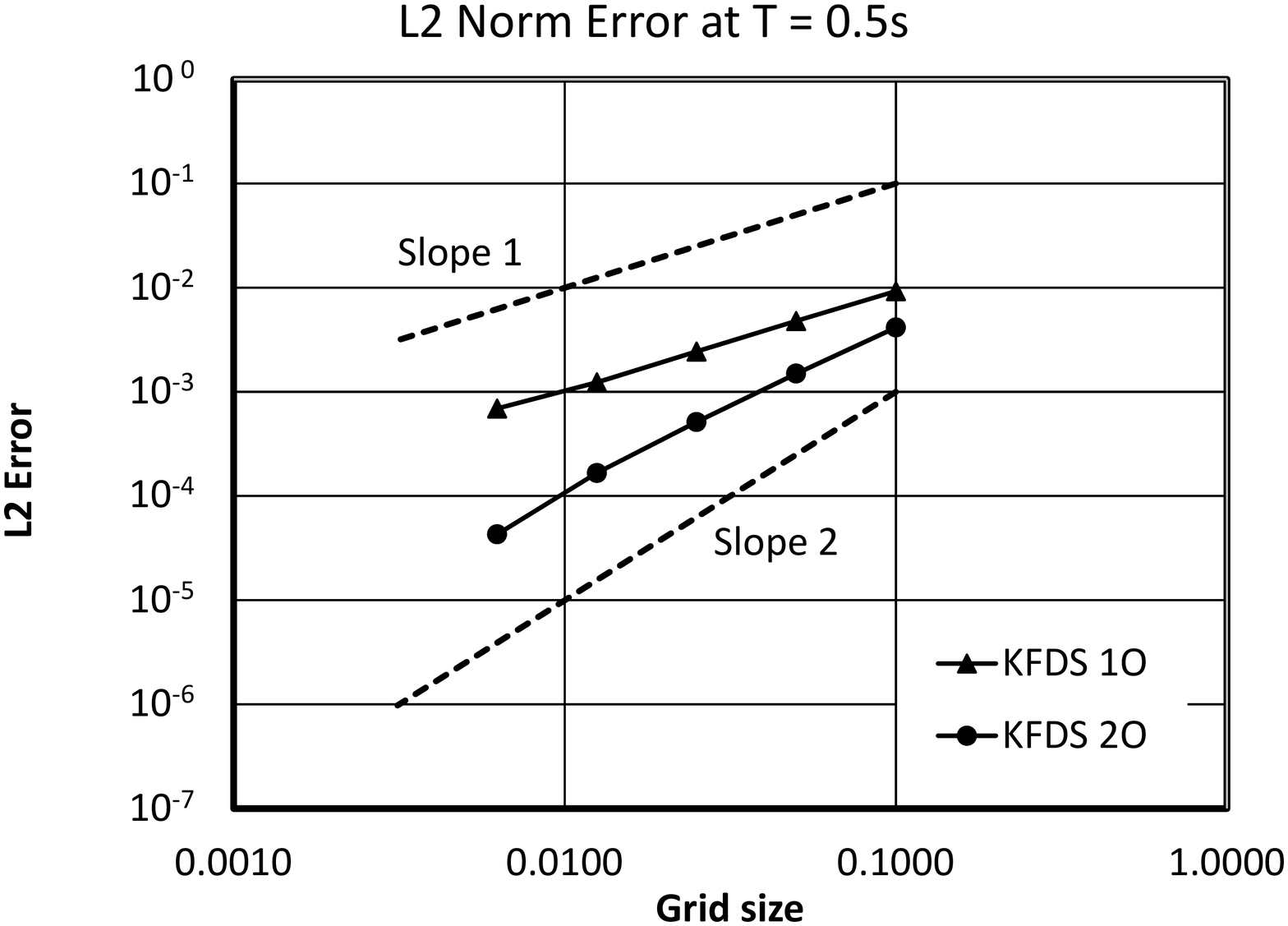}
\end{center} 
\caption{L2 norm errors for KFDS-1O and KFDS-2O schemes} 
\label{L2_Norm}  
\end{figure}

 The L1 and L2 norms obtained for each of the test cases of both first order accurate and second order accurate versions of KFDS scheme and are tabulated in tables 2 and 3 respectively. The log-log plots comparing the L1 and L2 norm errors with  slope 2 are shown in  figures[\ref{L1_Norm}] and [\ref{L2_Norm}] respectively.

\begin{table}[h!] 
\begin{center}
\begin{tabular}{|c|c|c|c|c|c|c|}
\hline {$\mathrm{N}$} & {Grid Spacing} & { L1 Error } & { EOC } & { L2 Error } &{ EOC } \\
\hline \hline 10 & 0.10000 & 0.011613777927 & & 0.009267137803 & \\
\hline 20 &0.05000& 0.006088680399 & 0.93164 & 0.004795147573 & 0.95055 \\
\hline 40 &0.02500 & 0.003097630000 & 0.97496 & 0.002436038922 & 0.97704 \\
\hline 80 &0.01250& 0.001566789511 & 0.98335 & 0.001234024541 & 0.98117 \\
\hline 160 &0.00625& 0.000838144804 & 0.90254 & 0.000692443990 & 0.83360 \\
\hline 
\end{tabular} 
\caption{EOC using L1 and L2 error norms for KFDS-1O with a smooth periodic test case.} 
\end{center}
\end{table} 

\begin{table}[h!] 
\begin{center}
\begin{tabular}{|l|l|l|l|l|l|}
\hline {$\mathrm{N}$}  & {Grid Spacing}&  { L1 Error } & { EOC } & { L2 Error } &{ EOC } \\
\hline \hline10 &0.10000& 0.004957979218 & & 0.004141851951 & \\
\hline 20 & 0.05000& 0.001549110966 & 1.67831 & 0.001495025225 & 1.47011 \\
\hline 40 &0.02500& 0.000482411631 & 1.68310 & 0.000513302408 & 1.54229 \\
\hline 80 &0.01250& 0.000136000286 & 1.82666 & 0.000166466154 & 1.62458 \\
\hline 160 &0.00625& 0.000035931520 & 1.92029 & 0.000042902870 & 1.95608 \\
\hline
\end{tabular}
\end{center}
\caption{EOC using L1 and L2 error norms for KFDS-2O with a smooth periodic test case.} 
\end{table} 

\clearpage

\subsection{Test cases for 1-D Euler equations}
    The initial conditions for the one dimensional Euler test cases are summarized in Table [\ref{1DEulerTestCases}]. 
		 
\begin{figure} 
\begin{center} 
\includegraphics[width=14.5cm]{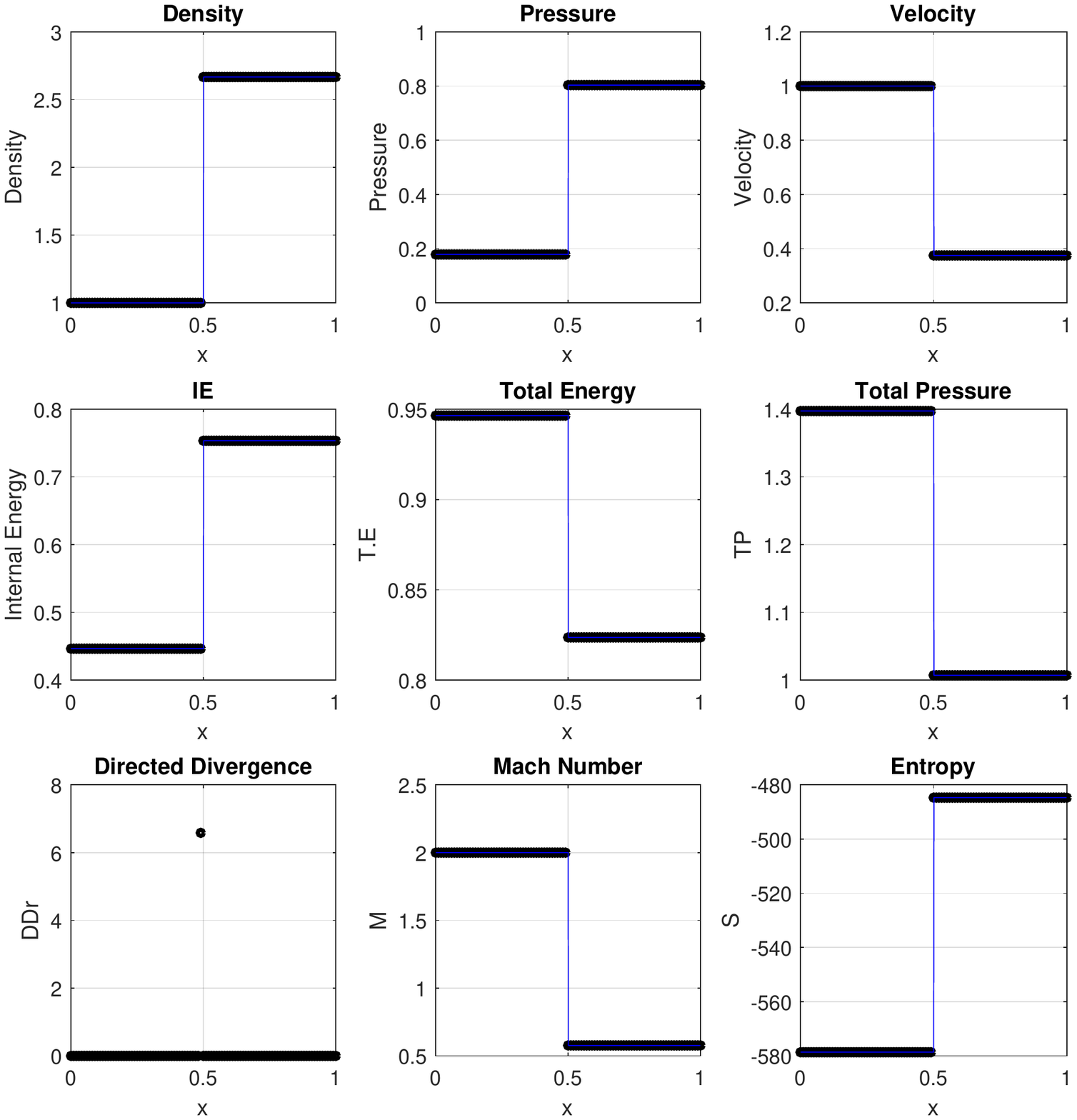}
\includegraphics[width=14.5cm]{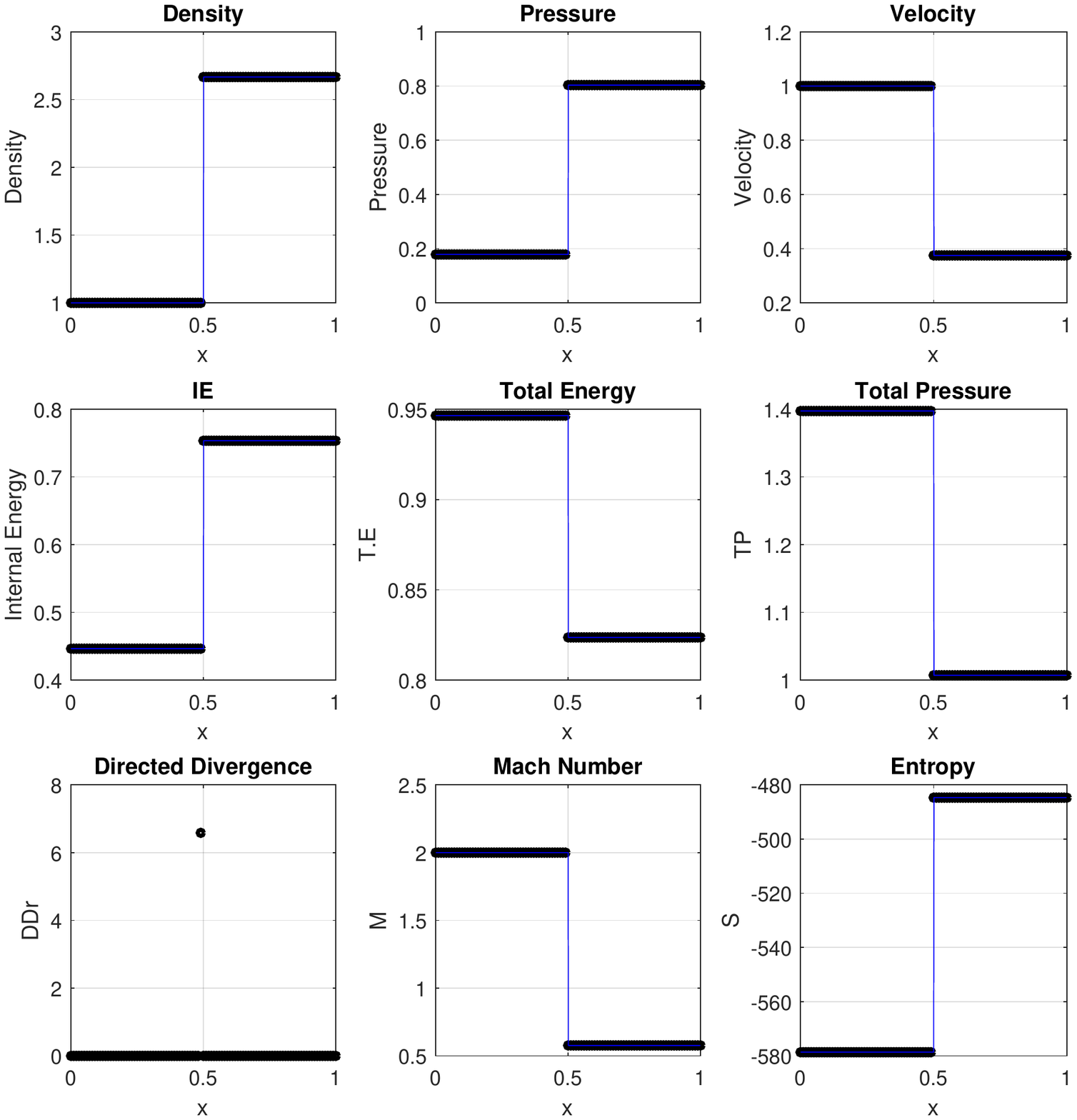}
\end{center} 
\caption{Test case 1:  (a) KFDS-A and  (b) KFDS-B} 
\label{1DEulerTC_1_LLF_RIC}  
\end{figure}

\begin{figure} 
\begin{center} 
\includegraphics[width=14.5cm,angle=0]{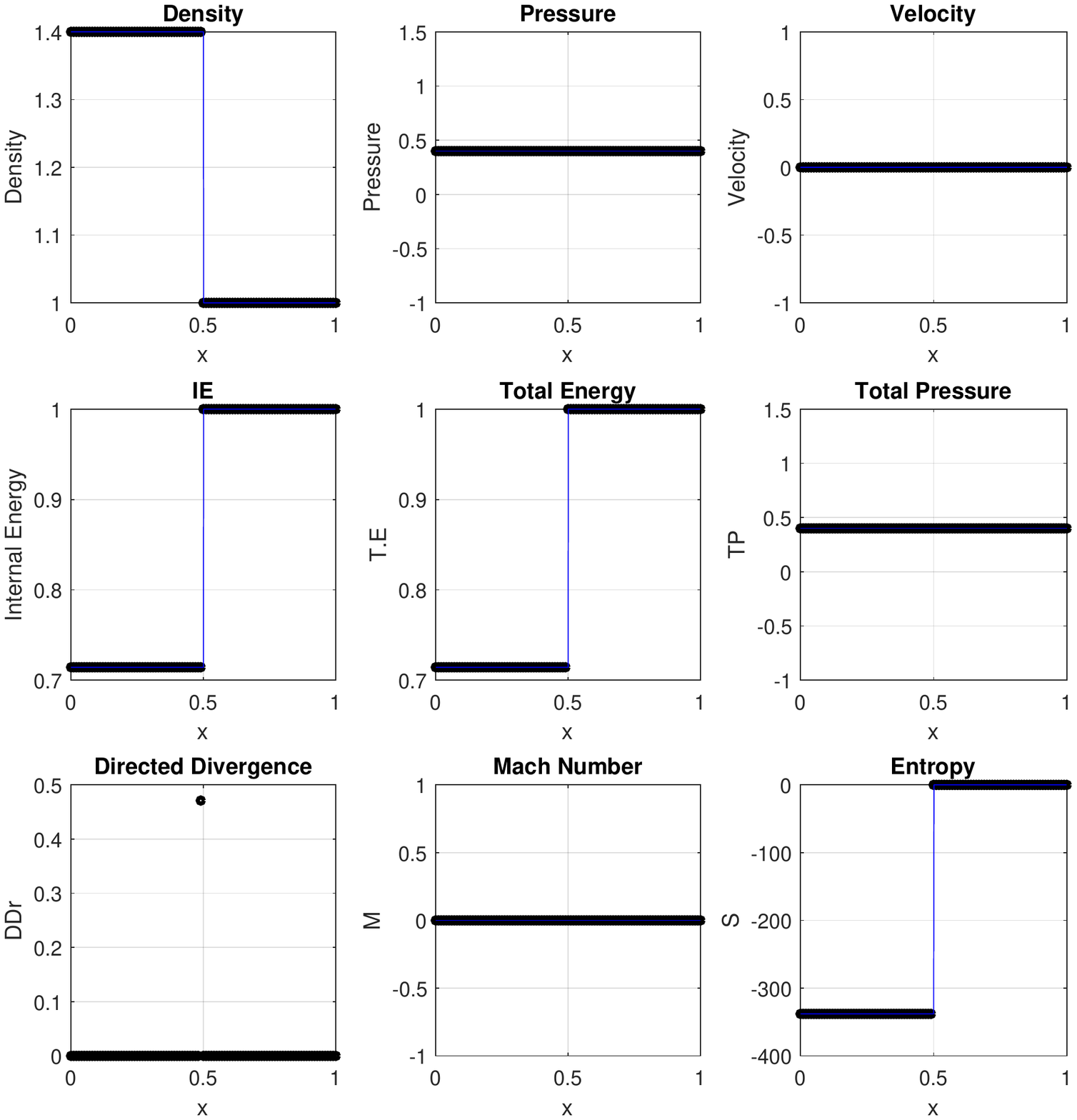}
\includegraphics[width=14.5cm,angle=0]{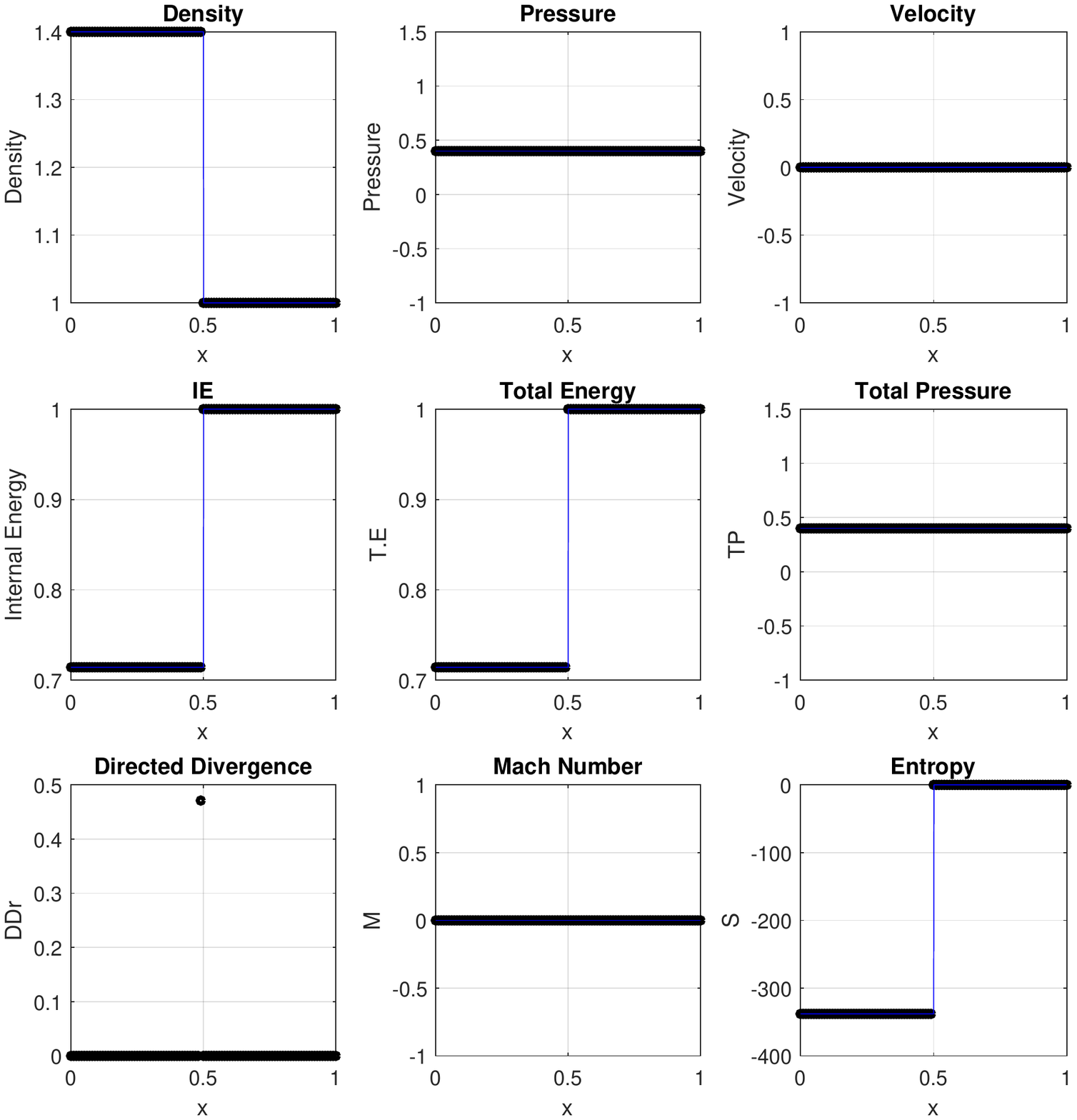}
\end{center} 
\caption{Test case 2:  (a) KFDS-A and  (b) KFDS-B} 
\label{1DEulerTC_2_LLF_RIC}  
\end{figure}

\begin{table}[!hbt]
\centering
\caption{1D Euler Equations: Test Case Matrix}
\label{1DEulerTestCases}
\begin{tabular}{|l|c|c|c|c|c|c|c|}
\hline S No.&Test Case & $\rho_{L}$ & $u_{L}$ & $p_{L}$ & $\rho_{R}$ & $u_{R}$ & $p_{R}$\\
\hline 1 & Steady shock &1.0 & 1.0 & 0.17875 & 2.6665 & 0.375 & 0.80357\\
\hline 2 & Steady contact discontinuity &1.4 & 0 & 0.4 & 1.0 & 0 & 0.4\\
\hline 3 & Shock tube  &1.0 & 0.75 & 1.0 &0.125 & 0 & 0.1\\
\hline 4 & Overheating  &1.0 & -2.0 & 0.4 & 1.0 & 2.0 & 0.4\\
\hline 5 & Toro Case 3  &1.0 & 0 & 1000 & 1.0 & 0 & 0.01\\
\hline 6 & Shock collision &5.99924 & 19.5975 & 460.894 & 5.99242 & -6.19633 & 46.0950\\
\hline 7 & Toro Case 5  &1.0 &  -19.5975 & 1000 & 1.0 & -19.59745 & 0.01\\
\hline
\end{tabular}
\end{table}
		The first two test cases involve  the steady shock and steady contact discontinuity respectively. The test results for the two test cases are shown in Figures [\ref{1DEulerTC_1_LLF_RIC}] and [\ref{1DEulerTC_2_LLF_RIC}]. The  KFDS scheme has the ability to capture steady shock waves and steady contact discontinuities exactly, without numerical diffusion. The introduction of additional numerical diffusion for smooth regions is regulated by the $D^{2}$-distance and entropy difference.  Thus, the exact discontinuity capturing ability of the scheme is undisturbed, as observed in the results.  

\begin{figure} 
\begin{center} 
\includegraphics[width=14.5cm,angle=0]{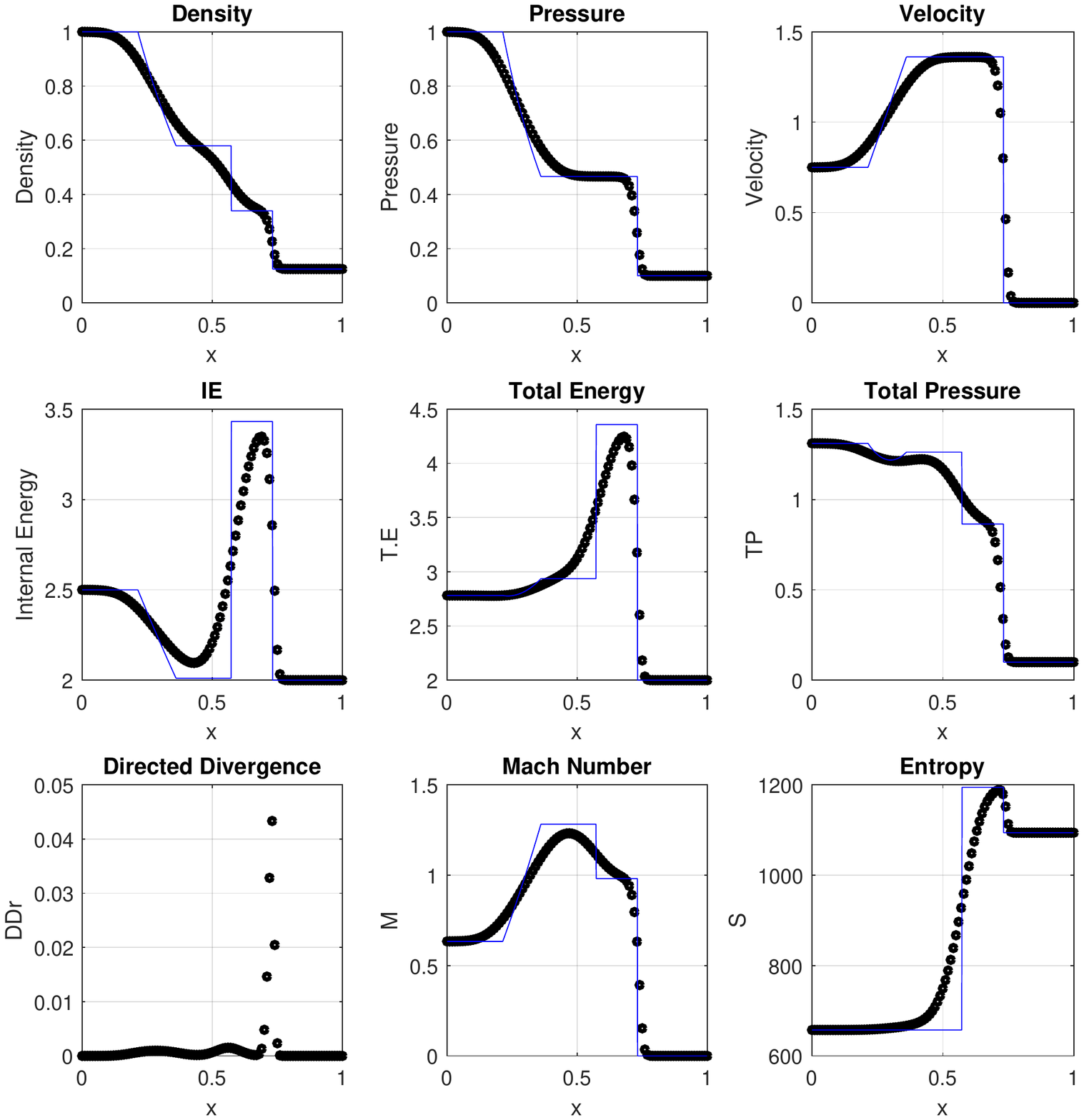}
\includegraphics[width=14.5cm,angle=0]{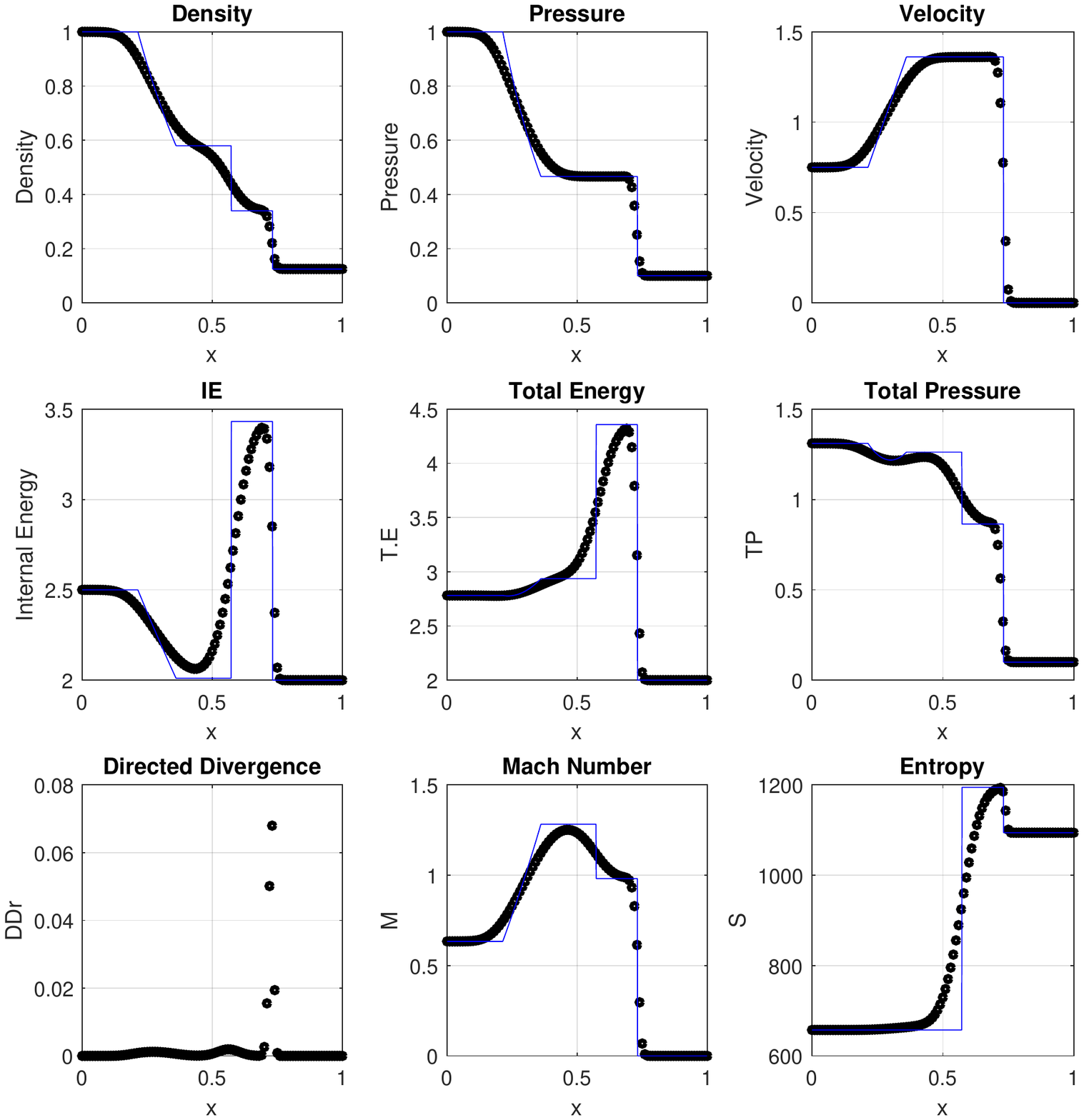}
\end{center} 
\caption{Test case 3: (a) KFDS-A and  (b) KFDS-B} 
\label{1DEulerTC_3_LLF}  
\end{figure}

The third test case is a classic shock tube test case wherein the gases are maintained at different thermodynamic states on either side of the diaphragm. Upon rupturing the diaphragm, the abrupt difference in the pressure, density and temperature results in the formation and evolution of an expansion wave, a contact discontinuity and a shock wave in the flow field.  While the flow is unsteady in nature, this test case assesses the accuracy of the scheme as well as its ability to capture each of the nonlinear and linear waves in the flow field.  Ideally, most low diffusive schemes would require an entropy fix to overcome the problems of unphysical expansion shocks that arise in the expansion regions.  The use of $D^{2}$-distance based additional diffusion has ensured an alternate way of overcoming this problem as is evident in the results [Fig.\ref{1DEulerTC_3_LLF}]. 

\begin{figure} 
\begin{center} 
\includegraphics[width=14.5cm,angle=0]{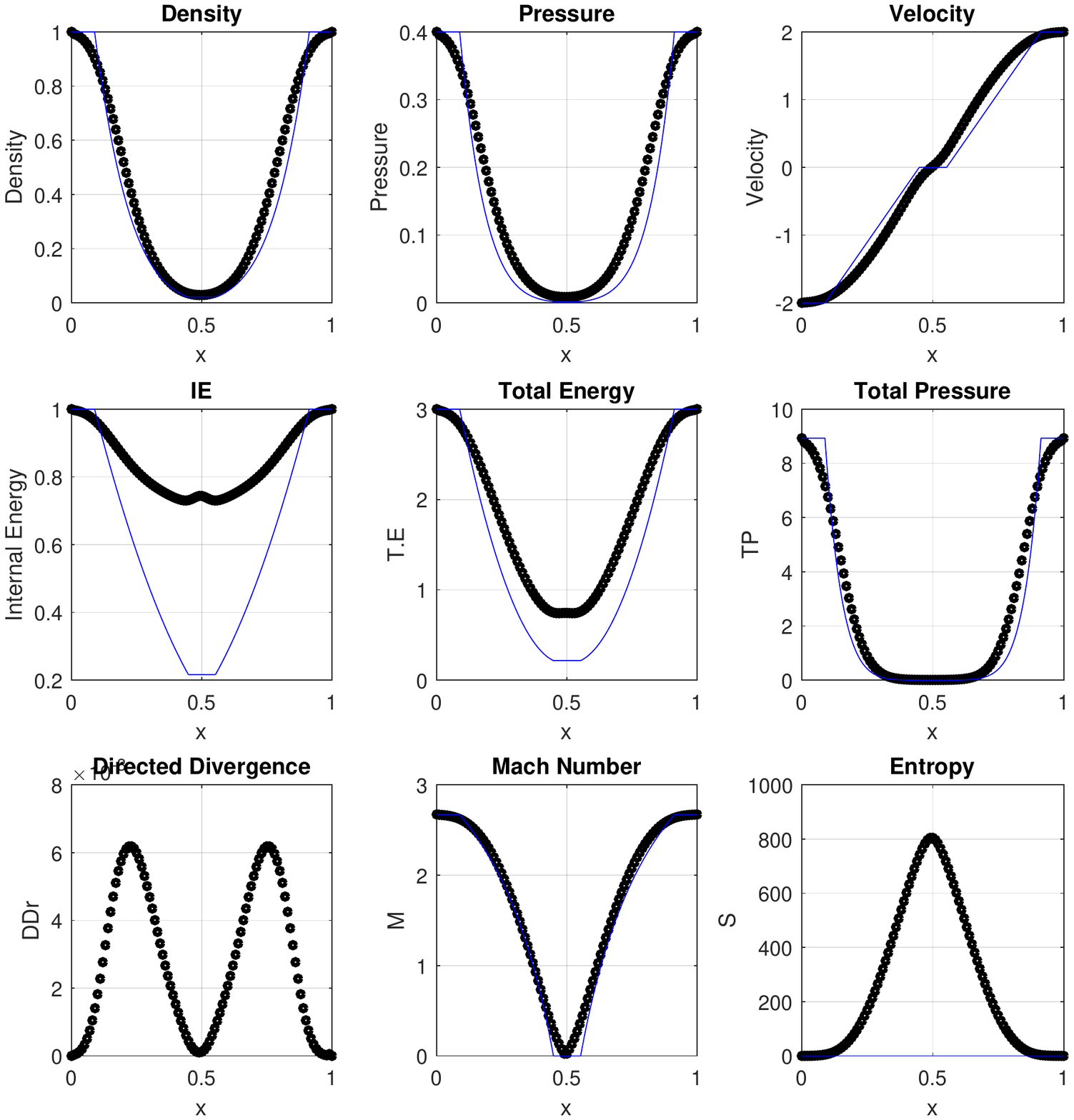}
\includegraphics[width=14.5cm,angle=0]{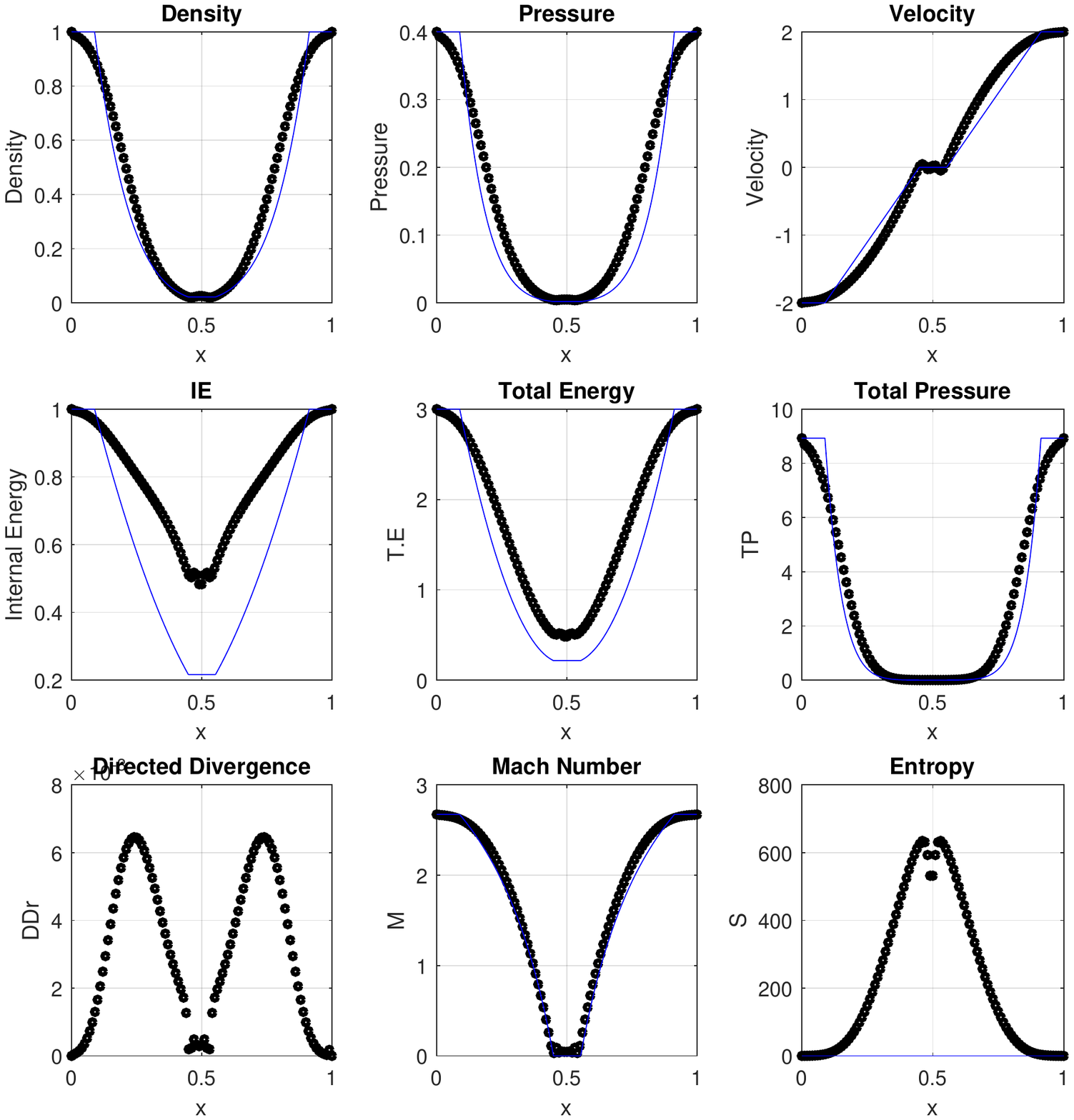}
\end{center} 
\caption{Test case 4: (a) KFDS-A and  (b) KFDS-B} 
\label{1DEulerTC_4_LLF}  
\end{figure}

\begin{figure} 
\begin{center} 
\includegraphics[width=14.5cm,angle=0]{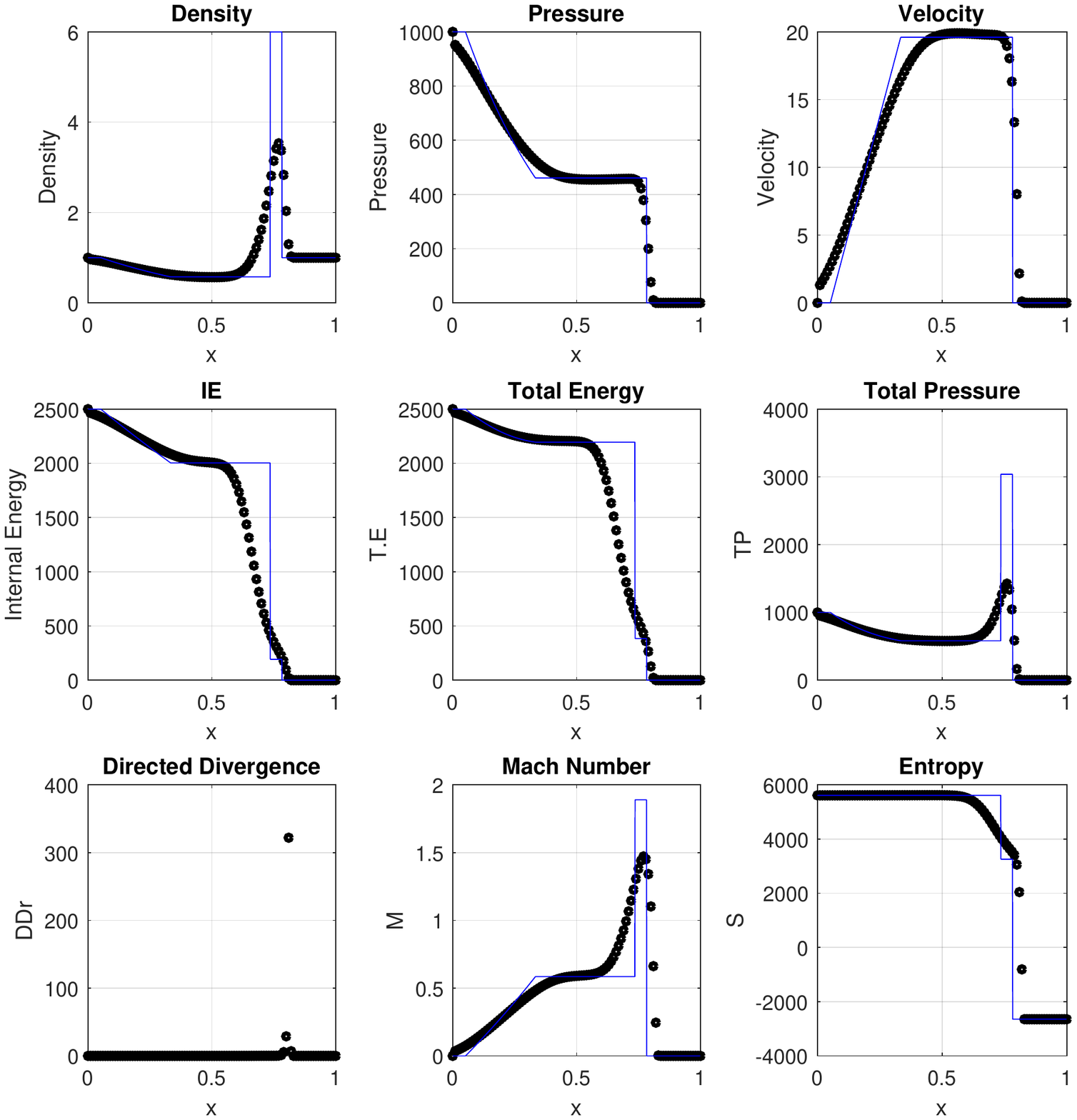}
\includegraphics[width=14.5cm,angle=0]{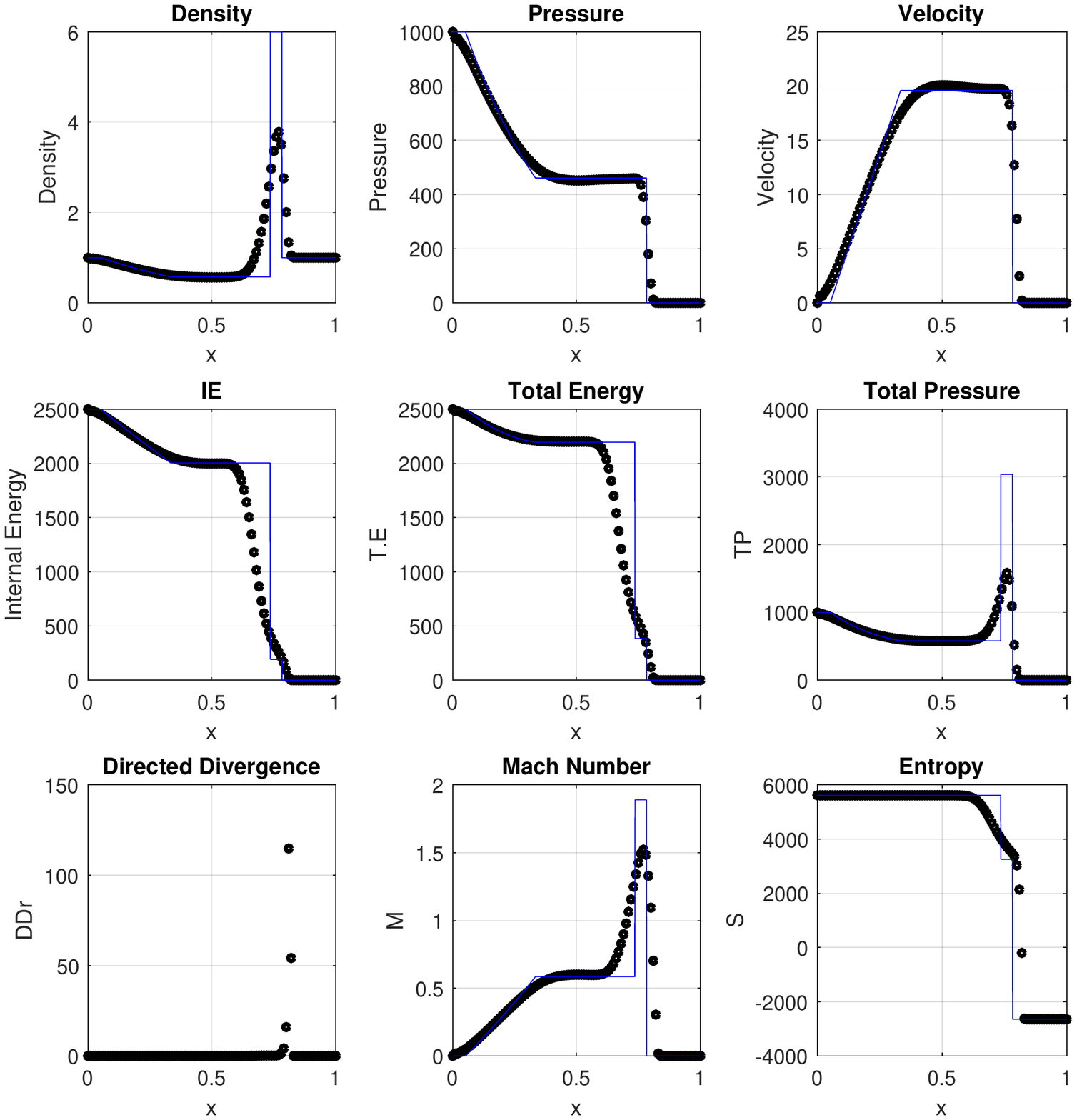}
\end{center} 
\caption{Test case 5: (a) KFDS-A and  (b) KFDS-B} 
\label{1DEulerTC_5_LLF}  
\end{figure}

\begin{figure} 
\begin{center} 
\includegraphics[width=14.5cm,angle=0]{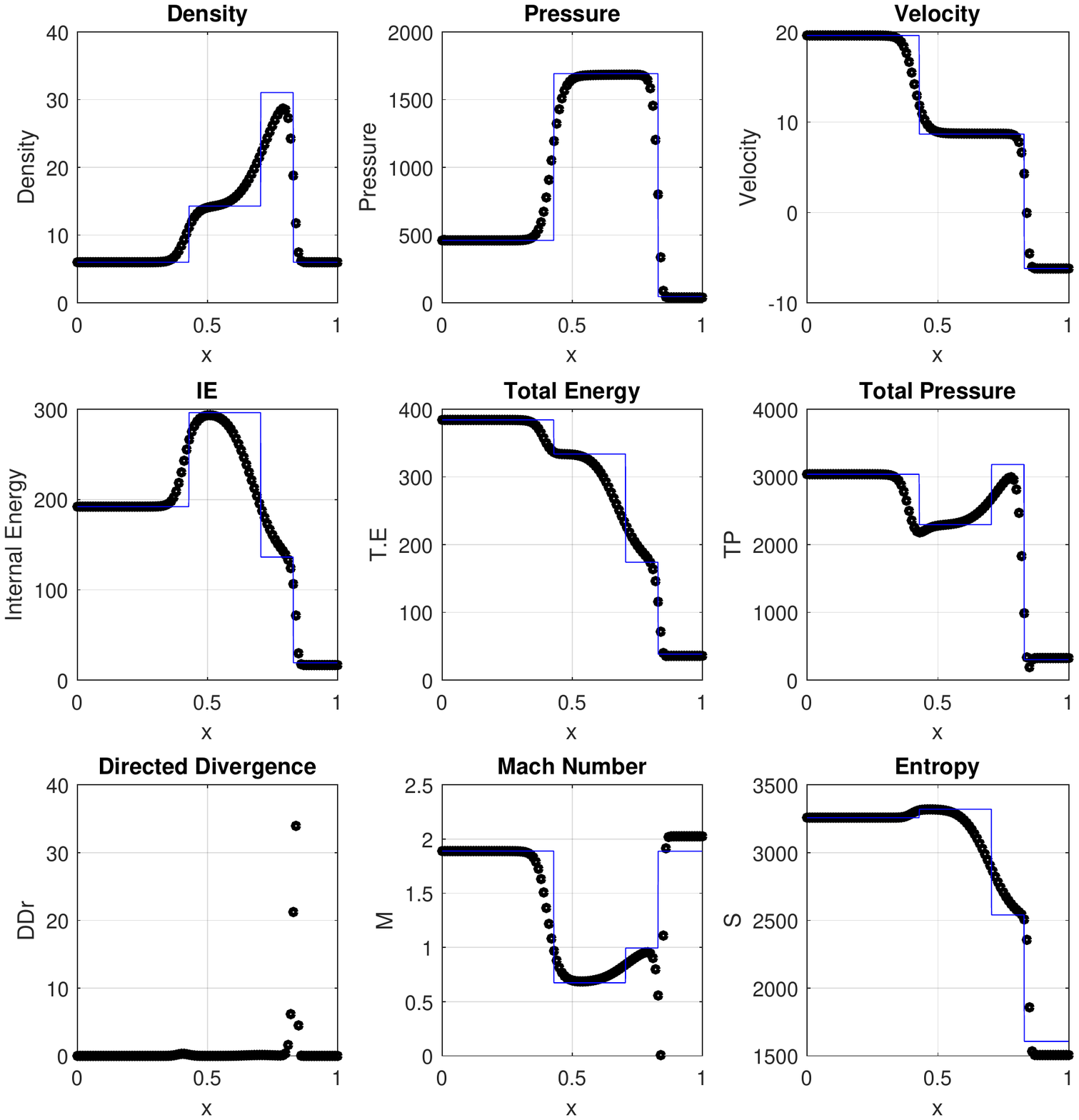}
\includegraphics[width=14.5cm,angle=0]{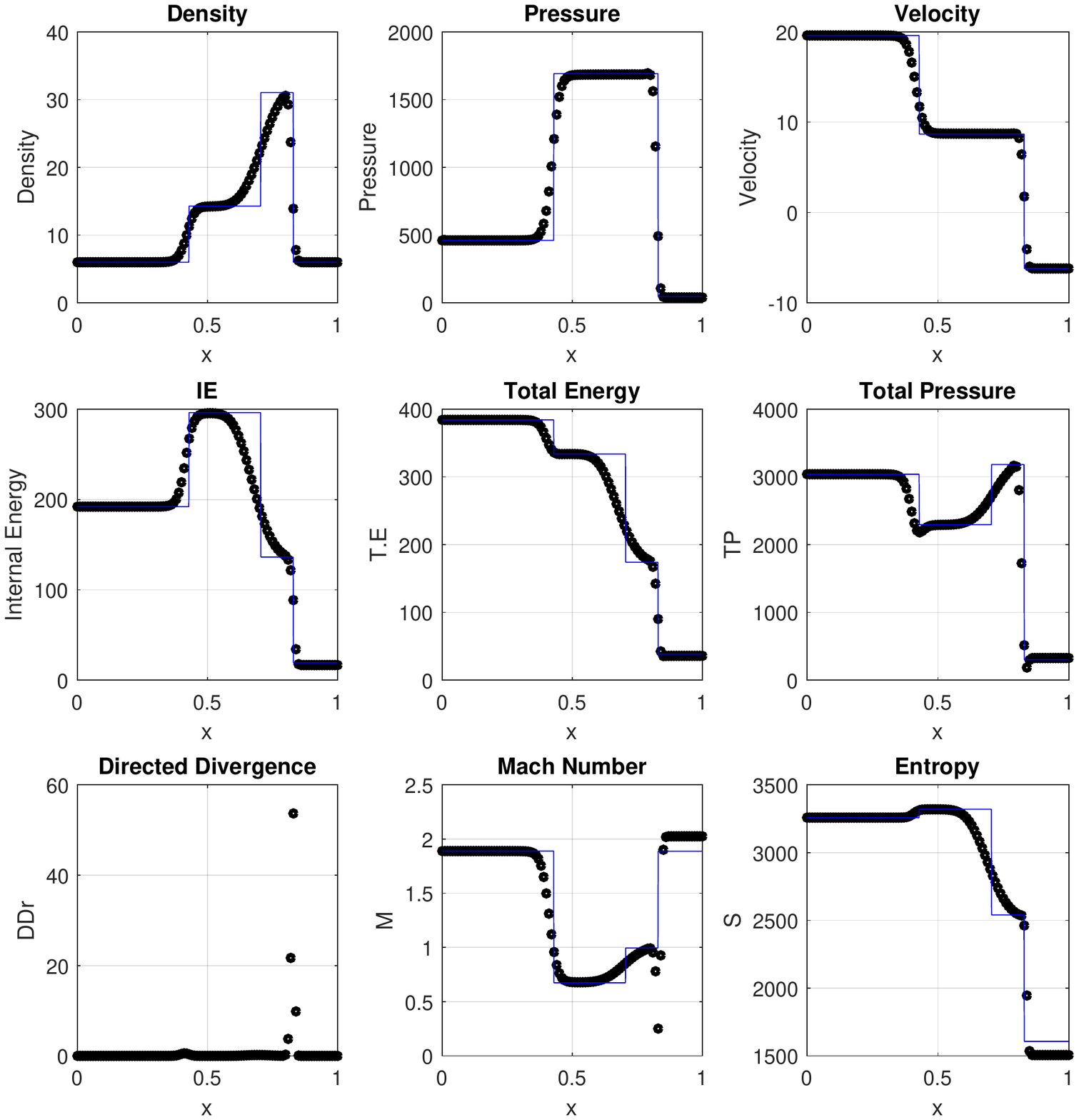}
\end{center} 
\caption{Test case 6:(a) KFDS-A and  (b) KFDS-B} 
\label{1DEulerTC_6_LLF}  
\end{figure}

\begin{figure} 
\begin{center} 
\includegraphics[width=15.0cm,angle=0]{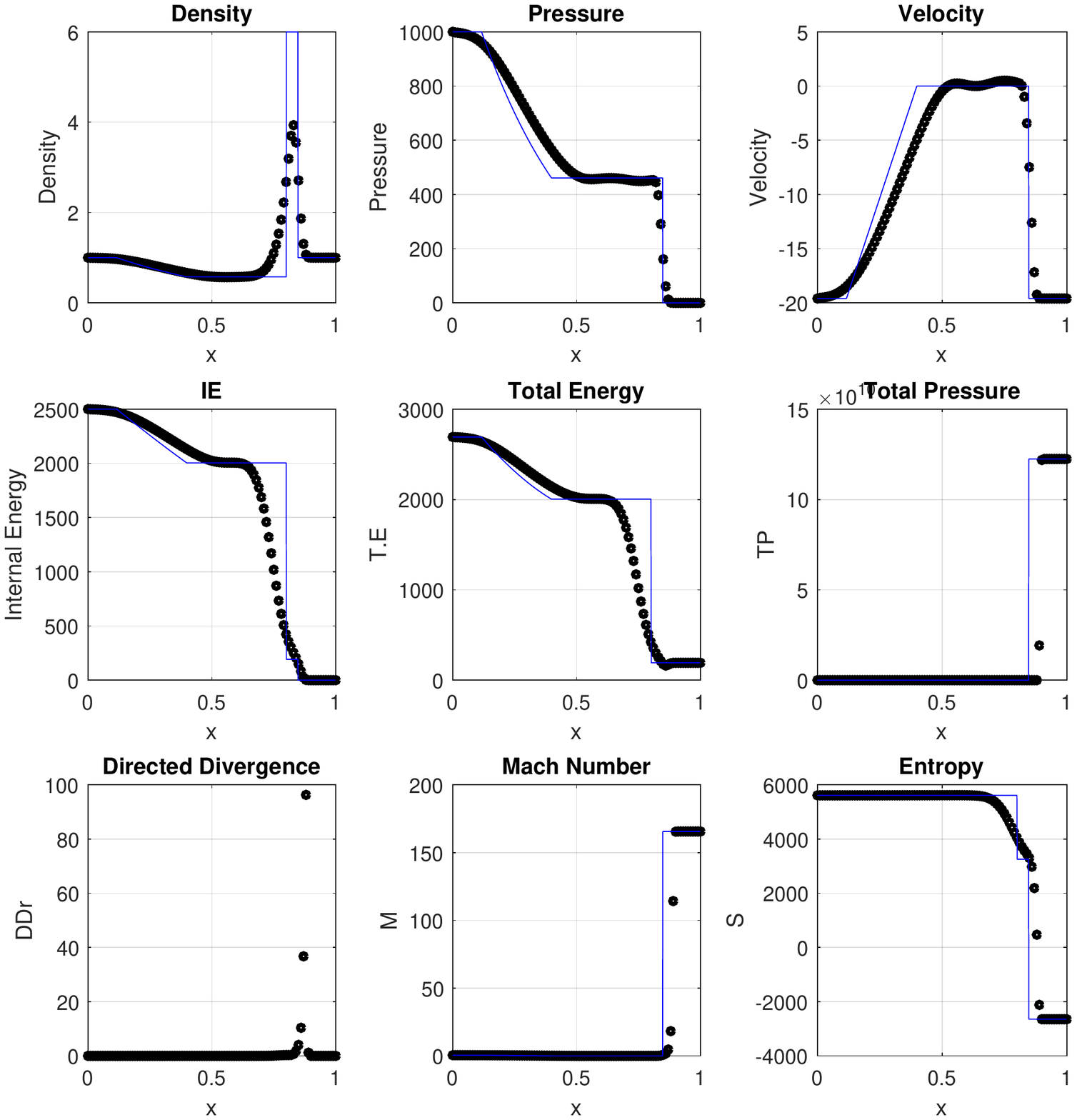}
\includegraphics[width=15.0cm,angle=0]{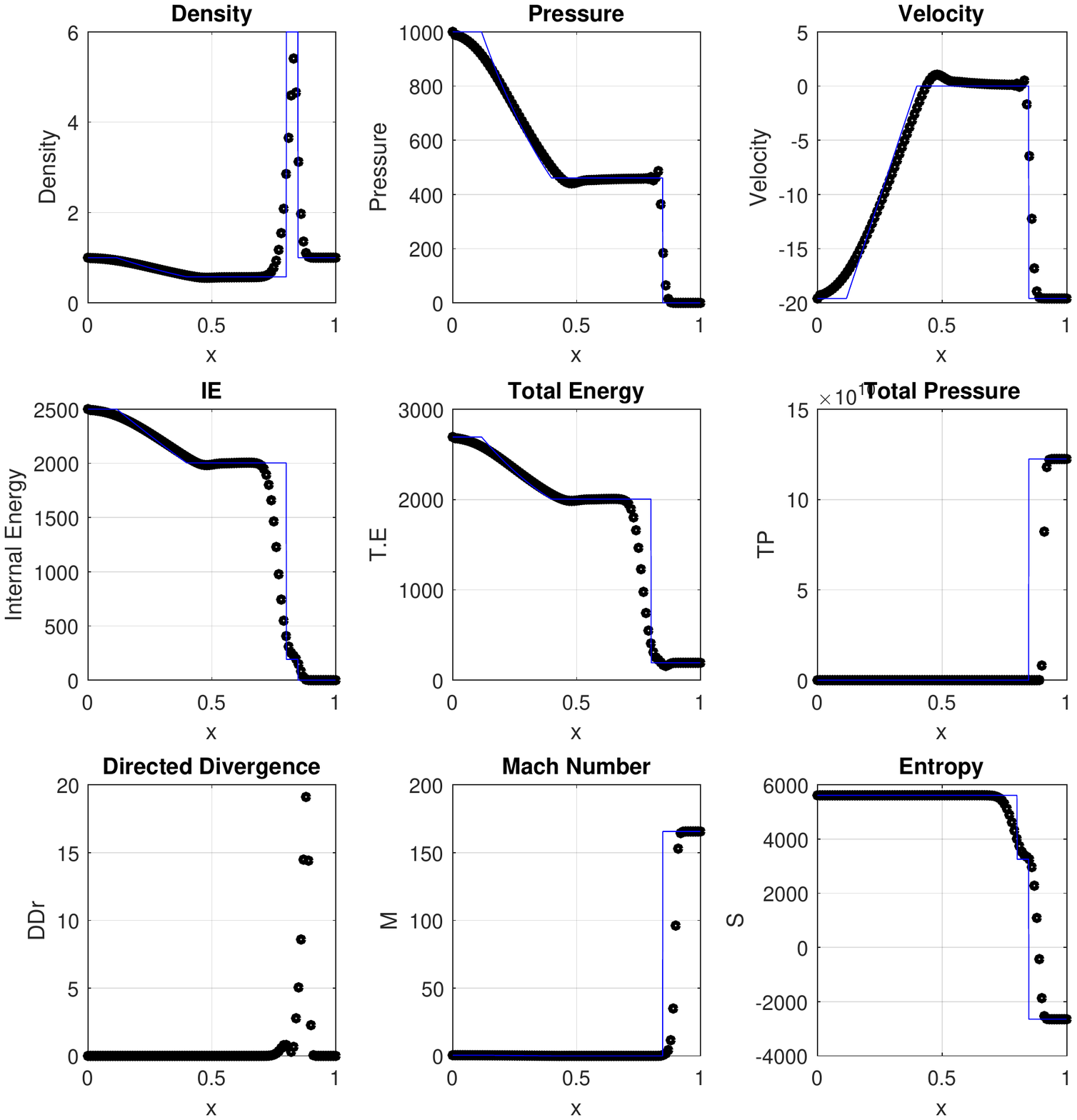}
\end{center} 
\caption{Test case 7:  (a) KFDS-A and  (b) KFDS-B} 
\label{1DEulerTC_7_LLF}  
\end{figure}

 The fourth test case is the well-known overheating problem where the initial strong velocity gradient results in the evolution of two expansion waves moving in the opposite directions, separated by a contact discontinuity. This case challenges the ability of the numerical scheme to provide physically relevant results.  Both the versions of KFDS scheme could generate stable results  [Fig.\ref{1DEulerTC_4_LLF}] without blowing up, but with a significant deviation in the internal energy near the contact discontinuity.  Test cases 5, 6 and 7 (Figures [\ref{1DEulerTC_5_LLF}], [\ref{1DEulerTC_6_LLF}] and [\ref{1DEulerTC_7_LLF}]) are meant to test the robustness of the scheme and its ability to handle very strong gradients in pressure.  Again it can be seen that both versions of KFDS are able to provide solutions without any  unphysical behaviour.  However, mild oscillations are observed in KFDS-B solutions in certain test cases and appear to die out with lower CFL numbers.
\clearpage
\newpage
\newpage
\newpage
\newpage
\newpage
\newpage

\clearpage
\newpage
\subsection{Test case for 2-D Euler equations} 
\subsubsection{Slip flow}
\begin{figure} 
\begin{center} 
\includegraphics[width=4cm,angle=0]{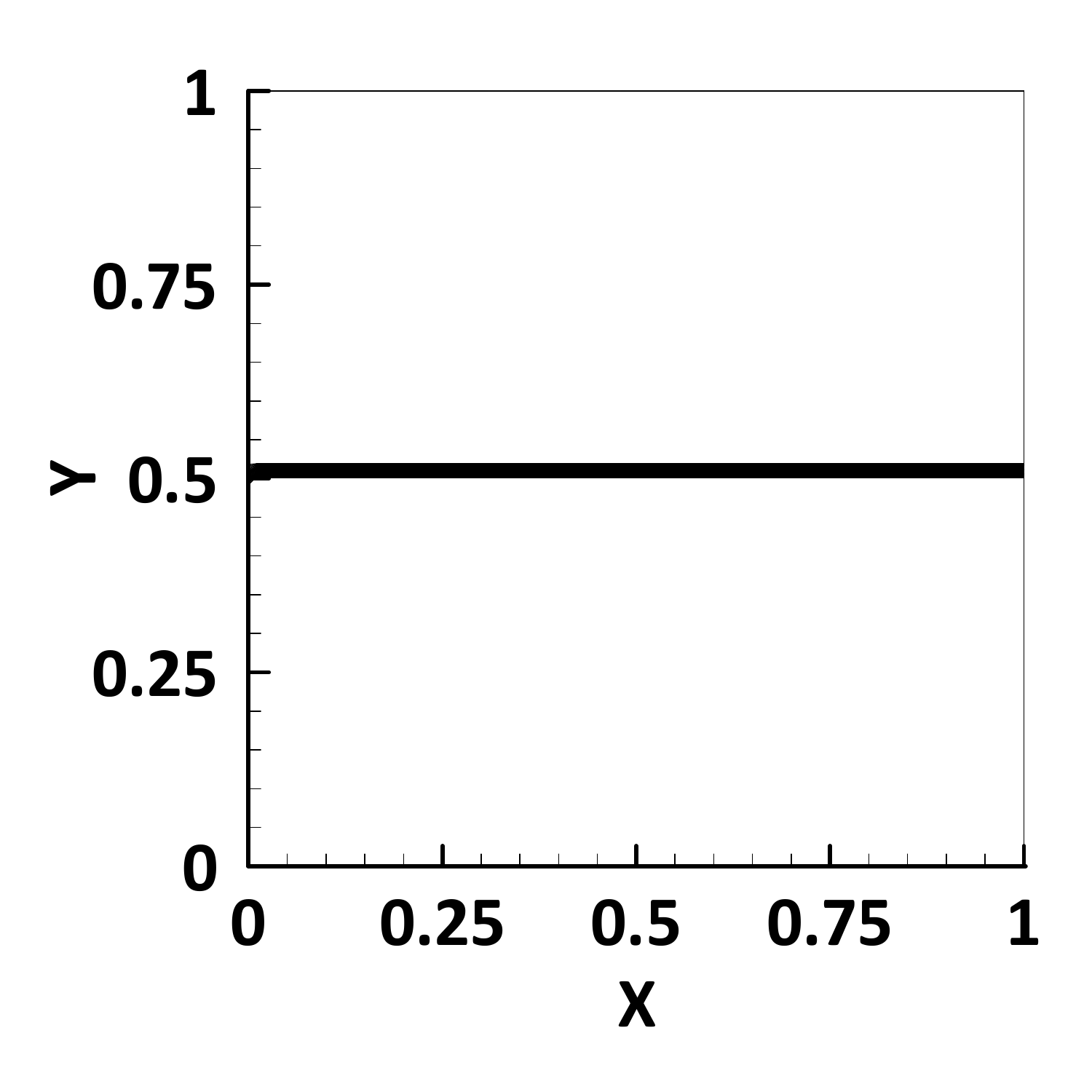}
\includegraphics[width=4cm,angle=0]{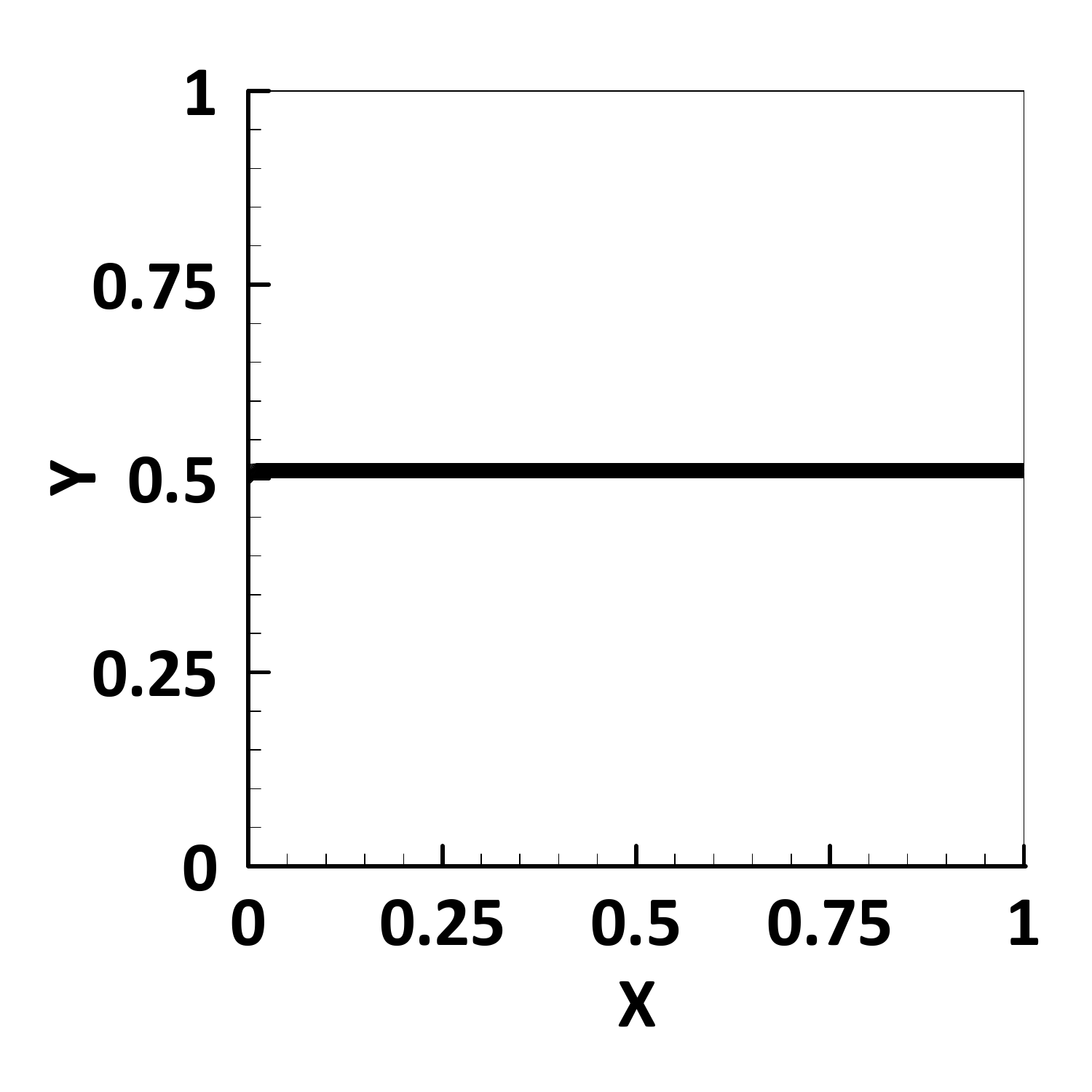}
\includegraphics[width=4cm,angle=0]{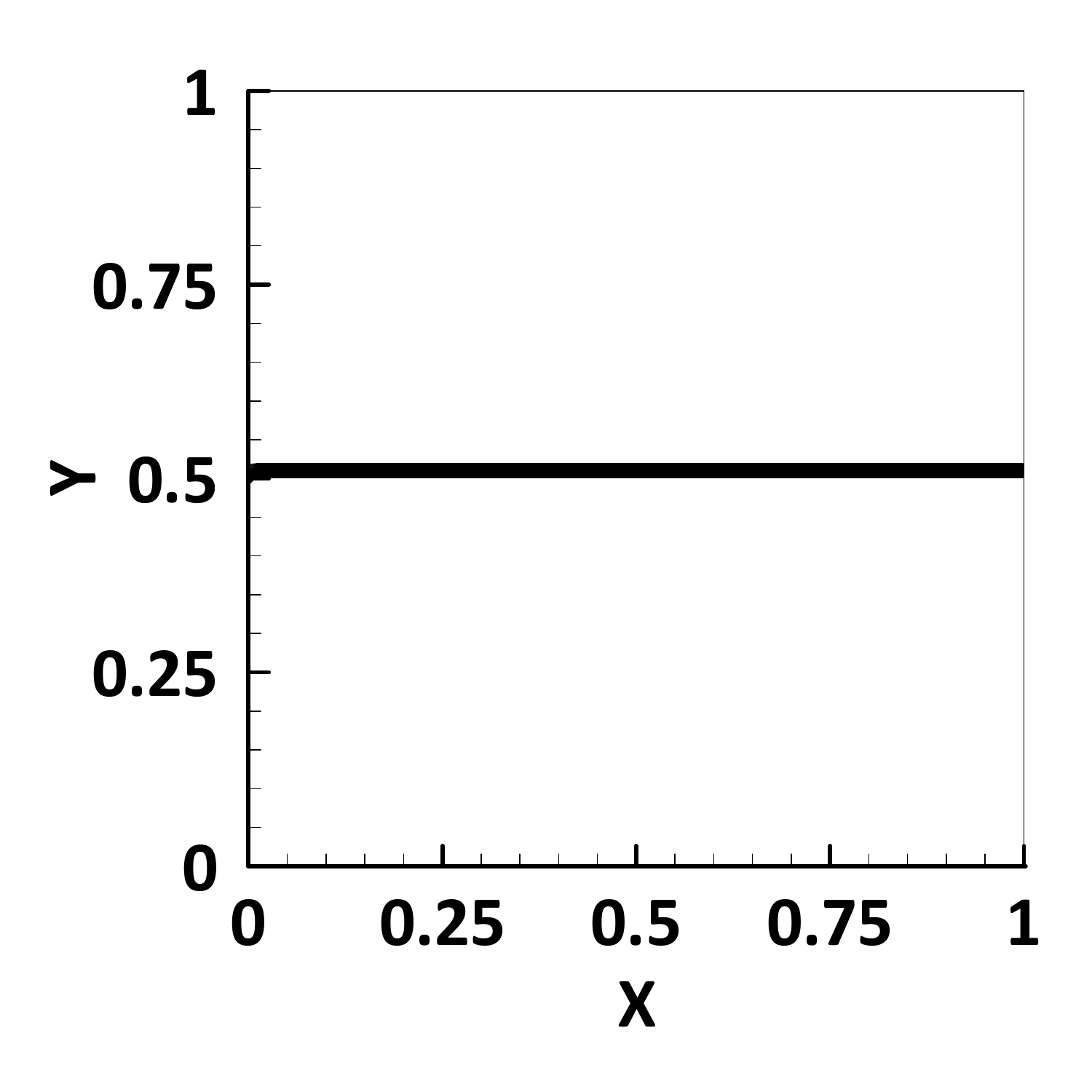}
\includegraphics[width=4cm,angle=0]{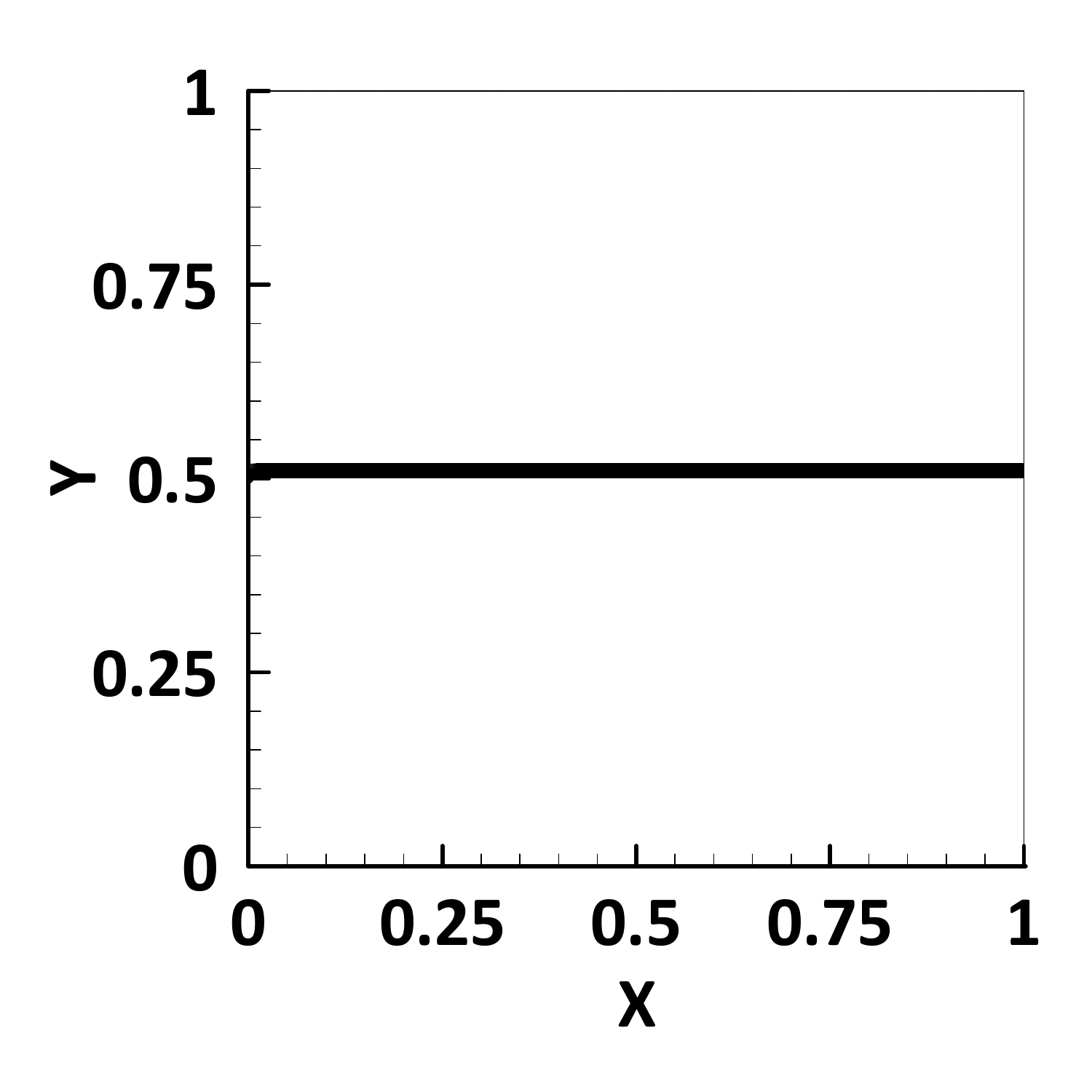}
\end{center} 
\caption{Test case: Slip Flow (20x20)- Mach contours (2.0: 0.025: 3.0)- (a)1O-KFDS-A  (b) 1O-KFDS-B (c) 2O-KFDS-A (d) 2O-KFDS-B}
\label{2DEulerTC_KFDS_SLIP}  
\end{figure}

The first of the test cases is a check on the ability to capture grid aligned contact discontinuities [\cite{Manna}]. The test case involves a uniform flow of two identical fluids with different speeds of u =2 and 3, which slip over each other. The computational domain [0,1][0,1] is discretised into a 20x20 grid and the domain is initialised with uniform pressure and density. The upper half of the left boundary is maintained at u=3 while the lower half of the left boundary is maintained at u = 2. The computational results for each version of the KFDS scheme are shown in Fig.[\ref{2DEulerTC_KFDS_SLIP}]. The basic ability of the KFDS scheme to capture grid aligned discontinuities exactly is retained. 

\subsubsection{Oblique shock reflection}

\begin{figure} 
\begin{center} 
\includegraphics[width=7.0cm,angle=0]{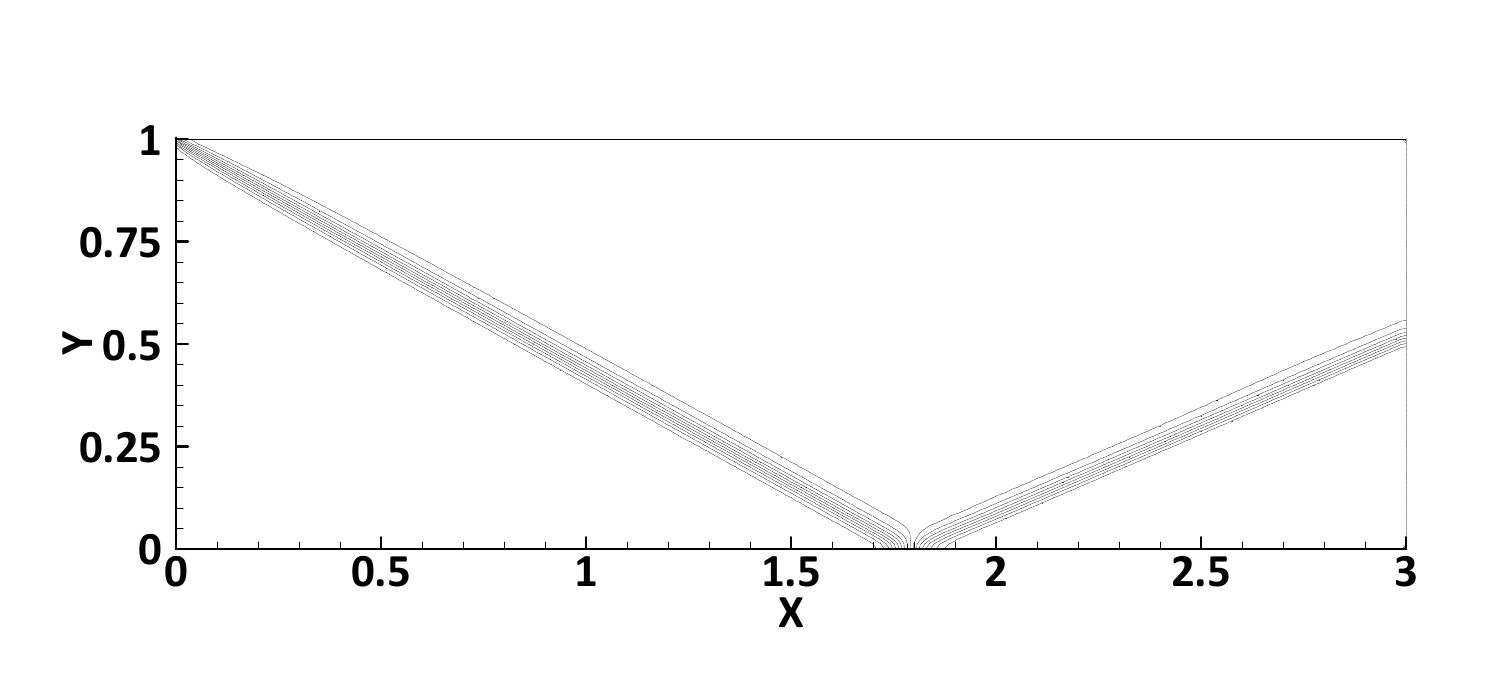}
\includegraphics[width=7.0cm,angle=0]{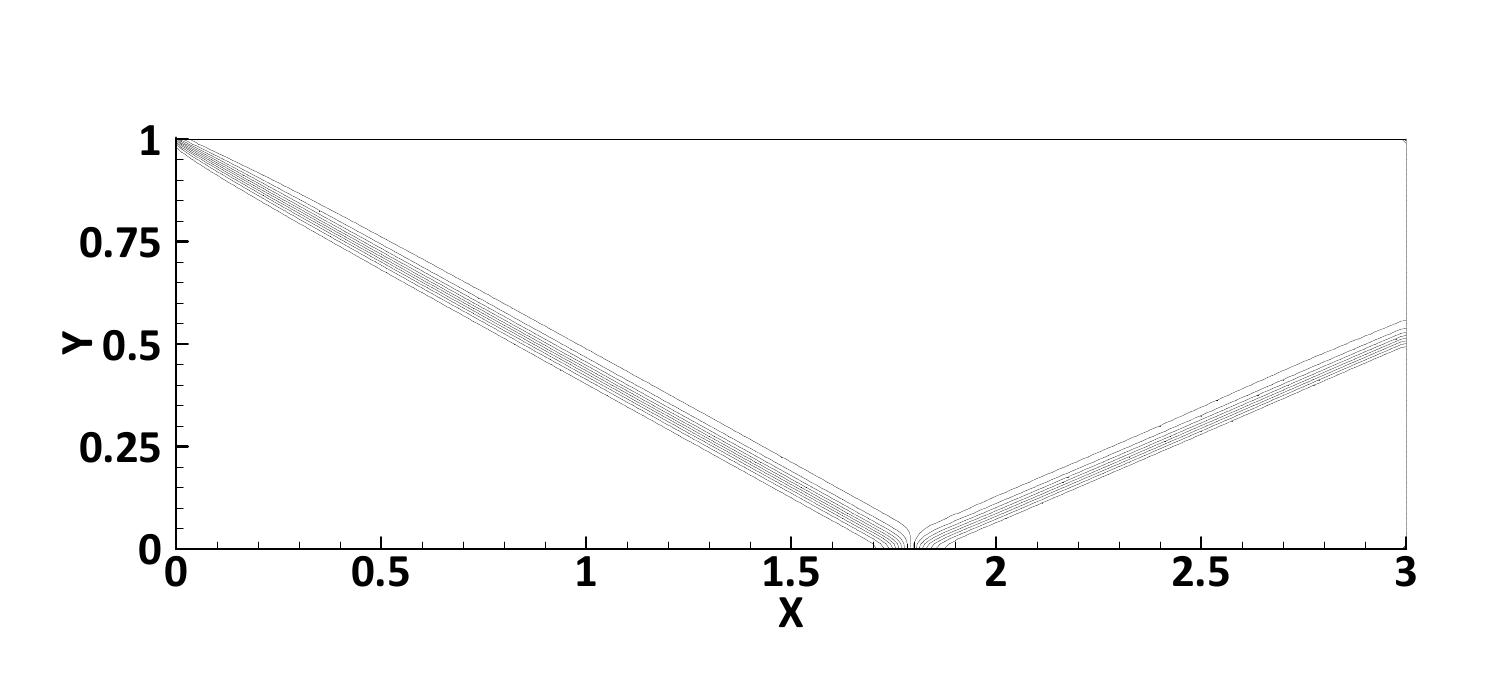}
\includegraphics[width=7.0cm,angle=0]{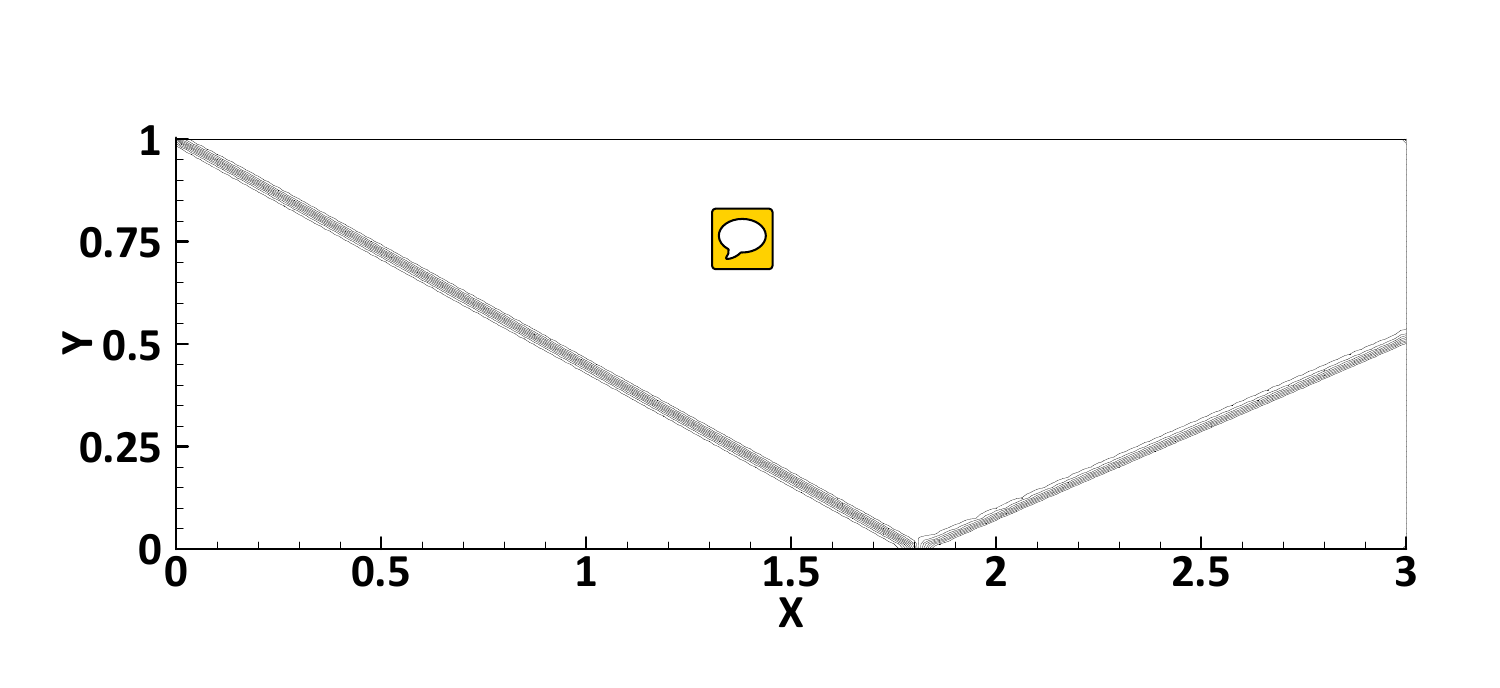}
\includegraphics[width=7.0cm,angle=0]{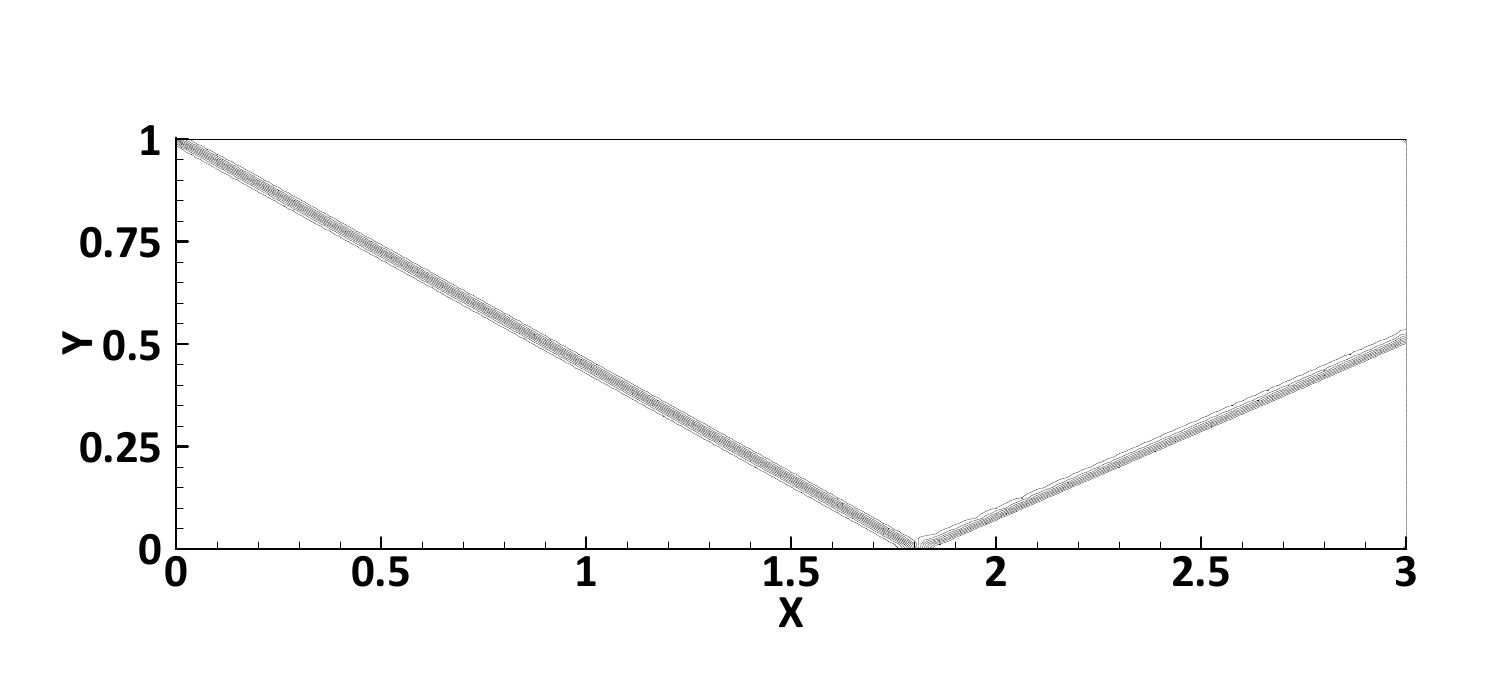}
\end{center} 
\caption{Test case: Oblique shock reflection (240x80)- pressure contours (0.7: 0.1: 2.9)- (a)1O-KFDS-A  (b) 1O-KFDS-B (c) 2O-KFDS-A (d) 2O-KFDS-B}
\label{2DEulerTC_KFDS_OSR}  
\end{figure}

 The computational domain for this test case [\cite{HartenWarming}] is rectangular:  $[0,3] \times[0,1]$.  An oblique shock wave is introduced from the top-left corner by means of inflow boundary conditions at the left boundary and the post-shock boundary conditions on the top side of the domain -    
$ Inflow: (\rho, u, v, p)_{0, y, t}=(1.0,2.9,0,1 / 1.4)$  and 
$ Post-shock \ conditions: (\rho, u, v, p)_{x, 1, t}=(1.69997,2.61934,-0.50633,1.52819)$ [\cite{Jin_Xin}].  The bottom side is maintained as a solid wall where flow tangency  boundary condition is applicable and at the right boundary the supersonic outflow boundary condition is prescribed.  The computational grid used for the numerical simulation is 240x80.  The test results are shown for both versions of KFDS scheme with first order and second order accuracy in Fig.[\ref{2DEulerTC_KFDS_OSR}]. The shock positions and the point of reflection match well with the results presented in [\cite{Jaisankar}]. There is a marked improvement in the shock crispness between the first order and second order accurate results.

\clearpage
\newpage

\subsubsection{Supersonic flow over a compression ramp}
\begin{figure} 
\begin{center} 
\includegraphics[width=7.5cm,angle=0]{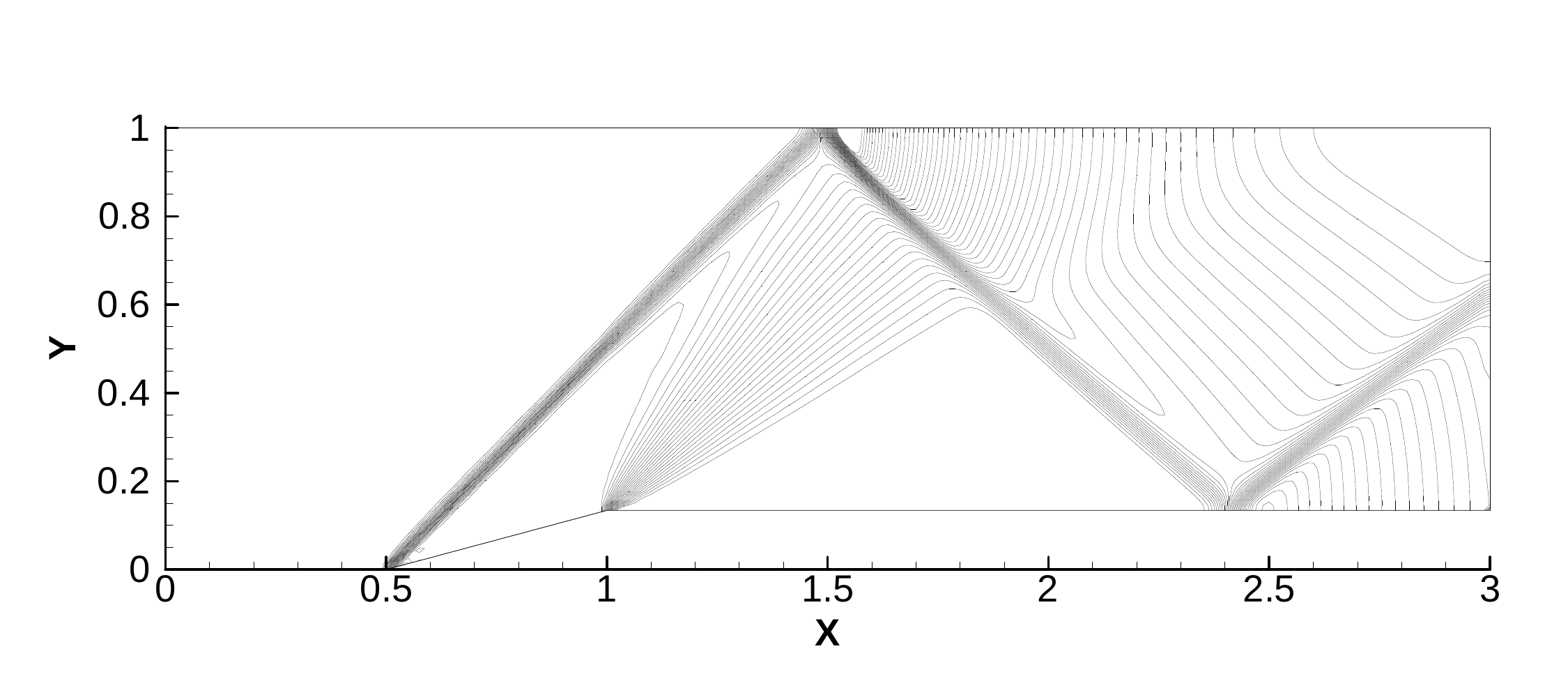}
\includegraphics[width=7.5cm,angle=0]{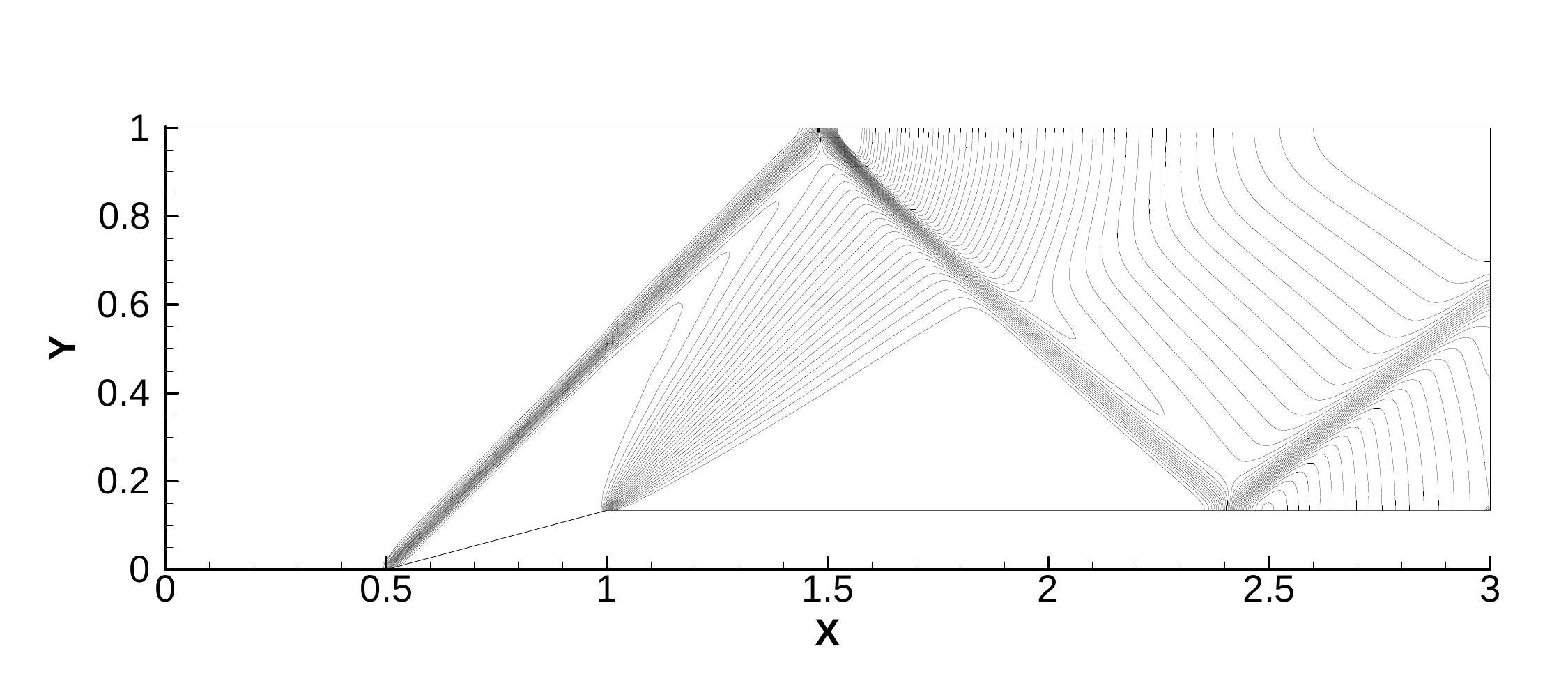}
\includegraphics[width=7.5cm,angle=0]{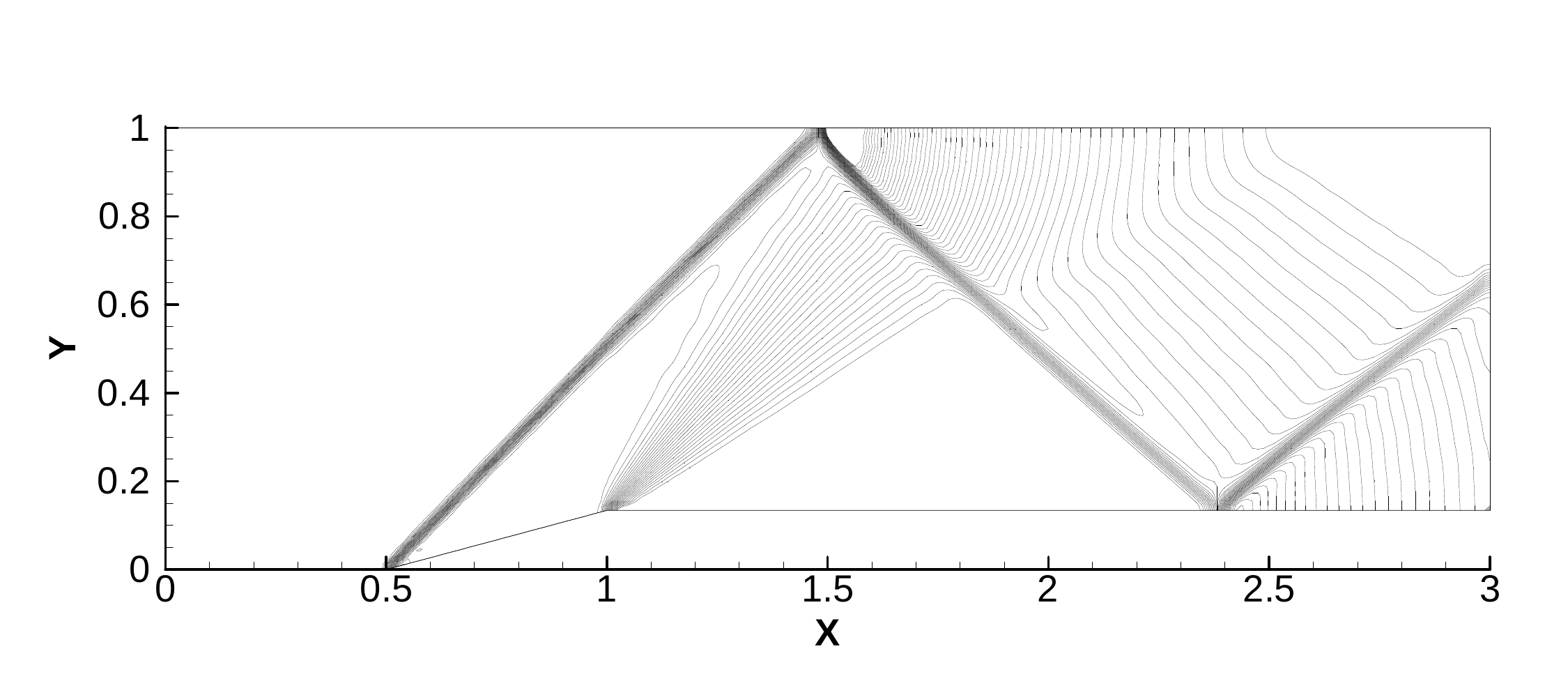}
\includegraphics[width=7.5cm,angle=0]{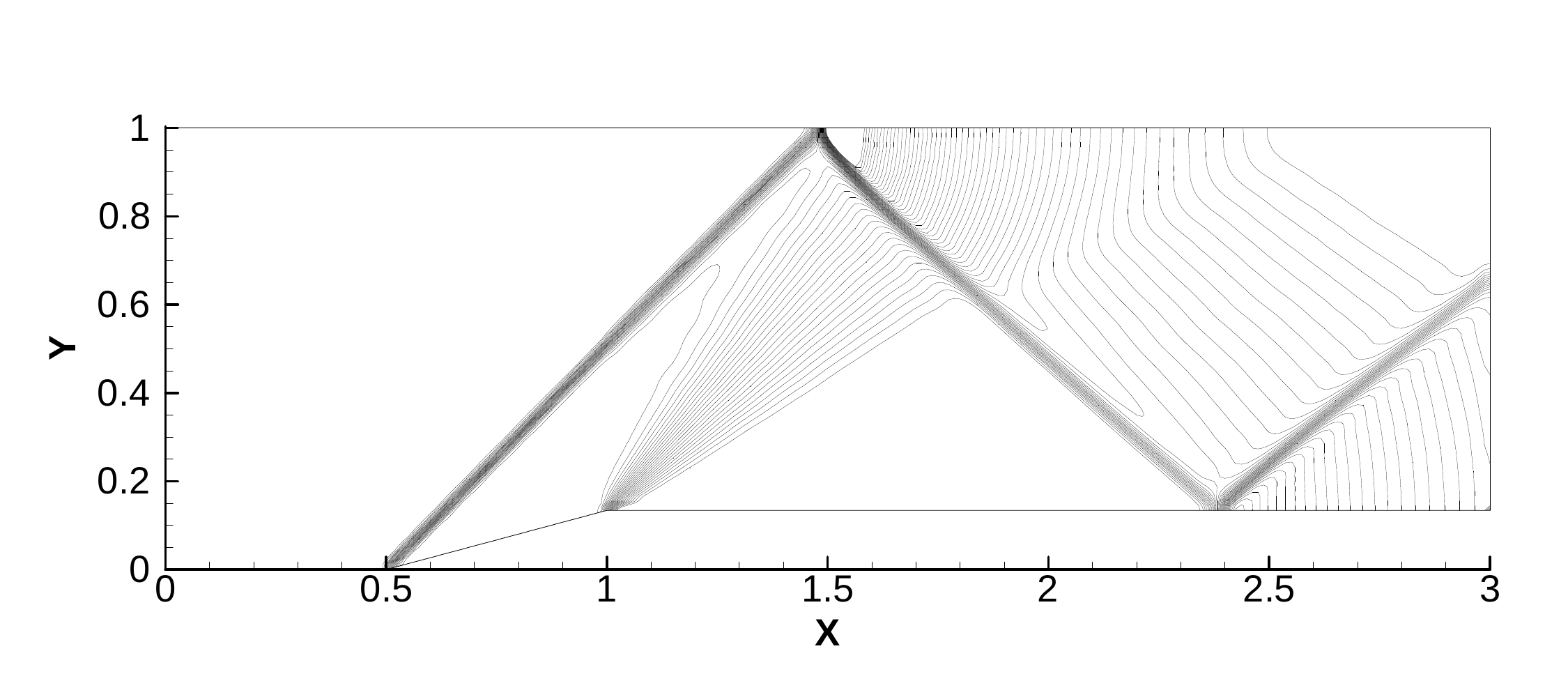}
\end{center} 
\caption{Test case: Supersonic flow over a ramp (240x80)- pressure contours 1.1 : 0.05 :3.8)- (a)1O-KFDS-A  (b) 1O-KFDS-B (c) 2O-KFDS-A (d) 2O-KFDS-B}
\label{2DEulerTC_KFDS_RAMP}  
\end{figure}

The computational domain for this test case \cite{Vanleer}  involves a $15^{o}$ ramp at the bottom wall of the  computational domain of size $[0,3] \times[0,1]$. A steady inflow of Mach 2 over the ramp results in the generation of an oblique shock wave   at the compression corner. This oblique shock hits the top wall and gets further reflected down the flow. This reflected shock wave interacts with the expansion wave generated from the upper corner of the ramp.  This test case involves evolution of both shock and expansion waves and their interactions.  The test results are shown for both the versions of KFDS with first order and second order accuracy in Fig.[\ref{2DEulerTC_KFDS_RAMP}].  No expansion shocks are seen, thus confirming the efficacy of the strategy of adding additional numerical diffusion based on $D^{2}$-distance.    

\subsubsection{Supersonic flow over a bump}

\begin{figure} 
\begin{center} 
\includegraphics[width=7.5cm]{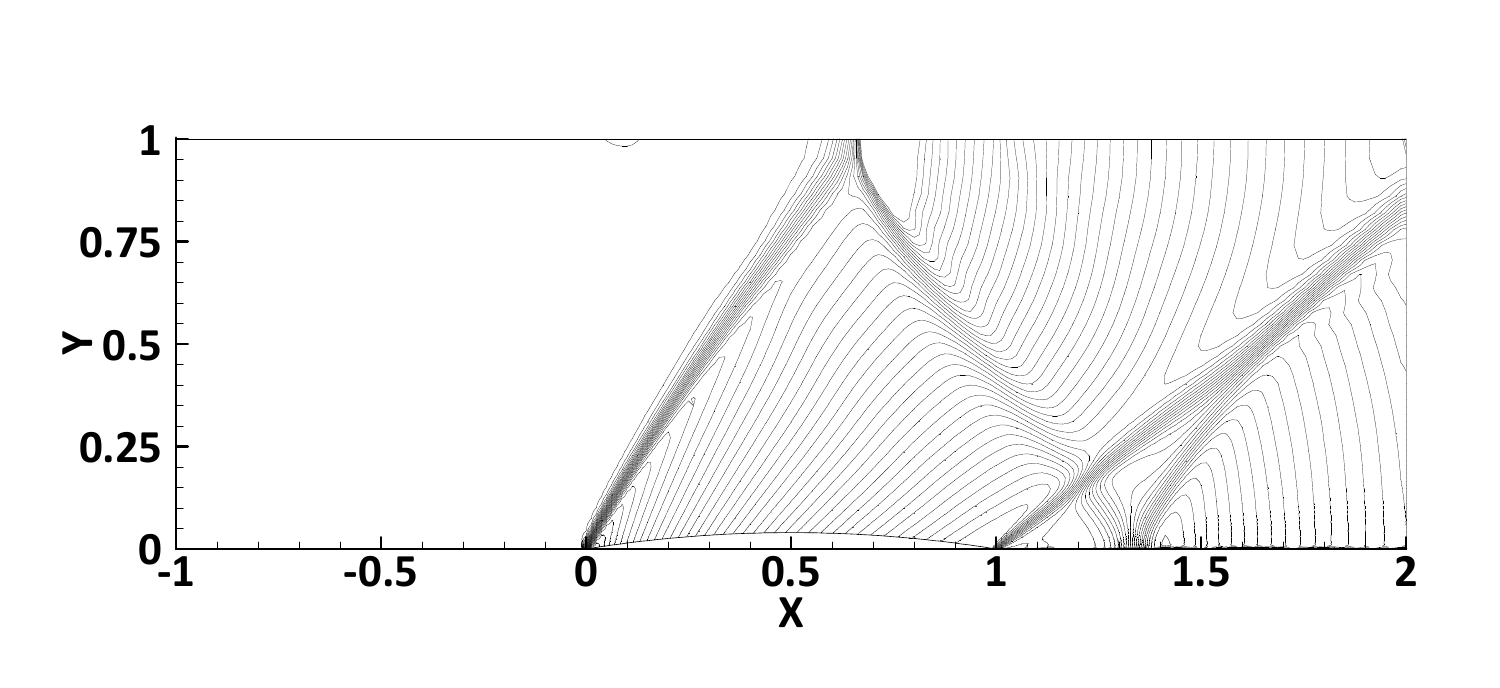}
\includegraphics[width=7.5cm]{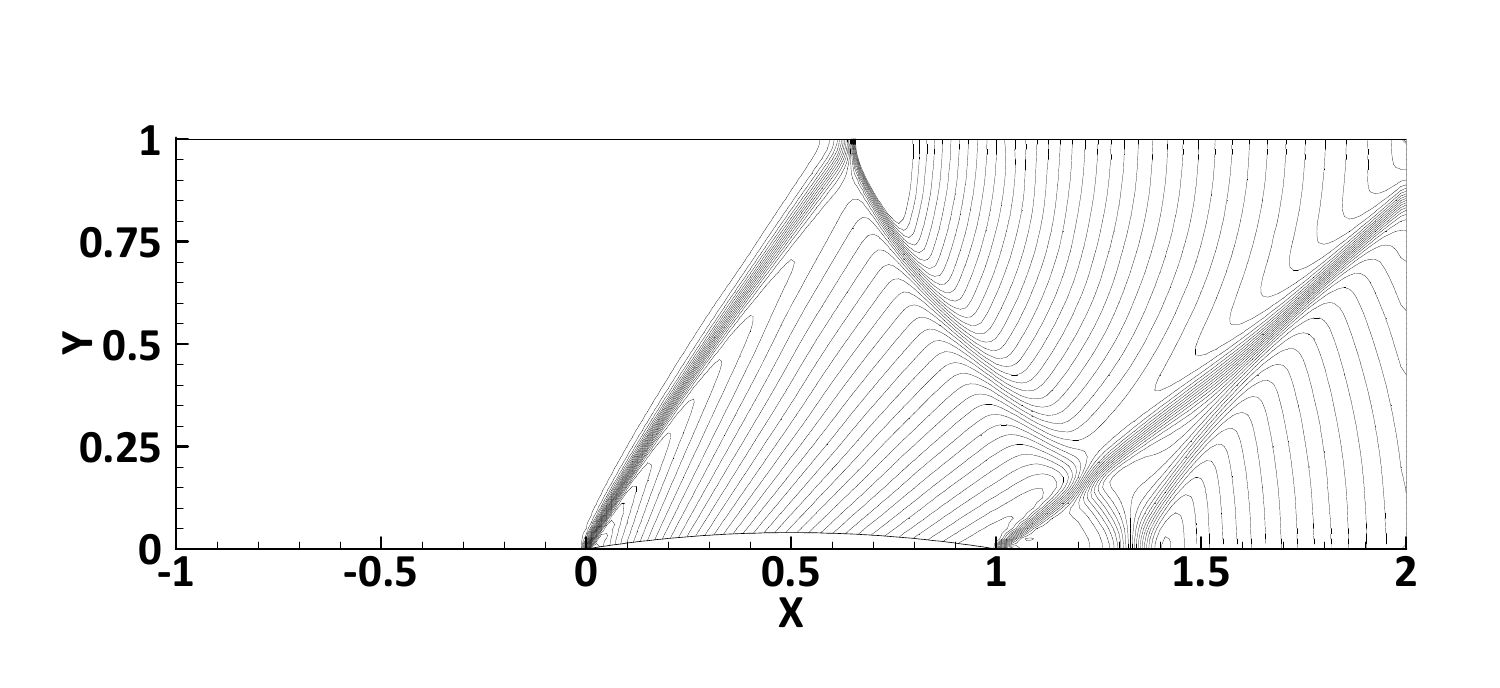}
\includegraphics[width=7.5cm]{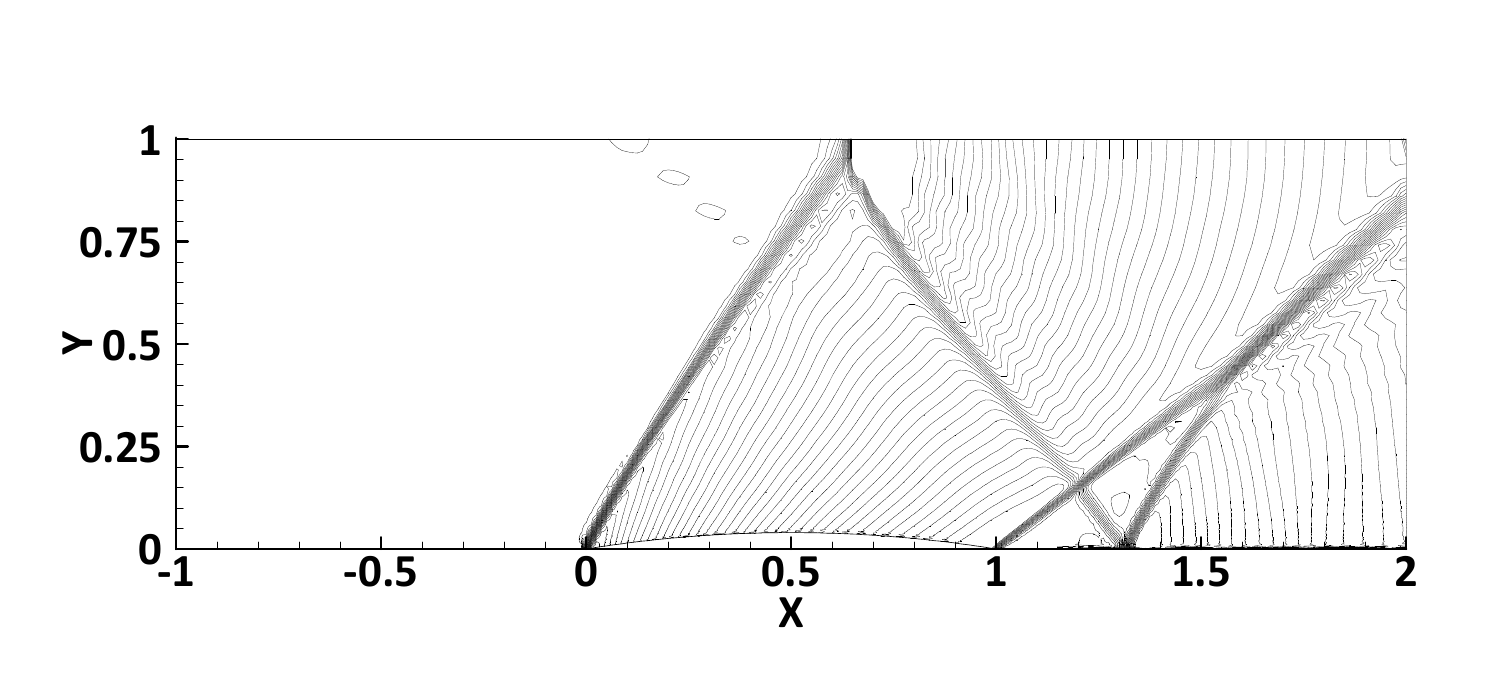}
\includegraphics[width=7.5cm]{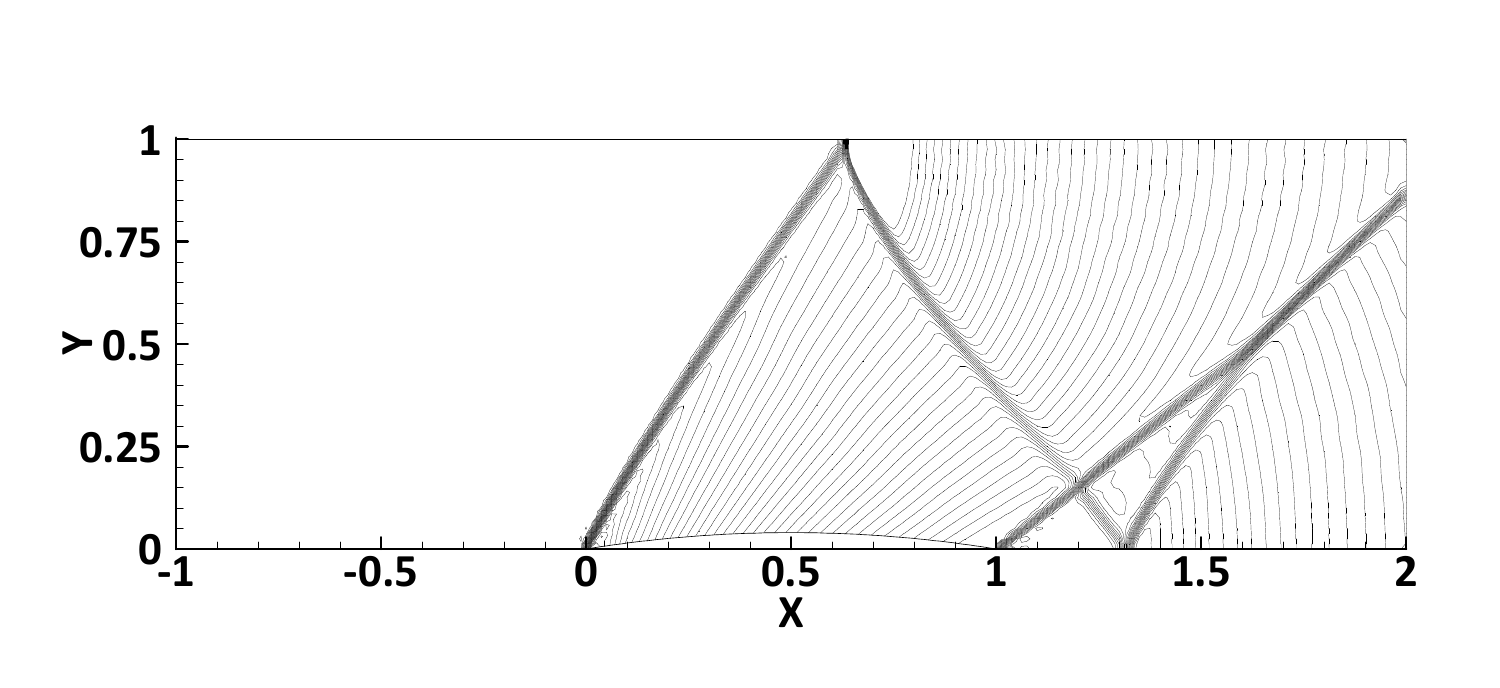}
\end{center} 
\caption{Test case: Supersonic flow over a bump (240x80) - pressure contours (0.7: 0.025: 1.5)- (a)1O-KFDS-A  (b) 1O-KFDS-B (c) 2O-KFDS-A (d) 2O-KFDS-B}
\label{2DEulerTC_KFDS_BUMP}  
\end{figure}

This test case [\cite{Ripley}] involves a channel having a 4\% thick circular arc bump on the bottom side of the test domain of [-1,2]x[0,1]. The bump has a chord length of 1 unit and is located at $x = 0.5 $.  The left side is marked as a supersonic inlet with free stream Mach number 1.4.  The bottom side is defined as free slip wall. The top side of the domain is defined as an inviscid wall, while the right side of the domain is marked as supersonic outlet. The results for each version of the scheme is shown in Fig.[\ref{2DEulerTC_KFDS_BUMP}]. Both versions of KFDS scheme capture the shock initiation, reflections and interactions, and the positions of reflections and interactions are captured with reasonable accuracy .  

\subsubsection{Hypersonic flow over a half cylinder}
\begin{figure} 
\begin{center} 
\includegraphics[width=4cm]{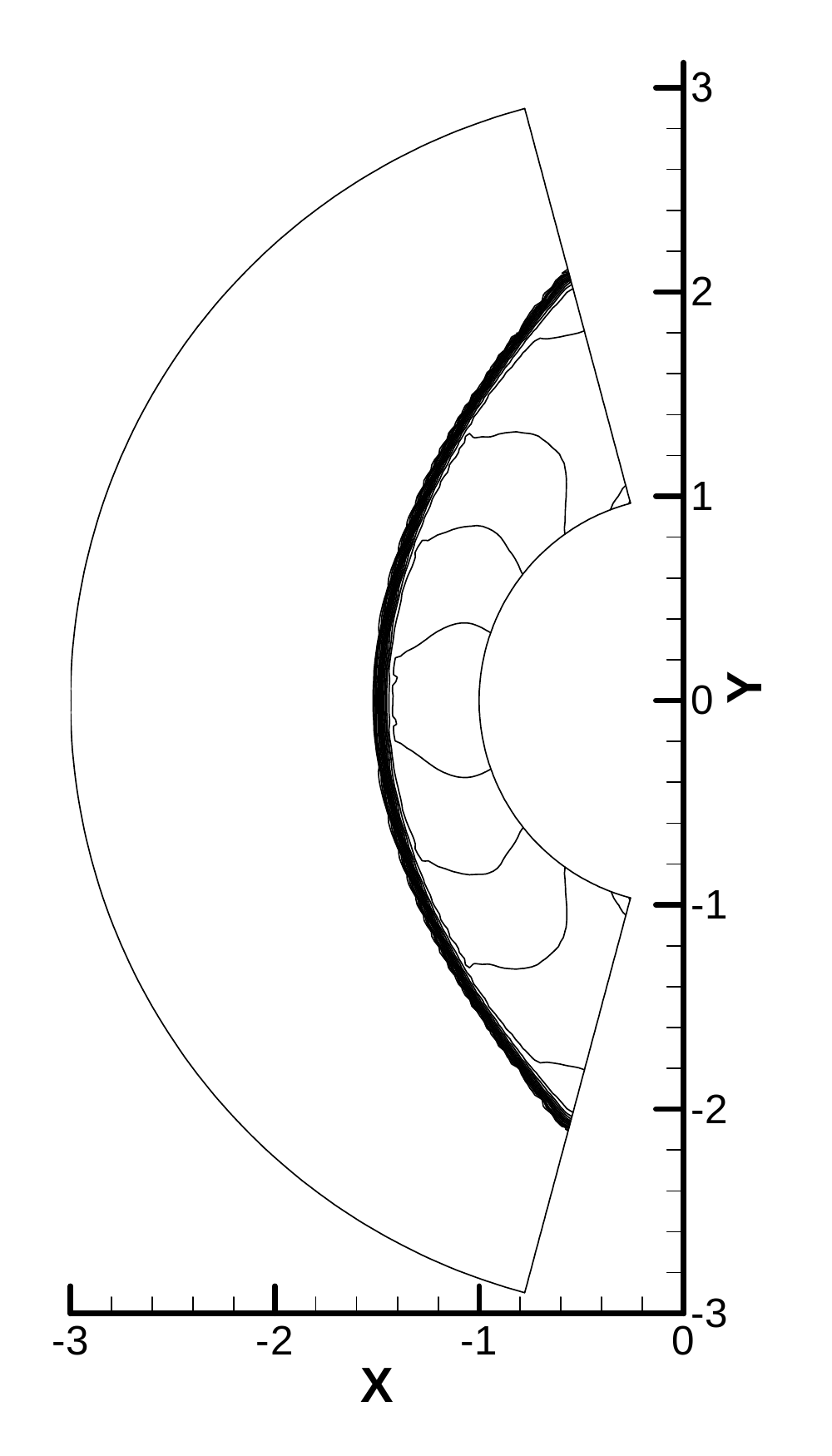}
\includegraphics[width=4cm]{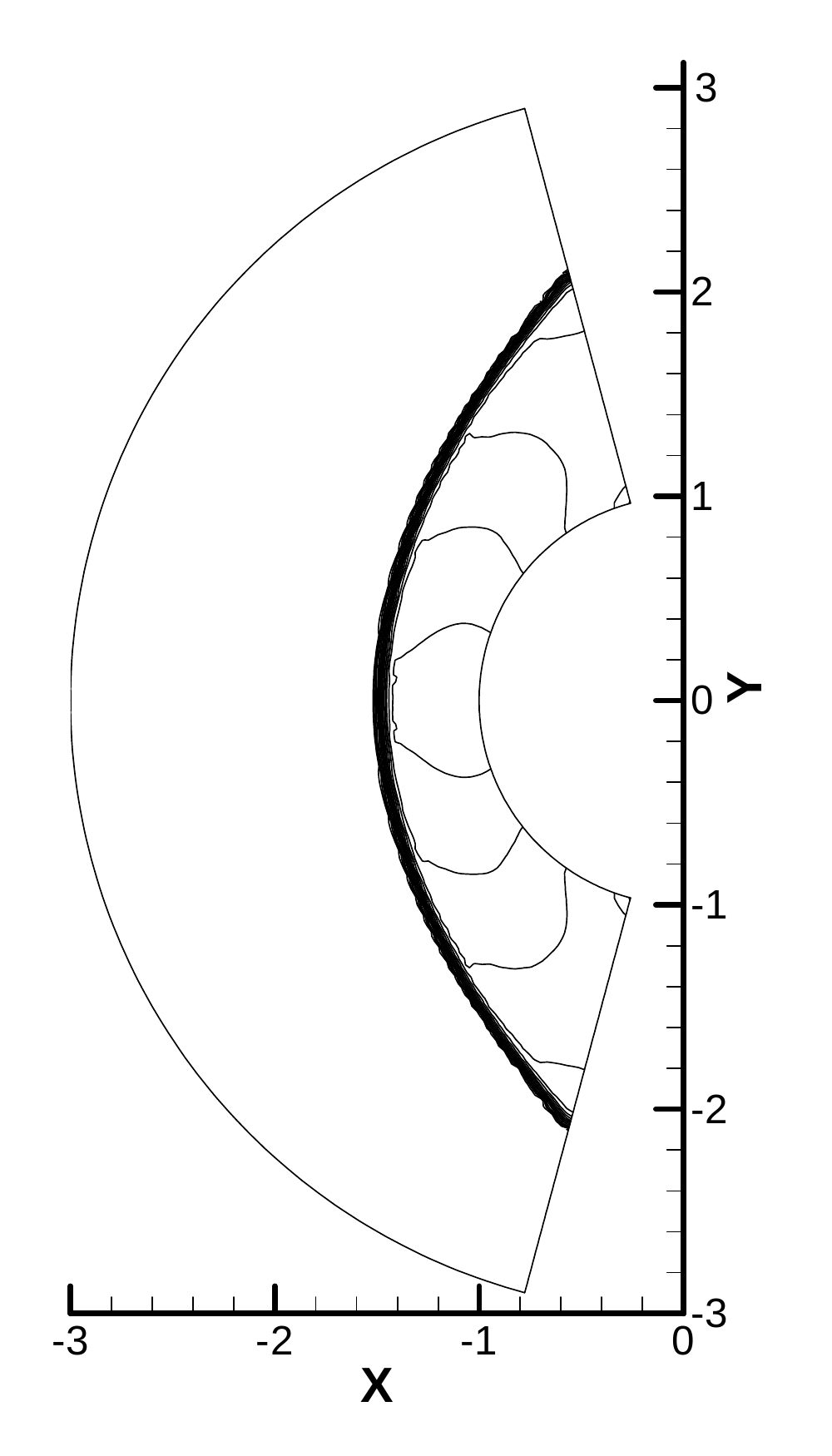}
\includegraphics[width=4cm]{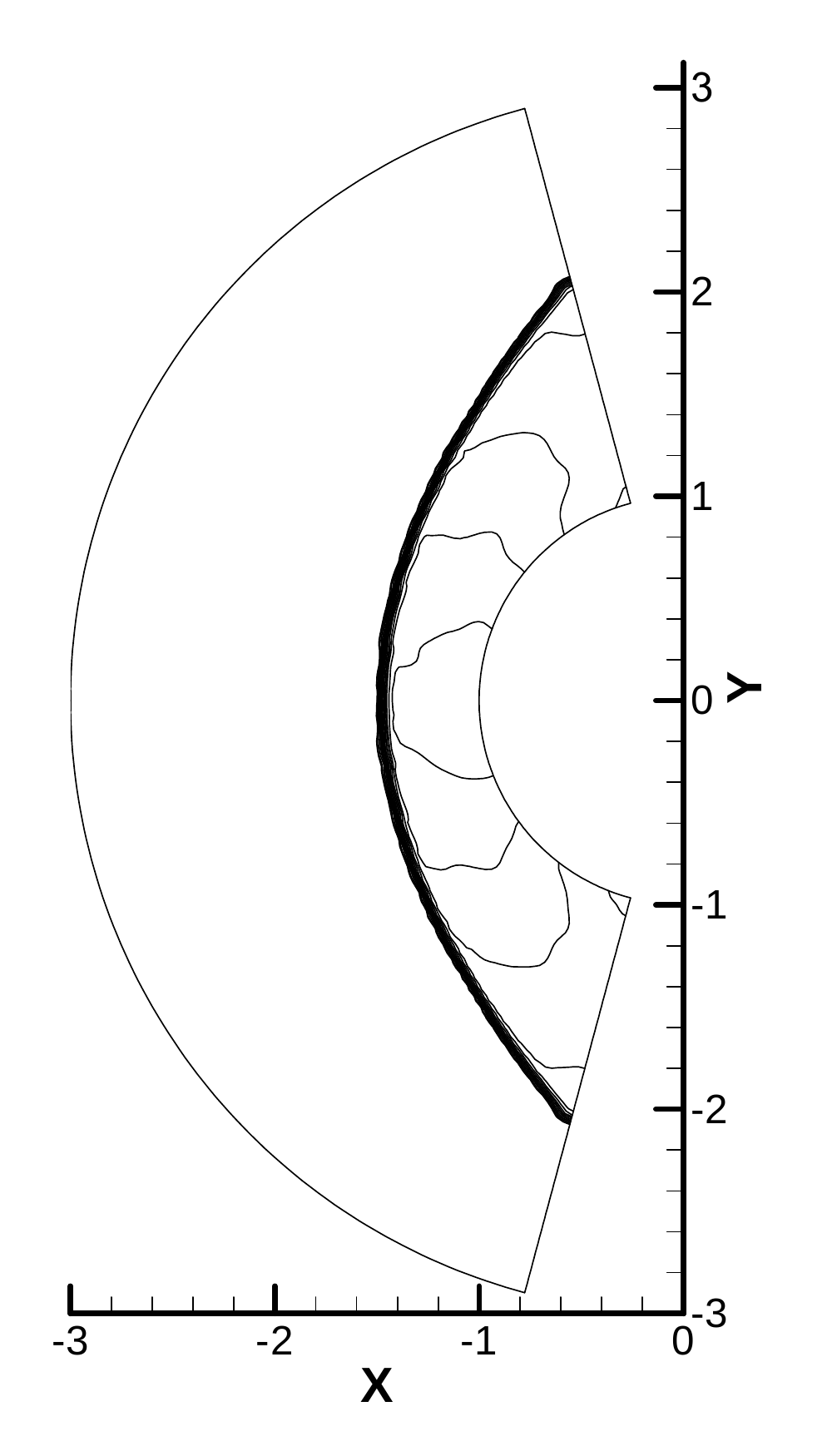}
\includegraphics[width=4cm]{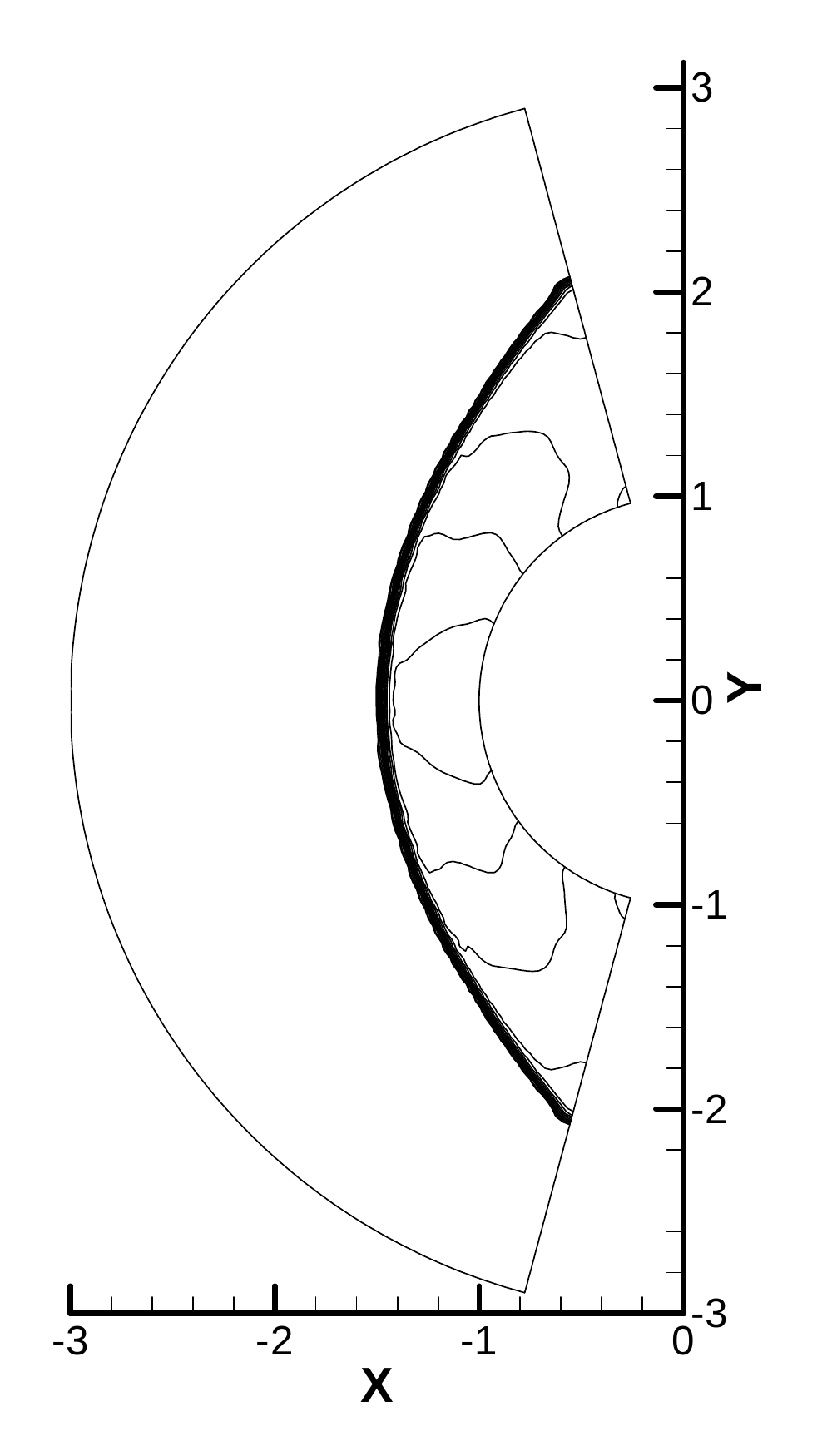}
\end{center} 
\caption{Testcase:Hypersonic flow over a half cylinder at Mach 6 - Mach contours (0.0:0.4:7.6)-  (160x80)- (a)1O-KFDS-A  (b) 1O-KFDS-B (c) 2O-KFDS-A (d) 2O-KFDS-B}
\label{2DEulerTC_KFDS_M6_CYL}  
\end{figure}

\begin{figure} 
\begin{center} 
\includegraphics[width=4cm]{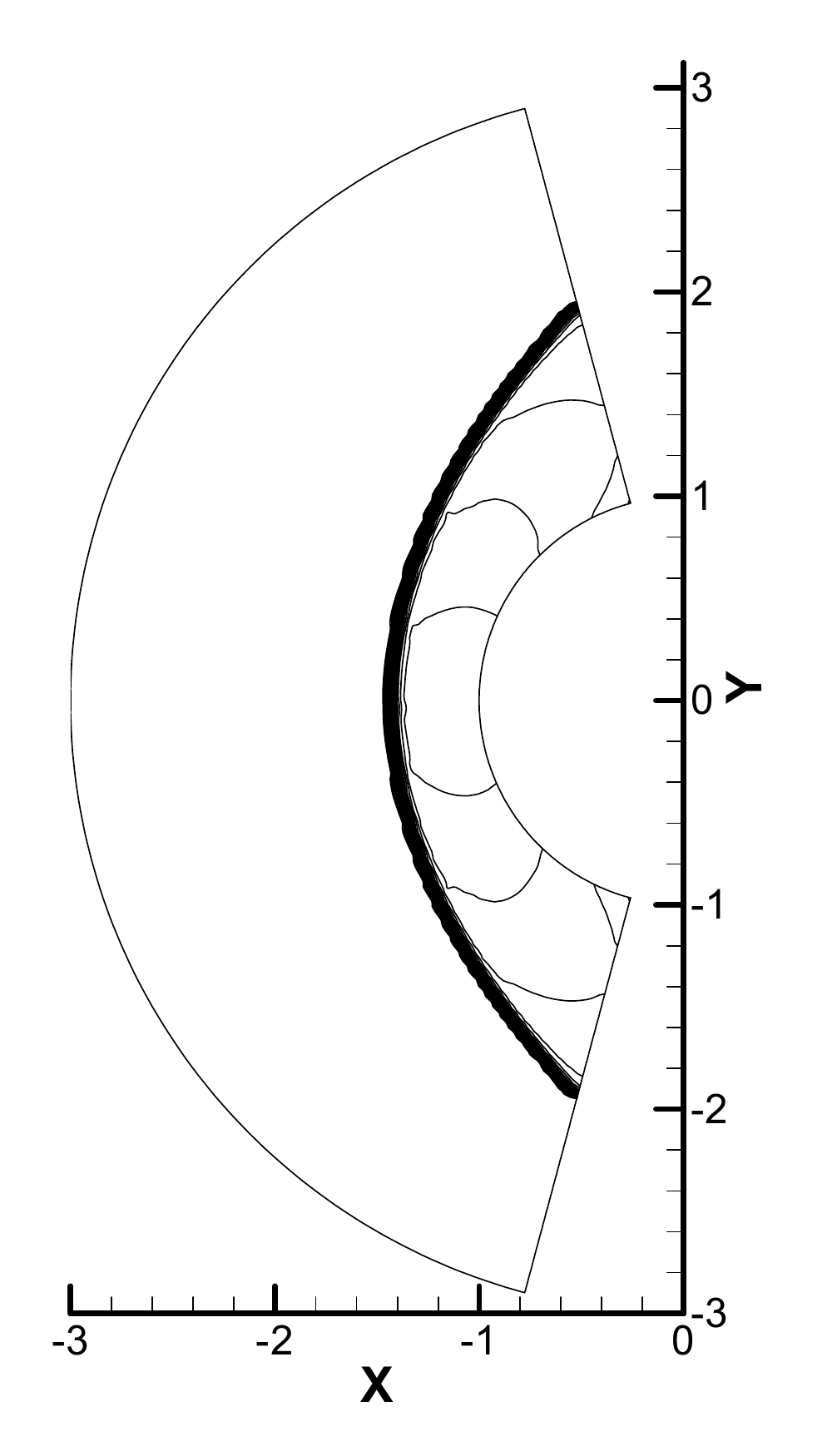}
\includegraphics[width=4cm]{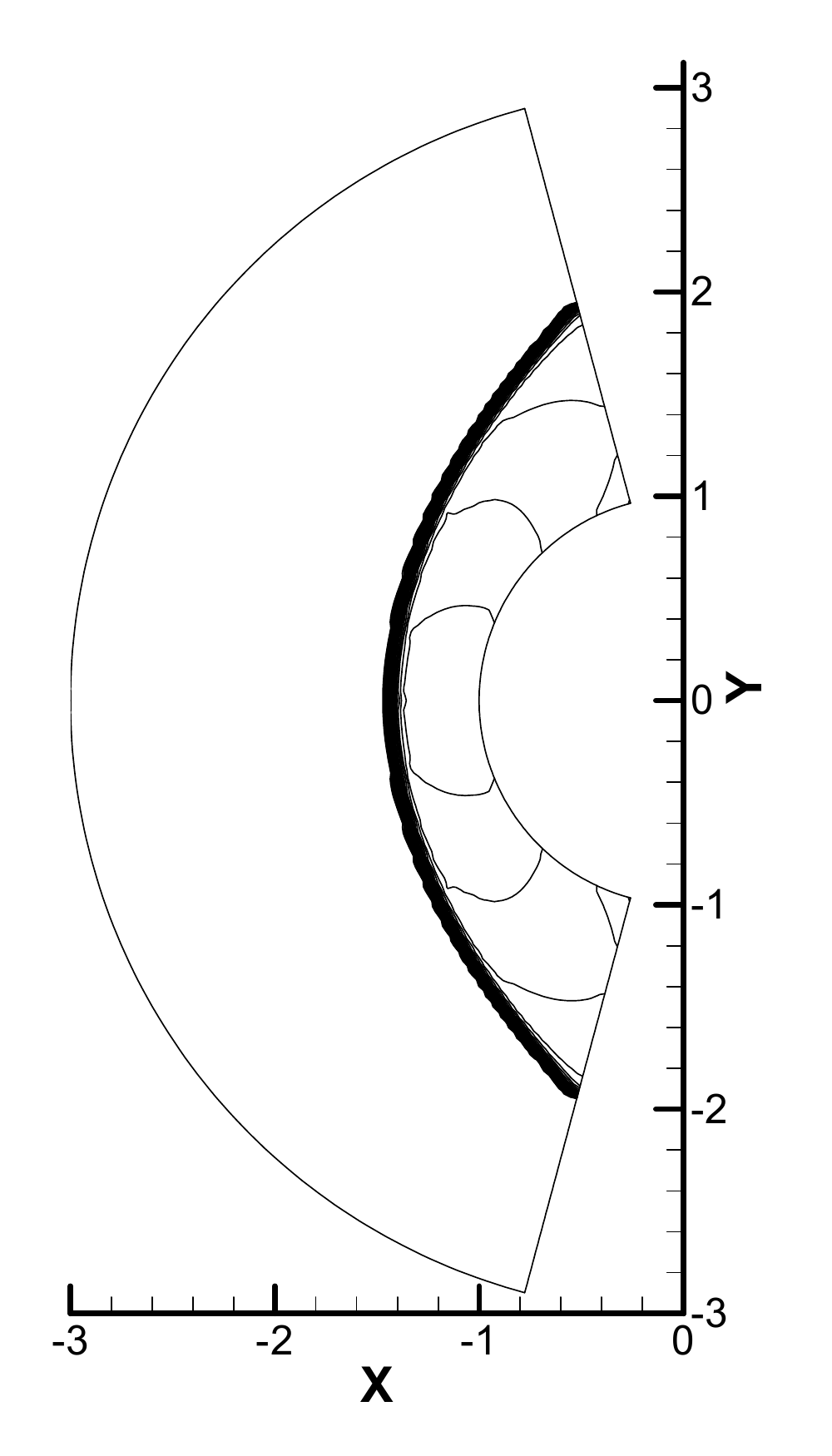}
\includegraphics[width=4cm]{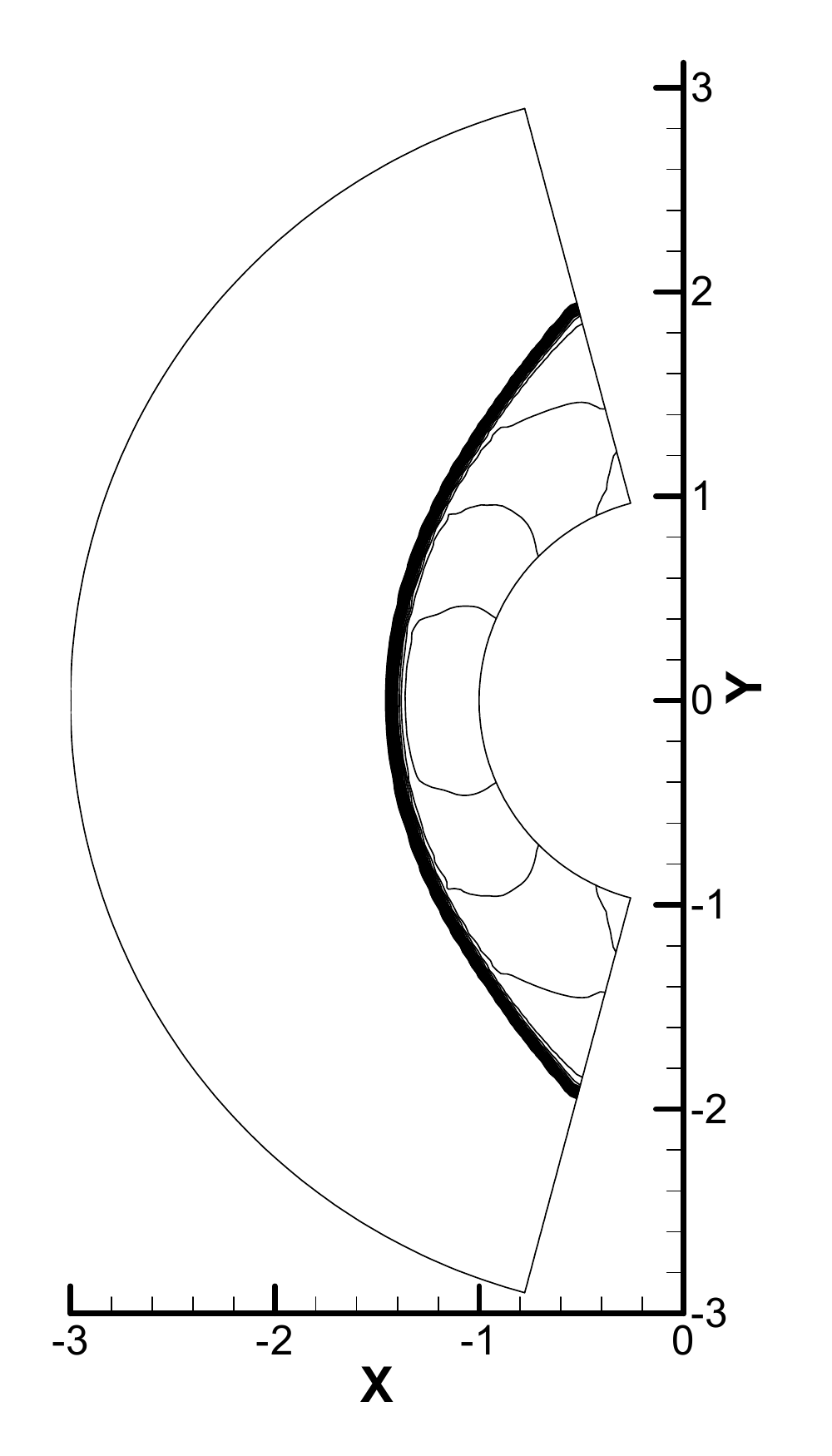}
\includegraphics[width=4cm]{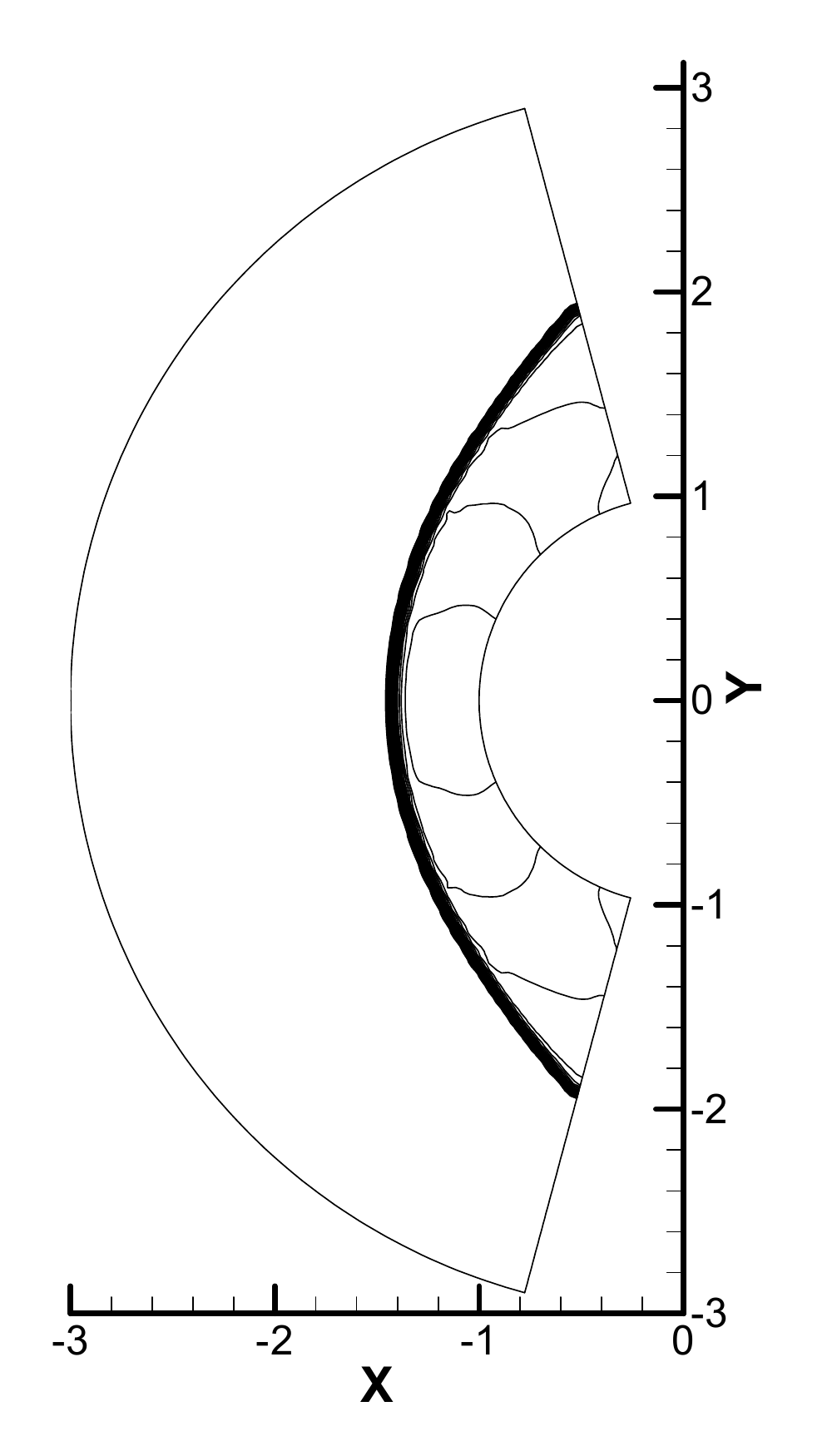}
\end{center} 
\caption{Testcase:Hypersonic flow over a half cylinder at Mach 20 - Mach contours (0.0:0.4:20.0)- (160x80)- (a)1O-KFDS-A  (b) 1O-KFDS-B (c) 2O-KFDS-A (d) 2O-KFDS-B}
\label{2DEulerTC_KFDS_M20_CYL}  
\end{figure}

 This test case involves  hypersonic inflows at Mach 6 and Mach 20 interacting with the leading half side of the cylinder. Ideally the flow would result in the evolution of a bow shock wave which is located upstream of the cylinder and is detached from it. The profile of the shock is dependent on the upstream Mach number. Incidentally some numerical schemes resolve these detached bow shock waves with anomalies known as  carbuncle shock waves, upstream of the bow shock on the stagnation line, and are reported as shock instabilities [\cite{Quirk}].  The density contours for both the versions of KFDS scheme for Mach 6 are shown in Fig.[\ref{2DEulerTC_KFDS_M6_CYL}]. Similarly, The density contours for both the versions of KFDS scheme for Mach 20 are shown in Fig.[\ref{2DEulerTC_KFDS_M20_CYL}]. As can be seen, both the schemes do not exhibit any form of shock instabilities and resolve the bow shocks with reasonable accuracy.

\clearpage 

\begin{figure} 
\begin{center} 
\includegraphics[width=7.5cm,angle=0]{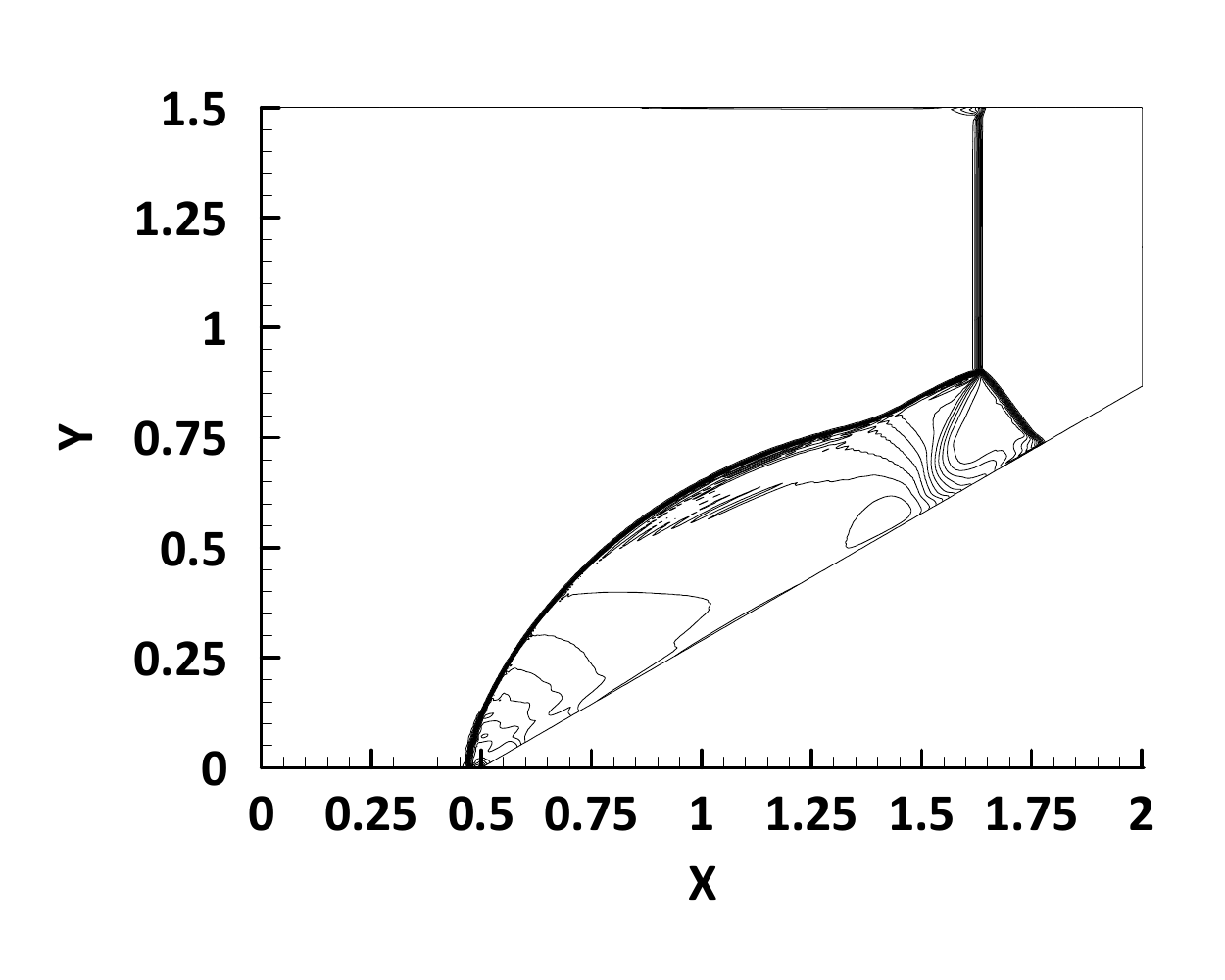}
\includegraphics[width=7.5cm,angle=0]{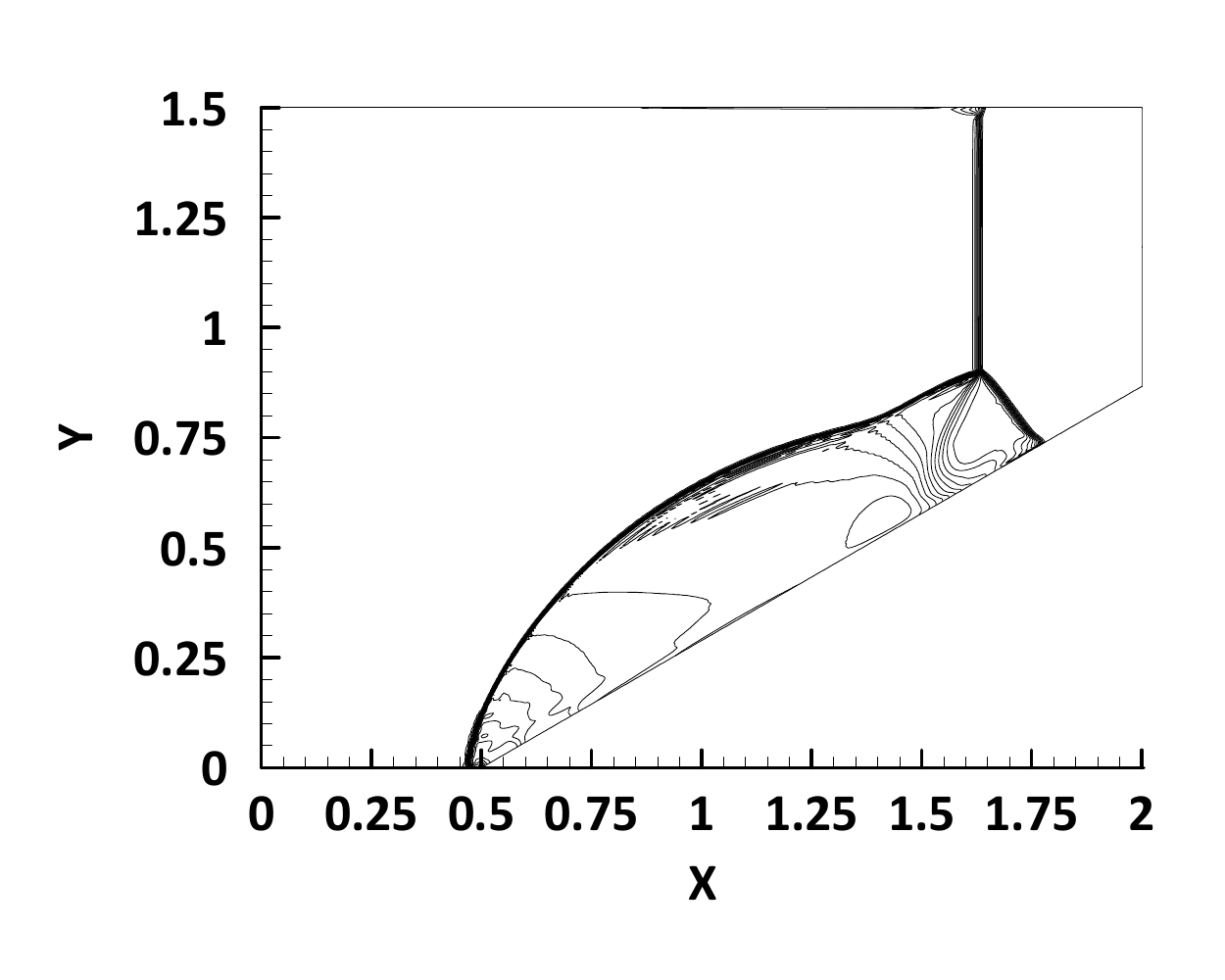}
\includegraphics[width=7.5cm,angle=0]{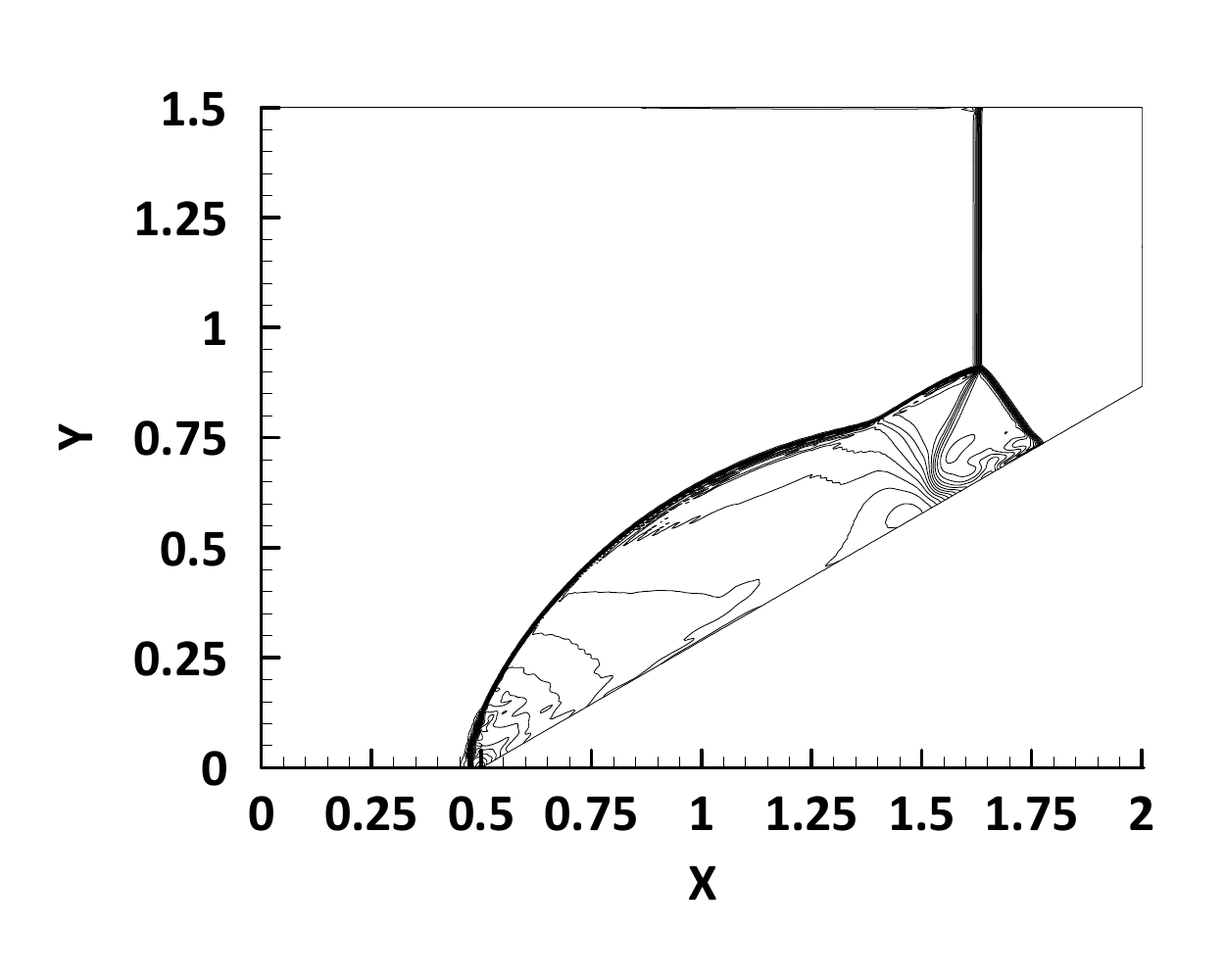}
\includegraphics[width=7.5cm,angle=0]{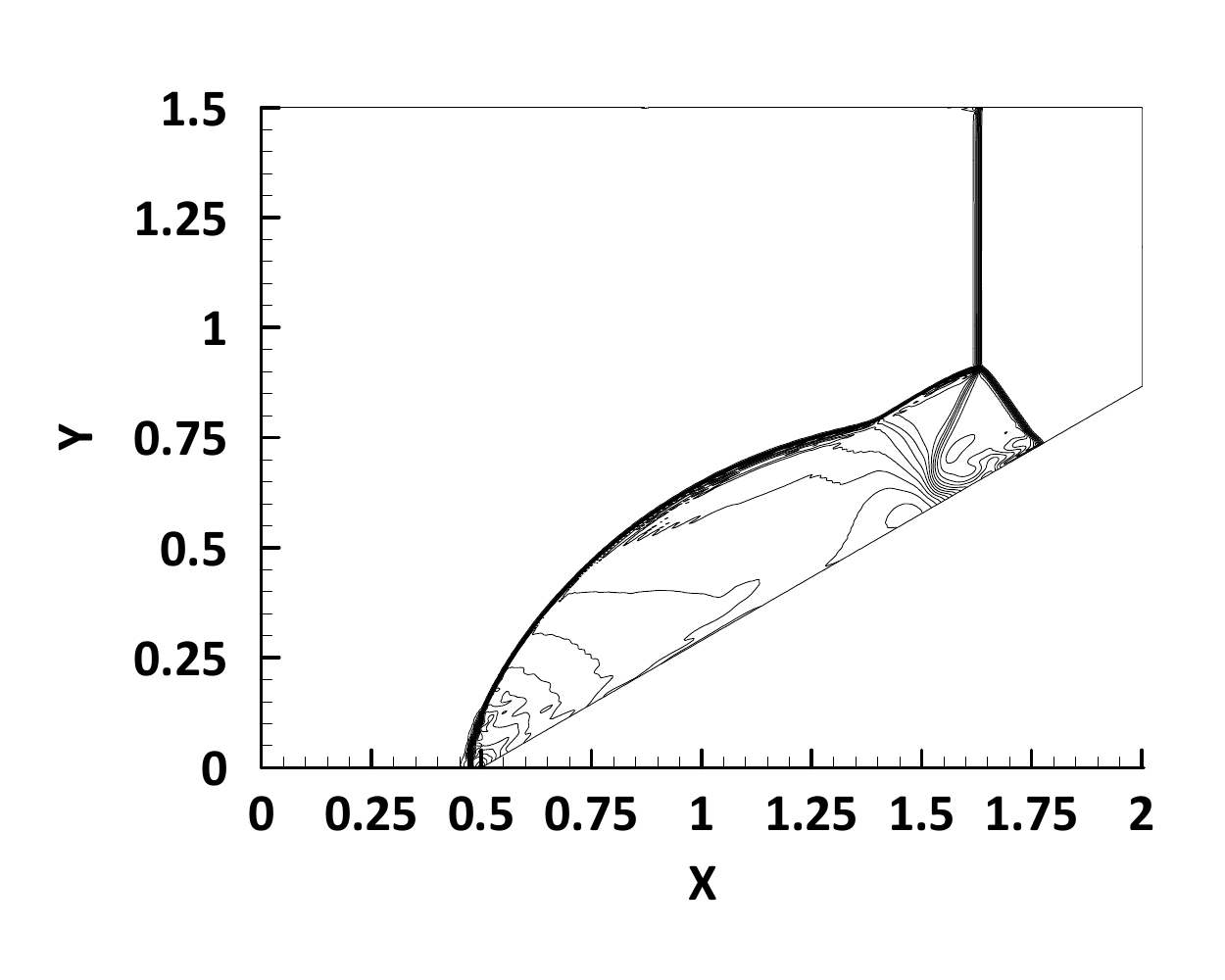}
\end{center} 
\caption{Test case: Double Mach reflection over a wedge - density contours (5.0:0.5:20.0) (400x400)- (a)1O-KFDS-A  (b) 1O-KFDS-B (c) 2O-KFDS-A (d) 2O-KFDS-B}
\label{2DEulerTC_KFDS_WEDGE}  
\end{figure} 
\clearpage
 \newpage
\subsubsection{Planar shock reflection over a wedge} 
This test case [\cite{Quirk}] comprises of a computational domain  $[0,2] \times[0,1.5]$  with a $30^{o}$ wedge positioned from $x = 0.5 $. The test case involves the interaction of a normal shock wave moving at Mach 5.5 with the wedge.  The initial shock is located at x = 0.25 and the computational domain to the the right of x=0.25 is initialized with stationary fluid with density 1.4 and pressure 1. The domain to the left of the shock is initialized with primitive variable values corresponding to the shock as obtained from the moving shock relations.  The supersonic nature of the inflow results in an evolution of an oblique shock at the root of the wedge which interacts with the moving normal shock and  gets reflected, thereby forming a triple point of shocks.  Essentially this test case evaluates the ability of the numerical scheme to handle shock instabilities and to provide a physically realistic solution. Many popular numerical schemes produce kinked mach stems near the triple point which are unphysical. 
 The results obtained from both the versions of KFDS scheme at t= 0.25s are shown in Fig.[\ref{2DEulerTC_KFDS_WEDGE}]. Both versions of KFDS scheme capture  the triple point well and do not produce unphysical kinked Mach stems. 

\begin{figure} 
\begin{center} 
\includegraphics[width=6cm,angle=0]{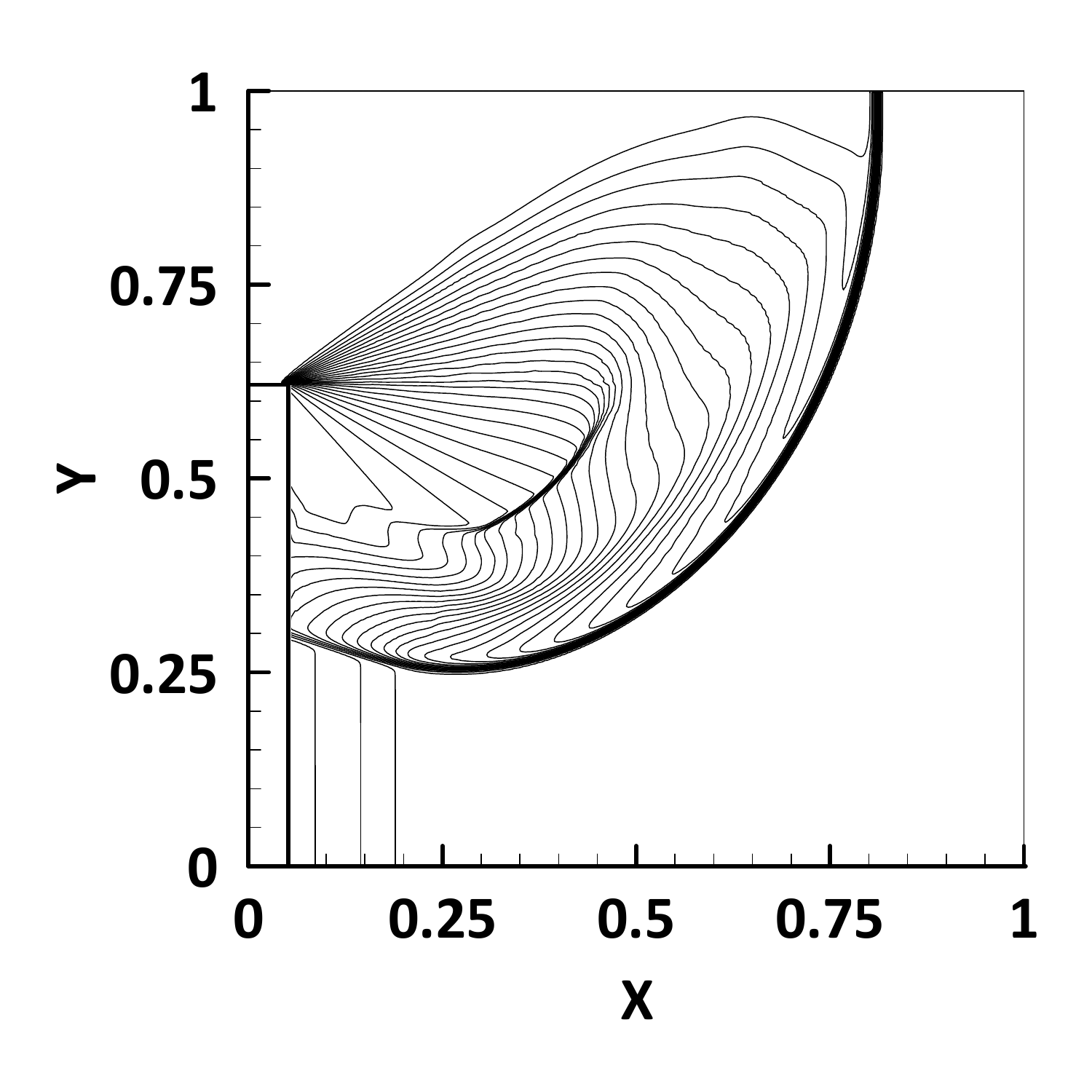}
\includegraphics[width=6cm,angle=0]{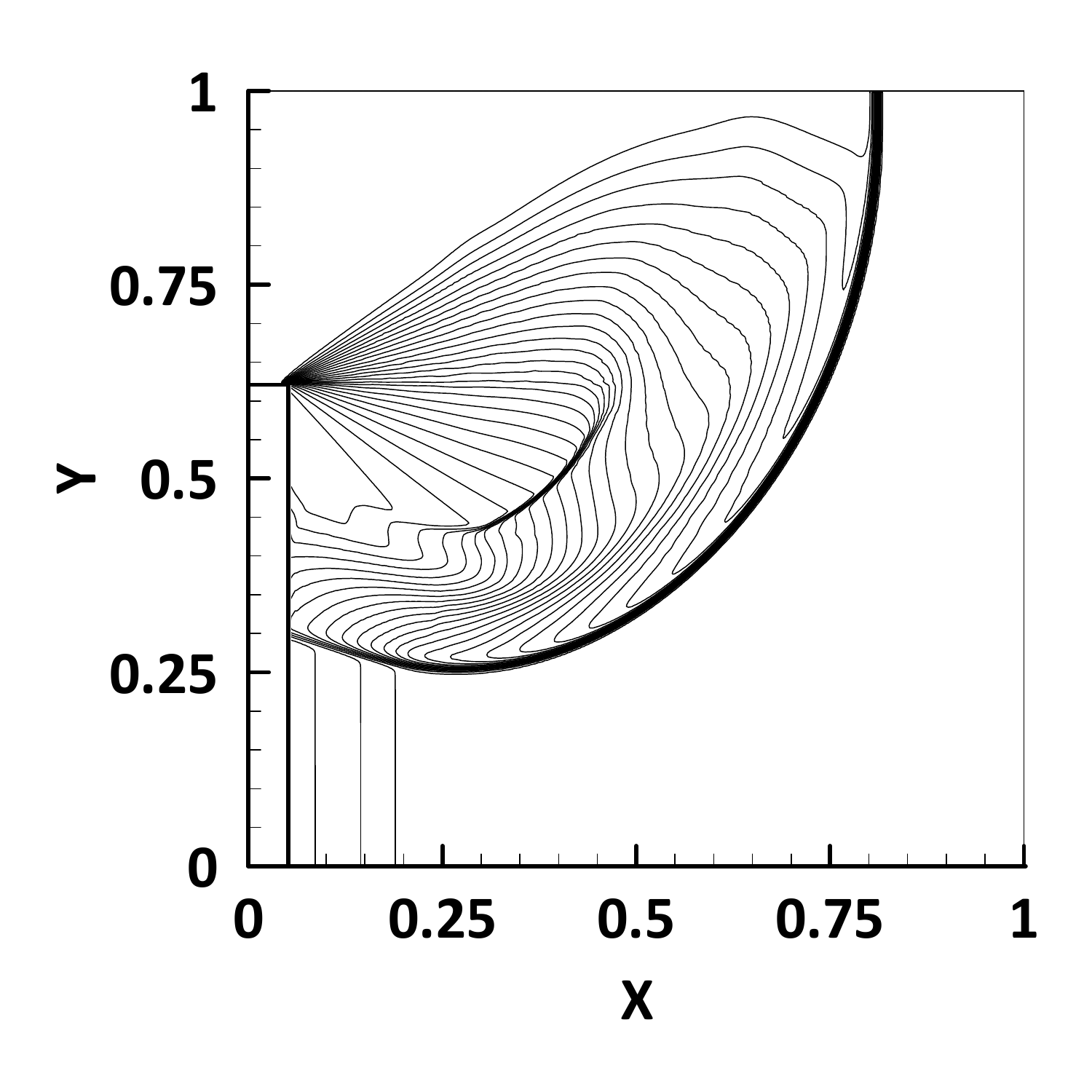}
\includegraphics[width=6cm,angle=0]{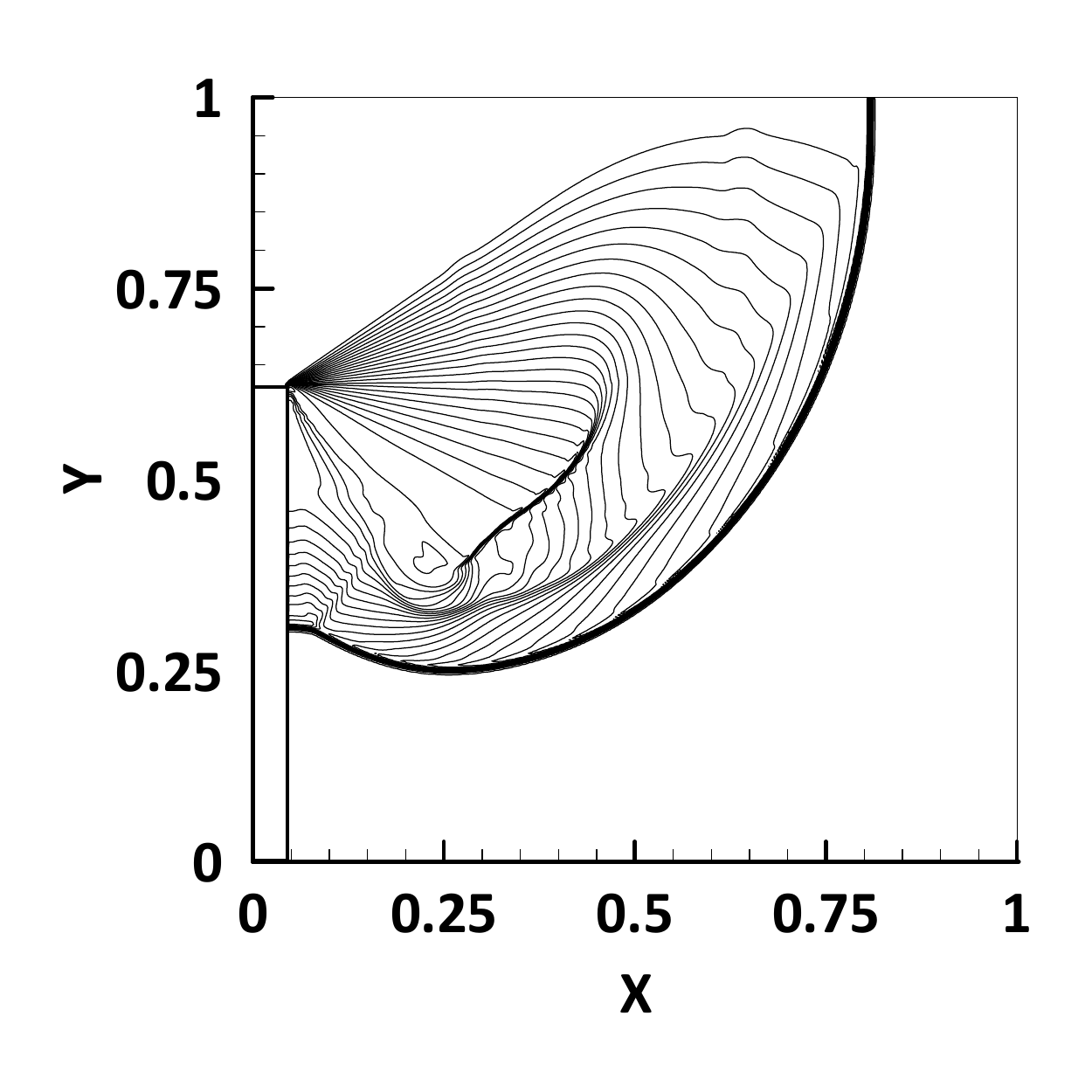}
\includegraphics[width=6cm,angle=0]{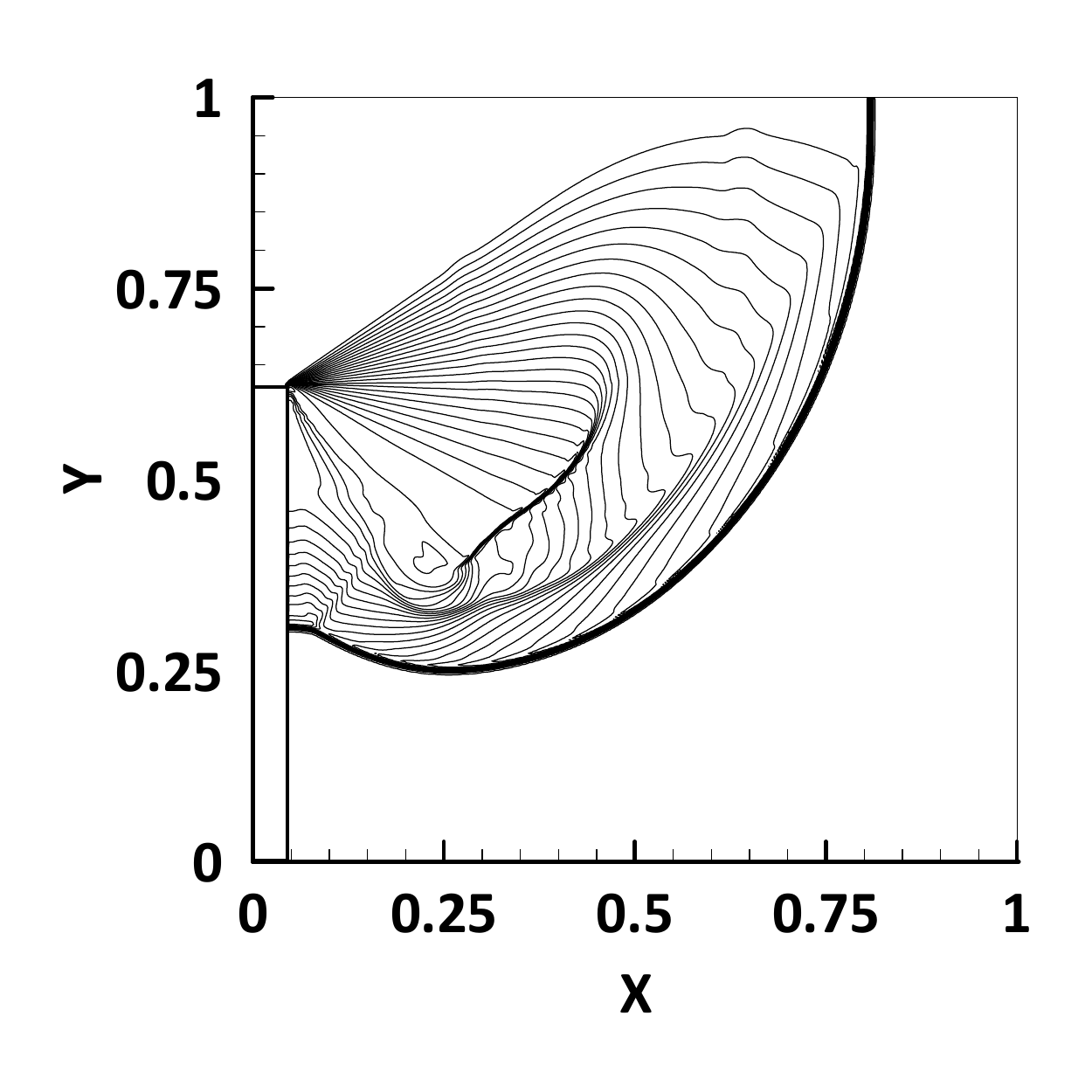}
\end{center} 
\caption{Test case: Shock Diffraction in a sudden expansion region- density contours (0.5:0.25:6.75) (400x400)- (a)1O-KFDS-A  (b) 1O-KFDS-B (c) 2O-KFDS-A (d) 2O-KFDS-B}
\label{2DEulerTC_KFDS_SHKDIFF}  
\end{figure} 

\subsubsection{Shock diffraction}
This test case [\cite{Quirk}] is important to assess the presence of shock instabilities and unphysical features in the strong expansion fan appearing with some numerical schemes. The test case involves the forward travel of a strong shock  with incident Mach Number 5.09 wherein the domain comprises of a sudden expansion around a $90^{0}$ corner.  The shock diffracts around the corner [X=0.05, Y=0.625]. At the corner, the sudden area expansion leads to a strong expansion fan.  Both the above nonlinear waves further interact.  For this test case we discretize the computational domain [0,1][0,1] into a 400x400 grid. The initial shock position is specified at x=0.05. The domain is initialized with the initial conditions $(\rho, u, v, p)=[1.4,0,0,1.0]$ to the left of the shock and with post shock conditions to the right of the shock (computed from moving shock relations).  The results obtained from the KFDS schemes at $t = 0.1561 s$ are shown in Fig.[\ref{2DEulerTC_KFDS_SHKDIFF}]. As can be observed, the both versions of KFDS scheme resolve the flow features arising due to the strong initial gradients well and do not produce any anomalies.  

\subsection{2D viscous flow test cases} 
\subsubsection{Blasius flow}
\begin{figure} 
\begin{center}  
\includegraphics[width=7.5cm]{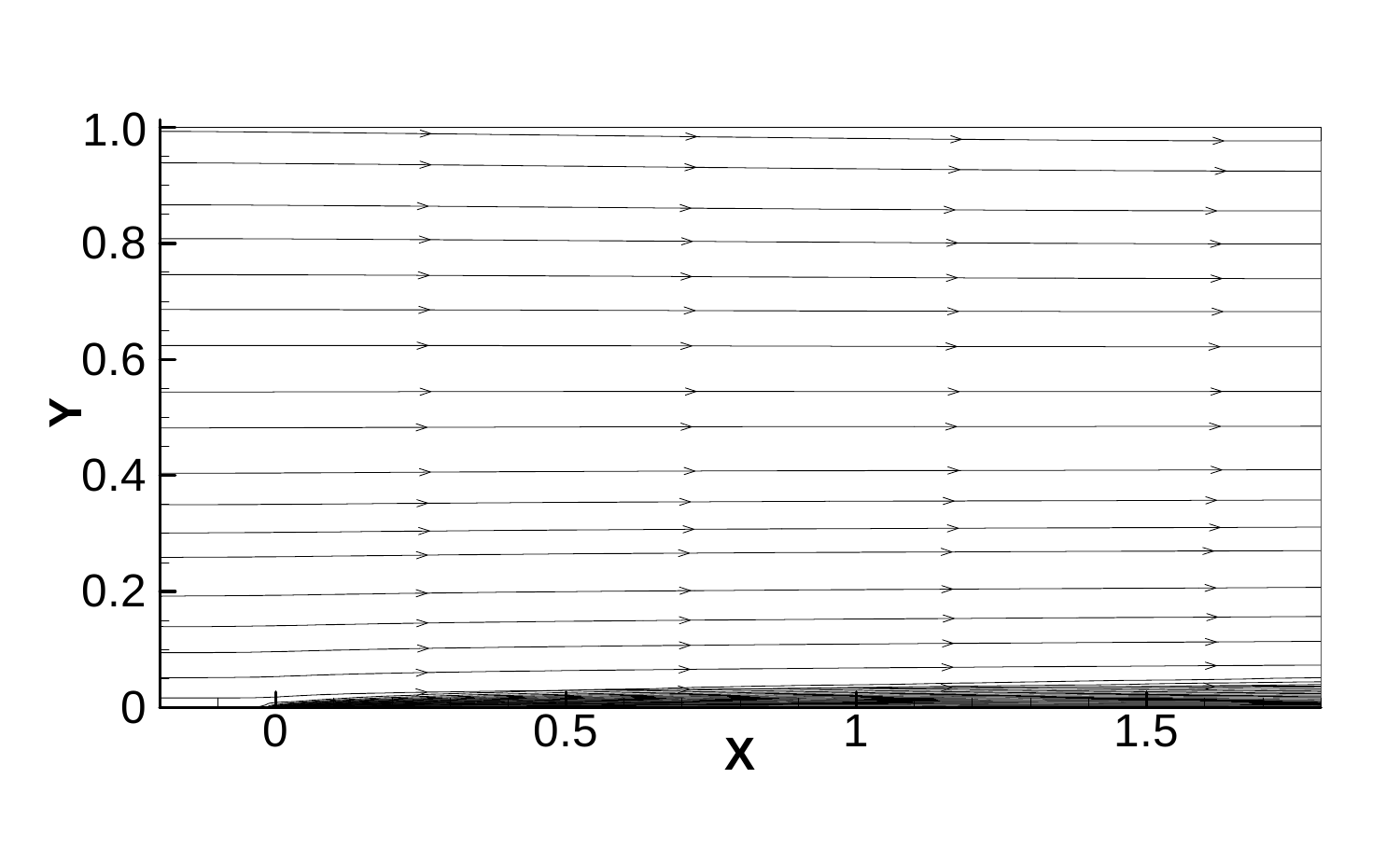}
\includegraphics[width=7.5cm]{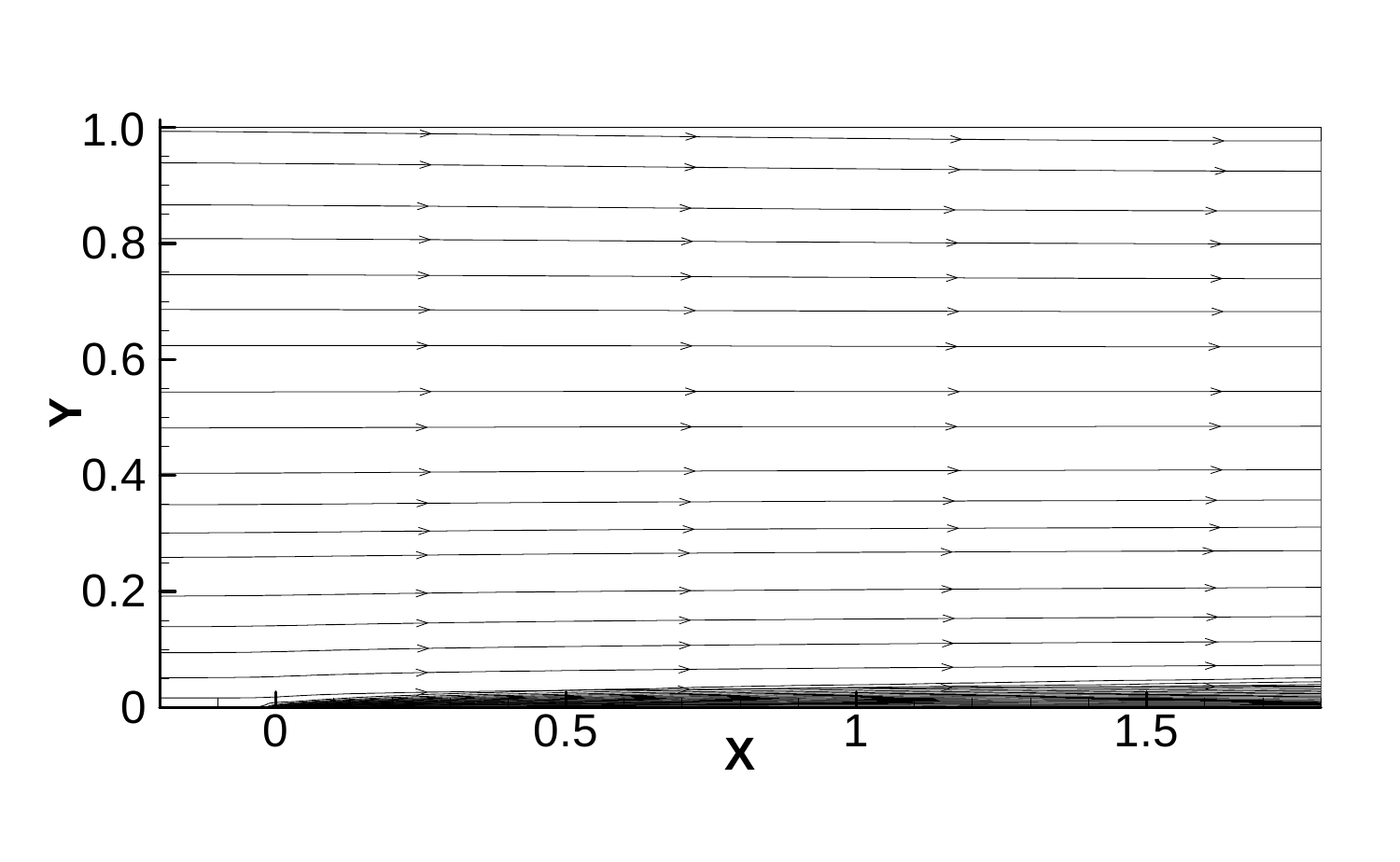}
\includegraphics[width=7.5cm]{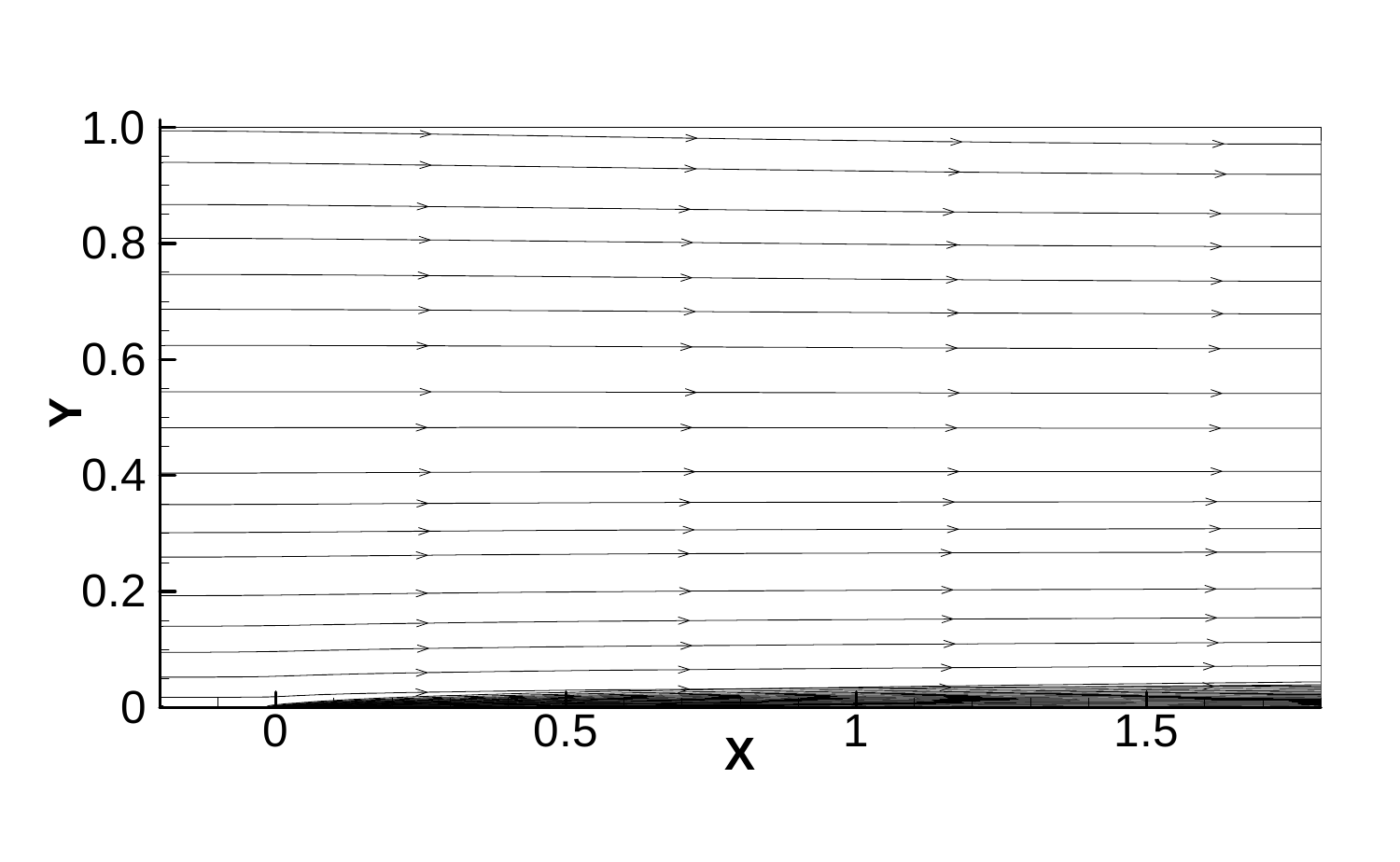}
\includegraphics[width=7.5cm]{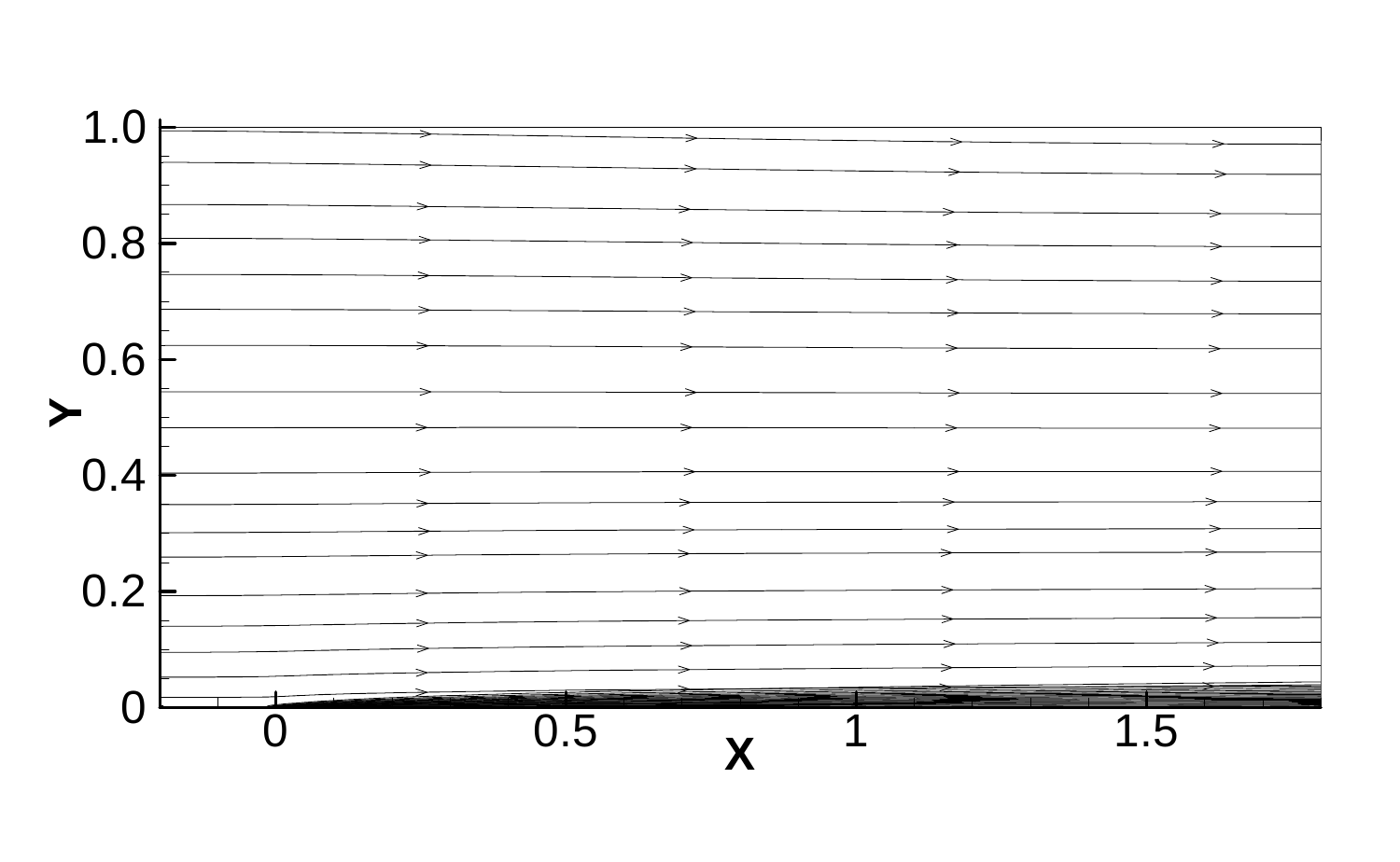}
\end{center} 
\caption{Test case: Blasius flow over a flat plate (105x65)- (a)1O-KFDS-A  (b) 1O-KFDS-B (c) 2O-KFDS-A (d) 2O-KFDS-B}
\label{2D_NS_TC_KFDS_Blasius}  
\end{figure}

\begin{figure} 
\begin{center} 
\includegraphics[width=6.5cm]{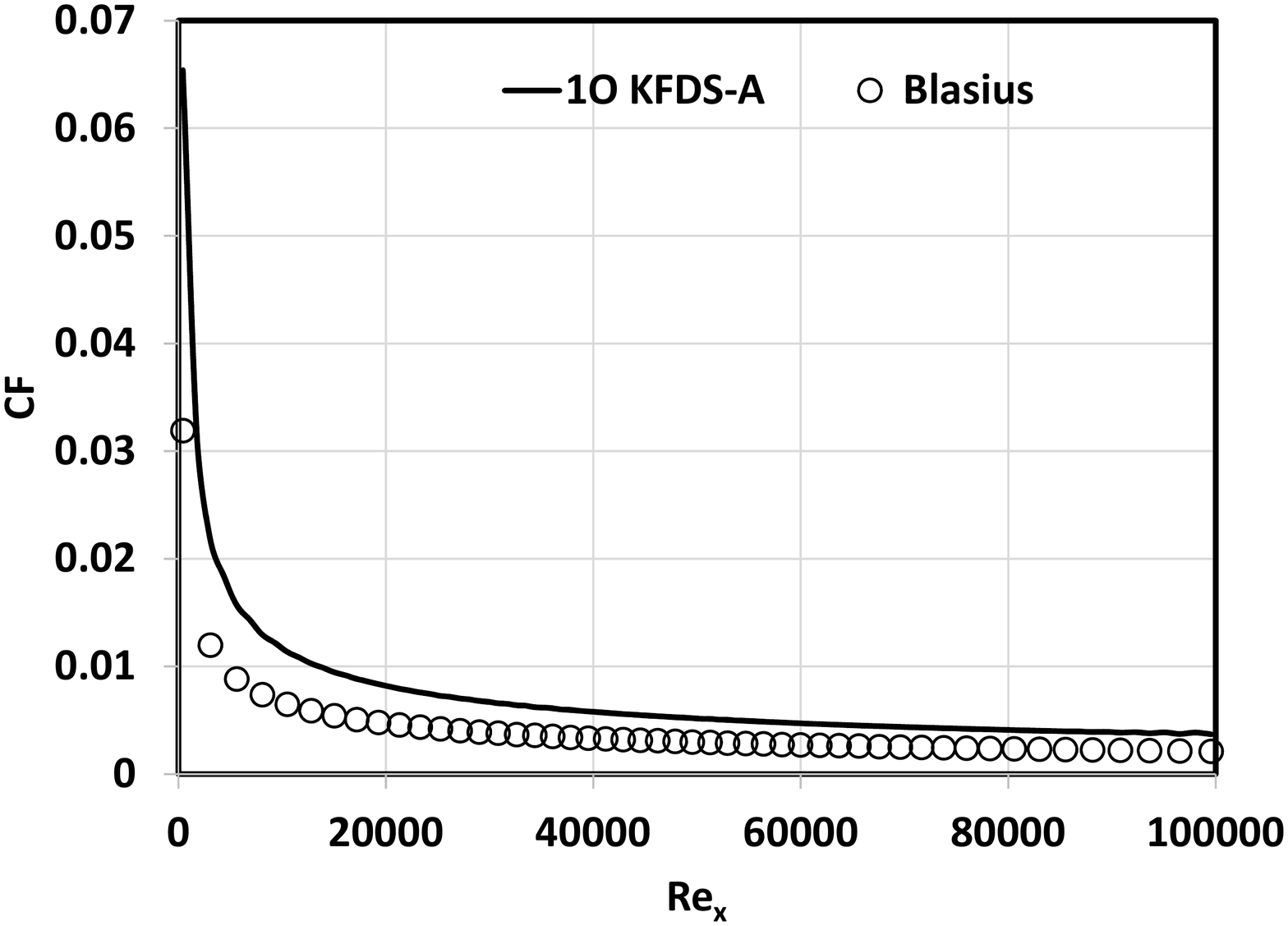}
\includegraphics[width=6.5cm]{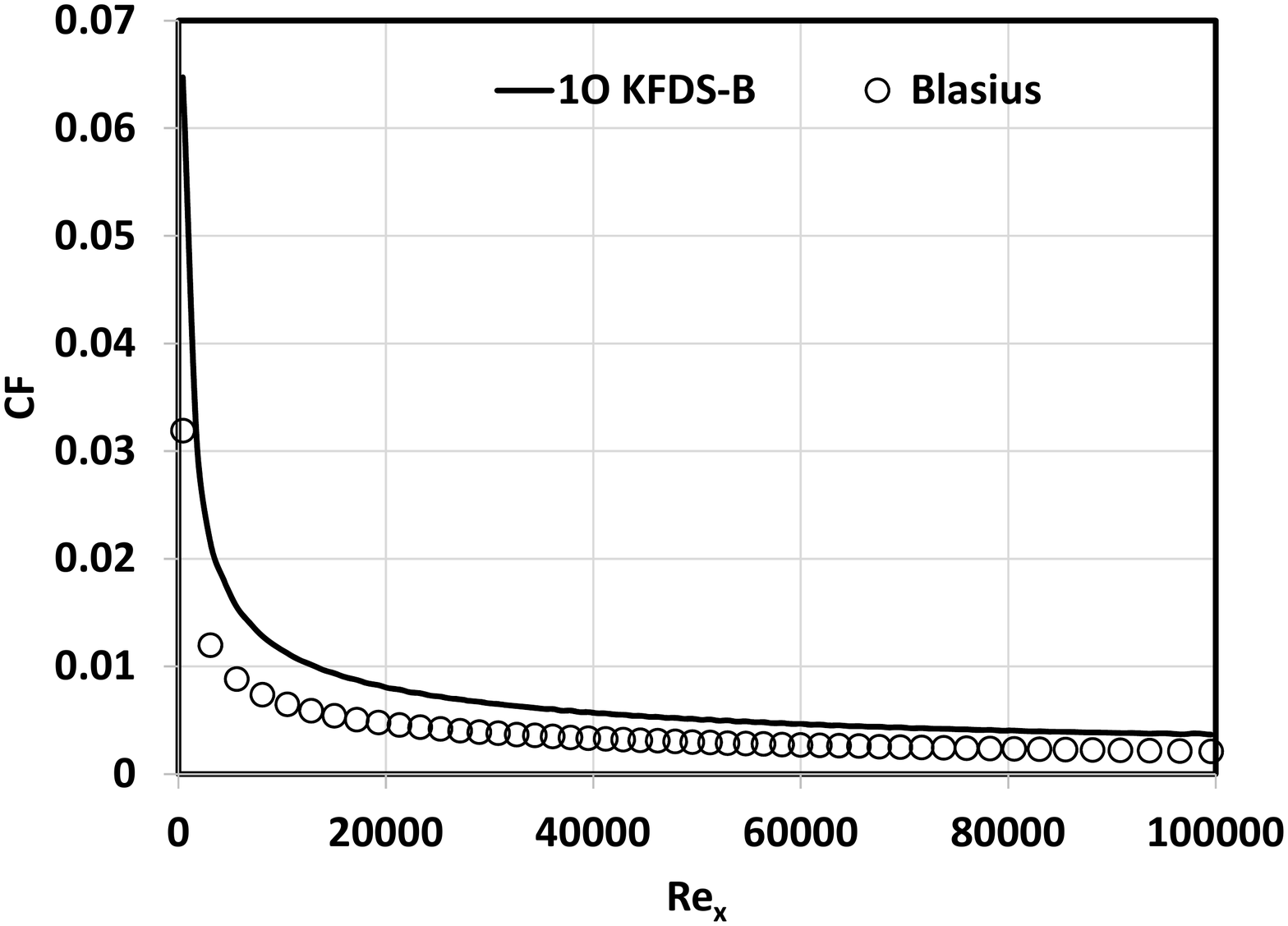}
\includegraphics[width=6.5cm]{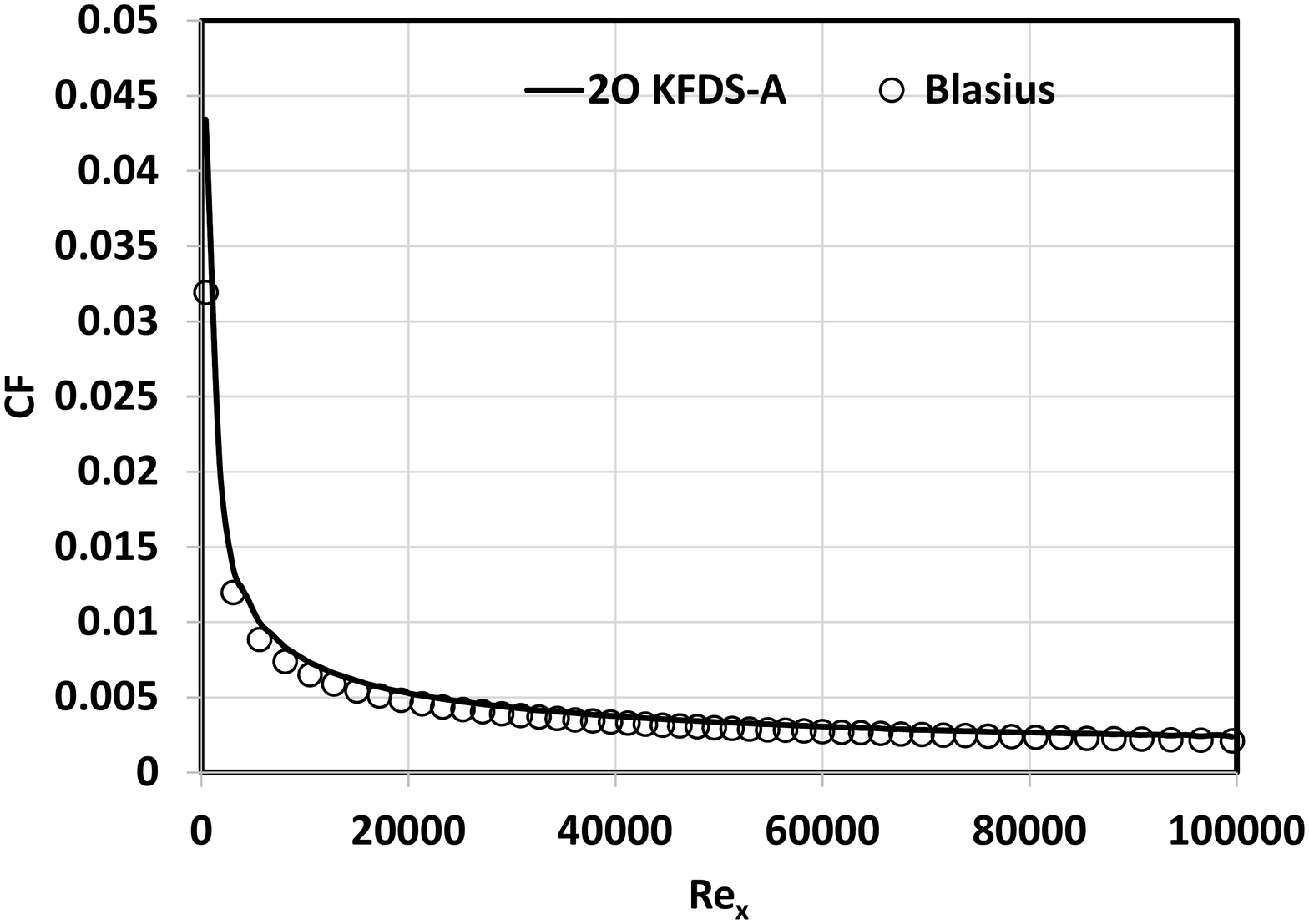}
\includegraphics[width=6.5cm]{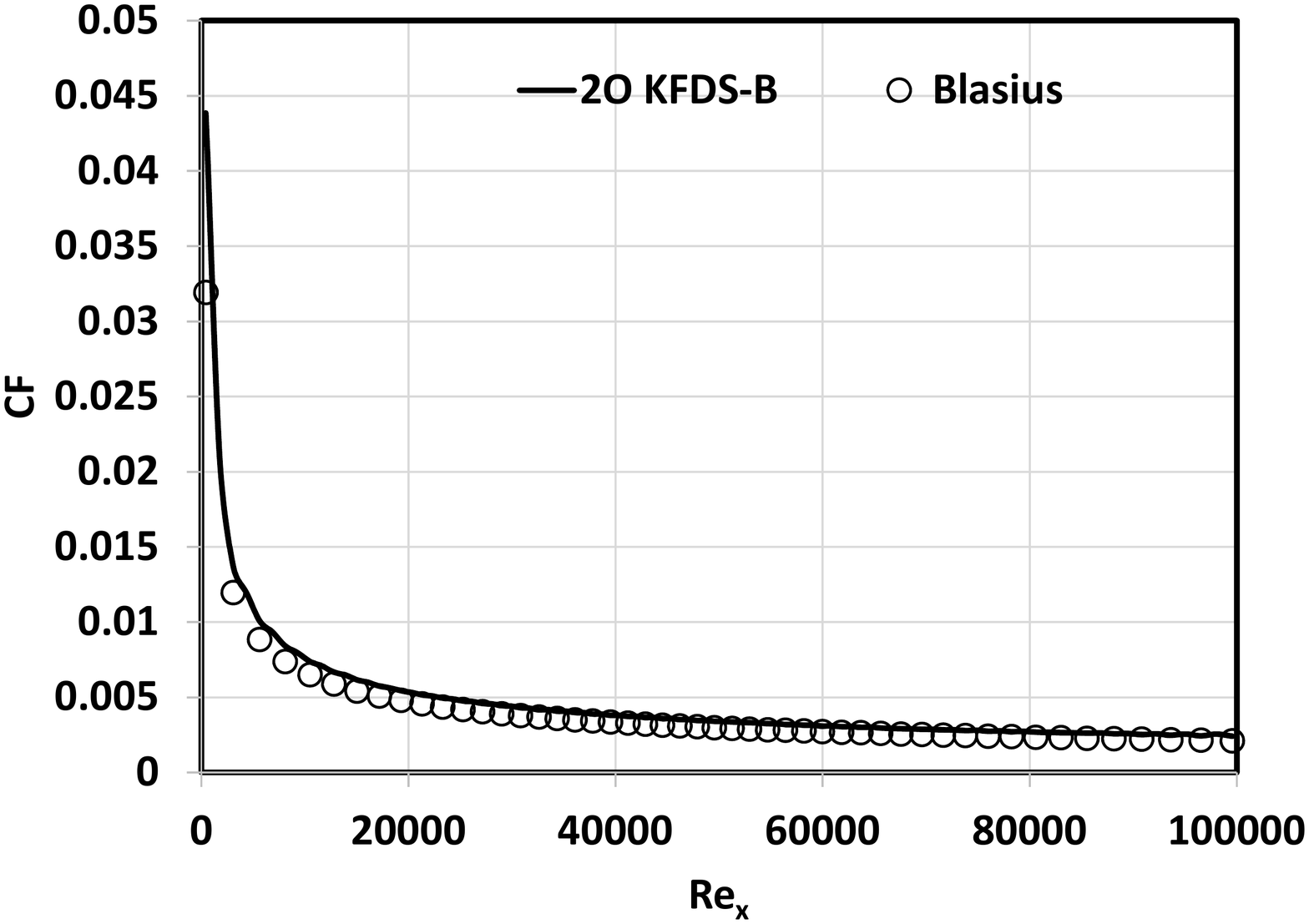}
\end{center} 
\caption{Test case: Blasius flow over a flat plate (105x65)- $C_f$ plots - (a)1O-KFDS-A  (b) 1O-KFDS-B (c) 2O-KFDS-A (d) 2O-KFDS-B}
\label{2D_NS_TC_KFDS_BlasiusCF}  
\end{figure}

\begin{figure} 
\begin{center} 
\includegraphics[width=6.0cm]{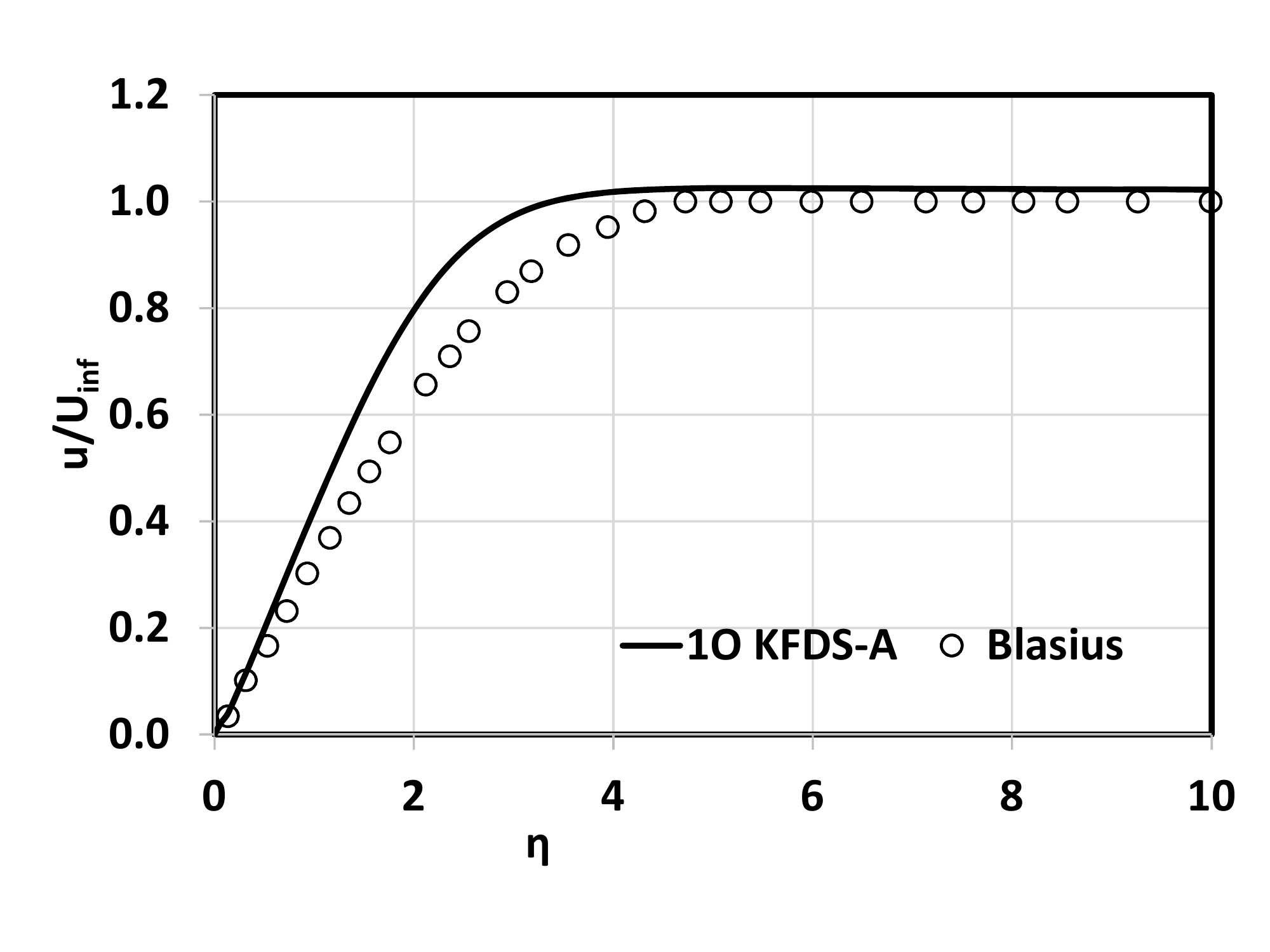}
\includegraphics[width=6.0cm]{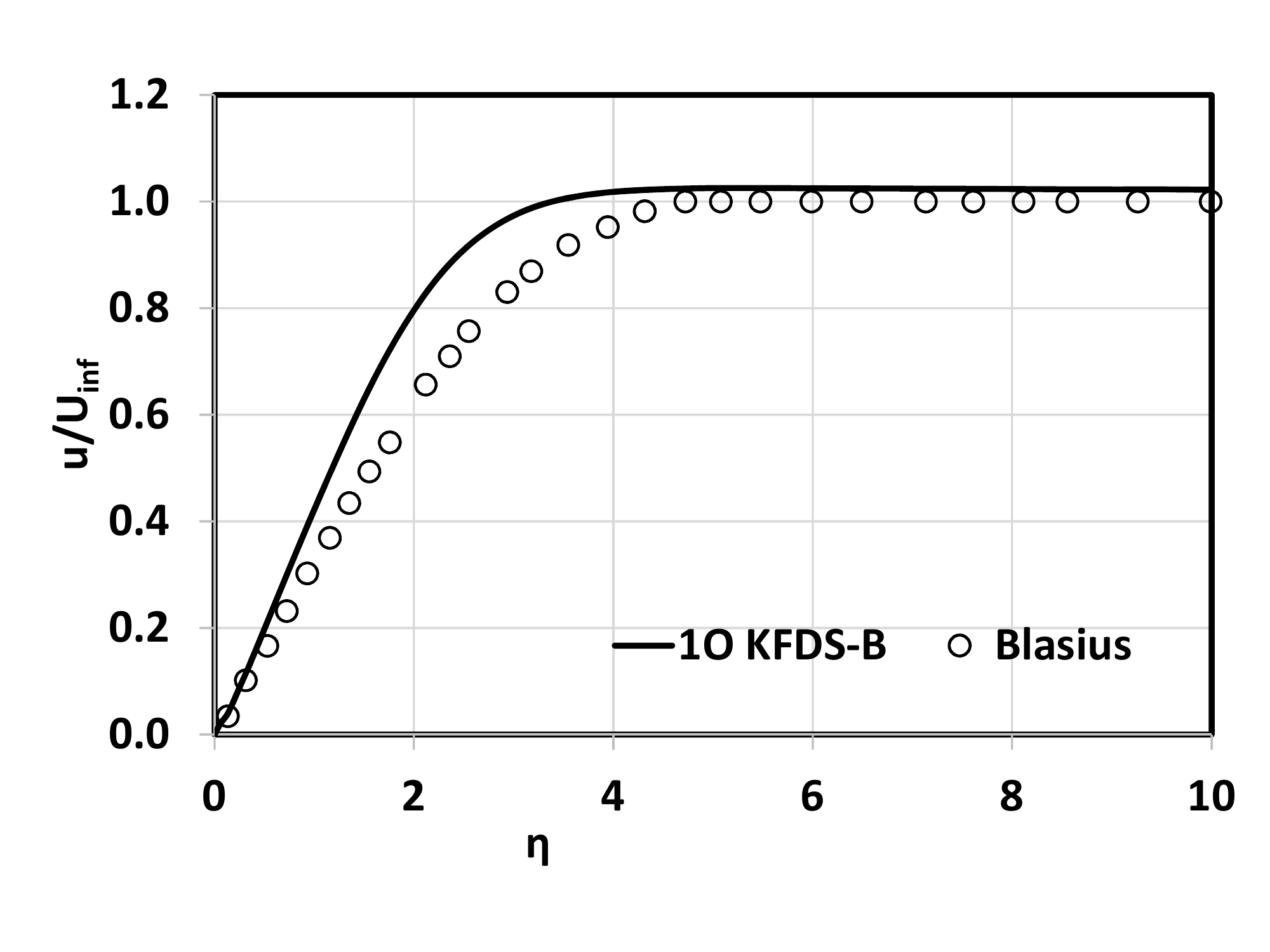}
\includegraphics[width=6.0cm]{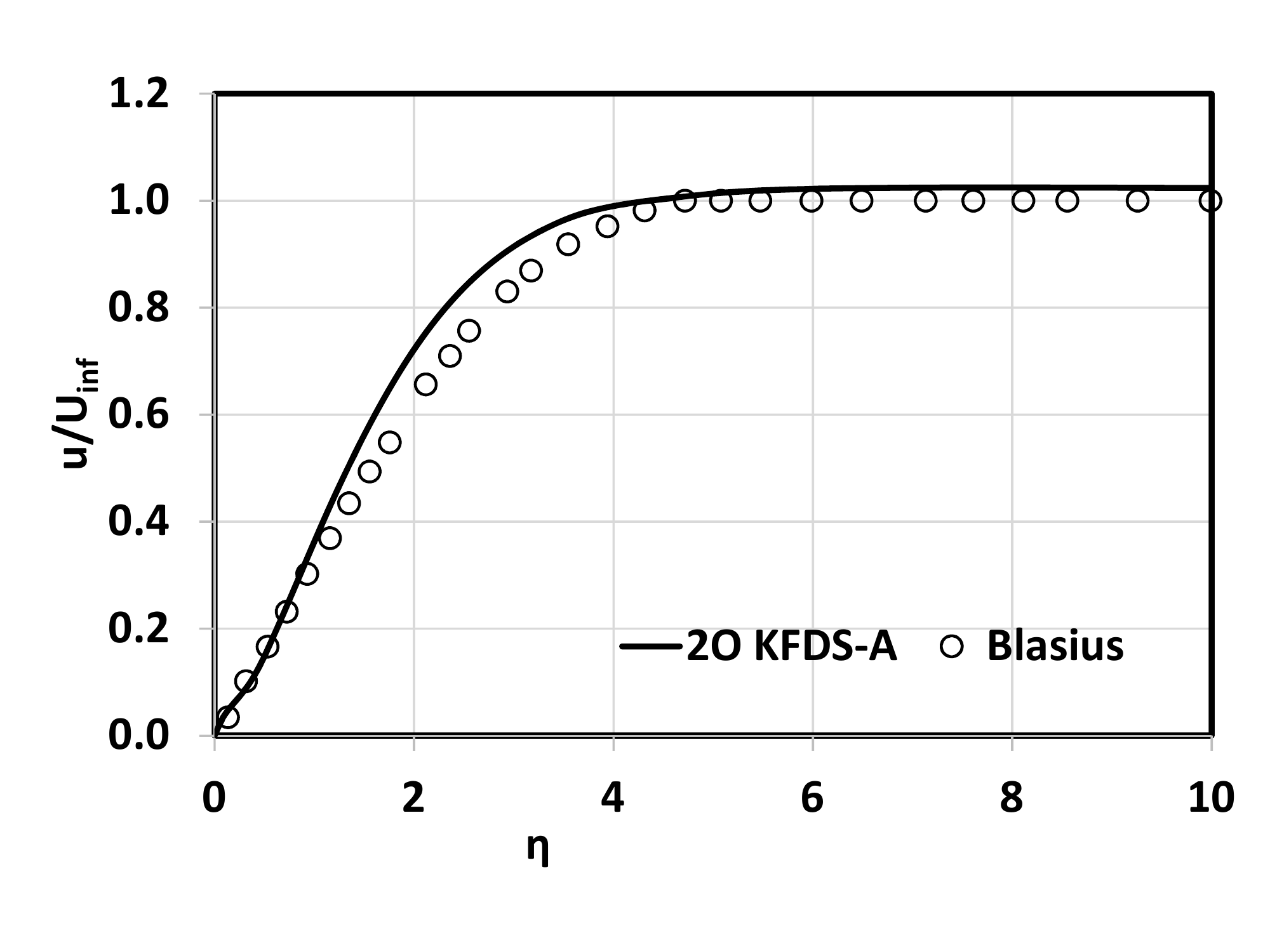}
\includegraphics[width=6.0cm]{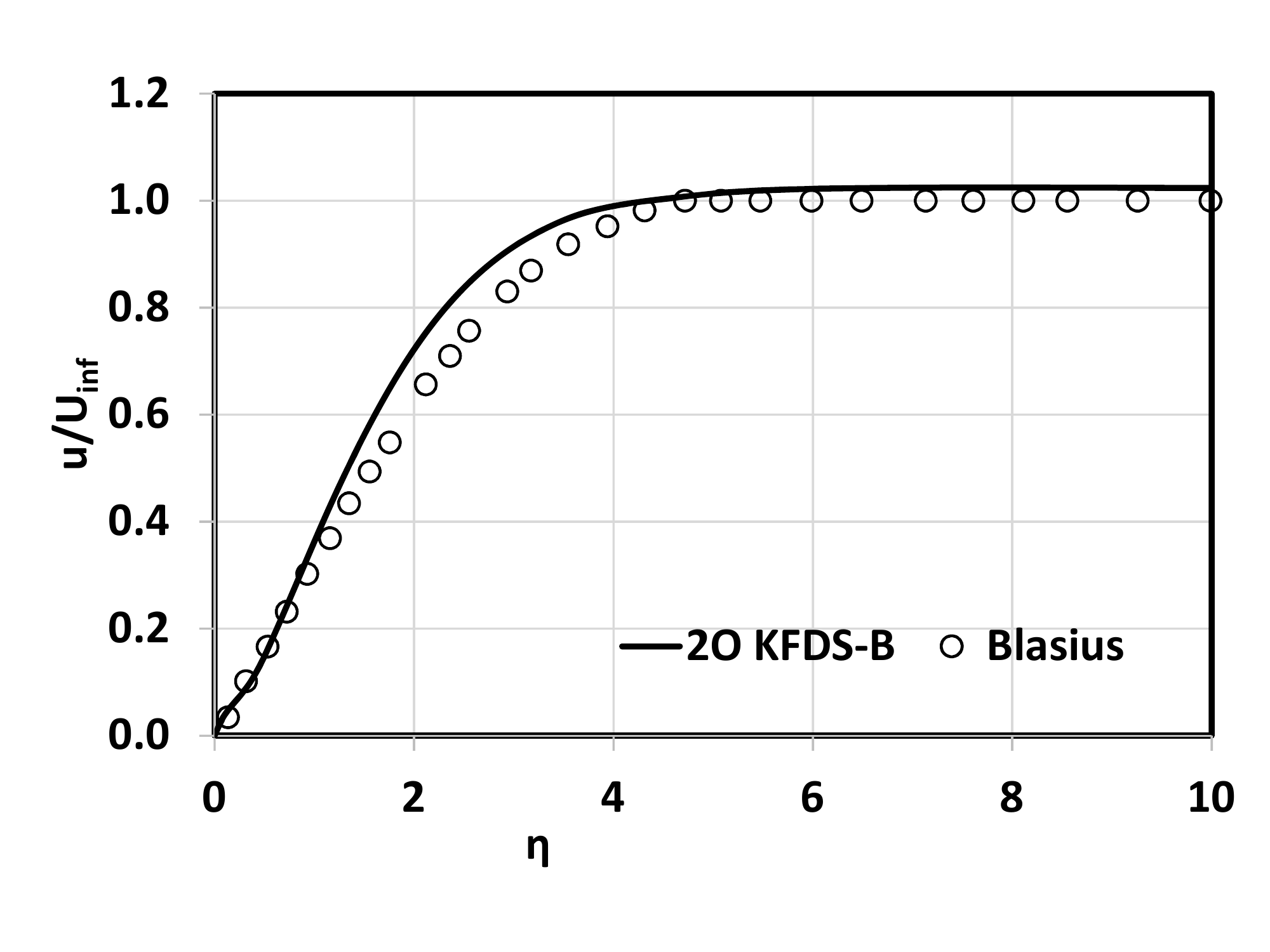}
\end{center} 
\caption{Test case: Blasius flow over a flat plate (105x65)- Velocity Profiles - (a)1O-KFDS-A  (b) 1O-KFDS-B (c) 2O-KFDS-A (d) 2O-KFDS-B}
\label{2D_NS_TC_KFDS_BlasiusV}  
\end{figure}

This test problem is essentially a validation for the viscous part in a numerical scheme. The test involves viscous laminar flow over a flat plate at zero angle of incidence.  The inlet has a freestream Mach number of 0.15 and Reynolds number 10000. The computational domain $[-0.2,1.8] \times[0,1]$ is rectangular with the flat plate on the bottom side from $x=0$ to $x=1.8$ defined as a viscous wall.  The inlet is defined with uniform total pressure and total temperature with zero vertical velocity component.  The top and the outlet are defined with free stream boundary conditions.  The remaining portion upstream of the plate is defined with symmetry boundary condition. The boundary layer evolves over the viscous wall and the grows in thickness in proportion to the length of the flat plate.  A stretched grid of $105 \times 65$, with a geometrically increasing ratio of $1.025$ is used.  The boundary layer profile for each version of KFDS scheme is shown in Fig.[\ref{2D_NS_TC_KFDS_Blasius}]. The skin friction distribution and the velocity profiles are shown in Fig.[\ref{2D_NS_TC_KFDS_BlasiusCF}] and Fig.[\ref{2D_NS_TC_KFDS_BlasiusV}] respectively.  Comparison with analytical solution (Blassius profile) is provided.  Second order versions of the KFDS schemes agree well with the analytical results, while the effect of numerical diffusion can be seen in the first order results.   

\clearpage

\newpage

\begin{figure} 
\begin{center} 
\includegraphics[width=6.5cm]{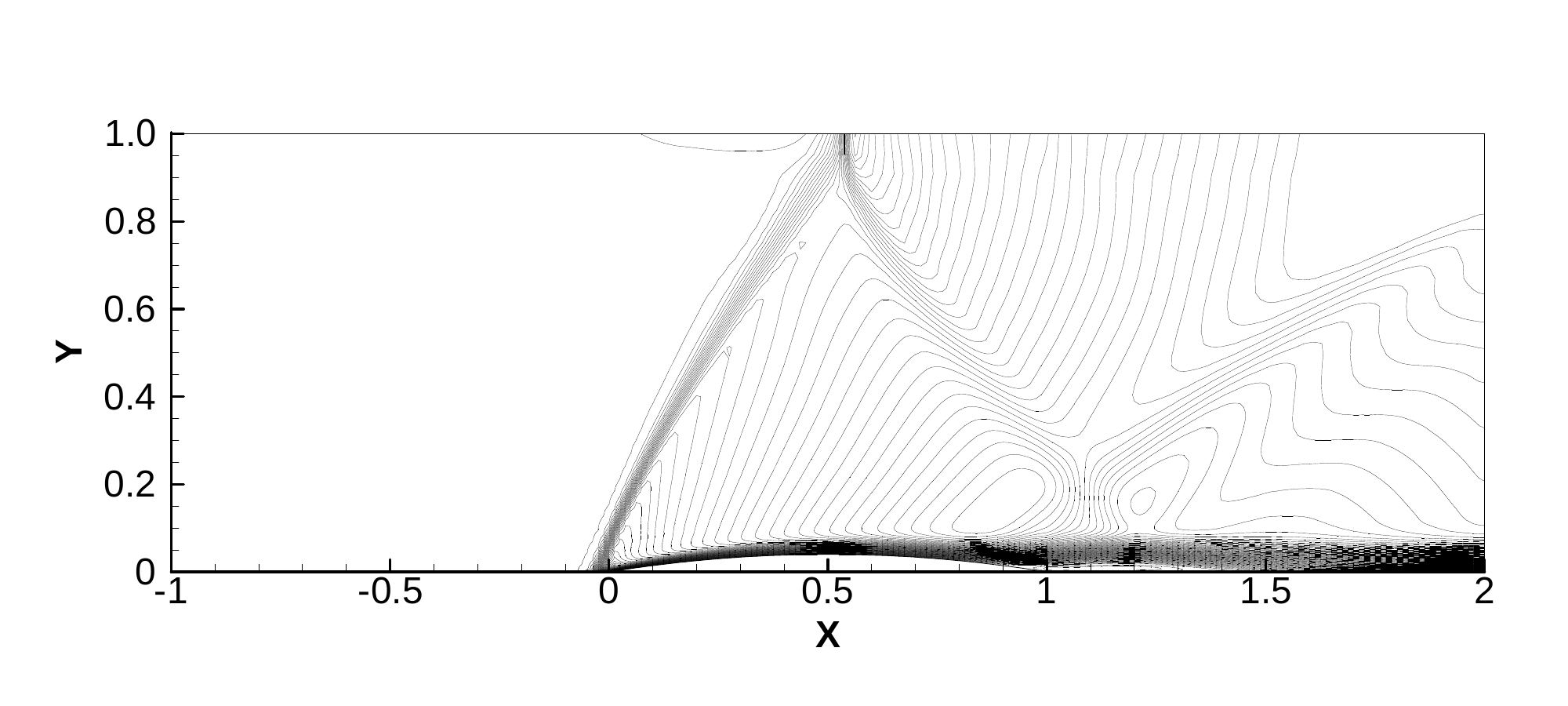}
\includegraphics[width=6.5cm]{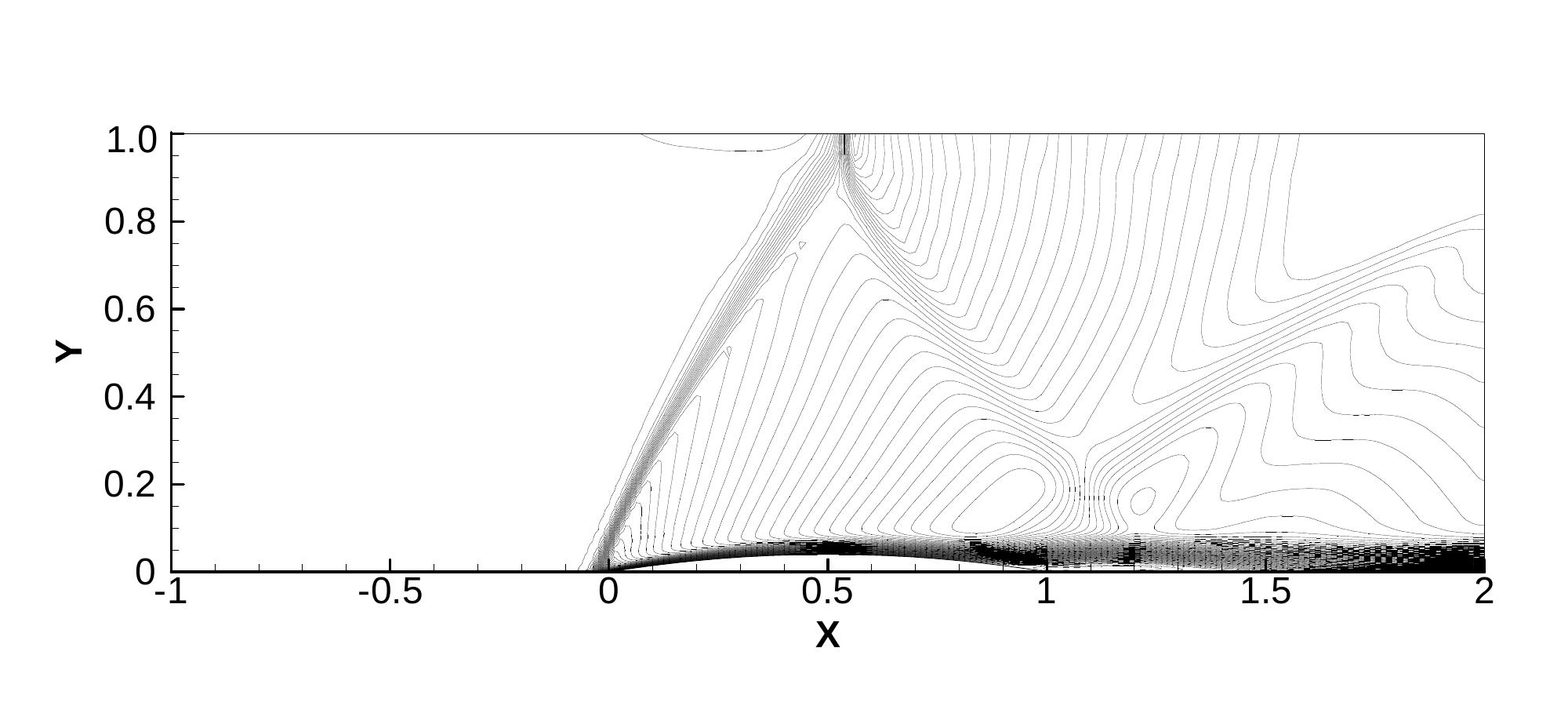}
\includegraphics[width=6.5cm]{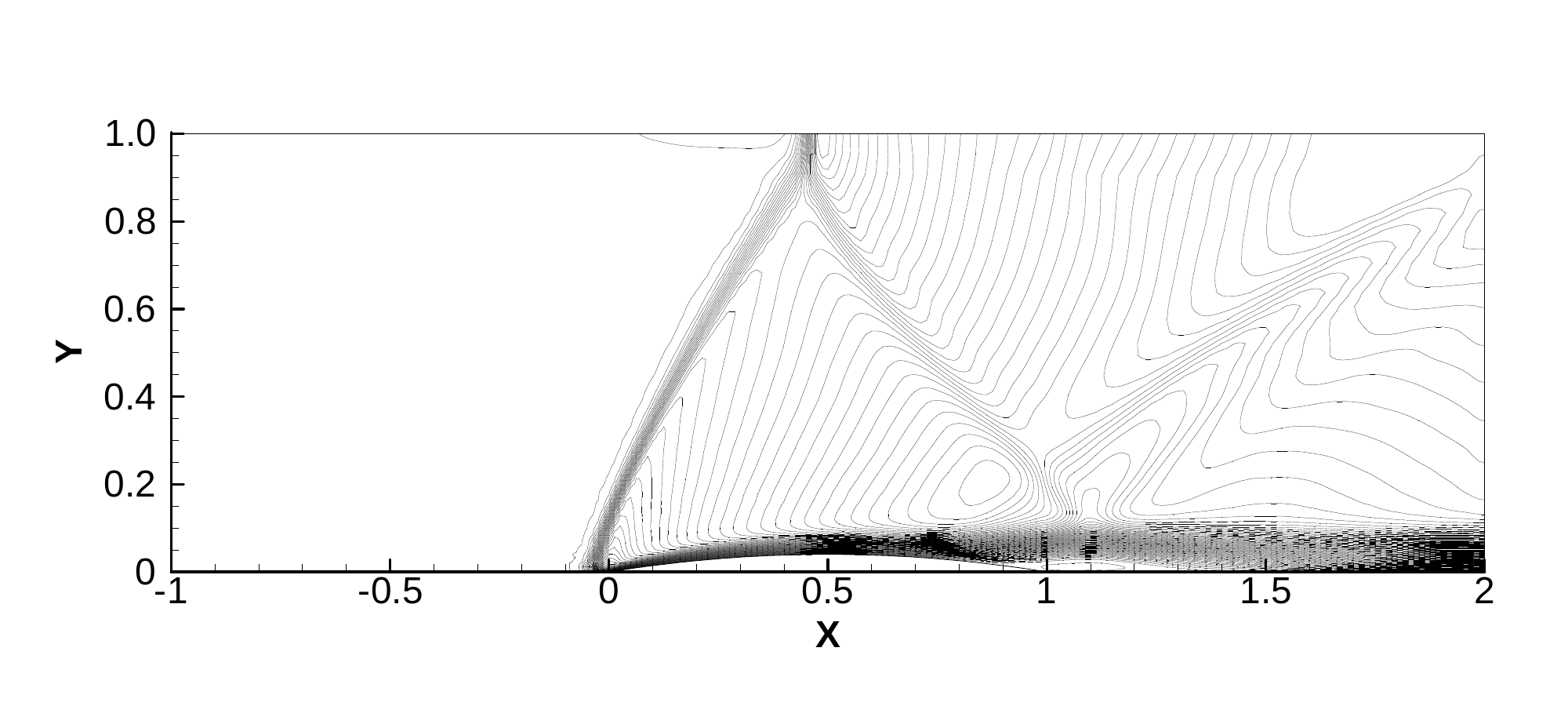}
\includegraphics[width=6.5cm]{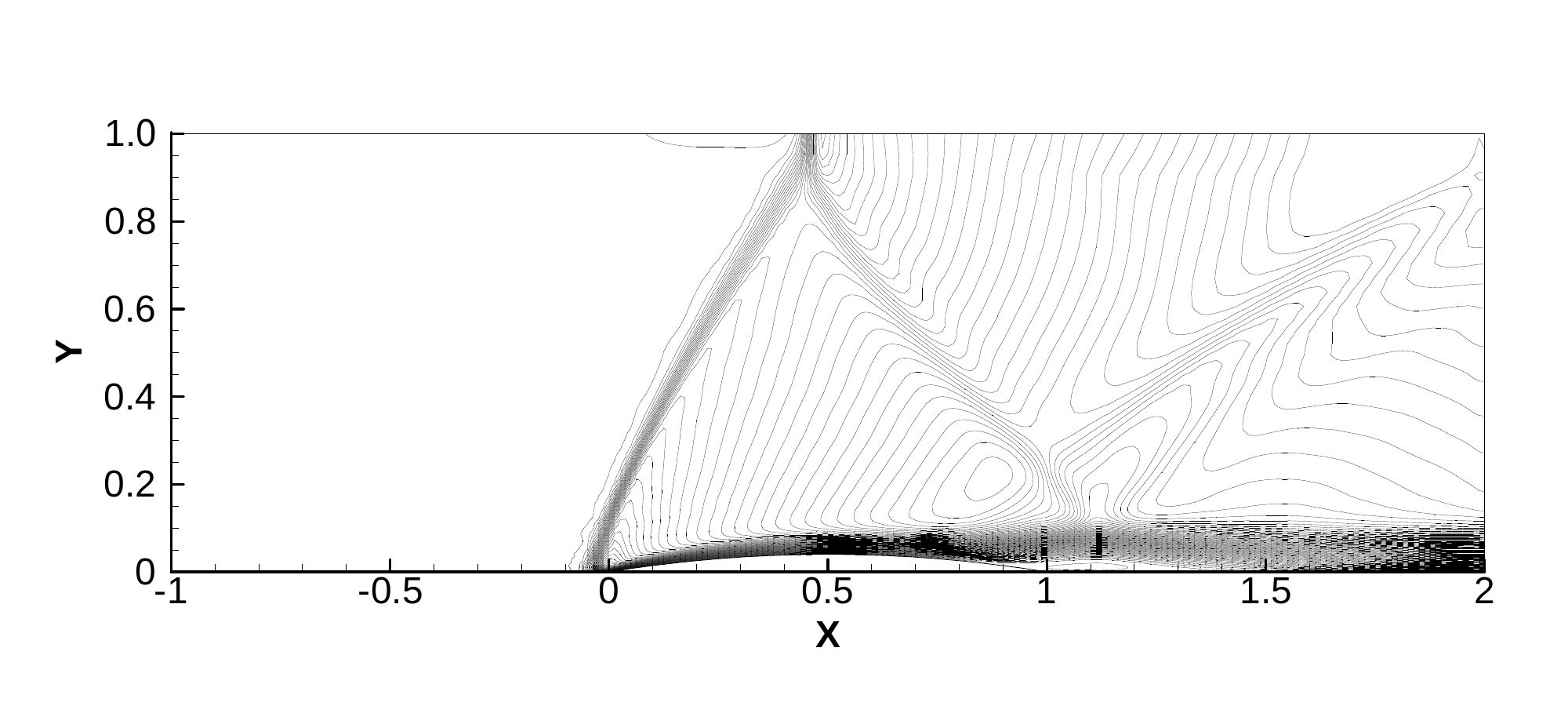}
\end{center} 
\caption{Test case: Supersonic viscous flow over a bump (240x80) -  Mach contours (0: 0.0215: 1.5)- (a)1O-KFDS-A  (b) 1O-KFDS-B (c) 2O-KFDS-A (d) 2O-KFDS-B}
\label{2D_NS_TC_KFDS_BUMP}  
\end{figure}

\subsubsection{Supersonic viscous flow over a bump}
\begin{figure} 
\begin{center} 
\includegraphics[width=6.5cm]{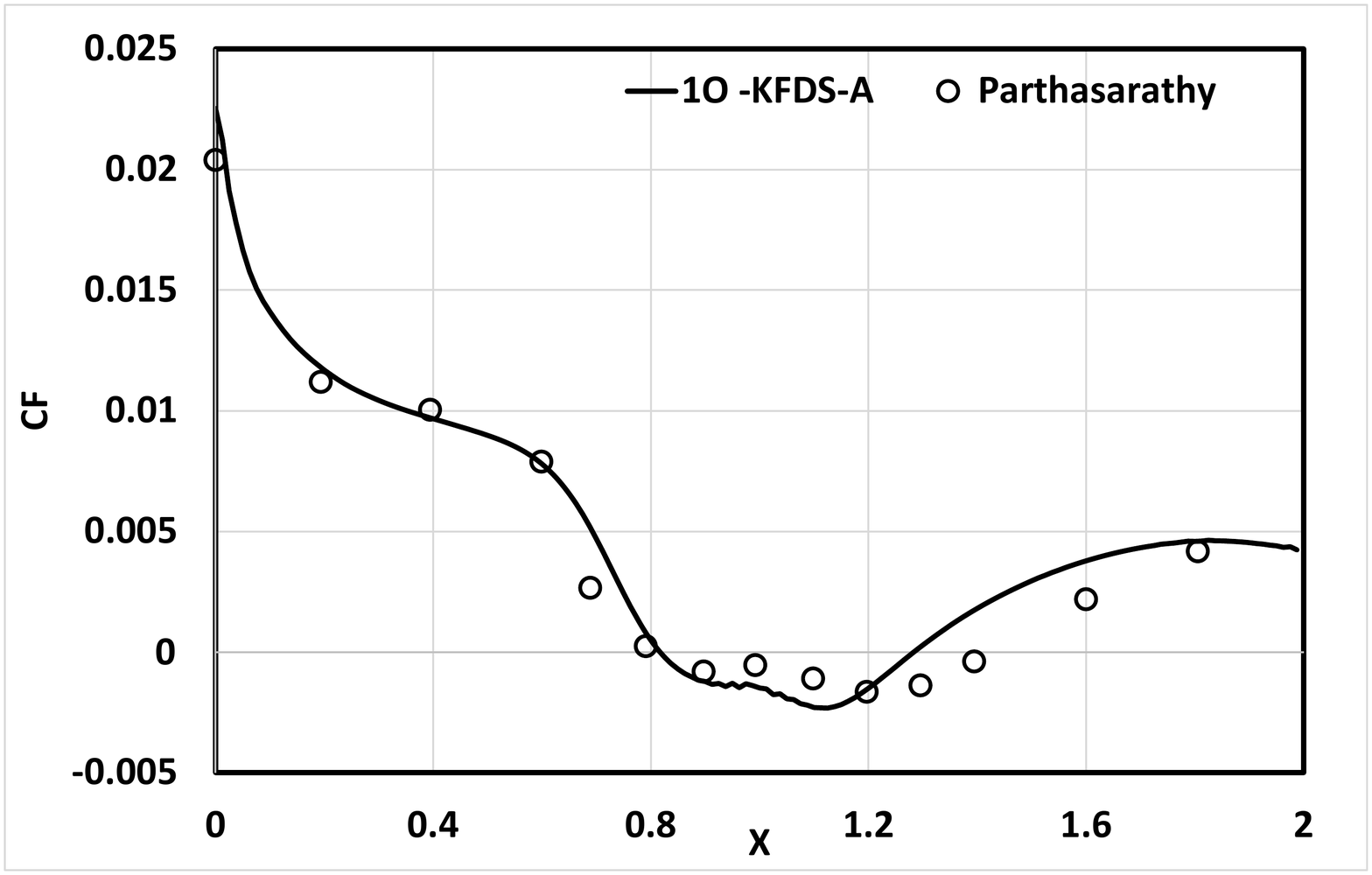}
\includegraphics[width=6.5cm]{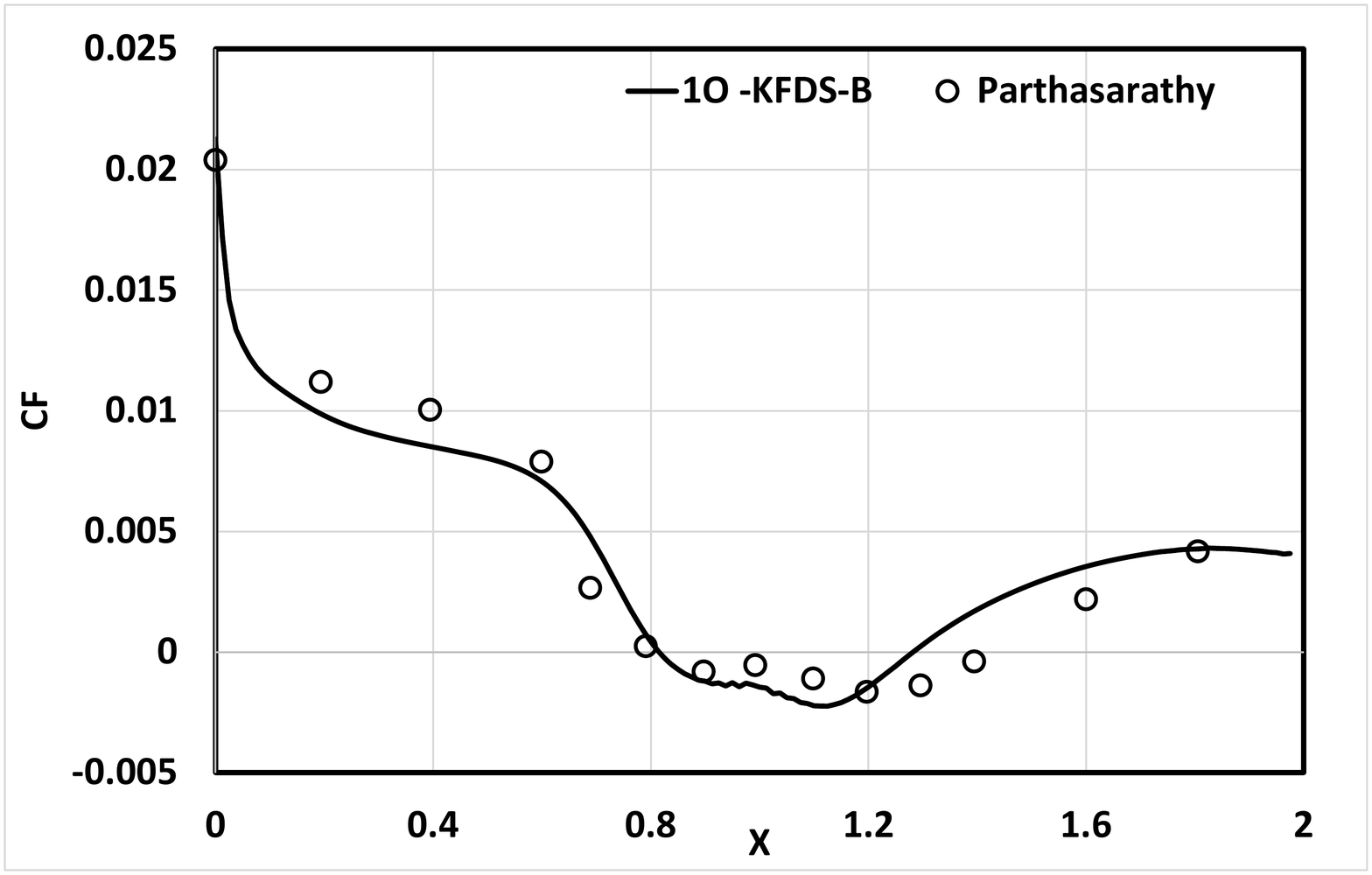}
\includegraphics[width=6.5cm]{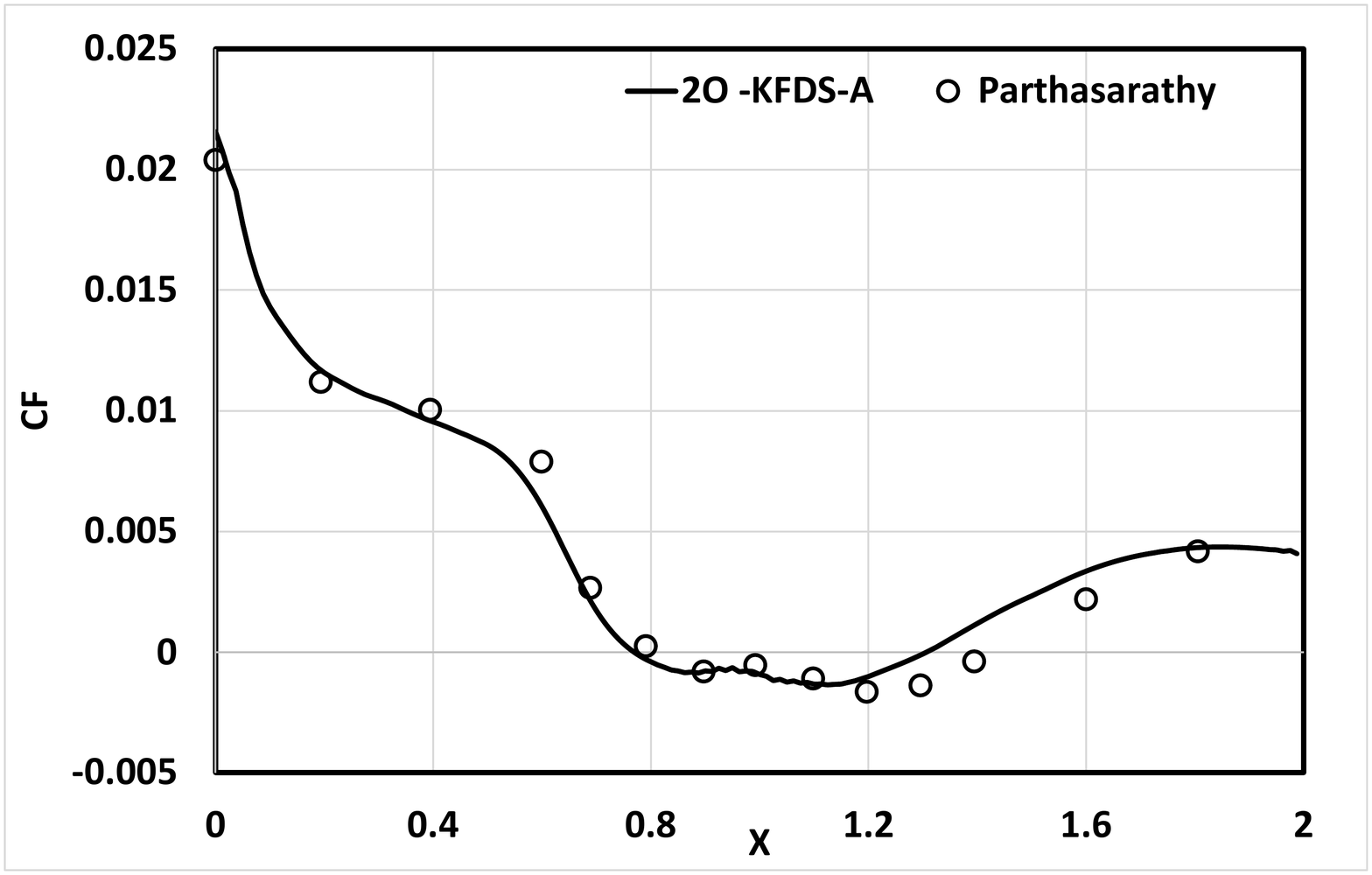}
\includegraphics[width=6.5cm]{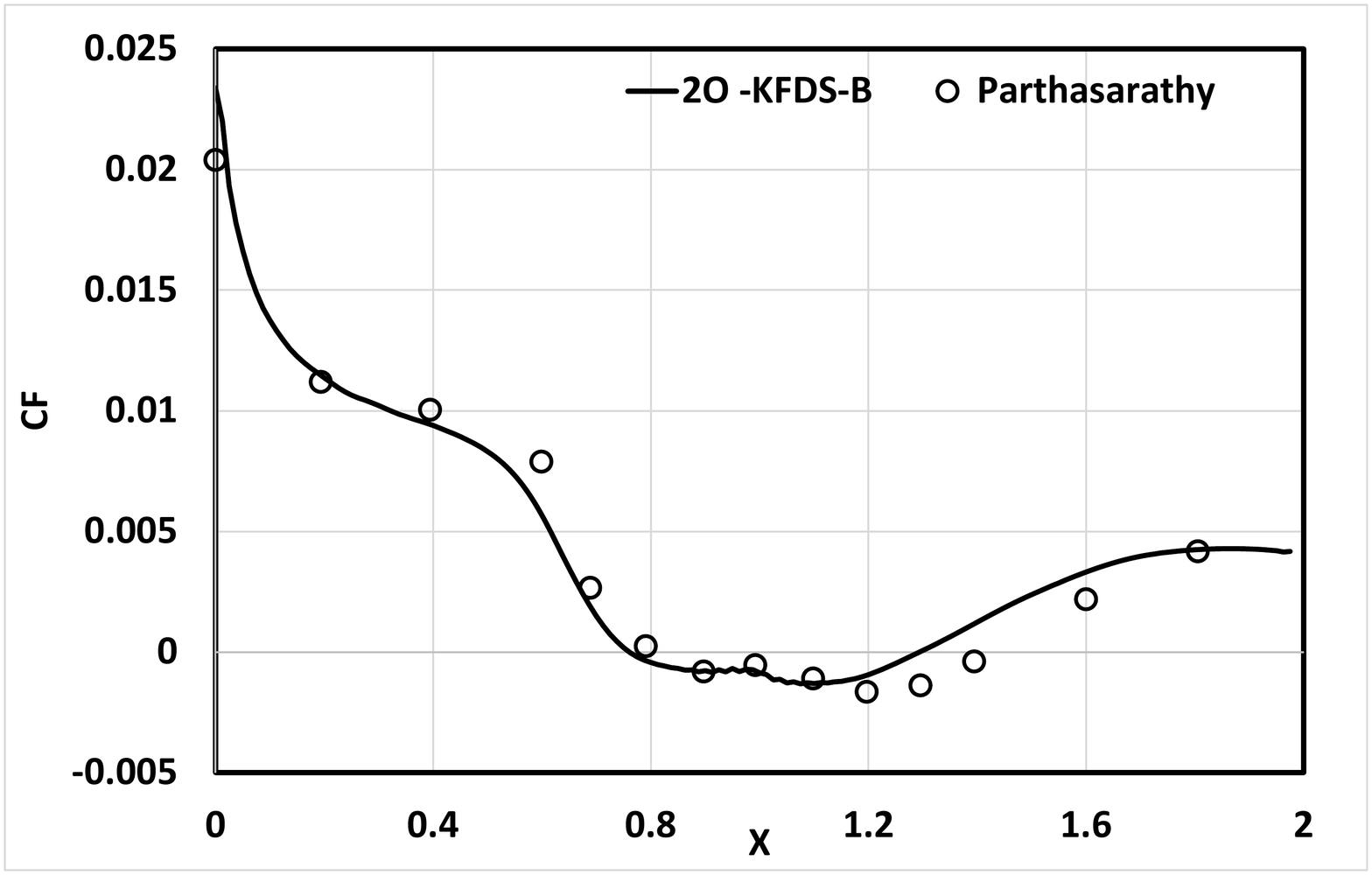}
\end{center} 
\caption{Test case: Supersonic viscous flow over a bump (240x80), $C_f$ plots - (a)1O-KFDS-A  (b) 1O-KFDS-B (c) 2O-KFDS-A (d) 2O-KFDS-B}
\label{2D_NS_TC_KFDS_BUMP_CF}  
\end{figure}

This test case [\cite{Parthasarathy}] involves a channel having a 4\% thick circular arc bump on the bottom side of the test domain [-1,2]x[0,1]. The bump has a chord length of 1 unit and is located at $x = 0.5$.  The left side is marked as a supersonic inlet with free stream Mach number 1.4 and Reynolds number 8000. Symmetry boundary condition is imposed on the bottom wall from $x =-1$ to $x=0$. The remaining portion of the bottom side is marked as a viscous wall.  The top side of the domain is defined as an inviscid wall, while the right side of the domain is marked as supersonic outlet.  A geometrically stretched grid of size $240 \times 80$, with a $4.5\%$ increase is used.  The results for each version of the KFDS scheme are  shown in Fig.[\ref {2D_NS_TC_KFDS_BUMP}]. The interaction of the reflected shock wave with separated flow region to the right of the bump results in a weak reflection as can be seen in the results. The skin friction coefficients for the test case are shown in Fig.[\ref{2D_NS_TC_KFDS_BUMP_CF}], are compared with those from the reference \cite{Parthasarathy} and show reasonable agreement.

\newpage
\subsubsection{Shock wave - boundary layer interaction}
\begin{figure} 
\begin{center} 
\includegraphics[width=7cm]{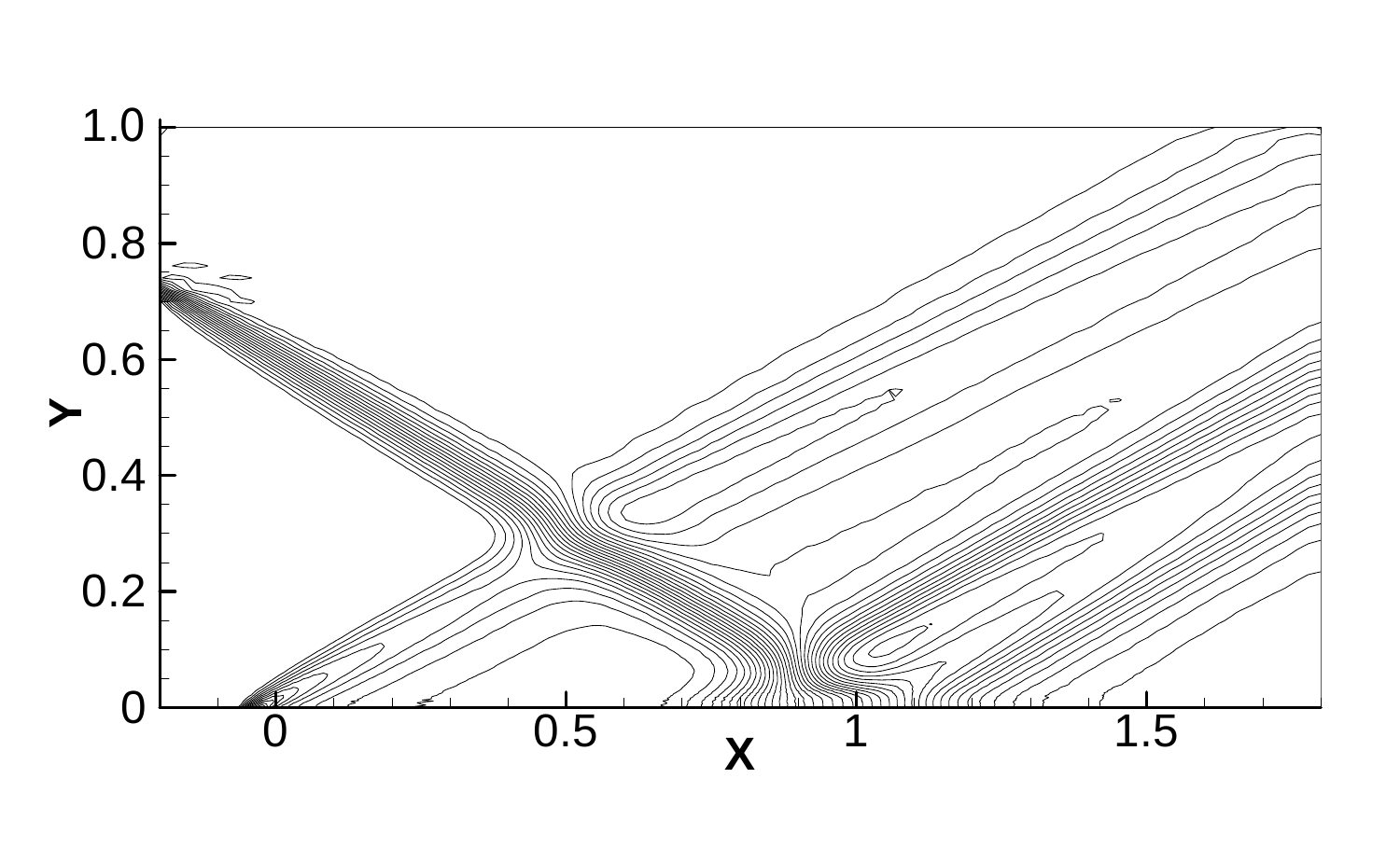}
\includegraphics[width=7cm]{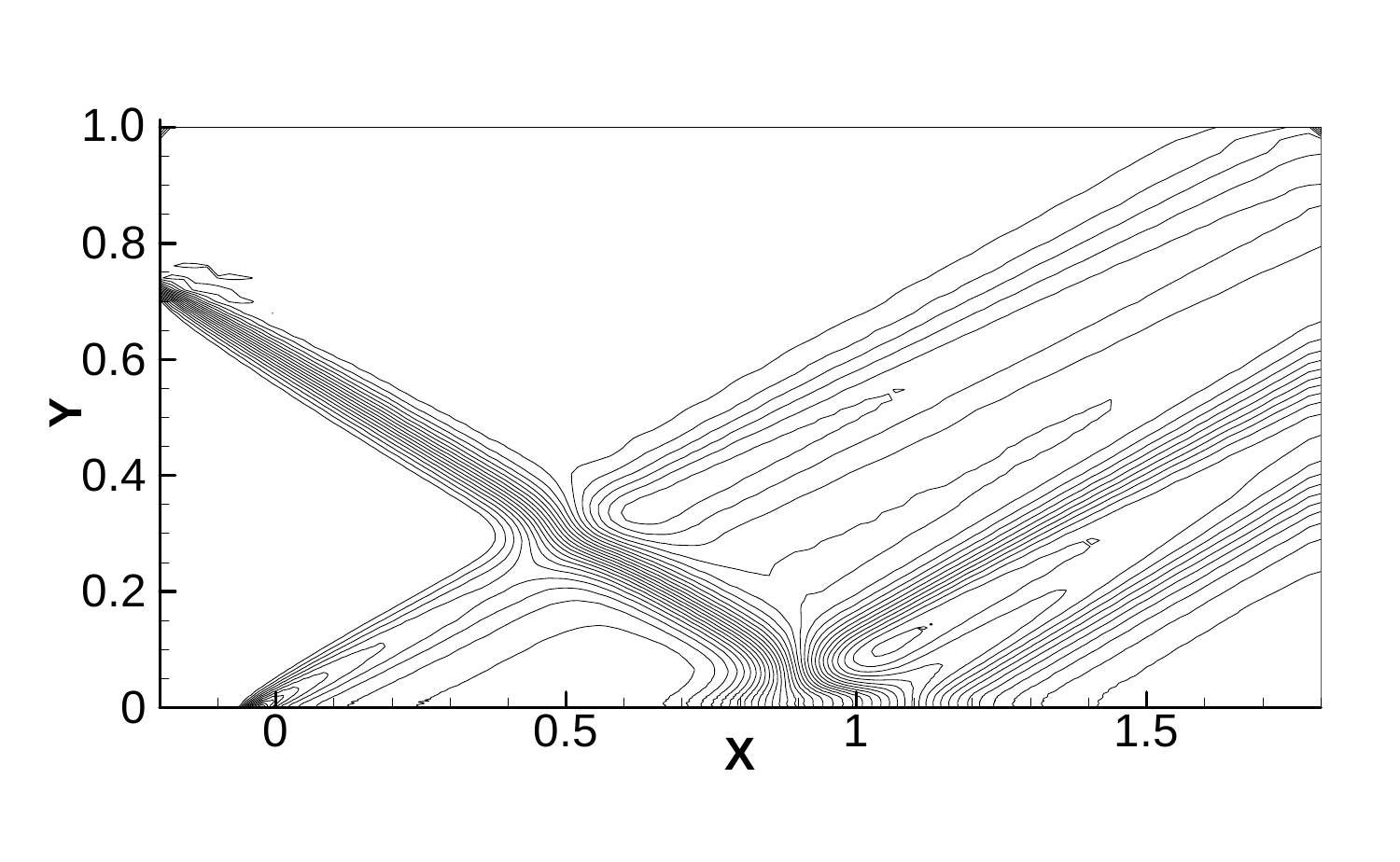}
\includegraphics[width=7cm]{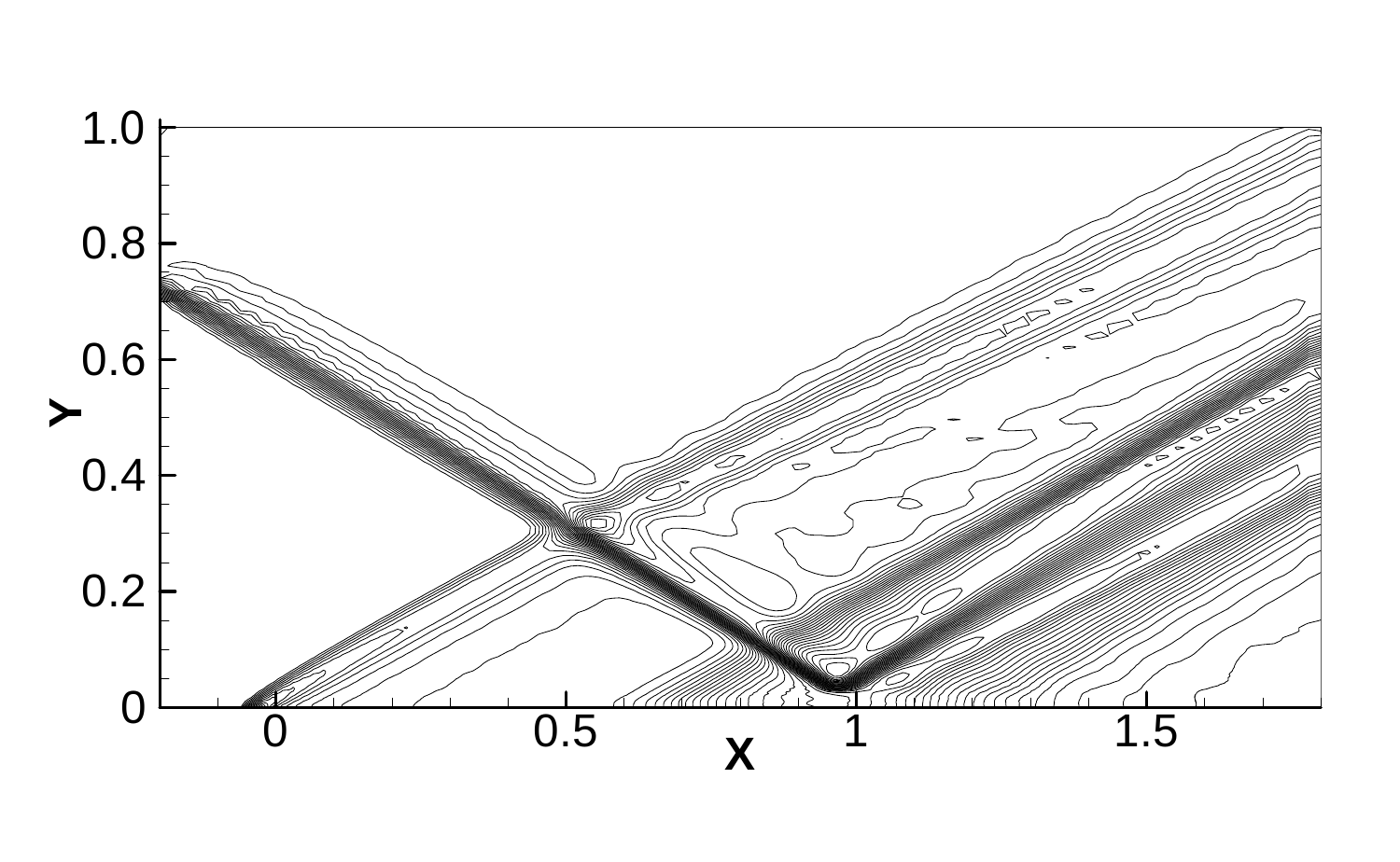}
\includegraphics[width=7cm]{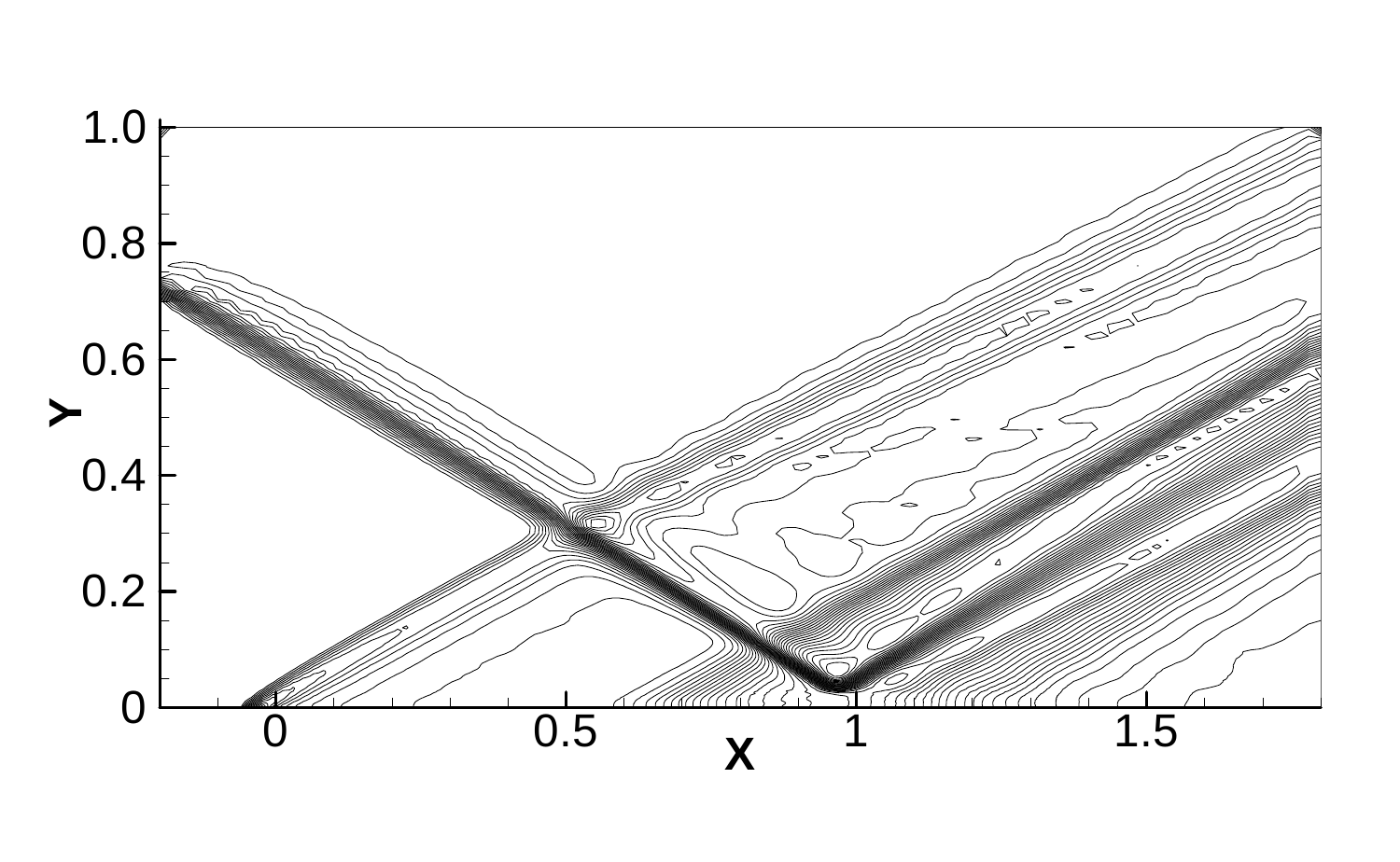}
\end{center} 
\caption{Test case: Shock wave - boundary layer interaction (141x121) - pressure contours (a)1O-KFDS-A  (b) 1O-KFDS-B (c) 2O-KFDS-A (d) 2O-KFDS-B}
\label{2D_NS_TC_KFDS_SWBLI_Press}  
\end{figure}

\begin{figure} 
\begin{center} 
\includegraphics[width=7cm]{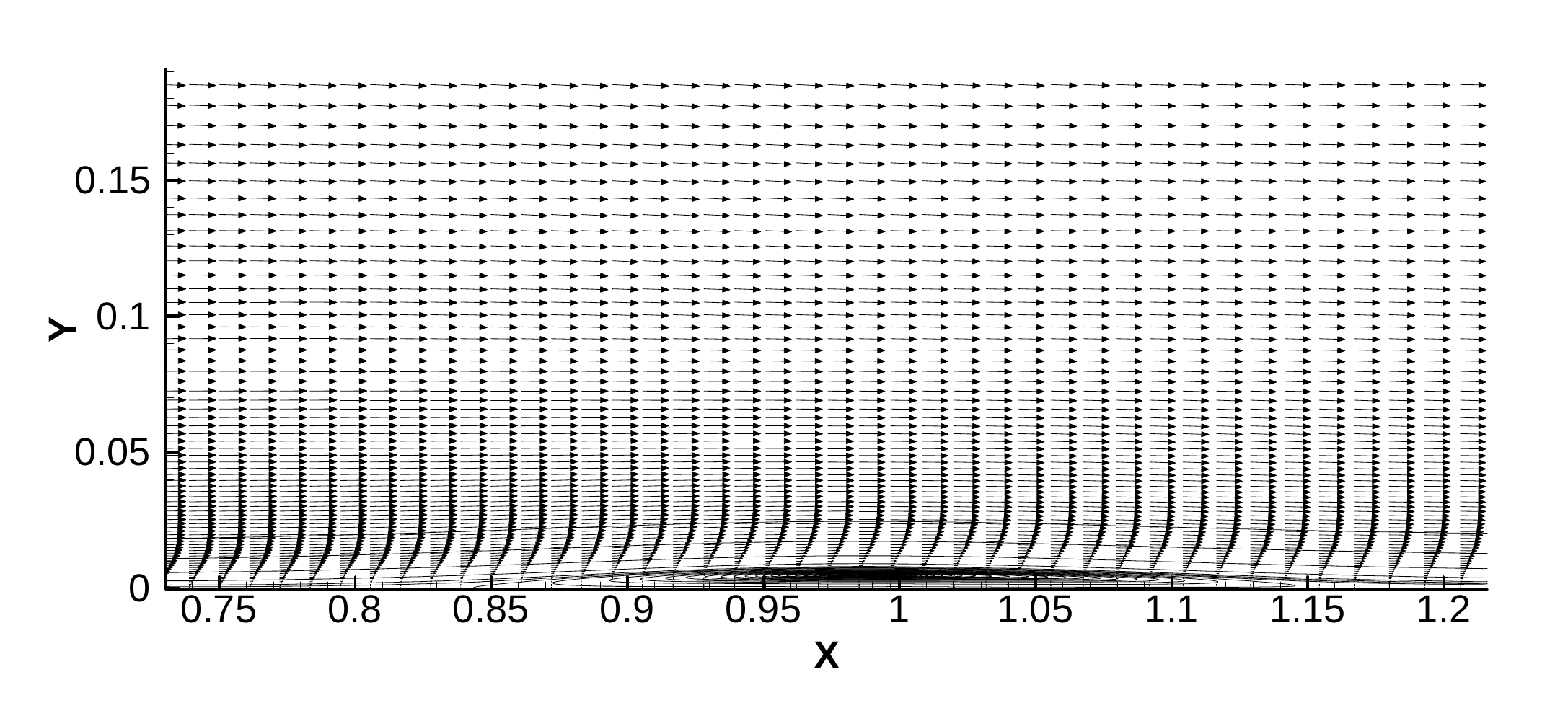}
\includegraphics[width=7cm]{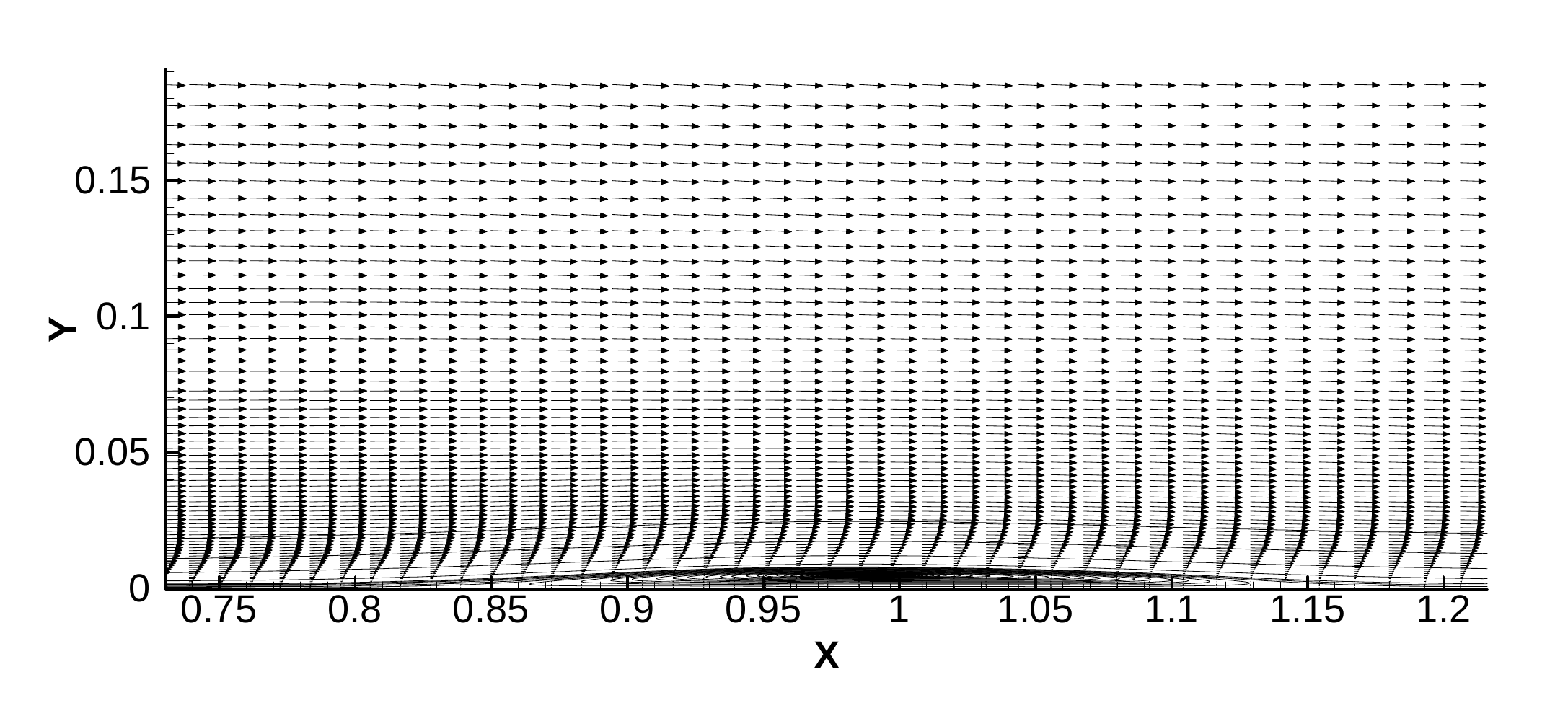}
\includegraphics[width=7cm]{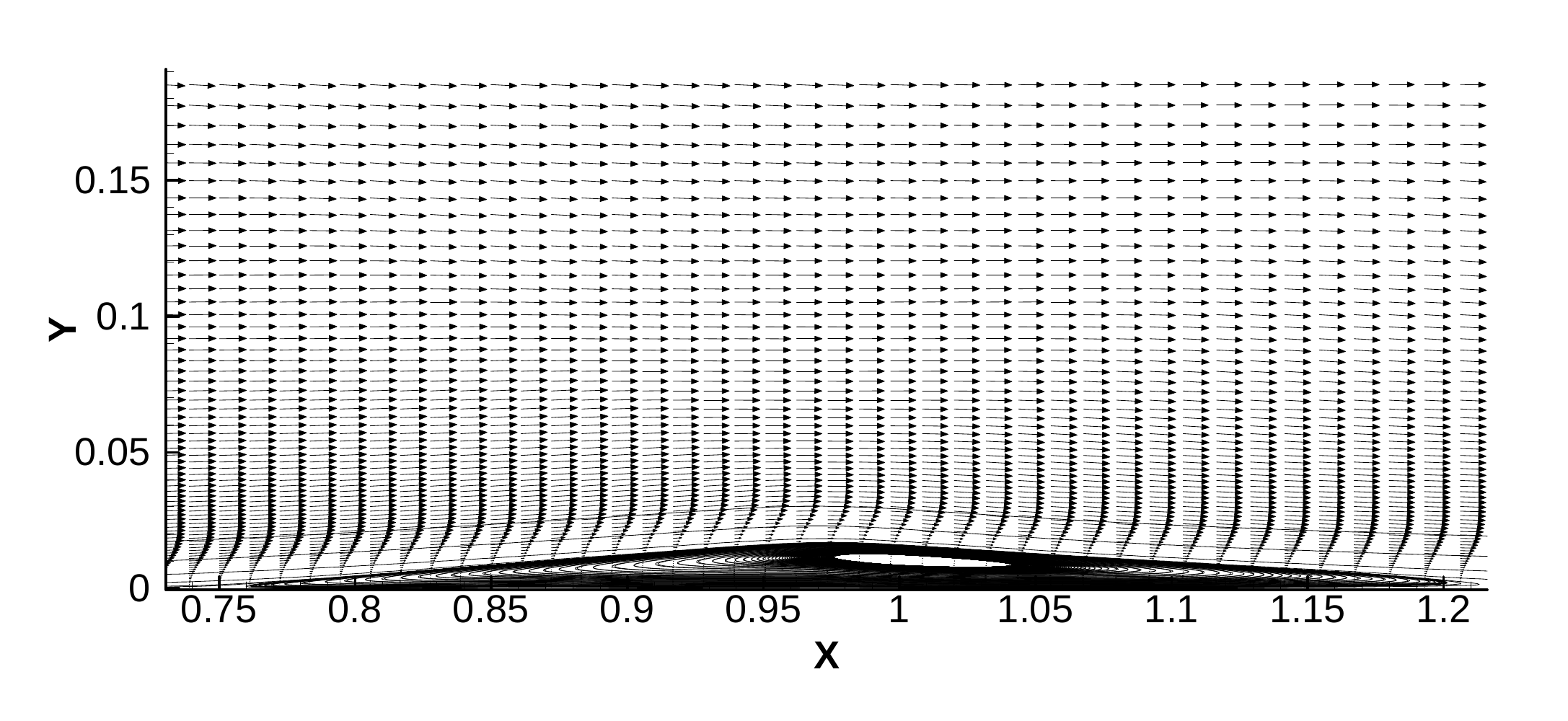}
\includegraphics[width=7cm]{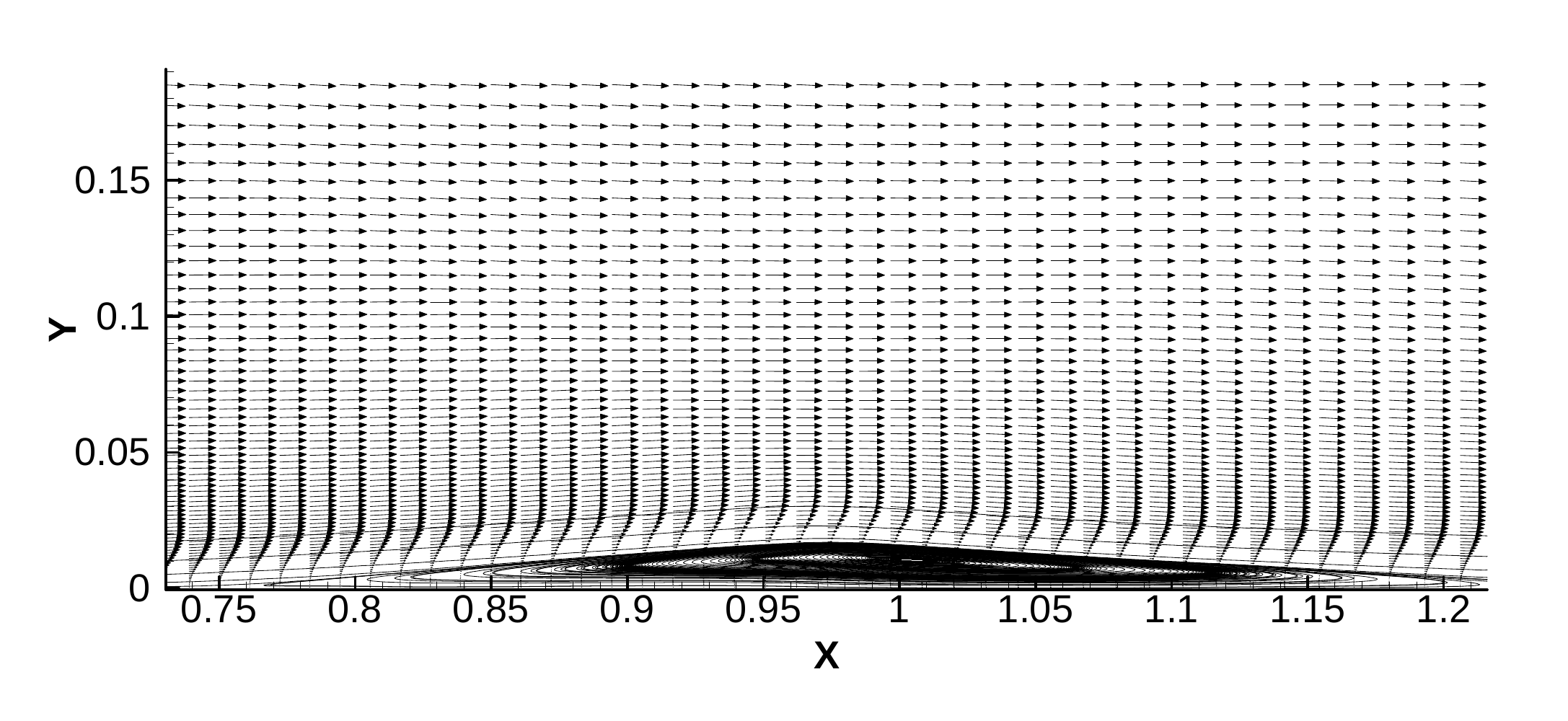}
\end{center} 
\caption{Test case: Shock wave - boundary layer interaction(141x121) - pressure contours (a)1O-KFDS-A  (b) 1O-KFDS-B (c) 2O-KFDS-A (d) 2O-KFDS-B}
\label{2D_NS_TC_KFDS_SWBLI_Stream}  
\end{figure}

\begin{figure} 
\begin{center} 
\includegraphics[width=6cm]{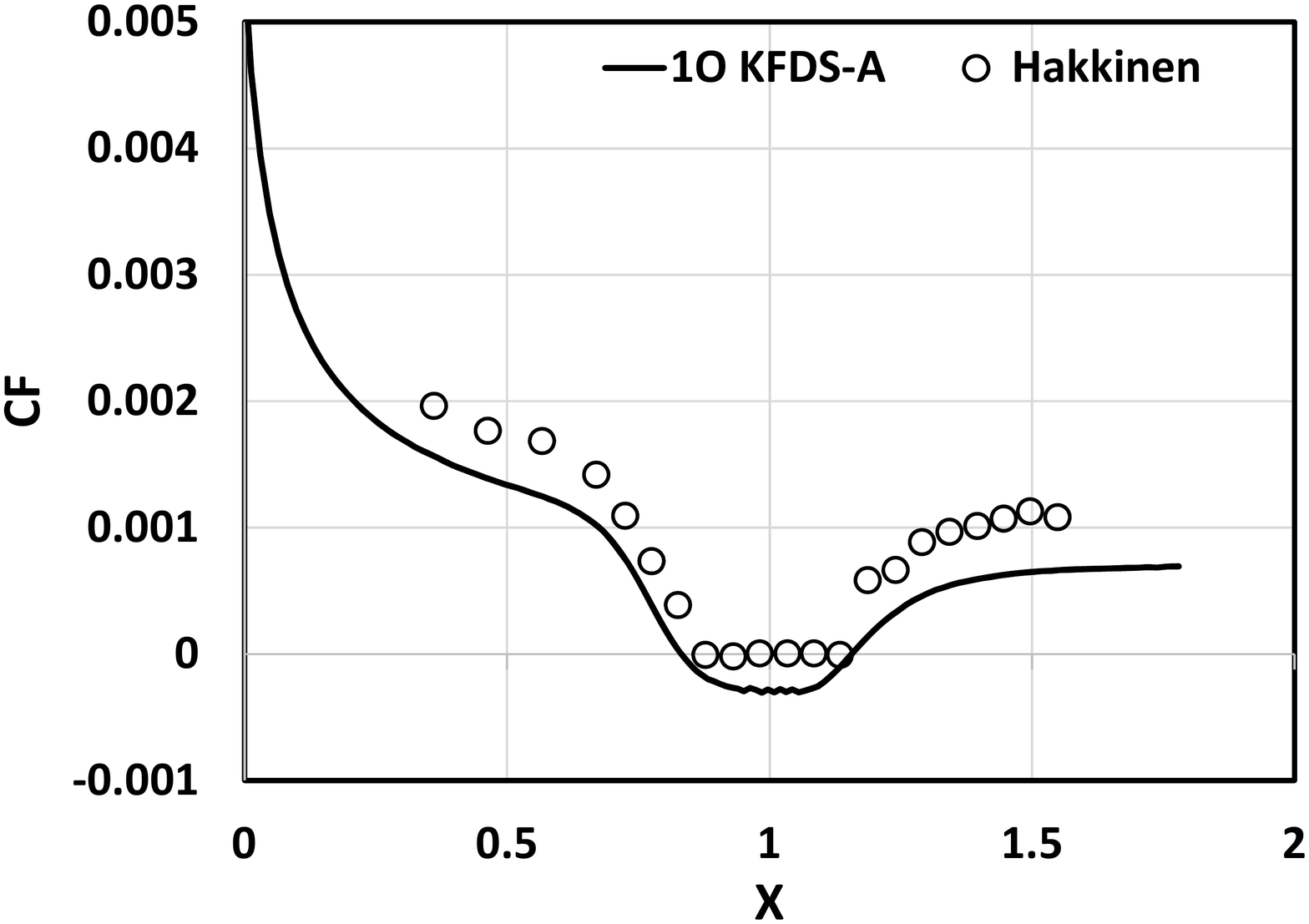}
\includegraphics[width=6cm]{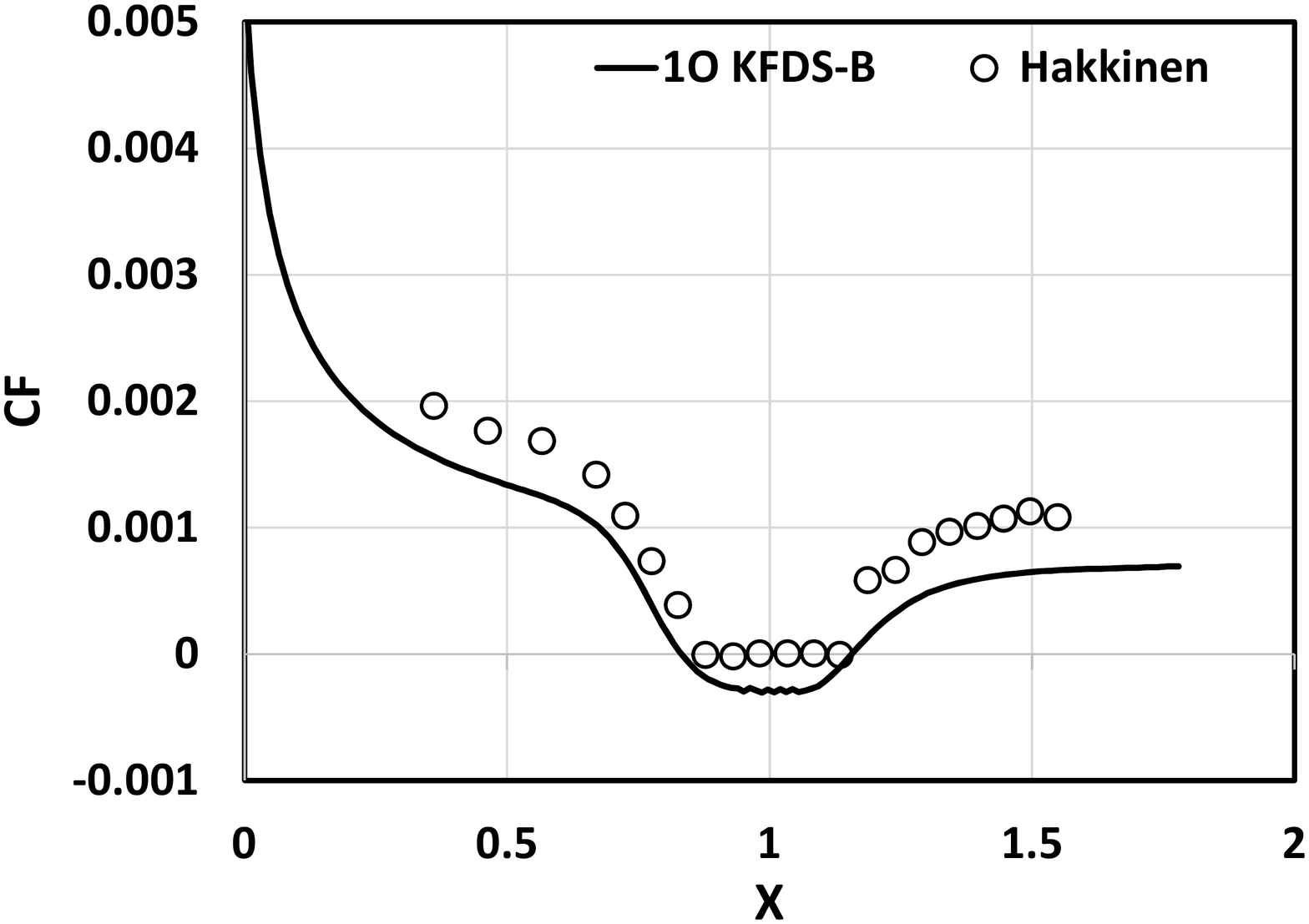}
\includegraphics[width=6cm]{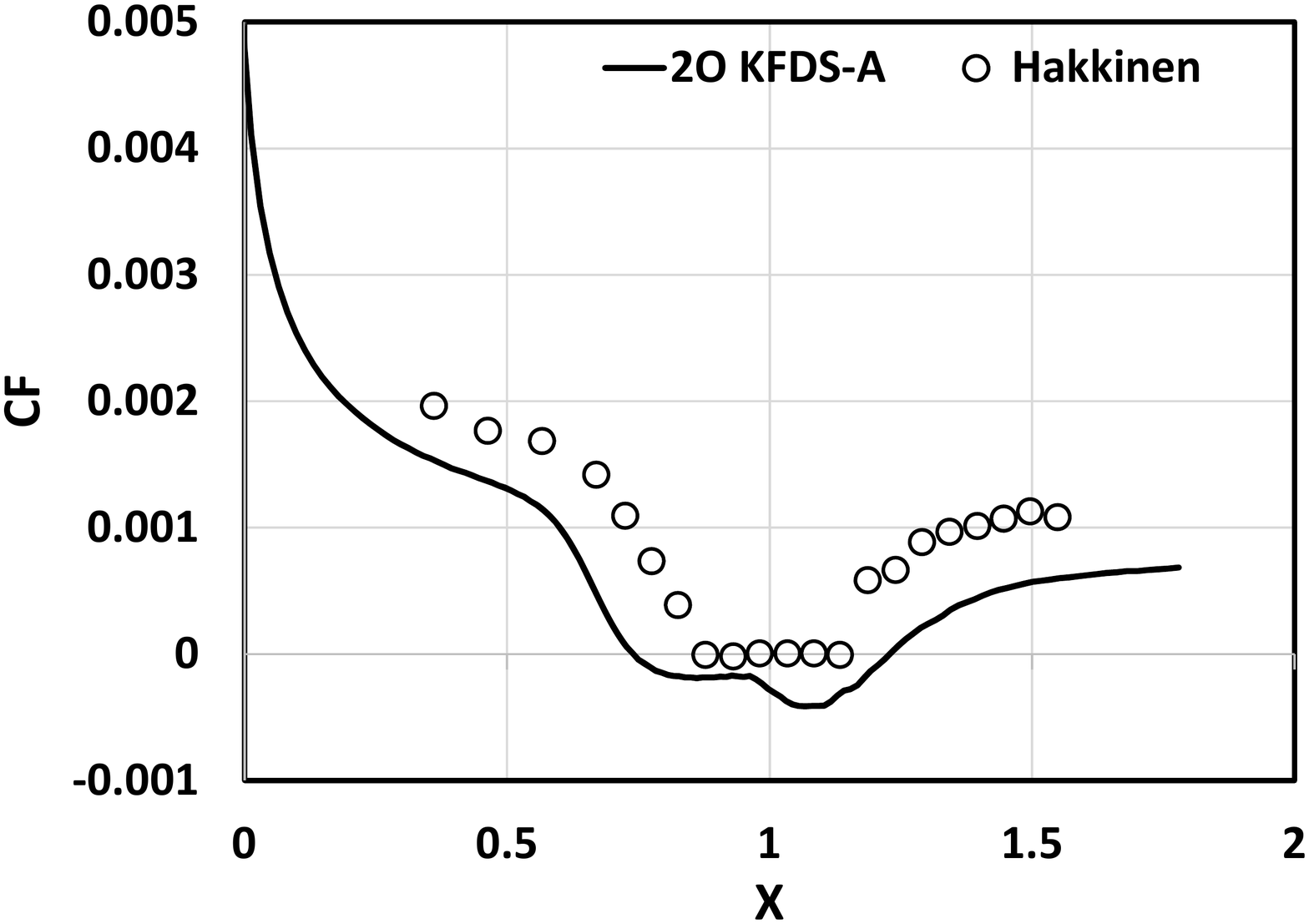}
\includegraphics[width=6cm]{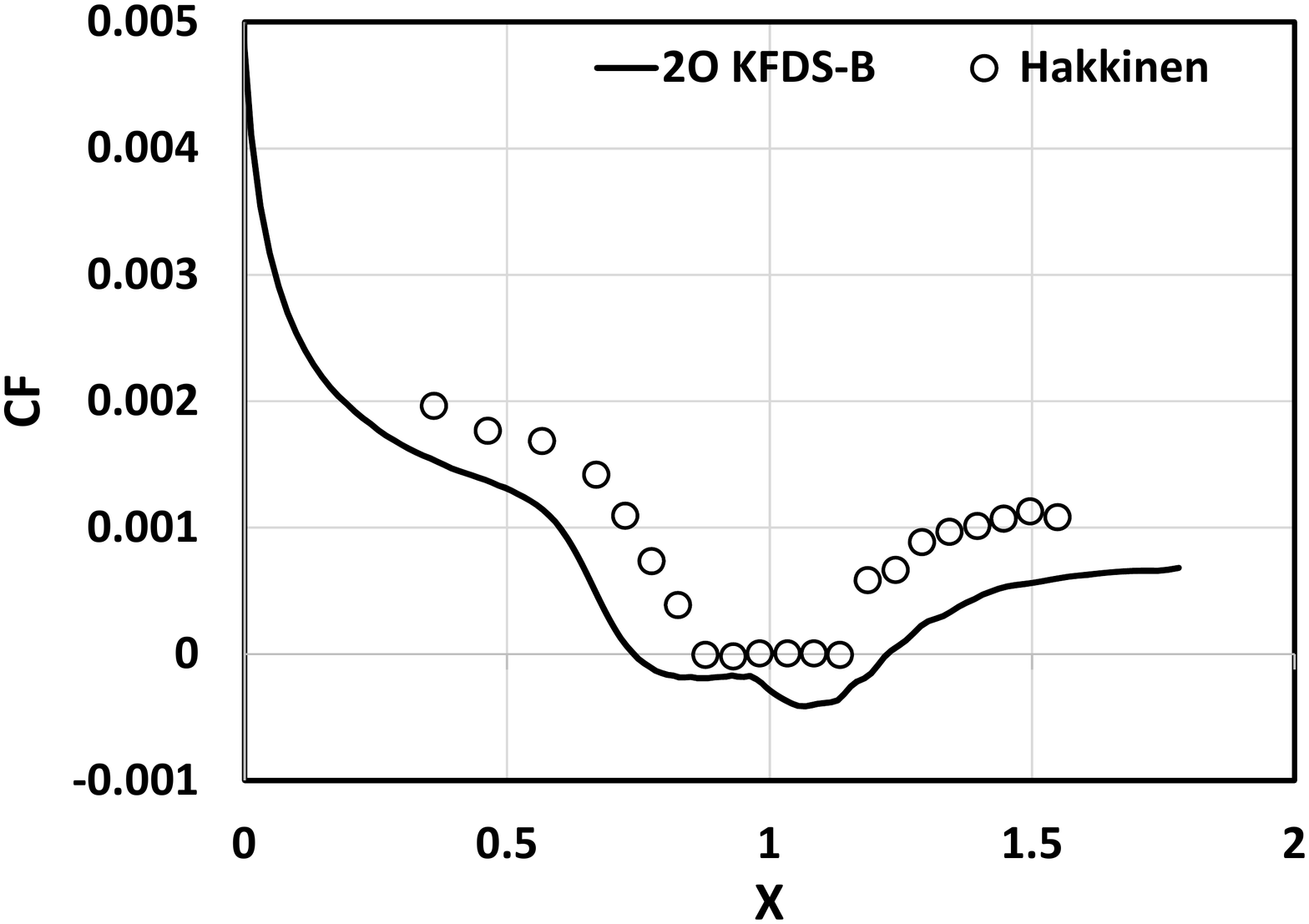}
\end{center} 
\caption{Test case: Shock wave - boundary layer interaction(141x121) - $C_f$ plots - (a)1O-KFDS-A  (b) 1O-KFDS-B (c) 2O-KFDS-A (d) 2O-KFDS-B}
\label{2D_NS_TC_KFDS_SWBLI_CF}  
\end{figure}

\begin{figure} 
\begin{center} 
\includegraphics[width=6cm]{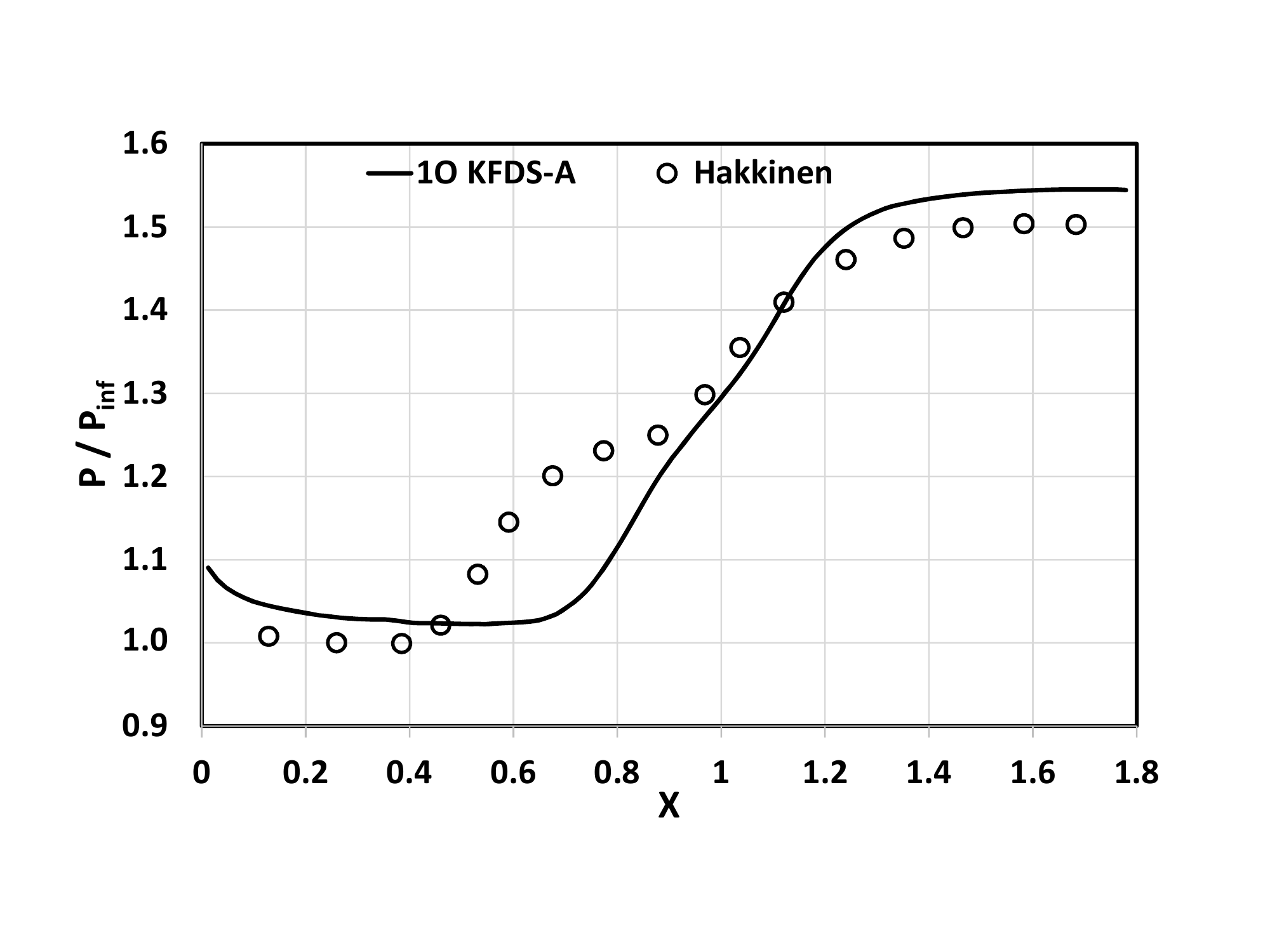}
\includegraphics[width=6cm]{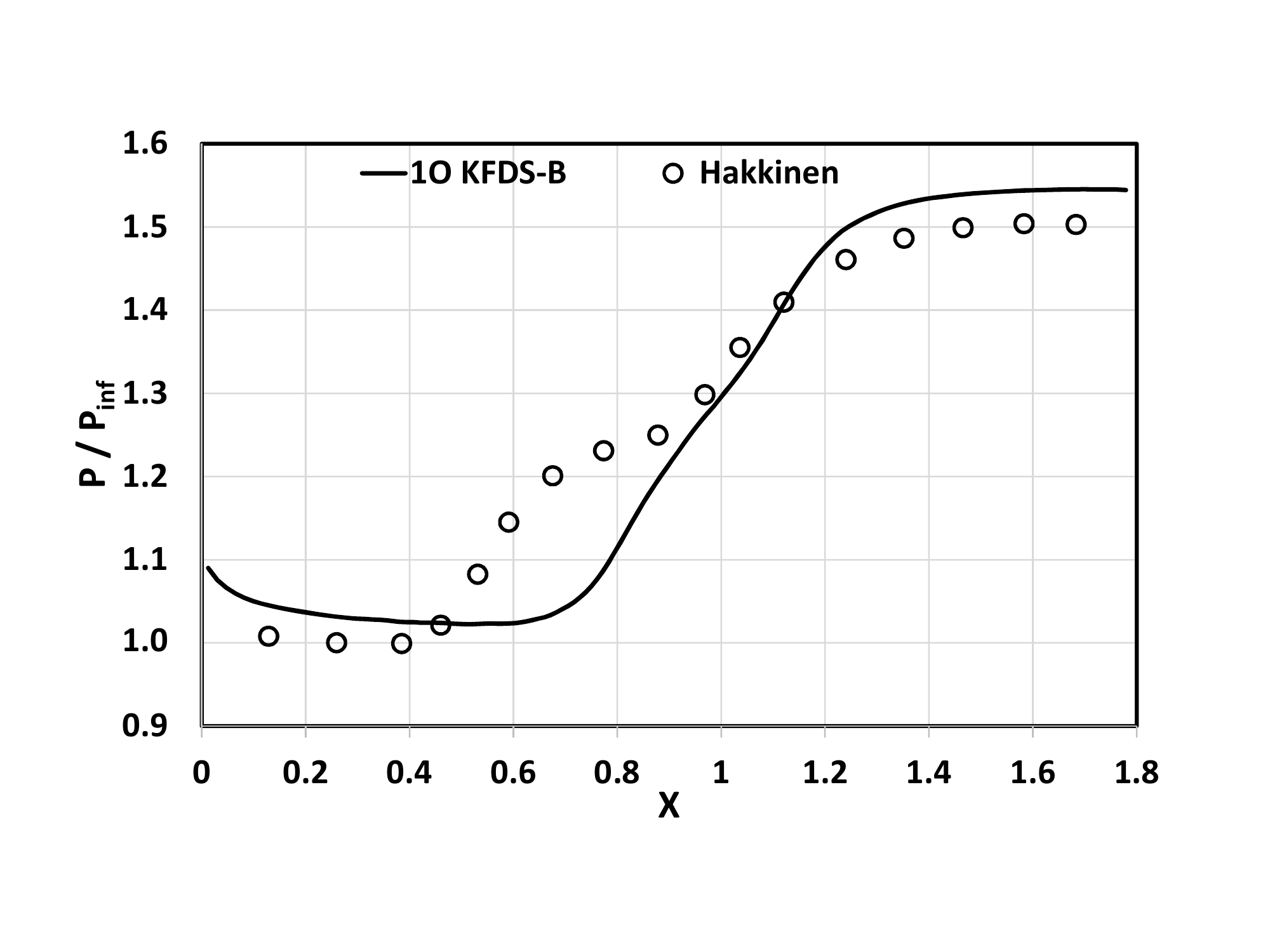}
\includegraphics[width=6cm]{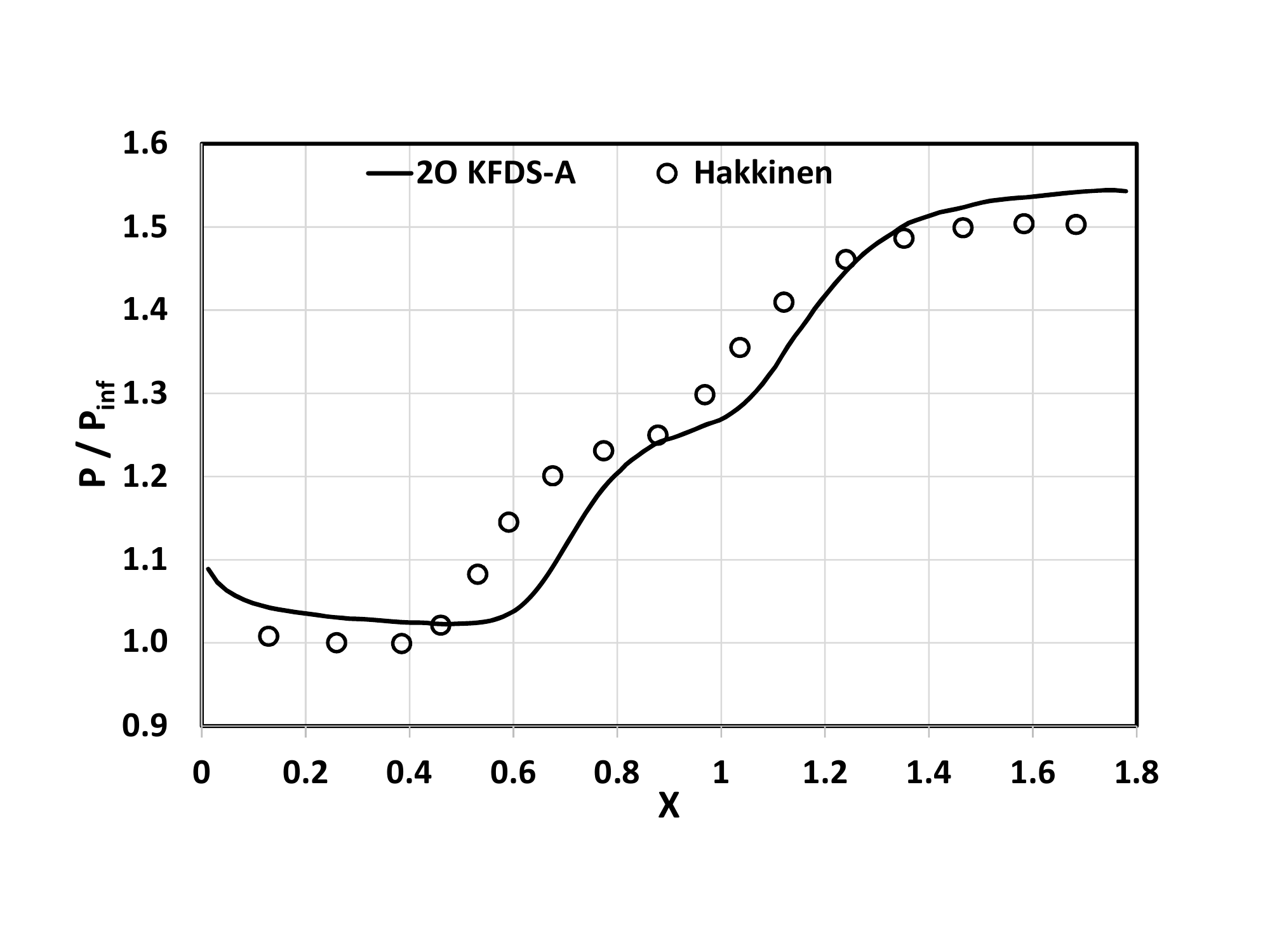}
\includegraphics[width=6cm]{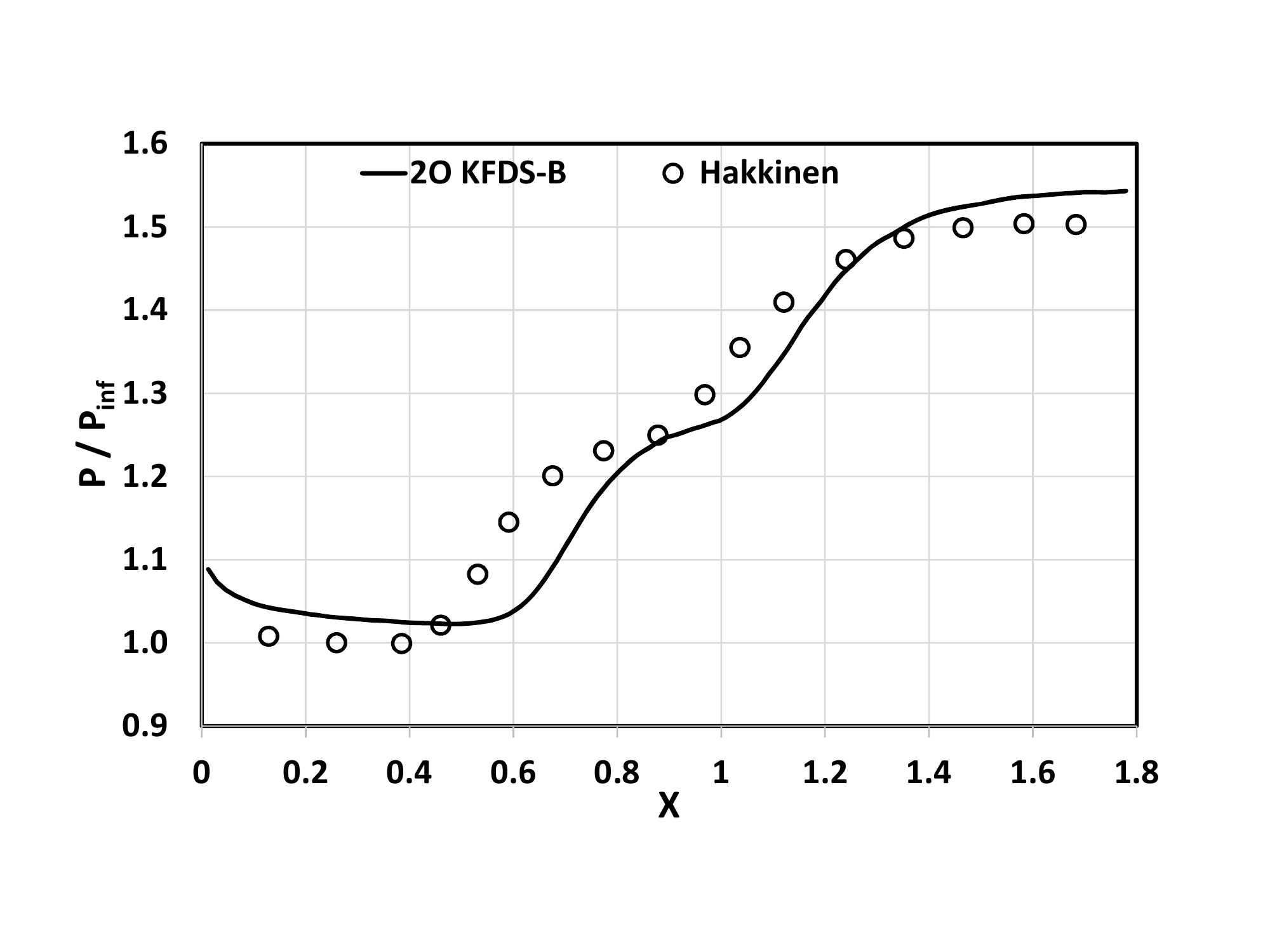}
\end{center} 
\caption{Test case: Shock wave - boundary layer interaction(141x121) - pressure coefficient plots - (a)1O-KFDS-A  (b) 1O-KFDS-B (c) 2O-KFDS-A (d) 2O-KFDS-B}
\label{2D_NS_TC_KFDS_SWBLI_P}  
\end{figure}

This test case [\cite{Degrez}] constitutes an interaction of an oblique shock wave with a laminar boundary layer that evolves on the bottom side of the computational domain [-0.2,1.8] x [0,1].  The interaction results in the formation of a streak of compression waves reflecting from the boundary layer accompanied by an adjacent expansion fan. Further, a recirculation zone in the form of a bubble develops on the surface around which the flow separates from the surface and gets reattached.  It is evident that a wide combination of flow features evolve in the solution and thus tests the capability of the numerical scheme to capture each phenomenon listed.  The inlet boundary comprises of a supersonic inflow with Mach number 2.0 and Reynolds number 100000 till y=0.765. The inlet region above y=0.765 and the topside of the domain is initialized with post-shock boundary conditions.  The right side of the domain is treated as a supersonic outlet.  A geometrically stretched grid of size $141 \times 121$, with a $4.5\%$ increase is used.  Fig.[\ref {2D_NS_TC_KFDS_SWBLI_Press}] presents the pressure contours for the numerical scheme. The scheme is able to resolve the incident and reflected shock wave along with the expansion regions. 
Fig.[\ref {2D_NS_TC_KFDS_SWBLI_Stream}] portrays the flow vectors and the streamlines closer to the bottom wall.  The bubble resulting from the flow separation and reattachment can be observed. The computed skin friction coefficients  and pressure coefficients are shown in Fig.[\ref {2D_NS_TC_KFDS_SWBLI_CF}] and Fig.[\ref {2D_NS_TC_KFDS_SWBLI_P}] respectively.  Both versions of KFDS schemes are able to capture the relevant flow features with reasonable accuracy.   

\clearpage 

\section{Conclusions}
A new discrete velocity model based kinetic flux difference splitting scheme is introduced and its capability to handle the complex flow features of compressible flows without the need of an entropy fix, yet retaining its basic ability to exactly capture grid aligned steady shocks and contact discontinuities, is demonstrated.  Enforcing the principle of flux equivalence across a steady discontinuity results in satisfying R-H conditions in the discretization, which leads to exact capturing of steady discontinuities.  The use of $D^{2}$-distance as a gradient sensor enables in precise addition of numerical diffusion only in the expansion regions and thus helps avoiding any entropy condition violations.  Various benchmark test cases for inviscid and viscous flows demonstrate the robustness and accuracy of the scheme.   

\section*{Acknowledgments} 
   The third author thanks the University of South Africa, Johannesburg for funding during his sabbatical leave, which facilitated the formulation of the new algorithm presented in this paper. The first author thanks Hindustan Aeronautics Limited, Bangalore for funding the author, which facilitated this research activity.

\clearpage 
\newpage 
\begin{appendix} 
\section{\bf \large Some important moments for DVBE}
\be
{\bf P} {\bf f}^{eq} = U
\ee

\be
{\bf P} \Lambda {\bf f}^{eq} =G(U)
\ee

\be 
G^{+}(U) = {\bf P}  \Lambda^{+} {\bf f}^{eq}
\ee
Therefore 
$$ \ba{rcl}{ G^{+}(U)}_{i} & = & \left[ 1 \ 1 \ 1 \right] \left[ \ba{ccc} \lambda^{+} & 0 & 0 \\ 0 & \lambda_{o}^{+} & 0 \\ 0 & 0 & 0 \ea \right]_{i} \left[ \ba{c} {\bf f_{+}}^{eq}\\ {\bf f_{o}}^{eq} \\ {\bf f_{-}}^{eq} \ea \right]_{i}\\[6mm]   
& = & \left \{ \lambda^{+} {\bf f_{+}}^{eq} +  \lambda_{o}^{+} {\bf f_{o}}^{eq} \right \}_{i} \\[2mm] 
\ea 
$$ 
which upon simplifying gives
\be 
{G^{+}(U)}_{i} = \left\{ \fr{1}{2} G(U) + \fr{\lambda}{2} U - \fr{(\lambda -  |\lambda_{o}|)}{2}{\bf f_{o}}^{eq} \right \}_{i}
\ee 
Similarly, starting with $G^{-}(u) = {\bf P} \Lambda^{-} {\bf f}^{eq}$, we get
$$ \ba{rcl}{ G^{-}(U)}_i & = & \left[ 1 \ 1 \ 1 \right] \left[ \ba{ccc} 0 & 0 & 0 \\ 0 & \lambda_{o}^{-} & 0 \\ 0 & 0 & \lambda^{-} \ea \right]_{i} \left[  \ba{c} {\bf f_{+}}^{eq}\\ {\bf f_{o}}^{eq} \\ {\bf f_{-}}^{eq} \ea \right]_{i} \\[6mm]   
& = &  \left\{ \lambda_{o}^{-} {\bf f_{o}}^{eq} + \lambda^{-} {\bf f_{-}}^{eq} \right \}_{i} \\[2mm] 
\ea 
$$ 
which upon simplifying gives 
\be 
{G^{-}(U)}_{i} = \left \{ \fr{1}{2} G(U) - \fr{\lambda}{2} U + \fr{(\lambda -  |\lambda_{o}|)}{2}{\bf f_{o}}^{eq} \right \}_{i}
\ee 
Further 
$$ \ba{rcl} {\bf P}| \Lambda| {\bf f}^{eq} _{i}& = & \left[ 1 \ 1 \ 1 \right] \left[ \ba{ccc}\lambda & 0 & 0 \\ 0 & \lambda_{o} & 0 \\ 0 & 0 & \lambda \ea \right] \left[ \ba{c} {\bf f_{+}}^{eq}\\ {\bf f_{o}}^{eq} \\ {\bf f_{-}}^{eq} \ea \right]_{i} \\[6mm]   
& = & \left \{ \lambda {\bf f_{+}}^{eq} +  \lambda_{o} {\bf f_{o}}^{eq} + \lambda {\bf f_{-}}^{eq} \right \}_{i} \\[2mm] 
& = & \left \{ \lambda U - (\lambda -  |\lambda_{o}|){\bf f_{o}}^{eq} \right \}_{i}\\
\ea 
$$ 
Similarly
\bea
{\bf P} {\bf f}^{CE} & = & {\bf P} {\bf f}^{eq}  - {\bf P} {\bf f_{v}}^{eq}  = U\\[2mm]
{\bf P} \Lambda {\bf f}^{CE} & = & {\bf P} \Lambda {\bf f}^{eq} -{\bf P} \Lambda {\bf f_{v}}^{eq} = G(U) -  G_{v}(U) 
\eea

\end{appendix} 

\newpage

\end{document}